\documentclass[lettersize,journal]{IEEEtran}
\usepackage{amsmath,amsfonts}
\usepackage{algorithmic}
\usepackage{array}
\usepackage{textcomp}
\usepackage{stfloats}
\usepackage{url}
\usepackage{verbatim}
\usepackage{graphicx}
\def\BibTeX{{\rm B\kern-.05em{\sc i\kern-.025em b}\kern-.08em
    T\kern-.1667em\lower.7ex\hbox{E}\kern-.125emX}}
\usepackage{balance}
\usepackage{array, calc, color, multicol}
\usepackage{mathtools}

\DeclarePairedDelimiter\floor{\lfloor}{\rfloor}
\usepackage{graphics, epstopdf, graphicx, epsfig, psfrag, epsf,subfigure}
\usepackage{amssymb, amsfonts, amsmath, times}
\usepackage{relsize}
\usepackage{multicol,balance, verbatim}
\usepackage{textcomp, mathrsfs, stmaryrd, setspace}
\usepackage{amssymb}
\usepackage{soul}
\usepackage[dvipsnames]{xcolor}
\usepackage{cite}
\usepackage{url}
\usepackage{lipsum}

\newcommand{\ie}{{\em i.e., }}
\newcommand{\Ie}{{\em I.e., }}
\newcommand{\eg}{{\em e.g., }}

\newtheorem{theorem}{Theorem}
\newtheorem{lemma}[theorem]{Lemma}

\newtheorem{corollary}[theorem]{Corollary}
\newtheorem{definition}{Definition}
\newtheorem{example}{Example}

\DeclareGraphicsExtensions{.eps, .pdf, .png, .jpg}   

\newcommand{\Nset}{\mathcal{N}}

\newcommand{\Eset}{\mathcal{E}}

\newcommand{\yas}[1]{\textcolor{magenta}{#1}}

\usepackage{lettrine}
\usepackage{amsmath}
\usepackage{adjustbox}
\usepackage{algorithm}
\usepackage{algorithmic}

\IEEEoverridecommandlockouts

\begin{document}
\title{Efficient Coded Multi-Party Computation at \\ Edge Networks}

\author{Elahe~Vedadi,~\IEEEmembership{Student~Member,~IEEE,}
       Yasaman~Keshtkarjahromi,~\IEEEmembership{Member,~IEEE,} ~~~~~~~~~~~~~~~~~
        Hulya~Seferoglu,~\IEEEmembership{Senior Member,~IEEE}
\IEEEcompsocitemizethanks{\IEEEcompsocthanksitem{The preliminary results of this paper were presented in part at the IEEE International Symposium on Information Theory (ISIT), 2022 \cite{AGE-CMPC}, and the IEEE 23rd International Workshop on Signal Processing Advances in Wireless Communication (SPAWC), 2022 \cite{PolyDot-CMPC}.}}

\IEEEcompsocitemizethanks{\IEEEcompsocthanksitem E. Vedadi and H. Seferoglu are with the Department of Electrical and Computer Engineering, University of Illinois Chicago. 
E-mails: evedad2@uic.edu, hulya@uic.edu. Y. Keshtkarjahromi is with Seagate Technology. E-mail: yasaman.keshtkarjahromi@seagate.com. 
}}

\markboth{IEEE Transactions on Information Forensics and Security}%
{Coded Privacy-Preserving Computation at
Edge Networks}

\maketitle


\begin{abstract}
Multi-party computation (MPC) is promising for designing privacy-preserving machine learning algorithms at edge networks.  An emerging approach is coded-MPC (CMPC), which advocates the use of coded computation to improve the performance of MPC in terms of the required number of workers involved in computations. The current approach for designing CMPC algorithms is to merely combine efficient coded computation constructions with MPC. We show that this approach fails short of being efficient; \eg entangled polynomial codes are not necessarily better than PolyDot codes in MPC setting, while they are always better for coded computation. Motivated by this observation, we propose a new construction;  Adaptive Gap Entangled (AGE) polynomial  codes for MPC. We show through analysis and simulations that MPC with AGE codes always perform better than existing CMPC algorithms in terms of the required number of workers as well as computation, storage, and communication overhead.
 \end{abstract}

\vspace{-15pt}

\section{\label{sec:introduction}Introduction} 
\IEEEPARstart{M}{assive} amount of data is generated at edge networks with the emerging Internet of Things (IoT). 
Indeed, the data generated by IoT devices is expected to reach 73.1 ZB by 2025, growing from 18.3 ZB in 2019 \cite{IDCReport}. This huge amount of data is expected to be processed in real-time in many time sensitive applications, which is extremely challenging if not impossible with existing centralized cloud due to limited bandwidth between the edge and centralized cloud \cite{DemocratizingNetworkEdge, EdgeComputingVideo, EdgeEatCloud}.

\begin{figure}[t!]
\vspace{-10pt}
\centering
{ \scalebox{.16}{\includegraphics{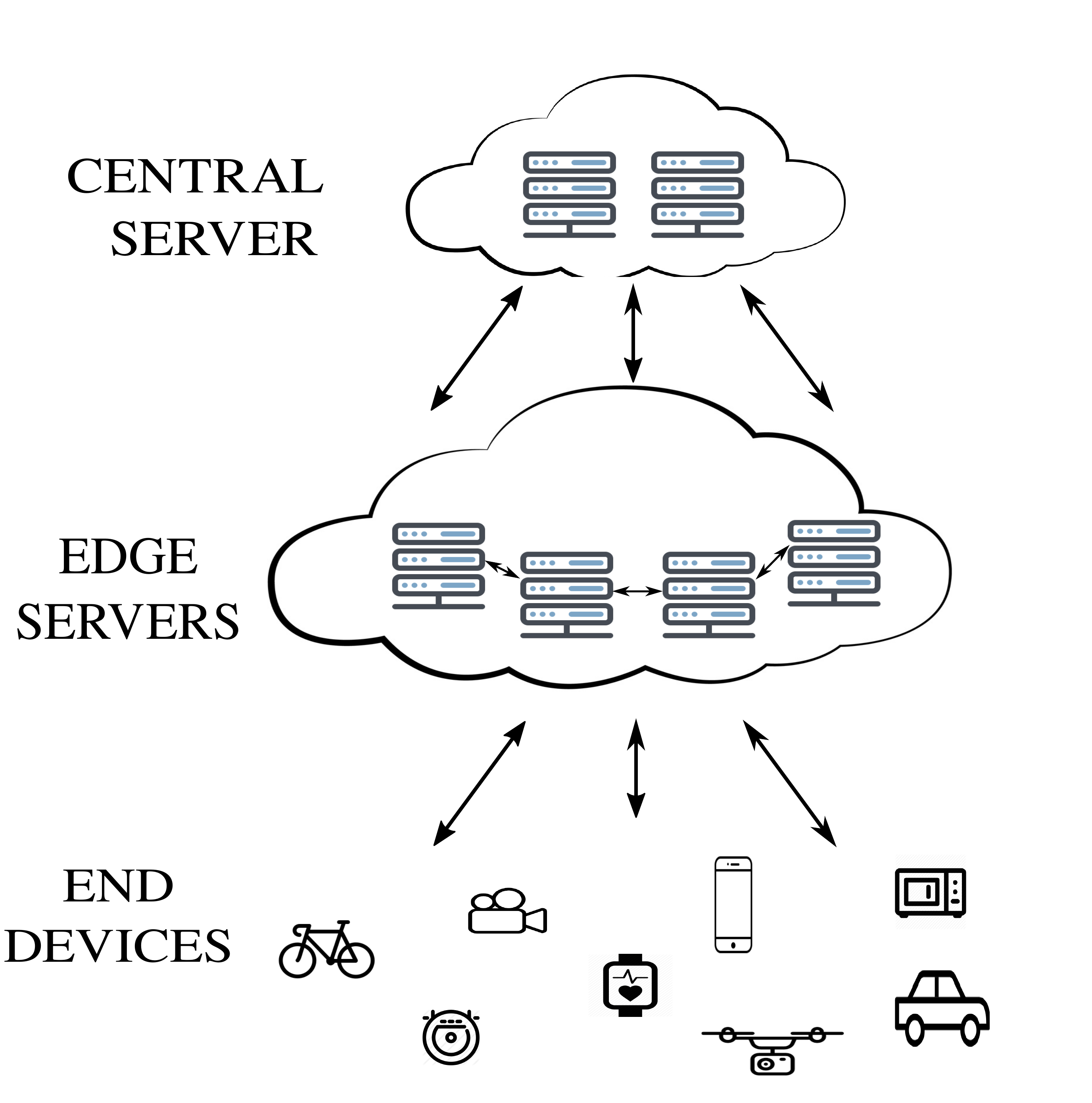}} }
\vspace{-10pt}
\caption{
An edge computing system. End devices generate and/or collect data, edge servers process data, and a central server collects the outcome of the processed data.}
\vspace{-15pt}
\label{fig:main_figure}
\end{figure}

We consider a distributed computing system at the edge, where data is generated and collected by end devices, Fig.~\ref{fig:main_figure}. The goal is to analyze this data through computationally-intensive machine learning algorithms to extract useful information. Computationally intensive aspects are distributively processed by the edge servers, and a central server collects the outcome of the processed data.  In this context, it is crucial to design efficient computation mechanisms at edge servers by taking into account the limited resources, including the number of edge serves, computing power, storage, and communication cost, while preserving privacy of data. 

Multi-party computation (MPC) is a privacy-preserving distributed computing framework \cite{scalableMPC}. In MPC, several parties (end devices in Fig. \ref{fig:main_figure}) have private data and the goal is to compute a function of data collectively with the participation of all parties (end devices and edge servers in Fig. \ref{fig:main_figure}), while preserving privacy, \ie each party only knows its own information. MPC can be categorized into cryptographic solutions \cite{Yao}, \cite{GMW} and information-theoretic solutions \cite{BGW}. In this paper, our focus is on the information-theoretic MPC solution; BGW (Ben-Or, Goldwasser and Widgerson) \cite{BGW} using Shamir's secret sharing scheme \cite{ShamirSS} thanks to its lower computational complexity and quantum safe nature \cite{10.1007/3-540-48405-1_4}. Despite its potential, BGW does not take into account the limited resources of edge devices.

An emerging approach is coded-MPC (CMPC), which advocates the use of coded computation \cite{SpeedUp-journal, Tradeoff-journal} to improve the performance of BGW in terms of the required number of workers involved in computations. However, the existing approach for designing CMPC algorithms \cite{8613446, Zhu2021ImprovedCF, 9333639} is to merely combine efficient coded computation constructions with MPC. This approach fails short of being efficient as it does not look at an important interaction between coded computation and MPC.

In this paper, we assume that end devices store/collect matrices, and the goal is to multiply these matrices in a privacy-preserving manner. 
We focus on matrix multiplication as these operations are the atomic functions computed over many iterations of several signal processing, machine learning, and optimization algorithms, such as gradient descent based algorithms, classification algorithms, etc. \cite{SpeedingUp, burges2005learning, zhang2004solving, bottou2010large}.

CMPC mechanisms based on Shamir's secret shares create a polynomial for each matrix, where a polynomial has two terms; \emph{coded} and \emph{secret}. The multiplication of matrices are performed by multiplying these polynomials, which create cross terms of \emph{coded} and \emph{secret} terms. Some of these cross terms are not used for reconstructing matrix multiplication from polynomials, so we refer them as \emph{garbage} terms. Our key observation in this paper is that the garbage terms, and designing the coded and secret terms by taking into account the garbage terms are crucial to reduce the required number of workers (edge servers in Fig.\ref{fig:main_figure}) in CMPC.

In fact, even if a code construction is optimized for coded computation, it may not perform well in CMPC due to the lack of the usage of garbage terms. For example, it is known that entangled polynomial codes always outperform PolyDot codes in terms of the number of required workers \cite{YuFundamentalLimits2018} for coded computation. However, we show in this paper that Entangled-CMPC does not always perform better than PolyDot-CMPC.  
%
%
%
Motivated by this observation, we propose a new construction; Adaptive Gap Entangled (AGE) polynomial codes for MPC setup. We show through analysis and simulations that MPC with AGE codes performs better than existing CMPC algorithms including Entangled-CMPC \cite{8613446}, SSMM \cite{Zhu2021ImprovedCF}, and GCSA-NA \cite{9333639} as well as our PolyDot-CMPC design in terms of the required number of workers as well as computation, storage, and communication overhead. {The main contributions of this paper are summarized in the following:}
{\begin{itemize}
\item We design PolyDot-CMPC, where we determine its secret terms by taking into account the garbage terms. We analyze the required number of workers by PolyDot-CMPC as compared to baselines. We show that PolyDot-CMPC reduces the required number of workers for several colluding workers as compared with baselines. In particular, we show that Entangled-CMPC does not always perform better than PolyDot-CMPC although it is always better for coded computation, according to \cite{YuFundamentalLimits2018}.    
\item We design Adaptive Gap Entangled (AGE) polynomial codes, where we determine both coded and secret terms by taking into account the garbage terms. We provide a theoretical analysis that AGE codes outperform existing CMPC algorithms \cite{8613446, Zhu2021ImprovedCF, 9333639} as well as our PolyDot-CMPC design in terms of the required number of workers. 
\item We analyze the storage, computation, and communication load requirements of AGE-CMPC and PolyDot-CMPC. We show that AGE-CMPC outperforms baselines in terms of these performance metrics. 
\item We provide a privacy analysis for AGE-CMPC and PolyDot-CMPC and show that both algorithms satisfy the privacy requirements that we define in Section \ref{sec:system}. 
\item We evaluate AGE-CMPC and PolyDot-CMPC via simulations and illustrate that AGE-CMPC outperforms the baselines in terms of the required number of workers, storage, computation, and communication load. 
\end{itemize}} 

{The structure of the rest of this paper is as follows. We give an overview of the related works in Section~\ref{sec:related}. In Section~\ref{sec:system}, we provide the system model. Section \ref{sec:PolyDotMPC} is dedicated to the detailed explanation of our PolyDot-CMPC framework. Section \ref{sec:AGEcodes-and-AGECMPC} includes the details of our proposed AGE codes and AGE-CMPC algorithm. Section \ref{sec:recoverythreshold,comp,com,S analysis-polydot} presents our analysis for computation, storage, communication overhead, and privacy of our proposed frameworks, PolyDot-CMPC and AGE-CMPC. We provide simulation results in Section \ref{sec:simulation}, and finally Section~\ref{sec:conc} concludes the paper.}

\vspace{-10pt}
\section{Related Work}\label{sec:related}
Coded computation advocates higher reliability and smaller delay in distributed computation by introducing redundancy \cite{SpeedingUp}. Significant effort is being put on constructing codes for fast and distributed matrix-vector multiplication \cite{SpeedingUp}, \cite{FerdinandAnytime}, matrix-matrix multiplication \cite{YuPolynomial2017, LeeHighDim2017, PolyDotMatDot, YuFundamentalLimits2018}, dot product and convolution of two vectors \cite{ShortDot}, \cite{DuttaCodedConvolution2017}, gradient descent \cite{TandonGradientCoding2017, HalbawiImprovingDistGradient2018, RavivGradientCyclic2018}, distributed optimization \cite{KarakusRedundancy2019}, Fourier transform \cite{YuCodedFourier2017}, and linear transformations \cite{YangComputeLinear2017}. As compared to this line of work, we consider privacy-preserving computation at edge networks.

Privacy is studied in coded computation. In \cite{8382305, SecureCoded, GASP}, the problem of matrix-matrix multiplication is considered for the case that a master possesses the input data and would like to perform multiplication on the data with the help of parallel workers, while the data is kept confidential from the workers. In \cite{bitar_trans_PRAC} and \cite{BPR17}, privacy is addressed for the same system model of master-worker setup, but for matrix-vector multiplication. As compared to this line of work, we focus on MPC, where there are multiple sources each having private input data, and the goal is that a master learns the result of the computation of a matrix multiplication with the help of parallel workers. The input data should be kept confidential from workers and the master. 

There is a line of work investigating CMPC. Lagrange Coded Computing  is designed  \cite{LCC1} in a coded computation setup for security and privacy. This work is extended for MPC setup \cite{LCC2}. 
The problem of limited memory at each party in MPC setup is addressed in \cite{PolynomCMPC}  by leveraging polynomial coded computation. This work is generalized using entangled polynomial codes for block-wise matrix multiplication \cite{8613446}. Secure multi-party batch matrix multiplication is considered in  \cite{9333639,Zhu2021ImprovedCF}, which modify the MPC system setup by employing the idea of noise alignment to reduce the communication load among workers. 
As compared to this line of work, we design PolyDot-CMPC and AGE-CMPC by taking into account the interaction between coded computation and MPC. In particular, we consider the garbage terms in our PolyDot-CMPC and AGE-CMPC design, where the garbage terms represent the interaction between coded computation and MPC. 

\vspace{-10pt}
\section{System Model and Motivation } \label{sec:system}

\textbf{Notations.} We denote the set of (i) natural numbers with $\mathbb{N}$, 
(ii) integers with $\mathbb{Z}$, and (iii) finite field with $\mathbb{F}$. 

\emph{Set of polynomial degrees:} The set of nonzero powers of a given polynomial $f(x) = \sum_{i=0}^n a_ix^i$  is denoted by $\mathbf{P}(f(x))$,  
 \begin{align}
    \mathbf{P}(f(x)) = \{i \in \mathbb{Z}: 0 \leq i \leq n,\; a_i \neq 0\}.
\end{align}

\emph{Set definitions and operations:} We use the following standard notations for arbitrary sets $\mathbf{A}$ and $\mathbf{B}$, where the elements of $\mathbf{A}$, $\mathbf{B}$ are integers, \ie  $a, b \in \mathbb{Z}$.  
 \begin{align}
   \mathbf{A} + \mathbf{B} = \{ a + b: a \in \mathbf{A},\; b \in \mathbf{B}\}, 
\end{align}
 \begin{align}
   \mathbf{A} + b = \{a+b: a \in \mathbf{A}\}.
\end{align}
 Furthermore, $|\mathbf{A}|$ stands for the cardinality of $\mathbf{A}$, $k|m$ means that $m$ is divisible by $k$, \ie $\mod\{m,k\}=0$. Finally, $\Omega_{a}^{b} =  \{a, \ldots, b\}$ refers to the set of integers between $a$ and $b$. 
 

\emph{Matrix splitting:} 
If a matrix $A$ is divided into $s$ row-wise and $t$ column-wise partitions, it is represented as
\begin{align}\label{eq:blockwise-part}
A = \left[ {\begin{array}{ccc}
   A_{0,0}&
   \ldots & A_{0,t-1}\\   
   \vdots&\ddots&\vdots \\ A_{s-1,0} &
   \ldots&A_{s-1,t-1}
  \end{array} } \right],
  \end{align}
  where for $A \in \mathbb{F}^{m \times m}$, $A_{j,i} \in \mathbb{F}^{\frac{m}{s} \times \frac{m}{t}}$ for $j \in \Omega_{0}^{s-1}$ and $i \in \Omega_{0}^{t-1}$.

\textbf{Setup.} We consider a system setup with $E$ end devices (sources), $N$ edge servers (workers), and a central server (master  node) as shown in Fig. \ref{fig:main_figure}. Each source node $e \in \Eset$, where $|\Eset|= E$, has private data $X_{e} \in \mathbb{F}^{\mu \times \nu}$. Each source node is connected to all worker nodes via device-to-device (D2D) links such as Wi-Fi Direct and offloads its data to worker nodes for privacy-preserving computation. Each worker node $W_n, n \in \Nset$ ($|\Nset| = N$) is connected to other worker nodes as well as the master node via D2D links. The source, worker, and  master nodes are all edge devices with limited resources. 

\textbf{Application.} The goal is to calculate a function of per source data; $Y = \gamma(X_{1}, \ldots, X_{E})$, while the privacy of data $X_{1}, \ldots, X_{E}$ is preserved. While function $\gamma(.)$ could be any polynomial function in MPC setup, we focus on matrix multiplication as (i) we would like to present our ideas in a simple way, and (ii) matrix multiplication forms an essential building block of many signal processing and machine learning algorithms (gradient descent, classification, etc.) \cite{SpeedingUp, burges2005learning, zhang2004solving, bottou2010large}. In particular, we consider $Y=\gamma(A, B) = A^TB$, where $X_1 = A$, $X_2 = B$, $A \in \mathbb{F}^{m \times m} $, $B \in \mathbb{F}^{m \times m}$. We note that we use square matrices from two sources for easy exposition, and it is straightforward to extend our results for more general matrices and larger number of sources. 
 
\textbf{Attack Model.}
We assume a semi-honest system model, where the sources, the worker nodes, and the master follow the defined protocols by our CMPC mechanisms, but they are curious about the private data. We assume that $z$ nodes ($z < \frac{N}{2}$) among  workers can collude to maximize the information that they can access. 
We design our CMPC mechanisms against $z$ colluding workers to provide privacy-preserving computation. 


\textbf{Privacy Requirements.} We define the privacy requirements from the perspective of source, worker, and master nodes. 

\textit{Source perspective:} Source nodes should not learn anything about the private data of any other source nodes. This requirement is satisfied in our system as there is no communication among the source nodes.
    
\textit{Worker perspective:} Each worker should not learn anything about the private data $X_{1}, \ldots, X_{E}$ from the perspective of information-theoretic security. Also, workers should not learn anything when the workers communicate with each other, \ie
    \begin{align}\label{privacy-for-each-w-1}
&\tilde{H}(X_{1},\ldots,X_{E}|\underset{n \in \Nset_c}{\bigcup} (\{G_{n'}(\alpha_n), n'\in \Omega_1^N\},\underset{e \in \Eset}{\cup} F_e(\alpha_n))) \nonumber \\
&=\tilde{H}(X_{1}, X_{2},...,X_{E}),
    \end{align}
    where $\tilde{H}$ denotes the Shannon entropy, $\alpha_n$ is a parameter from finite field $\mathbb{F}$, which is defined by worker $W_n$, $G_{n'}(\alpha_n)$ is the data each worker $W_n$ receives from another worker $W_{n'}$, $F_e(\alpha_n)$ is the data received by each worker $W_n$ from source node $e$, $n \in \Nset_c$, and $\Nset_c$ is a subset of $\Nset$ satisfying $|\Nset_c| \leq z$. 
    
    
\textit{Master perspective}: The master node should not learn anything more than the final result $Y$, \ie
    \begin{align}\label{privacy-for-master}
\tilde{H}(X_{1},\ldots, X_{E}|Y,\underset{n \in \Nset}{\bigcup}I(\alpha_n))= \tilde{H}(X_{1},\ldots, X_{E}|Y),
    \end{align}
where $I(\alpha_n)$ is the data received from $W_n$ by the master node. 

\section{PolyDot Coded MPC (PolyDot-CMPC)} \label{sec:PolyDotMPC} 

In this section, we present our PolyDot coded MPC  (PolyDot-CMPC) algorithm that employs PolyDot codes \cite{PolyDotMatDot} to create coded terms. 
Our design is based on leveraging the garbage terms that are not required for computing $Y=A^TB$ and reusing them in the secret terms.

\vspace{-10pt}
\subsection{PolyDot-CMPC\label{sec:PolyDot-CMPC}}

\textbf{Phase 1 - Sources Share Data with Workers.} 
We have two sources; source 1 and source 2, where they have matrices $A$ and $B$, respectively. They divide matrices $A \in \mathbb{F}^{m \times m}$ and $B \in \mathbb{F}^{m \times m}$ into $s$ row-wise and $t$ column-wise partitions\footnote{
We note that in PolyDot-CMPC, we exclude the case of no partitioning, \ie $s=t=1$; This case corresponds to BGW, where coding is not required, and thus is excluded from our CMPC setup.} as in (\ref{eq:blockwise-part}), where $s,t \in \mathbb{N}$, and $s|m$ and $t|m$ hold. {Using the splitted matrices $A_{i,j} \in A^T$ and $B_{k,l} \in B$, where $i,l \in \Omega_{0}^{t-1}$, $j,k \in \Omega_{0}^{s-1}$}, they generate polynomials $F_A(x)$ and $F_B(x)$.

Polynomials $F_A(x)$ and $F_B(x)$ consist of coded and secret terms, \ie $F_{i'}(x)=C_{i'}(x)+S_{i'}(x),\; i' \in \{A,B\}$, where $C_{i'}(x)$'s are the coded terms defined by PolyDot codes \cite{PolyDotMatDot}, and $S_{i'}(x)$'s are the secret terms that we construct. The coefficients of the coded terms correspond to splitted matrices. In other words, $A_{i,j}$ and $B_{k,l}$ become one of the coefficients of $C_A(x)$ and $C_B(x)$, respectively. The powers and degree of $C_A(x)$ and $C_B(x)$ are determined by PolyDot codes \cite{PolyDotMatDot}, which will be explicitly defined as part of Theorem \ref{th:S_i(x)andC_i(x)-strategy-polydot}. The coefficients of $S_A(x)$ and $S_B(x)$ are drawn randomly from the same finite field that matrices $A$ and $B$ are defined. It is crucial to determine the powers and degree of $S_A(x)$ and $S_B(x)$ as they dictate the number of workers required for privacy-preserving calculation of $Y=A^TB$. Next, we discuss how the powers of $S_A(x)$ and $S_B(x)$ are determined. 


Let $\mathbf{P}(C_A(x))$ and $\mathbf{P}(C_B(x))$ be the sets of the powers of the polynomials $C_A(x)$ and $C_B(x)$ with coefficients larger than zero. $\mathbf{P}(C_A(x))$ and $\mathbf{P}(C_B(x))$ are expressed as \cite{PolyDotMatDot}
\begin{align}\label{eq:polydot-p(CA)-th}
    \mathbf{P}(C_{A}(x)) = & \{i+tj \in \mathbb{N}: i \in \Omega_{0}^{t-1},\; j \in \Omega_{0}^{s-1}\} \nonumber \\
   & =  \{0,\ldots,ts-1\},
\end{align}
\begin{align}\label{eq:polydot-p(CB)-th}
    \mathbf{P}(C_B(x)) = & \{t(s-1-k)+l\theta' \in \mathbb{N}:  k \in \Omega_{0}^{s-1},\; l \in \Omega_{0}^{t-1}\} \nonumber \\
    = & \{tq'+l\theta' \in \mathbb{N}: q' \in \Omega_{0}^{s-1},\; l \in \Omega_{0}^{t-1}\},
\end{align}
where $s,t \in \mathbb{N}$, and $\theta' = t(2s-1)$. 

As seen from (\ref{eq:polydot-p(CA)-th}) and (\ref{eq:polydot-p(CB)-th}), $\mathbf{P}(C_A(x)C_B(x))$, the set of the powers of the polynomial $C_A(x)C_B(x)$ with coefficients larger than zero, is expressed as $\mathbf{P}(C_A(x)C_B(x))= \{i+t(s-1+j-k)+tl(2s-1) \in \mathbb{N}: i,l \in \Omega_{0}^{t-1},\; j,k \in \Omega_{0}^{s-1} \}$. 

\begin{algorithm}[t!]
\caption{Setting the powers of $S_A(x)$ and $S_B(x)$ in PolyDot-CMPC}\label{alg:Alg-SA(x)-SB(x)-polydot}
\begin{algorithmic}[1]
\STATE \textbf{Inputs:} Number of colluding workers $z$, number of row and column partitions; $s$ and $t$. \\
\underline{\textbf{Step 1: Set $\mathbf{P}(S_{A}(x))$}.} 
\STATE Determine all elements of $\mathbf{P}(S_{A}(x))$ starting from the minimum possible element satisfying C1 in (\ref{eq:non_eq-polydot-thrm4}). 
 \\
 \underline{\textbf{Step 2: Set $\mathbf{P}(S_B(x))$.}}
 \STATE Fix $\mathbf{P}(S_{A}(x))$ in C2 and find all elements of the subset of $\mathbf{P}(S_B(x))$ starting from the minimum possible element that satisfy C2. Call this subset as $\mathbf{P'}(S_B(x))$.\\
 \STATE Determine all elements of the subset of $\mathbf{P}(S_B(x))$ starting from the minimum possible element that satisfy C3. Call this subset as $\mathbf{P''}(S_B(x))$.\\
 \STATE Find the intersection of $\mathbf{P'}(S_B(x))$ and $\mathbf{P''}(S_B(x))$ to form $\mathbf{P}(S_B(x))$.
\end{algorithmic}
\end{algorithm}
\setlength{\textfloatsep}{7pt}

We know from \cite{PolyDotMatDot} that $Y_{i,l}=\sum_{j=0}^{s-1}A_{i,j}B_{j,l}$,  which are the coefficients of $x^{i+t(s-1)+tl(2s-1)}$ in $C_A(x)C_B(x)$, are the elements of the final result $Y=A^TB$. Therefore, we define $\{i+t(s-1)+tl(2s-1) \in \mathbb{N}: i,l \in \Omega_{0}^{t-1}\}$ as the set of \emph{important powers} of $C_A(x)C_B(x)$. In other words, these are the powers of the terms that are required to decode $Y=A^TB$. Thus, we determine the secret terms $S_A(x)$ and $S_B(x)$ such that the important powers of $C_A(x)C_B(x)$ do not overlap (do not have common terms) with garbage terms (the terms that are not used for decoding $Y=A^TB$) such as $\mathbf{P}(C_A(x)S_B(x))$, $\mathbf{P}(S_A(x)C_B(x))$, and $\mathbf{P}(S_A(x)S_B(x))$.
More precisely, the following conditions should hold: 
\begin{align}\label{eq:non_eq-polydot-thrm4}
    & \text{C1: } i+t(s-1)+tl(2s-1) \not\in \mathbf{P}(S_{A}(x))+\mathbf{P}(C_B(x)), \nonumber \\
   & \text{C2: } i+t(s-1)+tl(2s-1) \not\in \mathbf{P}(S_{A}(x))+\mathbf{P}(S_B(x)), \nonumber \\
    & \text{C3: } i+t(s-1)+tl(2s-1) \not\in \mathbf{P}(S_B(x))+\mathbf{P}(C_{A}(x)),
\end{align} where $i, l \in \Omega_{0}^{t-1}$ and $s,t \in \mathbb{N}$. Our algorithm that determines $\mathbf{P}(S_A(x))$ and $\mathbf{P}(S_B(x))$ to satisfy the conditions in (\ref{eq:non_eq-polydot-thrm4}) is provided in Algorithm \ref{alg:Alg-SA(x)-SB(x)-polydot}. Next, we show in Theorem \ref{th:S_i(x)andC_i(x)-strategy-polydot} that our PolyDot-CMPC mechanism, where the coded terms of its polynomials $F_A(x)$ and $F_B(x)$ are determined according to Algorithm \ref{alg:Alg-SA(x)-SB(x)-polydot}, satisfy the conditions in (\ref{eq:non_eq-polydot-thrm4}). 
\vspace{-1pt}
\begin{theorem}\label{th:S_i(x)andC_i(x)-strategy-polydot}
With the following design of $F_A(x)$ and $F_B(x)$ in PolyDot-CMPC, the conditions in (\ref{eq:non_eq-polydot-thrm4}) are satisfied.
{\begin{align}\label{eq:FA1andFA2PolyDotCMPC}
   F_A(x)= \bigg\{ \begin{array}{cc}
   F_{A_1}(x)  &  z > ts-t \text{ and } s,t \neq 1\\
   F_{A_2}(x) & z \leq ts-t \text{ or } t=1 \text{ or } s=1\\
\end{array}\end{align}} 
\begin{align}\label{eq:FAPolyDotCMPC}
     F_{A_1}(x) = & \clipbox{0.5 0 1 0}{$\underbrace{{\sum_{i=0}^{t-1}\sum_{j=0}^{s-1} A_{i,j}x^{i+tj}}}_{\triangleq C_A(x)}$}+\clipbox{-3 0 11 0}{$\underbrace{{\sum_{w=0}^{ts-t-1}\sum_{l=0}^{p-1}\bar{A}_{(w+\theta'l)}x^{ts+\theta'l+w}}\hspace{1em}}_{\triangleq S_{A_1}(x)}$} \nonumber \\
      &\clipbox{220 0 -2 0}{$\underbrace{{\phantom{\hspace{22em}}+ \sum_{u=0}^{z-1-pt(s-1)}\bar{A}_{(u+t(s-1)+\theta'(p-1))} x^{ts+\theta'p+u}}}$},
\end{align}
\begin{align}\label{eq:FA2PolyDotCMPC}
F_{A_2}(x) = & \underbrace{\sum_{i=0}^{t-1}\sum_{j=0}^{s-1} A_{i,j}x^{i+tj}}_{\triangleq C_A(x)} + \underbrace{\sum_{u=0}^{z-1}\bar{A}_{u}x^{ts+\theta' p+u}}_{\triangleq S_{A_2}(x)},
\end{align}
{\begin{align}\label{eq:FBPolyDotCMPC}
   F_B(x)= \bigg\{ \begin{array}{cc}
   F_{B_1}(x)  &  z>\tau \text{ or } t=1 \text{ or } s=1\\
   F_{B_2}(x) & \frac{\tau+1}{2}< z \leq \tau \text{ and } s,t \neq 1\\
   F_{B_3}(x) & z \leq \frac{\tau+1}{2} \text{ and } s,t \neq 1
\end{array}\end{align} 
\begin{align}\label{eq:FB1PolyDotCMPC}
F_{B_1}(x)= \underbrace{\sum\limits_{k=0}^{s-1}\sum\limits_{l=0}^{t-1} B_{k,l}x^{\varpi+\theta'l}}_{\triangleq C_B(x)}+ \underbrace{\sum\limits_{r=0}^{z-1} \bar{B}_rx^{ts+\theta'(t-1)+r}}_{\triangleq S_{B_1}(x)},
\end{align} 
\begin{align}\label{eq:FB2PolyDotCMPC}
&F_{B_2}(x) = \clipbox{0.5 0 0.25 0}{$\underbrace{\sum\limits_{k=0}^{s-1}\sum\limits_{l=0}^{t-1} B_{k,l}x^{\varpi+\theta'l}}_{\triangleq C_B(x)}$}+\clipbox{1 0 10.7 0}{$\underbrace{\sum\limits_{d=0}^{\tau-z}\sum\limits_{l'=0}^{p'-1} \bar{B}_{(\theta'l'+d)}x^{ts+\theta'l'+d}\hspace{1em}}_{\triangleq S_{B_2}(x)}$} \nonumber\\&+ \clipbox{220 0 -2 0}{$\underbrace{\phantom{\hspace{22em}}\sum\limits_{v=0}^{z-1-p'(\tau-z+1)}\bar{B}_{(v+\tau-z+1+\theta'(p'-1))}x^{ts+\theta'p'+v}}$},
\end{align}}
\begin{align}\label{eq:FB3PolyDotCMPC}
       F_{B_3}(x) = & \underbrace{\sum_{k=0}^{s-1}\sum_{l=0}^{t-1} B_{k,l}x^{\varpi+\theta'l}}_{\triangleq C_B(x)} + \underbrace{\sum_{v=0}^{z-1}\bar{B}_{v}x^{ts+v}}_{\triangleq S_{B_3}(x)},
\end{align}
where $p = \min \{ \floor{\frac{z-1}{ts-t}},t-1\}$, $\varpi=t(s-1-k)$, {$\tau=\theta'-ts-t$, $p'=\min \{\floor{\frac{z-1}{\tau-z+1}},t-1\}$}. {Moreover,} $\bar{A}_{(w+\theta'l)}$, $\bar{A}_{(u+t(s-1)+\theta'(p-1))}$, and $\bar{A}_{u}$, are selected independently and uniformly at random in $\mathbb{F}^{\frac{m}{t} \times \frac{m}{s}}$, and $\bar{B}_r$, {$\bar{B}_{(\theta'l'+d)}$, $\bar{B}_{(v+\tau-z+1+\theta'(p'-1))}$}, and $\bar{B}_{v}$ are chosen independently and uniformly at random in $\mathbb{F}^{\frac{m}{s} \times \frac{m}{t}}$. 
\end{theorem}
{\em Sketch of Proof:}
{To prove this theorem, we first determine $\mathbf{P}(S_A(x))$ and $\mathbf{P}(S_B(x))$, based on the set of rules that are described in Algorithm \ref{alg:Alg-SA(x)-SB(x)-polydot} (which clearly satisfy the conditions in (\ref{eq:non_eq-polydot-thrm4})), then derive $F_{A}(x)$ and $F_{B}(x)$, accordingly. The proof is provided in Appendix A in {the supplemental materials}.}
\hfill $\Box$

After source 1 and source 2 determine $F_A(x)$ and $F_B(x)$, respectively, they calculate $F_A(\alpha_n)$ and $F_B(\alpha_n)$, where $\alpha_n$ is a constant associated with worker $W_n$ and known by all the workers in the system. Then, source 1 sends $F_A(\alpha_n)$ to worker $W_n$, and source 2 sends $F_A(\alpha_n)$ to worker $W_n$ for actual matrix multiplication computations. 

\vspace{-1pt}
\textbf{Phase 2 - Workers Compute and Communicate.} The second phase consists of workers processing data received from the sources and sharing the results with each other. In this phase, each worker $W_n$ calculates $H(\alpha_n) = F_{A}(\alpha_n)F_B(\alpha_n)$, where ${H}(x)$ is defined as: 
\begin{align}\label{eq:HxAGECMPC}
    {H}(x) = \sum_{n=0}^{\deg(F_A(x))+\deg(F_B(x))} {H}_nx^n = F_{A}(x)F_B(x),
\end{align}
where ${H}_u = \sum_{j=0}^{s-1}A_{i,j}B_{j,l}$ are the coefficients that are required for calculating $A^TB$, \ie $u = si+(s-1)+\theta l$ for $i,l \in \Omega_{0}^{t-1}$. 
Each worker $W_n$ has the knowledge of one point from ${H}(x)$ through calculation of $H(\alpha_n)=F_{A}(\alpha_n)F_B(\alpha_n)$. By applying Lagrange interpolation on (\ref{eq:HxAGECMPC}), there exist $r_n^{(i,l)}$'s {$ \in \mathbb{F}$} such that
\begin{align}\label{lagrangeinterpolpolydot}
    H_u=\sum_{j=0}^{s-1} A_{i,j}B_{j,l} = \sum_{n=1}^{N} r_n^{(i,l)} {H}(\alpha_n).
\end{align} Thus, each worker $W_n$ multiplies $r_n^{(i,l)}$'s with $H(\alpha_n)$ and shares them with the other workers, securely. In particular, for each worker $W_n$, there are $t^2$ coefficients of $r_n^{(i,l)}$. Therefore, each worker $W_n$ creates a polynomial $G_n(x)$ with the first $t^2$ terms allocated to multiplication of $r_n^{(i,l)}$ with $H(\alpha_n)$ and the last $z$ terms allocated to random coefficients to keep $H(\alpha_n)$ confidential from $z$ colluding workers:
\begin{align} \label{eq:GnAGECMPC}
    G_n(x) = \sum_{i=0}^{t-1}\sum_{l=0}^{t-1}r_n^{(i,l)}{H}(\alpha_n)x^{i+tl}+ \sum_{w=0}^{z-1} R^{(n)}_wx^{t^2+w},
\end{align}
where $R^{(n)}_w, w \in \Omega_{0}^{z-1}$ are chosen independently and uniformly at random from $\mathbb{F}^{\frac{m}{t} \times \frac{m}{t}}$.
Each worker $W_n$ sends $G_n(\alpha_{n'})$ to other workers ${W_{n'}}, n' \in \Nset, n'\neq n$. After all the data exchanges, each worker $W_{n'}$ has the knowledge of $G_n(\alpha_{n'})$, $\forall n$, which sums them up and sends it to the master in the last phase. The following equation represents the polynomial that is equal to the summation of $G_n(x)$:
\begin{align}
    I(x) = \sum_{n=1}^{N} G_n(x),
\end{align}
which can be equivalently written as:
\begin{align}\label{eq:IxAGECMPC}
    I(x)= & \sum_{i=0}^{t-1}\sum_{l=0}^{t-1}\sum_{n=1}^{N}r_{n}^{(i,l)}{H}(\alpha_n)x^{i+tl}+ \sum_{w=0}^{z-1}\sum_{n=1}^{N} R^{(n)}_wx^{t^2+w} \nonumber \\
    = &  \sum_{i=0}^{t-1}\sum_{l=0}^{t-1}\sum_{j=0}^{s-1}A_{i,j}B_{j,l}x^{i+tl} + \sum_{w=0}^{z-1}\sum_{n=1}^{N} R^{(n)}_wx^{t^2+w}.
\end{align}
\textbf{Phase 3 - Master Node Reconstructs $Y = A^TB$.} As seen in (\ref{eq:IxAGECMPC}), the coefficients for the first $t^2$ terms of $I(x)$ represent the components of the matrix $Y=A^TB$. On the other hand, the degree of $I(x)$ is $t^2+z-1$, therefore, the master can reconstruct $I(x)$ and extract $Y=A^TB$ after receiving $I(\alpha_n)$ from $t^2+z$ workers.

\begin{theorem} \label{th:N_PolyDot}
{The required number of workers for multiplication of two massive $A$ and $B$ employing PolyDot-CMPC, in a privacy preserving manner while there exist $z$ colluding workers in the system and due to the resource limitations each worker is capable of working on at most $\frac{1}{st}$ fraction of each input matrix, is expressed as follows}
\begin{align} \label{eq:N-PolyDot-DMPC}
N_{\text{PolyDot-CMPC}} =
\begin{cases}
   \psi_1, & ts<z \;\text{ or } t=1 
   \\
   \psi_2, & ts-t < z \leq ts \text{ and } t,s \neq 1\\
   \psi_3, & ts-2t < z \leq ts-t \text{ and } t,s \neq 1\\
   \psi_4, &  \upsilon' < z \leq ts-2t \text{ and } t,s \neq 1\\
   \psi_5, & z \leq \upsilon' \text{ and } t,s \neq 1\\
   \psi_6, & s=1 \text{ and } t \ge z \text{ and } t \neq 1
\end{cases}
\end{align}
where $\psi_1 = (p+2)ts+\theta'(t-1)+2z-1$, $\psi_2 = 2ts+\theta'(t-1)+3z-1$, $\psi_3 = 2ts+\theta'(t-1)+2z-1$, $\psi_4 = (t+1)ts+(t-1)(z+t-1)+2z-1$, $\psi_5 = \theta't+z$, and $\psi_6=t^2+2t+tz-1$,  $s|m$, and $t|m$ are satisfied, $p=\min\{\floor{\frac{z-1}{\theta'-ts}},t-1\}$, $\theta'= 2ts-t$ and $\upsilon'=\max\{ts-2t-s+2,\frac{ts-2t+1}{2}\}$. 
\end{theorem}
{\em Sketch of Proof:}
{The required number of workers in CMPC is equal to the number of terms in polynomial $H(x)=F_A(x)F_B(x)$ with non-zero coefficients \cite{PolynomCMPC}, \ie
\begin{align}
    N_{\text{PolyDot-CMPC}}=&|\mathbf{P}(H(x))| \nonumber\\
    =&|\mathbf{P}((C_A(x)+S_A(x))(C_B(x)+S_B(x)))| \nonumber \\
    =&|(\mathbf{P}(C_A(x))+\mathbf{P}(C_B(x))) \cup\nonumber \nonumber \\ &(\mathbf{P}(C_A(x))+\mathbf{P}(S_B(x))) \cup\nonumber\\
    & (\mathbf{P}(S_A(x))+\mathbf{P}(C_B(x))) \cup\nonumber\\
    &(\mathbf{P}(S_A(x))+\mathbf{P}(S_B(x)))|.
\end{align}
Therefore, to prove this theorem, we first determine $\mathbf{D}_1 = \mathbf{P}(C_A(x))+\mathbf{P}(C_B(x))$, $\mathbf{D}_2  =\mathbf{P}(C_A(x))+\mathbf{P}(S_B(x))$, $\mathbf{D}_3=\mathbf{P}(S_A(x))+\mathbf{P}(C_B(x))$, and $\mathbf{D}_4=\mathbf{P}(S_A(x))+\mathbf{P}(S_B(x))$. Then, we calculate $|\mathbf{D}_1 \cup \mathbf{D}_2 \cup \mathbf{D}_3 \cup \mathbf{D}_4|$.
The detailed proof is provided in Appendix B in {the supplemental materials}. } \hfill $\Box$

\vspace{-10pt}
\subsection{PolyDot-CMPC in Perspective}\label{sec:polydotcmpc-perspective}
This section provides the theoretical analysis on the performance of PolyDot-CMPC as compared with the baselines, 
Entangled-CMPC \cite{8613446}, SSMM \cite{Zhu2021ImprovedCF} and GCSA-NA \cite{9333639}\footnote{GCSA-NA is constructed for batch matrix multiplication. However, by considering the number of batches as one, it becomes a fair baseline to compare PolyDot-CMPC.
}, in terms of the required number of workers.

\begin{lemma}\label{lemma: regions where N_polydot<N_entangled}
PolyDot-CMPC requires less number of workers than Entangled-CMPC when the system parameters satisfy one of the following requirements:
\begin{enumerate}
    \item  $z>ts,\; p<\frac{t-1}{s},\; t\neq 1$
    \item $ts-s<z\leq ts,\; t-1>s,\; s,t \neq 1$
    \item $(t-1)^2<z<t(t-1),\; s=t-1,\; s,t \neq 1$
    \item $ts-t-\min\{0,1-\frac{2s-5}{t-3}\}<z\leq ts-s,\; t>3,\; s \neq 1$
    \item $s=2,\; t=3,\;z=4$
    \item $t=2,\; s=2,\; z=1,2$ 
    \item $\max\{st-t-s-\frac{2}{t-2}, ts-2t\} < z \leq ts-t,\; t>2,\; t\ge s,\; s \neq 1$
   \item $t<s\leq 2t,\; ts-s<z\leq ts-t,\; s,t \neq 1$ 
     \item $t=2,\; 3\leq s \leq 4,\; 2(s-2)<z\leq 2(s-1)$ 
    \item $st-2t < z \leq ts-s,\; t>2,\; t< s\leq 2t$
    \item $s>2t,\; ts-2t<z\leq ts-t,\; s,t \neq 1$ 
    \item $2t\ge s,\; \max\{ts-2t-s+2, \frac{ts-2t+1}{2}\} < z \leq \min\{st-2t, 2ts-t^2+t-2s+1\},\; s,t \neq 1$ 
    \item $s>2t,\; ts-s<z\leq ts-2t,\; t\neq 1,2$ 
    \item $4<s<z<2s-4,\; t=2$ 
    \item $ts-2t-s+2<z<ts-s,\; 2t<s,\; s,t \neq 1$
    \item $st-2s-t-\frac{1}{t-1}<z\leq \max\{ts-2t-s+2, \frac{ts-2t+1}{2}\},\; s,t \neq 1$. 
\end{enumerate}
In all other regions for the values of the system parameters $s, t$, and $z$, PolyDot-CMPC requires the same or larger number of workers.
\end{lemma}
{\em Sketch of Proof:}
{
The comparison between the number of workers required by PolyDot-CMPC and Entangled-CMPC is derived directly from comparing (\ref{eq:N-PolyDot-DMPC}) and $N_{\text{Entangled-CMPC}}$ in Theorem 1 in \cite{8613446}. The detailed analysis is provided in Appendix C.A of the supplemental materials.
} \hfill $\Box$
\begin{lemma}\label{lemma: regions where N_polydot<N_ssmm}
PolyDot-CMPC requires less number of workers than SSMM when the system parameters satisfy one of the following requirements:
\begin{enumerate}
    \item $z > \max\{ts,ts-t+\frac{pts}{t-1}\}, t\neq 1$
    \item $\frac{t-1}{t-2}(st-t)<z \leq ts$.
\end{enumerate}
In all other regions for the values of the system parameters $s, t$, and $z$, PolyDot-CMPC requires the same or larger number of workers.
\end{lemma}
{\em Sketch of Proof:}
The comparison between the number of workers required by PolyDot-CMPC and SSMM is derived directly from comparing (\ref{eq:N-PolyDot-DMPC}) and  {$N_{\text{SSMM}}$ provided in Theorem 1 in \cite{Zhu2021ImprovedCF}}. 
The detailed proof is provided in Appendix C.B {in the supplemental materials}. \hfill $\Box$

\begin{lemma}\label{lemma: regions where N_polydot<N_gcsana}
PolyDot-CMPC requires less number of workers than GCSA-NA when the system parameters satisfy one of the following requirements:
\begin{enumerate}
    \item $z>ts, p<\frac{t-1}{s}, t\neq 1$
    \item $s<t, ts-t<z \leq \min\{ts, t(t-1)-1\}$
    \item $z \leq ts-t$
    \item $s=1, t>z, t\neq 2$.
\end{enumerate}
In all other regions for the values of the system parameters $s, t$, and $z$, PolyDot-CMPC requires the same or larger number of workers.
\end{lemma}
{\em Sketch of Proof:}
The comparison between the number of workers required by PolyDot-CMPC and GCSA-NA is derived directly from comparing (\ref{eq:N-PolyDot-DMPC}) and {$N_{\text{GCSA-NA}}$ for one matrix matrix multiplication provided in Table 1 in \cite{9333639}.} 
The proof is provided in Appendix C.C {in the supplemental materials}. \hfill $\Box$

As seen from Lemma \ref{lemma: regions where N_polydot<N_entangled}, PolyDot-CMPC, a CMPC method based on PolyDot codes, outperforms Entangled-CMPC, a CMPC method based on entangled polynomial codes, for a range of values of system parameters. This observation is surprising as it is known that entangled polynomial codes constantly outperforms PolyDot codes for coded computation design \cite{YuFundamentalLimits2018}. This result shows that the design of secret terms jointly with the coded terms is crucial to reduce the required number of workers. Motivated by this observation, we design a new code construction that is optimized for CMPC. The details of our new construction is provided in the next section.

\vspace{-10pt}
\section{Adaptive Gap Entangled Polynomial Codes}\label{sec:AGEcodes-and-AGECMPC}

In this section, we introduce Adaptive Gap Entangled polynomial (AGE) codes and present our CMPC design with AGE codes; AGE-CMPC.
\vspace{-10pt}
\subsection{AGE Codes} \label{sec:AGECodes}

We consider the generalized formulation  \cite{YuFundamentalLimits2018} for coded computation of matrices $A$ and $B$ and create the coded term, 
\begin{align}\label{eq:generalEntangled}
    C_A(x) = & \sum\limits_{i=0}^{t-1}\sum\limits_{j=0}^{s-1}A_{i,j}x^{j\alpha+i\beta}, \nonumber \\
    C_B(x) = & \sum\limits_{k=0}^{s-1}\sum\limits_{l=0}^{t-1}B_{k,l}x^{(s-1-k)\alpha+\theta l} 
\end{align} where $\alpha, \beta, \theta \in \mathbb{N}$, $A_{i,j} \in A^T$ and $B_{k,l} \in B$. Several codes that have been designed for coded computation can be considered as the special case of (\ref{eq:generalEntangled}) by considering different values of $(\alpha,\beta,\theta)$. 
%
%
For example, PolyDot codes \cite{PolyDotMatDot} correspond to $(\alpha,\beta,\theta)=(t,1,t(2s-1))$, while generalized PolyDot codes \cite{Dutta2018AUC} and entangled polynomial codes \cite{YuFundamentalLimits2018} follow $(\alpha,\beta,\theta)=(1,s,ts)$, where $t$ is the number of column-wise partitions and $s$ is the number of row-wise partitions of matrices $A$ and $B$. The common goal of these codes is to reduce the degree of  $C_A(x)C_B(x)$ multiplication, which reduces the number of required workers in coded computation. 

On the other hand, in our PolyDot-CMPC construction and analysis, we observed that minimizing the degree of $C_A(x)C_B(x)$ is not necessarily good for CMPC (although it is for coded computation) to reduce the required number of workers. Our key observation is that if we keep the degree of $C_A(x)C_B(x)$ higher, we can potentially create gaps in the powers of $C_A(x)$ and $C_B(x)$. This would actually be better to align the garbage terms of $C_A(x)C_B(x)$ with the garbage terms coming from $C_A(x)S_B(x)$, $S_A(x)C_B(x)$, and $S_A(x)S_B(x)$ multiplications, which would reduce the degree of $H(x)$ and this is important to reduce the number of required workers (see the sketch of proof of Theorem \ref{th:N_PolyDot}) in CMPC.  

Thus, we construct new codes by considering $(\alpha,\beta,\theta)=(1,s,ts+\lambda)$ in (\ref{eq:generalEntangled}), where $\lambda$ is an integer in the range of $0\leq \lambda\leq z$, 
which we optimize to achieve the minimum required number of workers for CMPC\footnote{Note that $\lambda\ge 0$ is required for decodability, and $\lambda >z$ does not result in a more efficient AGE-CMPC, so we consider $0\leq \lambda\leq z$ range. The proof is provided in Appendix H in the supplemental materials.}. 
We note that different values of $\lambda$ results in different number of gaps in $C_B(x)$ and thus different number of garbage terms. The value of $\lambda$ will be determined adaptively based on the optimum number of workers required by CMPC. We call this code design \emph{``Adaptive Gap Entangled polynomial (AGE)"} codes.
Next, we prove the decodability of our AGE codes.

\begin{theorem}\label{th:decodabilityofAGEcodes}
AGE codes 
guarantee the decodability of $Y=A^TB$ from the polynomial $C_Y(x)=C_A(x)C_B(x)$.
\end{theorem}

{\em Sketch of Proof:}
{The components of $Y=A^TB$, \ie $Y_{i,l}=\sum_{j=0}^{s-1}A_{i,j}B_{j,l}$ are the coefficients of $x^{s-1+si+(ts+\lambda)l}$ in $C_Y(x)=C_A(x)C_B(x)$. To prove the decodability of AGE codes, we prove that the terms with the powers of $(s-1)+si+(ts+\lambda)l, i,l \in \Omega_{0}^{t-1}$, \ie the set of important powers, (i) do not have repetitive elements, \ie the set $\{(s-1)+si+(ts+\lambda)l: i,l \in \Omega_{0}^{t-1}\}$ consists of $t^2$ distinct elements, and (ii) do not have overlap with any other terms, \ie the set $\{(s-1)+si+(ts+\lambda)l: i,l \in \Omega_{0}^{t-1}\}$ and $\{j+is+(s-1-k)+(ts+\lambda)l: i,l \in \Omega_{0}^{t-1}, j,k \in \Omega_{0}^{s-1}, j\neq k\}$ do not overlap.}
{The proof is provided in Appendix D}. 
\hfill $\Box$
\vspace{-17pt}
\subsection{AGE-CMPC} \label{sec:AGE-CMPC}
\textbf{Phase 1 - Sources Share Data with Workers.} 
The operation of this phase is similar to the operation of phase 1 of PolyDot-CMPC detailed in Section \ref{sec:PolyDot-CMPC}. The only differences are how the coded terms $C_{A}(x)$, $C_{B}(x)$ and the secret terms $S_{A}(x)$, $S_{B}(x)$ are constructed. From (\ref{eq:generalEntangled}), $\mathbf{P}(C_A(x))$ and $\mathbf{P}(C_B(x))$, the set of all powers in the polynomials $C_A(x)$ and $C_B(x)$ with non-zero coefficients, are as follows:
\begin{align}\label{eq:AGE-p(CA)-th}
    \mathbf{P}(C_{A}(x)) = & \{j+si: i \in \Omega_{0}^{t-1},
    j \in \Omega_{0}^{s-1}\} 
    =  \{0,\ldots,ts-1\},
\end{align}
\begin{align}\label{eq:AGE-p(CB)-th}
    \mathbf{P}(C_B(x)) = & \{(s-1-k)+l(ts+\lambda): 
    k \in \Omega_{0}^{s-1},\;l \in \Omega_{0}^{t-1}\},
\end{align}
%
where $s,t \in \mathbb{N}$ and $\lambda \in \Omega_{0}^{z}$. In AGE-CMPC, $S_A(x)$ and $S_B(x)$ are defined such that $\mathbf{P}(C_A(x)S_B(x))$, $\mathbf{P}(S_A(x)C_B(x))$, and $\mathbf{P}(S_A(x)S_B(x))$ do not have common terms with the important powers of $(s-1)+si+(ts+\lambda)l$ for $i,l \in \Omega_{0}^{t-1}$. 
%
%
%
The reason is that $\{s-1-k+j+is+(ts+\lambda)l:\;i,l \in \Omega_{0}^{t-1},\; j,k \in \Omega_{0}^{s-1}\;s,t \in \mathbb{N}\}$ is the set of powers of polynomial $C_A(x)C_B(x)$, from which 
$\{(s-1)\alpha+i\beta+\theta l:\;i,l \in \Omega_{0}^{t-1},\;s,t \in \mathbb{N}\}$ is the set that is required to have no overlap with the other terms, called garbage terms, for successful recovery of $Y$. 
For this purpose, the following conditions should be satisfied: 
\begin{align}\label{eq:non_eq-AGE-conditions}
    & \text{C4: } (s-1)+si+(ts+\lambda)l \not\in \mathbf{P}(S_B(x))+\mathbf{P}(C_A(x)), \nonumber \\
   & \text{C5: } (s-1)+si+(ts+\lambda)l \not\in \mathbf{P}(S_A(x))+\mathbf{P}(C_B(x)), \nonumber \\
    & \text{C6: } (s-1)+si+(ts+\lambda)l \not\in \mathbf{P}(S_A(x))+\mathbf{P}(S_B(x)).
\end{align}

\begin{algorithm}[t!]
\caption{Setting the powers of $S_A(x)$ and $S_B(x)$ in AGE-CMPC}\label{alg:Alg-SA(x)-SB(x)}
\begin{algorithmic}[1]
\STATE \textbf{Inputs:} Number of colluding workers $z$, number of row and column partitions; $s$ and $t$, parameter $\lambda \in \mathbb{N}$. \\
\underline{\textbf{Step 1: Determining $\mathbf{P}(S_B(x))$.}}
\STATE
Set the elements of $\mathbf{P}(S_B(x))$ as $z$ consecutive elements starting from the maximum important power plus 1.\\
 \underline{\textbf{Step 2: Determining $\mathbf{P}(S_A(x))$.}}
 \STATE Fix $\mathbf{P}(S_B(x))$ in C5 and find all elements of $\mathbf{P}(S_A(x))$ starting from the minimum possible element that satisfies C5. 
\end{algorithmic}
\end{algorithm}
\setlength{\textfloatsep}{5pt}

\vspace{-10pt}
Our strategy for determining $\mathbf{P}(S_A(x))$ and $\mathbf{P}(S_B(x))$, {summarized in Algorithm \ref{alg:Alg-SA(x)-SB(x)}}, is as follows. First, we set the elements of $\mathbf{P}(S_{B}(x))$ as $z$ consecutive elements starting from the maximum important power plus one, \ie $s-1+s(t-1)+(ts+\lambda)(t-1)$ plus one; $\mathbf{P}(S_B(x))=\{ts+(ts+\lambda)(t-1),\dots,ts+(ts+\lambda)(t-1)+z-1\}$ or equivalently:
$\mathbf{P}(S_B(x))= \{ts+\theta(t-1)+r, r\in \Omega_0^{z-1}, \theta=ts+\lambda\}$. 
We note that the elements of $\mathbf{P}(C_A(x))$ and $\mathbf{P}(S_A(x))$ are powers of polynomials, so they are non-negative. Therefore, by starting the elements of $\mathbf{P}(S_B(x))$ from the maximum important power plus one, C4 and C6 are satisfied. Then, we find all elements of the subset of $\mathbf{P}(S_A(x))$, starting from the minimum possible element, that satisfies C5 in (\ref{eq:non_eq-AGE-conditions}). Using this strategy, we can determine $S_A(x)$ and $S_B(x)$ as
\begin{align}\label{eq:S-A}
   S_A(x)= \bigg\{ \begin{array}{cc}
   S_{A_1}(x) 
     &  z > \lambda, \text{ and } t \neq 1\\
   S_{A_2}(x) & {z \leq \lambda}, \text{ or } t=1,
\end{array}\end{align} where $S_{A_1}(x) = \sum_{w=0}^{\lambda-1} \sum_{l=0}^{q-1}   \bar{A}_{(w+\theta l)}x^{ts+\theta l+w}  +  \sum_{u=0}^{z-1-q\lambda}   \bar{A}_{(u+\lambda+\theta(q-1))}  x^{ts+\theta q+u}$, $S_{A_2}(x)$ $=$ $\sum_{u=0}^{z-1}$ $\bar{A}_{u}x^{ts+u}$, $\bar{A}_{(w+\theta l)}$, $\bar{A}_{(u+\lambda+\theta(q-1))}$, and $\bar{A}_{u}$ are chosen independently and uniformly at random in $\mathbb{F}^{\frac{m}{t} \times \frac{m}{s}}$, $\theta=ts+\lambda$ and $q=\min\{\floor{\frac{z-1}{\lambda}},t-1\}$, and 
\begin{align}\label{eq:S-B}
   S_{B}(x)=\sum_{r=0}^{z-1} \bar{B}_rx^{ts+\theta(t-1)+r},
\end{align} where $\bar{B}_r$ is chosen independently and uniformly at random in $\mathbb{F}^{\frac{m}{s} \times \frac{m}{t}}$. 
\begin{theorem}\label{th:FA-FB-AGE-thrm}
The polynomials $S_{A}(x)$ and $S_{B}(x)$ defined in (\ref{eq:S-A}) and (\ref{eq:S-B}) satisfy the conditions in (\ref{eq:non_eq-AGE-conditions}). 
\end{theorem}
{\em Sketch of Proof:}
{To prove this theorem, we determine $\mathbf{P}(S_A(x))$ and $\mathbf{P}(S_B(x))$, based on our strategy that is described in Algorithm \ref{alg:Alg-SA(x)-SB(x)} and show that they satisfy the conditions in (\ref{eq:non_eq-AGE-conditions}). In other words, We first show that $\mathbf{P}(S_B(x))=\{ts+(ts+\lambda)(t-1),\ldots,ts+(ts+\lambda)(t-1)+z-1\}$ in (\ref{eq:S-B}) satisfies C4 in (\ref{eq:non_eq-AGE-conditions}). Then, we fix $\mathbf{P}(S_{B}(x))$ in C6 of (\ref{eq:non_eq-AGE-conditions}), and find $\mathbf{P}(S_A(x))$ that satisfies C5 and C6.
The detailed proof is provided in Appendix E {in the supplemental materials}.} \hfill $\Box$

In phase 1, source 1 and source 2 find the optimum $\lambda$ based on the optimization problem explained in Algorithm \ref{alg:Alg-AGE-CMPC}, then create $F_A(x)=C_A(x)+S_A(x)$ and $F_B(x)=C_B(x)+S_B(x)$, respectively, using the obtained optimum $\lambda$, and share $F_{A}(\alpha_n)$ and $F_B(\alpha_n)$ with each worker $W_n$. Due to using $z$ random terms in constructing $F_{A}(x)$ and $F_B(x)$, no information about $A$ and $B$ is revealed to any workers.

\textbf{Phase 2 - Workers Compute and Communicate.} This phase is the same as phase 2 of PolyDot-CMPC detailed in Section \ref{sec:PolyDot-CMPC}.

\textbf{Phase 3 - Master Node Reconstructs $Y = A^TB$.} This phase is the same as phase 3 of PolyDot-CMPC detailed in Section \ref{sec:PolyDot-CMPC}.

\begin{theorem} \label{th:N_AGE}
The total number of workers required to compute $Y = A^TB$ using AGE-CMPC, when there exist $z$ colluding workers and each worker can work on at most $\frac{1}{st}$ 
fraction of data from each source due to the computation or storage constraints, is expressed as
\begin{align}\label{eq:eq:N-AGE-CMPC-optimization}
    N_{\text{AGE-CMPC}} =\begin{cases} \displaystyle\min_{\lambda} \Gamma(\lambda) & t \neq 1\\
    2s+2z-1 & t=1
    \end{cases}
\end{align} 
where $\Gamma(\lambda)$ is defined as
\begin{align}\label{eq:N-AGE-CMPC}
\Gamma(\lambda)=\begin{cases}
   \Upsilon_1(\lambda), & z>ts-s,\; \lambda=0\\
   \Upsilon_2(\lambda), & z \leq ts-s,\; \lambda=0\\
   \Upsilon_3(\lambda), & \lambda=z\\
   \Upsilon_4(\lambda), & z>ts,\; 0 < \lambda < z\\
   \Upsilon_5(\lambda), & z\leq ts,\; 0< \lambda < z,\; ts<\lambda+s-1\\
   \Upsilon_6(\lambda), & \lambda+s-1<z\leq ts,\;0 < \lambda < z,\; q\lambda \geq s \\
   \Upsilon_7(\lambda), & \lambda+s-1<z\leq ts,\;0 < \lambda < z,\; q\lambda < s\\
   \Upsilon_8(\lambda), & z \leq \lambda+s-1\leq ts,\;0 < \lambda < z,\; q\lambda \geq s\\
   \Upsilon_9(\lambda), & z \leq \lambda+s-1 \leq ts,\;0 < \lambda < z,\; q\lambda < s,
\end{cases}
\end{align}
and $\Upsilon_1(0)=2st^2+2z-1$, $\Upsilon_2(0)=st^2+3st-2s+t(z-1)+1$, $\Upsilon_3(z)=2ts+(ts+z)(t-1)+2z-1$, $\Upsilon_4(\lambda)=(q+2)ts+\theta(t-1)+2z-1$, $\Upsilon_5(\lambda)=3ts+\theta(t-1)+2z-1$, $\Upsilon_6(\lambda)=2ts+\theta(t-1)+(q+2)z-q-1$, $\Upsilon_7(\lambda)= \theta(t+1)+q(z-1)-2\lambda +z+ts+\min\{0, z+s(1-t)-\lambda q-1\}$, $\Upsilon_8(\lambda)=2ts+\theta(t-1)+3z+(\lambda+s-1)q-\lambda-s-1$, $\Upsilon_9(\lambda)=\theta(t+1)+q(s-1)-3\lambda+3z-1\nonumber+\min\{0,ts-z+1+\lambda q-s\}$, $s \geq 1$, $t \geq 2$, $s|m$, $t|m$ are satisfied, $q=\min\{\floor{\frac{z-1}{\lambda}},t-1\}$ and $\theta = ts+\lambda$.
\end{theorem}
{\em Sketch of Proof:}
{Similar to the proof of Theorem \ref{th:N_PolyDot}, to calculate the number of required workers, we calculate its equivalent term, $|\mathbf{D}_1 \cup \mathbf{D}_2 \cup \mathbf{D}_3 \cup \mathbf{D}_4|$, where $\mathbf{D}_1 = \mathbf{P}(C_A(x))+\mathbf{P}(C_B(x))$, $\mathbf{D}_2  =\mathbf{P}(C_A(x))+\mathbf{P}(S_B(x))$, $\mathbf{D}_3=\mathbf{P}(S_A(x))+\mathbf{P}(C_B(x))$, and $\mathbf{D}_4=\mathbf{P}(S_A(x))+\mathbf{P}(S_B(x))$. The proof is provided in Appendix F {in the supplemental materials}}. \hfill $\Box$

{\textbf{AGE-CMPC in a Nutshell.} Algorithm \ref{alg:Alg-AGE-CMPC} provides an overview of AGE-CMPC operation. Next, we provide an example to illustrate AGE-CMPC operation.} 

\begin{algorithm}[t!]
\caption{Operation of AGE-CMPC}\label{alg:Alg-AGE-CMPC}
\begin{algorithmic}[1]
\STATE \textbf{Inputs:}  
Matrices $A, B$, number of colluding workers $z$, number of row and column partitions; $s$ and $t$.
\STATE  \textbf{Parameters known by all workers:} Chosen parameters $\alpha_n$, Lagrange interpolation coefficients $r_1^{(i,l)}, \ldots, r_N^{(i,l)}$. \\
 \underline{\textbf{Phase 0: Calculation of $\lambda^*$}}
 \STATE Calculate $\lambda^*$ by solving the optimization problem in (\ref{eq:eq:N-AGE-CMPC-optimization}). \\
 \underline{\textbf{Phase 1: Sources Share Data with Workers.}}
 \STATE Sources $1$ and $2$ determine $F_A(x)$ and $F_B(x)$, respectively using  $\lambda^*$.
 \STATE Sources $1$ and $2$ send $F_A(\alpha_n)$ and $F_B(\alpha_n)$, respectively to worker $W_n$. \\
 \underline{\textbf{Phase 2: Workers Compute and Communicate.}}
 \STATE Worker $W_n$ computes ${H}(\alpha_n)=F_A(\alpha_n)F_B(\alpha_n)$, and $G_n(x)$ according to (\ref{eq:GnAGECMPC}). 
 \STATE Worker $W_n$ sends $G_n(\alpha_{n'})$ to worker $W_{n'}$.
\STATE Worker $W_{n'}$ computes $I(\alpha_{n'}) = \sum_{n=1}^{N} G_n(\alpha_{n'})$ and sends it to the master. \\
 \underline{\textbf{Phase 3: Master Node Reconstructs $Y = A^T B$.}}
\STATE The master reconstructs $I(x)$ when it receives results from $t^2+z$  workers.
\STATE The master calculates $A^TB$ from the coefficients of the first $t^2$ terms of $I(x)$ according to (\ref{eq:IxAGECMPC}).
 \end{algorithmic}
\end{algorithm}

\begin{example}\label{ex:AGECMPC} \emph{AGE-CMPC.} 
Let us consider a scenario with $s=t=z=2$ for two sources (Source $1$ and Source $2$) that have matrices $A$ and $B$. The sources partition the matrices to $st=4$ sub-matrices; \ie $s=2$ row-wise and $t=2$ column-wise partitions. These sub-matrices will be multiplied with the help of a number of workers, where $z=2$ workers are adversaries.   

For this purpose, as it is mentioned in the first step of Algorithm \ref{alg:Alg-AGE-CMPC} (in phase $0$), the optimization problem in (\ref{eq:eq:N-AGE-CMPC-optimization}) will be solved to determine $\lambda^*$ (the optimum $\lambda$ that minimizes the required number of workers, $N_{\text{AGE-CMPC}}$). The solution of (\ref{eq:eq:N-AGE-CMPC-optimization}) is $N_{\text{AGE-CMPC}} = 17$ for $\lambda^* = 2$ 
when $s=t=z=2$. This means that 17 workers are required by AGE-CMPC to guarantee privacy. We note that the required number of workers by Entangled-CMPC  \cite{8613446} is $N_{\text{Entangled-CMPC}}=19$. As seen, AGE-CMPC reduces the required number of workers as compared to Entangled-CMPC. 

In phase 1, source $1$ and source $2$ first calculate $C_A(x)$, $C_B(x)$, $S_A(x)$, and $S_B(x)$ according to (\ref{eq:generalEntangled}), (\ref{eq:S-A}), and (\ref{eq:S-B}) for $\lambda=\lambda^*=2$: $C_A(x) = A_{0,0} + A_{0,1}x + A_{1,0}x^2 + A_{1,1}x^3$ and $C_B(x) = B_{0,0}x + B_{1,0} + B_{0,1}x^7 + B_{1,1}x^6$, $S_A(x) = \bar{A}_0x^4 + \bar{A}_1x^5$ and $S_B(x) = \bar{B}_0x^{10} + \bar{B}_1x^{11}$. Then these sources create $F_A(x)=C_A(x)+S_A(x)$ and $F_B(x)=C_B(x)+S_B(x)$ accordingly: $F_A(x)=A_{0,0} + A_{0,1}x + A_{1,0}x^2 + A_{1,1}x^3+\bar{A}_0x^4 + \bar{A}_1x^5$, $F_B(x)=B_{0,0}x + B_{1,0} + B_{0,1}x^7 + B_{1,1}x^6+ \bar{B}_0x^{10} + \bar{B}_1x^{11}$. At the end of phase 1, the sources collaborate to create each $\alpha_n \in \mathbb{F}$, for $n \in \Omega_1^{17}$, randomly and then each source sends its private data, $F_A(\alpha_n)$, $F_B(\alpha_n)$ to worker $W_n$.

In phase 2, each worker $W_n$, $n \in \Omega_1^{17}$ computes ${H}(\alpha_n)=F_A(\alpha_n)F_B(\alpha_n)=A_{0,0}B_{1,0}+(A_{0,0}B_{0,0}+A_{0,1}B_{1,0})x+(A_{0,1}B_{0,0}+A_{1,0}B_{1,0})x^2+A_{1,0}B_{0,0}x^3+(A_{1,1}B_{0,0}+\bar{A}_{0}B_{1,0})x^4+(\bar{A}_{0}B_{0,0}+\bar{A}_{1}B_{1,0})x^5+(A_{0,0}B_{1,1}+\bar{A}_{1}B_{0,0})x^6+(A_{0,0}B_{0,1}+A_{0,1}B_{1,1})x^7+(A_{0,1}B_{0,1}+A_{1,0}B_{1,1})x^8+(A_{1,0}B_{0,1}+A_{1,1}B_{1,1})x^9+(A_{0,0}\bar{B}_{0}+A_{1,1}B_{0,1}+\bar{A}_0B_{1,1})x^{10}+(A_{0,0}\bar{B}_{1}+A_{0,1}\bar{B}_{0}+\bar{A}_0B_{0,1}+\bar{A}_1B_{1,1})x^{11}+(A_{0,1}\bar{B}_{1}+A_{1,0}\bar{B}_{1}+\bar{A}_1B_{0,1})x^{12}+(A_{1,0}\bar{B}_{1}+A_{1,1}\bar{B}_{0})x^{13}+(A_{1,1}\bar{B}_{1}+\bar{A}_{0}\bar{B}_{0})x^{14}+(\bar{A}_{0}\bar{B}_{1}+\bar{A}_{1}\bar{B}_{0})x^{15}+(\bar{A}_{1}\bar{B}_{1})x^{16}$. Then, it computes $G_n(x)$ according to (\ref{eq:GnAGECMPC}): $G_n(x)=r_n^{(0,0)}{H}(\alpha_n)+r_n^{(0,1)}{H}(\alpha_n)x^2+r_n^{(1,0)}{H}(\alpha_n)x+r_n^{(1,1)}{H}(\alpha_n)x^3+ R^{(n)}_0x^{4}+R^{(n)}_1x^{5}$. Next, worker $W_n$ sends $G_n(\alpha_{n'})$ to all other workers $W_{n'}$. Finally, worker $W_{n'}$ computes $I(\alpha_{n'}) = \sum_{n=1}^{17} G_n(\alpha_{n'})$ and sends it to the master.

In phase 3, the master reconstructs $I(x)$ when it receives $I(\alpha_{n'})$ results from $6$ workers, as the degree of $I(x)$ is 5 and is equal to
\begin{align}
     I(x) &= 
     (A_{0,0}B_{0,0} + A_{0,1}B_{1,0})+ (A_{1,0}B_{0,0} + A_{1,1}B_{1,0}) x \nonumber \\
     & + (A_{0,0}B_{0,1} + A_{0,1}B_{1,1}) x^2+ (A_{1,0}B_{0,1} + A_{1,1}B_{1,1})x^3 \nonumber \\
     &+\sum_{n=1}^{17} R^{(n)}_0x^4+\sum_{n=1}^{17} R^{(n)}_1x^5 \nonumber
  \end{align}
according to (\ref{eq:IxAGECMPC}). In the last step, the master calculates $A^TB$ from the coefficients of the first $4$ terms of $I(x)$:
 \begin{align}
   A^TB = \left[{\begin{array}{cc}
   A_{0,0}B_{0,0} + A_{0,1}B_{1,0} & A_{0,0}B_{0,1} + A_{0,1}B_{1,1}\\
   A_{1,0}B_{0,0} + A_{1,1}B_{1,0} & A_{1,0}B_{0,1} + A_{1,1}B_{1,1}
   \end{array} } \right]. \nonumber
 \end{align}
\hfill $\Box$
\end{example}

\vspace{-15pt}
\subsection{AGE-CMPC in Perspective}\label{sec:AGECMPCvsOthers}
In this section, we compare AGE-CMPC with the baselines in terms of the required number of workers. 
\begin{lemma}\label{corol:AGE-Vs-Ent-SSMM-GC-Poly}
$N_{\text{AGE-CMPC}}$ is always less than or equal to the number of workers required by Entangled-CMPC \cite{8613446}, SSMM \cite{Zhu2021ImprovedCF}, GCSA-NA (for one matrix multiplication) \cite{9333639}, and PolyDot-CMPC.
\end{lemma}

{\em Sketch of Proof:}
{
The comparison between the required number of workers by AGE-CMPC and Entangled-CMPC, SSMM, GCSA-NA, and PolyDot-CMPC is derived directly from comparing (\ref{eq:N-AGE-CMPC}) with $N_{\text{Entangled-CMPC}}$ from Theorem 1 in \cite{8613446}, $N_{\text{SSMM}}$ provided in Theorem 1 in \cite{Zhu2021ImprovedCF}, $N_{\text{GCSA-NA}}$ for one matrix multiplication proposed in Table 1 in \cite{9333639}, and $N_{\text{PolyDot-CMPC}}$ provided in (\ref{eq:N-PolyDot-DMPC}), respectively. The detailed comparisons are provided in Appendix G.A, G.B, G.C, and G.D, respectively, in the supplemental materials. 
} 
\hfill $\Box$

\vspace{-5pt}
\section{Computation, Storage, and Communication Requirements and Privacy Guarantee of the Coded MPC Methods}\label{sec:recoverythreshold,comp,com,S analysis-polydot} 

{In this section, we provide the theoretical analysis of the computation, storage, and communication overhead required by the coded MPC methods including Entangled-CMPC \cite{8613446} and our designed PolyDot-CMPC and AGE-CMPC. We will also prove the privacy guarantee of PolyDot-CMPC and AGE-CMPC at the end of this section.}

\vspace{-15pt}
\subsection{Computation Overhead} 

We define the computation overhead as the total number of scalar multiplications performed by each worker. We do not consider additions in the analysis as the computation complexity of addition is negligible as compared with multiplication. 
\begin{corollary} \label{cor:computation_all_methods-polydot}
The total computation overhead per worker to compute $Y=A^TB$ using coded MPC methods of Entangled-CMPC, PolyDot-CMPC and AGE-CMPC 
is expressed as
\begin{align} \label{eq:polydot_computation_cost}
    \xi
    =  \frac{m^3}{st^2}+m^2+N
    (t^2+z-1)\frac{m^2}{t^2},
\end{align} where $m$ is the number of rows/columns of matrices $A$ and $B$, $s$ and $t$ are the number of row-wise and column-wise partitions, respectively, $z$ is the number of colluding workers, and $N$ is the required number of workers by each method.
\end{corollary} 
{\em Proof:} Based on Phase 1 of coded MPC (Entangled-CMPC, PolyDot-CMPC and AGE-CMPC), each worker computes $H(\alpha_n)=F_A(\alpha_n)F_B(\alpha_n)$. 
In coded MPC, $F_A(\alpha_n) \in \mathbb{F}^{ \frac{m}{t} \times \frac{m}{s}}$ and $F_B(\alpha_n) \in \mathbb{F}^{\frac{m}{s} \times \frac{m}{t}}$, so $\frac{m^3}{st^2}$ scalar multiplications are computed. 

After computing $H(\alpha_n)$, each worker $W_n$ needs to compute polynomial $G_n(x)$ for $N$ different points; $\alpha_{n'}$, $n' \in \Omega_1^N$, following {(\ref{eq:GnAGECMPC})}\footnote{$G_n(\alpha_{n'})$, for $n' \in \Omega_1^N\; , n'\neq n$, is required to be calculated to be sent to the other workers and $G_n(\alpha_n)$ is required to be calculated for the calculation of $I(\alpha_n)$ in {(\ref{eq:IxAGECMPC})}.}. For this purpose, worker $W_n$ first  multiplies $r_n^{i,l}$ for $i,l \in \Omega_0^{t-1}$ with $H(\alpha_n)\in \mathbb{F}^{\frac{m}{t} \times \frac{m}{t}}$. This requires $t^2\frac{m^2}{t^2}=m^2$ scalar multiplications. Then,  $r_n^{i,l}H(\alpha_n)$ is multiplied with $\alpha_{n'}^{i+tl}$ for all $n' \in \Omega_1^N$ workers. This requires $N(t^2-1)\frac{m^2}{t^2}$ scalar multiplications. To calculate the second part of $G_n(\alpha_{n'})$, $n' \in \Omega_1^N$, $W_n$ multiplies $\alpha_{n'}^{t^2+w}$ with random matrices $R^{(n)}_w \in \mathbb{F}^{\frac{m}{t} \times \frac{m}{t}}$, for $w \in \Omega_0^{z-1}$. This requires $Nz\frac{m^2}{t^2}$ scalar multiplications. In total, each worker $W_n$ computes $m^2+N(t^2+z-1)\frac{m^2}{t^2}$ scalar multiplications to obtain $G_n(\alpha_{n'})$'s. Then worker $W_n$ adds $G_n(\alpha_{n'})$, for $n' \in \Omega_1^N$, where the complexity of addition is negligible. 

By summing up the number of scalar multiplications computed by each worker $W_n$, the computation overhead of coded MPC becomes $\frac{m^3}{st^2}+m^2+N(t^2+z-1)\frac{m^2}{t^2}$ per worker. This concludes the proof. \hfill $\Box$

\vspace{-15pt}
\subsection{Storage Overhead}
We define the storage overhead as the total number of scalar parameters that should be stored in all phases of coded MPC at each worker.\footnote{We note that it is possible to delete some of the data after each phase once they are not needed for future steps, but we do not consider deleting data for easy exposition.} These parameters include the received parameters from the other workers as well as those that are computed and stored to be used in the next computations.
\begin{corollary} \label{cor:Storage_all_methods-polydot}
The total storage overhead per worker to compute $Y=A^TB$ using coded MPC methods of Entangled-CMPC, PolyDot-CMPC and AGE-CMPC is expressed as
\begin{align} \label{eq:polydot_storage}
    \sigma
    = (2N
    +z+1)\frac{m^2}{t^2}+\frac{2m^2}{st}+t^2, 
\end{align} where $m$ is the number of rows/columns of matrices $A$ and $B$, $s$ and $t$ are the number of row-wise and column-wise partitions, respectively, $z$ is the number of colluding workers, and $N$ is the required number of workers by each method.
\end{corollary}
{\em Proof:} 
Based on Phase 1 of coded MPC (Entangled-CMPC,
PolyDot-CMPC and AGE-CMPC), each worker $W_n$ receives $F_A(\alpha_n)$ and $F_B(\alpha_n)$ each with the size of $\frac{m}{t} \times \frac{m}{s}$ from the sources. This requires storing $\frac{2m^2}{st}$ scalar parameters.

In Phase 2, each worker $W_n$ stores $H(\alpha_n)$ with the size of $\frac{m}{t} \times \frac{m}{t}$ computed by multiplying $F_A(\alpha_n)$ with $F_B(\alpha_n)$. This requires storing $\frac{m^2}{t^2}$ scalar parameters.

Next, each worker $W_n$ creates the polynomial $G_n(x)$ to calculate different points of it. For this purpose, $W_n$ needs to store the coefficients of this polynomial. According to {(21)}, the random variables $r_n^{i,l},\; i,l \in \Omega_0^{t-1}$, with the total number of $t^2$ scalar parameters are stored. In addition, the random matrices $R^{(n)}_w \in \mathbb{F}^{\frac{m}{t} \times \frac{m}{t}}$ for $w \in \Omega_0^{z-1}$ are stored. In total, this requires storing $t^2+z\frac{m^2}{t^2}$ scalar parameters.

After creating $G_n(x)$, worker $W_n$ needs to compute it at points $\alpha_{n'},\; n' \in \Omega_1^N$, where $G_n(\alpha_{n'}),\; n' \in \Omega_1^N,\; n'\neq n$, will be sent to the other workers and $G_n(\alpha_n)$ is stored for the calculation of $I(\alpha_n)$ in {(22)}. 
Also, worker $W_n$ receives $G_{n'}(\alpha_n)$ from the other workers, which will be stored in its storage. As $G_{n}(x) \in \mathbb{F}^{\frac{m}{t} \times \frac{m}{t}}$, in total, this step requires storing $(2N-1)\frac{m^2}{t^2}$ scalar parameters.

Finally, worker $W_n$ needs to store $I(\alpha_n)$. As $I(\alpha_n) \in \mathbb{F}^{\frac{m}{t} \times \frac{m}{t}}$, this requires storing $\frac{m^2}{t^2}$ scalar parameters. 

By summing up the number of scalar parameters required to be stored by each worker $W_n$, the storage overhead of coded MPC becomes $(2N+z+1) \frac{m^2}{t^2}+\frac{2m^2}{st}+t^2$.
This concludes the proof. \hfill $\Box$
\vspace{-10pt}
\subsection{Communication Overhead} 
We define the communication overhead as the total number of scalar parameters that are exchanged among all workers in Phase 2. Note that there are other data transmissions; from sources to workers in Phase 1, and from workers to the master in Phase 3. We do not include these communications in the communication overhead calculation as they are negligible as compared to the data exchange among workers in Phase 2.
\begin{corollary}
\label{cor:communication_worker_worker_all_methods-polydot}
Communication overhead to compute $Y=A^TB$, using coded MPC methods of Entangled-CMPC, PolyDot-CMPC, and AGE-CMPC is expressed as
\begin{align} \label{eq:polydot_comm}
    \zeta
    = N
    (N
    -1)\frac{m^2}{t^2},
\end{align} where $m$ is the number of rows/columns of matrices $A$ and $B$, $t$ is the number of column-wise partitions, and $N$ is the required number of workers by each method.
\end{corollary}
{\em Proof:} In Phase 2 of coded MPC, each worker $W_n, n \in \Nset$ sends $G_n(\alpha_{n'})$ to worker $W_{n'},\; n' \in \Omega_1^N,\; n'\neq n$.  
As $G_n(\alpha_{n'}) \in \mathbb{F}^{\frac{m}{t} \times \frac{m}{t}}$, the communication overhead among workers is equal to $N(N-1)\frac{m^2}{t^2}$. This concludes the proof. \hfill $\Box$ 
\vspace{-10pt}
\subsection{Privacy Analysis}
\begin{theorem}\label{thm:privacy-AGE-Poly}
Our proposed PolyDot-CMPC and AGE-CMPC algorithms satisfy the privacy constraints (\ref{privacy-for-each-w-1}) and (\ref{privacy-for-master}) stated in Section \ref{sec:system}.
\end{theorem}
{\em Proof:} In order to prove this theorem we use Corollary 6, Lemma 7, and Corollary 8 in the proof of Theorem 3 in \cite{PolynomCMPC}, as well as the following lemma which is a generalized version of Lemma 7 in \cite{PolynomCMPC} in terms of the degree of polynomials $\mathbf{U}_i(x)$ and $\mathbf{R}_i(x)$ for $i \in \Omega_1^m$ with a change of notation for the sake of consistency with the remaining parts of this paper.
\begin{lemma}\label{lem:privacyproofs}
Let us consider $m$ polynomials $\mathbf{U}_1(x), \mathbf{U}_2(x), \dots, \mathbf{U}_m(x)$ of arbitrary degree $n$, where their coefficients are chosen from an arbitrary joint distribution in $\mathbb{F}^{\mu \times \nu}$. Let $\mathcal{A}$ denotes the order set of those coefficients. Consider the polynomials 
\begin{align}
&\mathbf{T}_1(x) = \mathbf{U}_1(x)+\mathbf{R}_1(x), \nonumber \\
&\mathbf{T}_2(x) = \mathbf{U}_2(x)+\mathbf{R}_2(x), \nonumber \\
\vdots \nonumber \\
&\mathbf{T}_m(x) = \mathbf{U}_m(x)+\mathbf{R}_m(x),
\end{align}
where for $i \in \Omega_1^m$, $\mathbf{R}_i(x): \mathbb{F} \rightarrow \mathbb{F}^{\mu \times \nu}$ is a polynomial with $z$ distinct terms, where the $z$ coefficients are chosen independently and uniformly at random from $\mathbb{F}^{\mu \times \nu}$. Then, $I(\mathcal{A};\mathbf{\widetilde{T}})=0$, where $\mathbf{\widetilde{T}} $ is defined as 
\begin{align}\label{eq:T-tilde}
\mathbf{\widetilde{T}} = \left[ {\begin{array}{ccc}
   \mathbf{\widetilde{T}}_1(\beta_1)&
   \ldots & \mathbf{\widetilde{T}}_1(\beta_z)\\   
   \vdots&\ddots&\vdots \\ \mathbf{\widetilde{T}}_m(\beta_1) &
   \ldots&\mathbf{\widetilde{T}}_m(\beta_z)
  \end{array} } \right],
  \end{align}
  for some arbitrary values $\beta_1, \beta_2, \dots, \beta_z \in \mathbb{F}$.
\end{lemma}
{\em Proof:} The proof is similar to the proof of Lemma 7 in \cite{PolynomCMPC}, since the proof is valid for any degree of polynomials $\mathbf{U}_i(x)$ and $\mathbf{R}_i(x)$, as far as $\mathbf{R}_i(x)$ contains $z$ distinct terms. Below is the detailed proof. First, let us define
\begin{align}
\mathbf{\widetilde{U}} = \left[ {\begin{array}{ccc}
   \mathbf{\widetilde{U}}_1(\beta_1)&
   \ldots & \mathbf{\widetilde{U}}_1(\beta_z)\\   
   \vdots&\ddots&\vdots \\ \mathbf{\widetilde{U}}_m(\beta_1) &
   \ldots&\mathbf{\widetilde{U}}_m(\beta_z)
  \end{array} } \right],
  \end{align}
  \begin{align}
  \mathbf{\widetilde{R}} = \left[ {\begin{array}{ccc}
   \mathbf{\widetilde{R}}_1(\beta_1)&
   \ldots & \mathbf{\widetilde{R}}_1(\beta_z)\\   
   \vdots&\ddots&\vdots \\ \mathbf{\widetilde{R}}_m(\beta_1) &
   \ldots&\mathbf{\widetilde{R}}_m(\beta_z)
  \end{array} } \right].
  \end{align}
  For any $\mathbf{T,U,} \in \mathbb{F}^{m\mu \times z\nu}$ we have
  \begin{align}\nonumber
&\text{Pr}(\mathbf{\widetilde{U}}=\mathbf{U}|\mathbf{\widetilde{T}}=\mathbf{T}) \\ \nonumber
& \buildrel \rm (a) \over = \frac{\text{Pr}(\mathbf{\widetilde{T}}=\mathbf{T}|\mathbf{\widetilde{U}}=\mathbf{U})\text{Pr}(\mathbf{\widetilde{U}}=\mathbf{U})}{\sum_{\mathbf{U}_j \in \mathbb{F}^{m\mu \times z\nu}}\text{Pr}(\mathbf{\widetilde{T}}=\mathbf{T}|\mathbf{\widetilde{U}}=\mathbf{U}_j)\text{Pr}(\mathbf{\widetilde{U}}=\mathbf{U}_j)} \nonumber \\
& = \frac{\text{Pr}(\mathbf{\widetilde{R}}=\mathbf{T}-\mathbf{U}|\mathbf{\widetilde{U}}=\mathbf{U})\text{Pr}(\mathbf{\widetilde{U}}=\mathbf{U})}{\sum_{\mathbf{U}_j \in \mathbb{F}^{m\mu \times z\nu}}\text{Pr}(\mathbf{\widetilde{R}}=\mathbf{T}-\mathbf{U}_j|\mathbf{\widetilde{U}}=\mathbf{U}_j)\text{Pr}(\mathbf{\widetilde{U}}=\mathbf{U}_j)} \nonumber\\
& \buildrel \rm (b) \over = \frac{\text{Pr}(\mathbf{\widetilde{R}}=\mathbf{T}-\mathbf{U})\text{Pr}(\mathbf{\widetilde{U}}=\mathbf{U})}{\sum_{\mathbf{U}_j \in \mathbb{F}^{m\mu \times z\nu}}\text{Pr}(\mathbf{\widetilde{R}}=\mathbf{T}-\mathbf{U})\text{Pr}(\mathbf{\widetilde{U}}=\mathbf{U}_j)} \nonumber \\
& = \frac{\text{Pr}(\mathbf{\widetilde{U}}=\mathbf{U})}{\sum_{\mathbf{U}_j \in \mathbb{F}^{m\mu \times z\nu}}\text{Pr}(\mathbf{\widetilde{U}}=\mathbf{U}_j)} 
 = \text{Pr}(\mathbf{\widetilde{U}}=\mathbf{U}),
  \end{align}
where (a) comes from Bayesian Rule, and (b) is resulted from Corollary 6 in \cite{PolynomCMPC}, that says each row of matrix $\mathbf{\widetilde{R}}$ has a uniform distribution over $\mathbb{F}^{\mu \times z\nu}$, so $\mathbf{\widetilde{R}}$ has a uniform distribution over $\mathbb{F}^{m\mu \times z\nu}$, and consequently $\text{Pr}(\mathbf{\widetilde{R}}=\mathbf{T}-\mathbf{U})=\text{Pr}(\mathbf{\widetilde{R}}=\mathbf{T}-\mathbf{U}_j)=\frac{1}{|\mathbb{F}|^{m\mu z\nu}}$. Thus, we have $\tilde{H}(\mathbf{\widetilde{U}}|\mathbf{\widetilde{T}})=\tilde{H}(\mathbf{\widetilde{U}})$, and $I(\mathbf{\widetilde{U}};\mathbf{\widetilde{T}})=0$. Moreover, from the definition of $\mathcal{A}$, $\mathcal{A} \rightarrow \mathbf{\widetilde{U}} \rightarrow \mathbf{\widetilde{T}}$ is a Markov chain, and as a result according to data processing inequality we have $I(\mathcal{A};\mathbf{\widetilde{T}}) \leq I(\mathbf{\widetilde{U}};\mathbf{\widetilde{T}})=0$. This concludes the proof of Lemma \ref{lem:privacyproofs}. 
\hfill $\Box$

{Therefore, from Lemma \ref{lem:privacyproofs} we conclude that Corollary 8 in \cite{PolynomCMPC} is valid for any polynomial with arbitrary powers of the coded and secret terms. Thus, polynomial $\mathbf{F}_{\mathbf{Y},b,t,k}(x)$, in Corollary 8 in \cite{PolynomCMPC}, can be substituted with  $F_A(x)$ and $F_B(x)$ in (\ref{eq:FA1andFA2PolyDotCMPC}), (\ref{eq:FBPolyDotCMPC}) in PolyDot-CMPC or $F_A(x)$ and $F_B(x)$ constructed using (\ref{eq:generalEntangled}), (\ref{eq:S-A}), and (\ref{eq:S-B}) in AGE-CMPC. The rest of the proof of Theorem \ref{thm:privacy-AGE-Poly} follows directly from the proof of Theorem 3 in \cite{PolynomCMPC}.\hfill $\Box$
}
\vspace{-5pt}

\begin{figure}[t!]
\centering

\scalebox{.16}{ \includegraphics{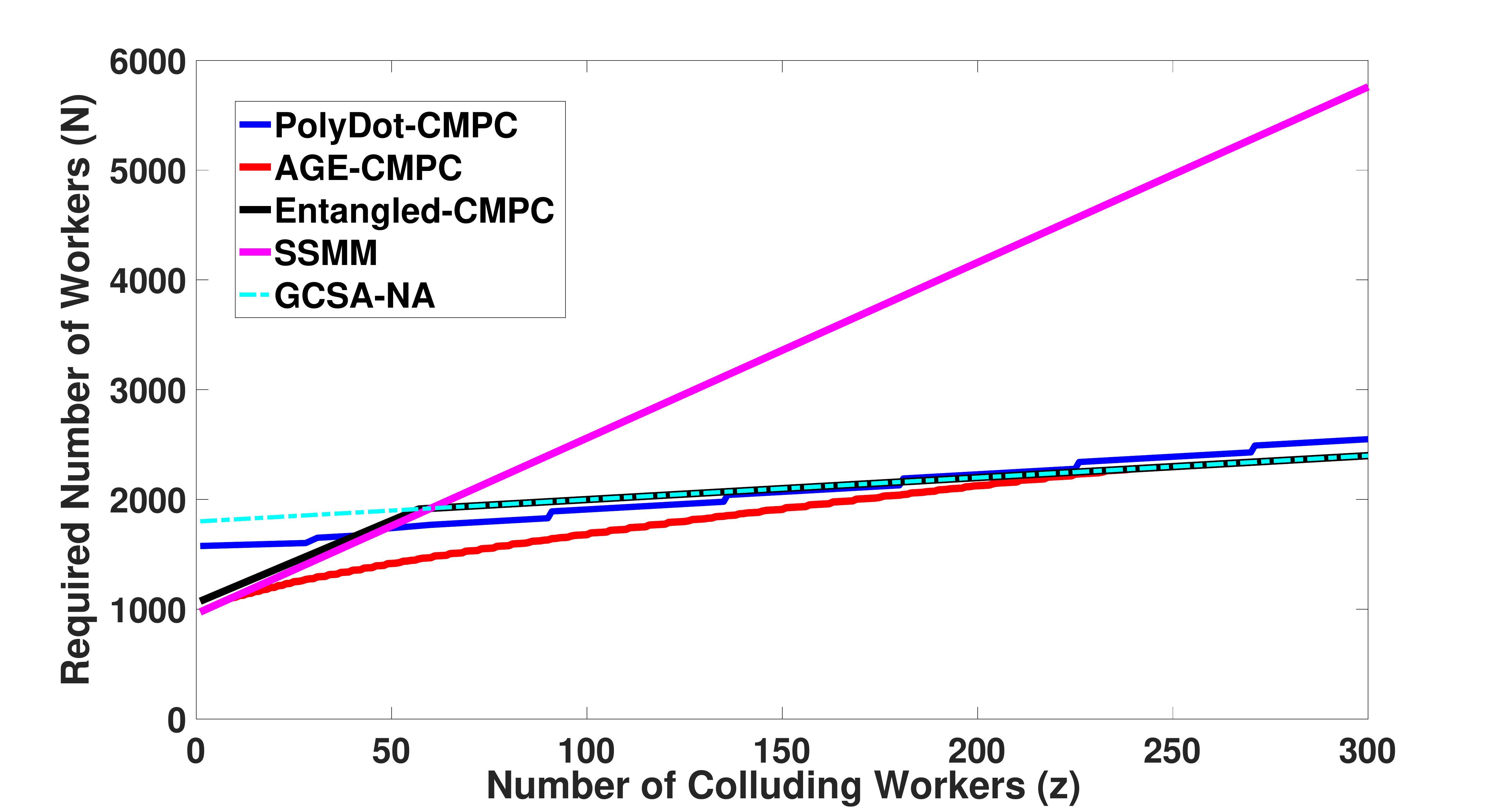}}
\vspace{-10pt}
\caption{Required number of workers versus number of colluding workers. The parameters are set to $s=4,\; t=15$ and $1 \leq z \leq 300$.}
\vspace{-5pt}
\label{fig:NvsZ-comp}
\vspace{-5pt}
\end{figure}

\begin{figure}[t!]
\centering
\scalebox{.16}{ \includegraphics{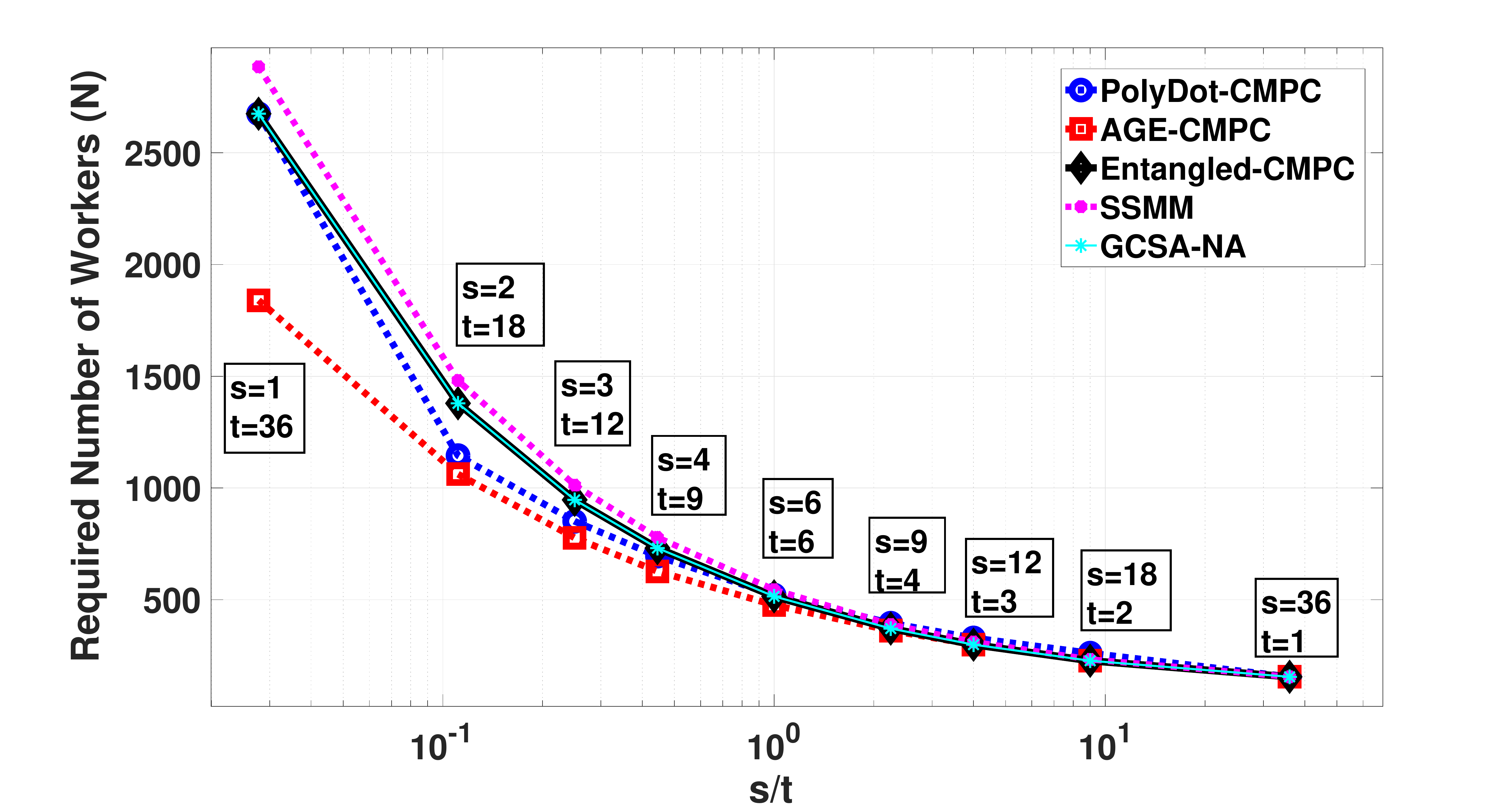}} 
\vspace{-5pt}
\caption{Required number of workers.}
\vspace{-1pt}
\label{fig:N-AGE}
\vspace{1pt}
\end{figure}

\vspace{-5pt}
\section{Performance Evaluation}\label{sec:simulation} 

We evaluated the performance of our algorithms, PolyDot-CMPC and AGE-CMPC, and compared them with the baselines, (i) Entangled-CMPC \cite{8613446}, (ii) SSMM \cite{Zhu2021ImprovedCF}, and (iii) GCSA-NA {for one matrix multiplication} \cite{9333639}. 


{Fig.~\ref{fig:NvsZ-comp} shows the number of workers required for computing $Y=A^TB$ versus the number of colluding workers, where the matrices $A$ and $B$ are divided into $s=4$ row-wise and $t=15$ column-wise partitions, and the number of colluding workers varies in the range of $1 \leq z \leq 300$. As seen, AGE-CMPC requires less number of workers than all other methods for any number of colluding workers, which confirms Lemma \ref{corol:AGE-Vs-Ent-SSMM-GC-Poly}. For small number of colluding workers, \ie $1 \leq z \leq 48$, SSMM \cite{Zhu2021ImprovedCF} is the second best choice. PolyDot-CMPC performs better than all the baselines, excluding AGE-CMPC,  when $49 \leq z \leq 180$. On the other hand, GCSA-NA \cite{9333639} and Entangled-CMPC \cite{8613446} have similar performance and perform better than SSMM and PolyDot-CMPC when $181 \leq z \leq 300$. These results confirm Lemmas \ref{lemma: regions where N_polydot<N_entangled}, \ref{lemma: regions where N_polydot<N_ssmm}, and \ref{lemma: regions where N_polydot<N_gcsana} as PolyDot-CMPC performs better than the baselines, excluding AGE-CMPC, for a range of colluding workers.} 

\begin{figure*}[ht]
\centering
\subfigure[Computation]{ \scalebox{0.6}{\includegraphics[width=0.52\textwidth]{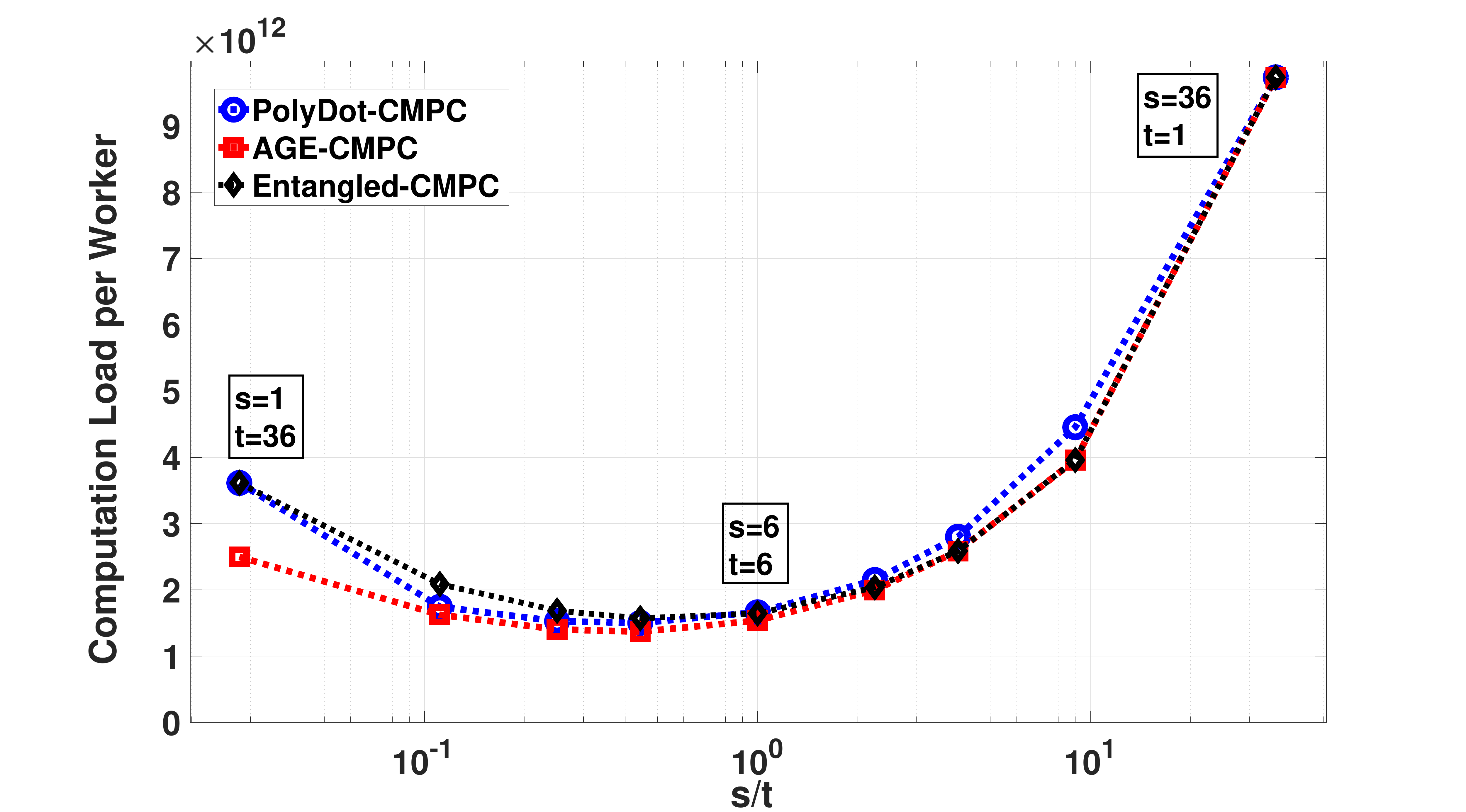}} } 
\subfigure[Storage]{ \scalebox{0.6}{\includegraphics[width=0.52\textwidth]{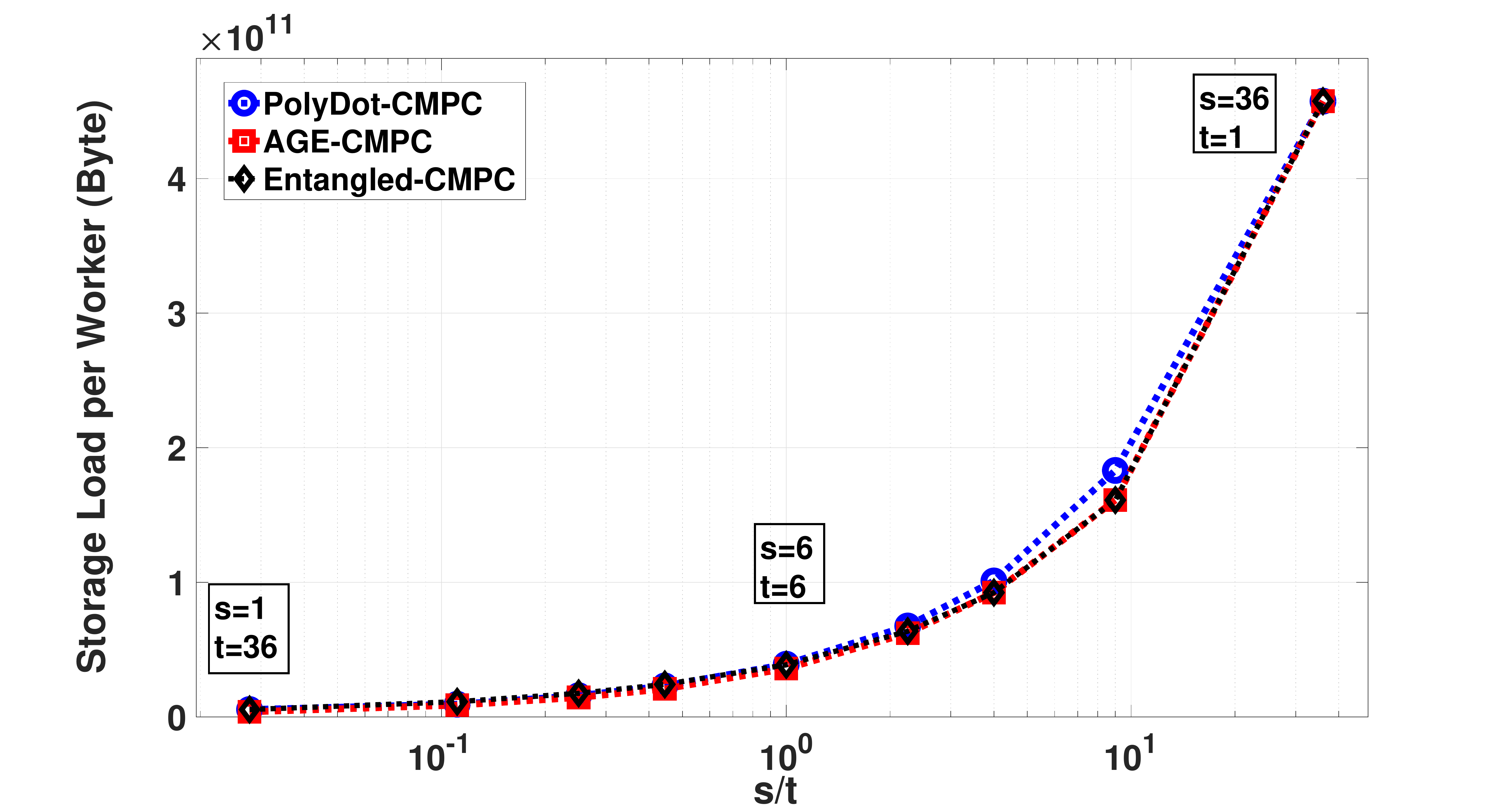}} } 
%
%
\subfigure[Communication]{ \scalebox{0.6}{\includegraphics[width=0.52\textwidth]{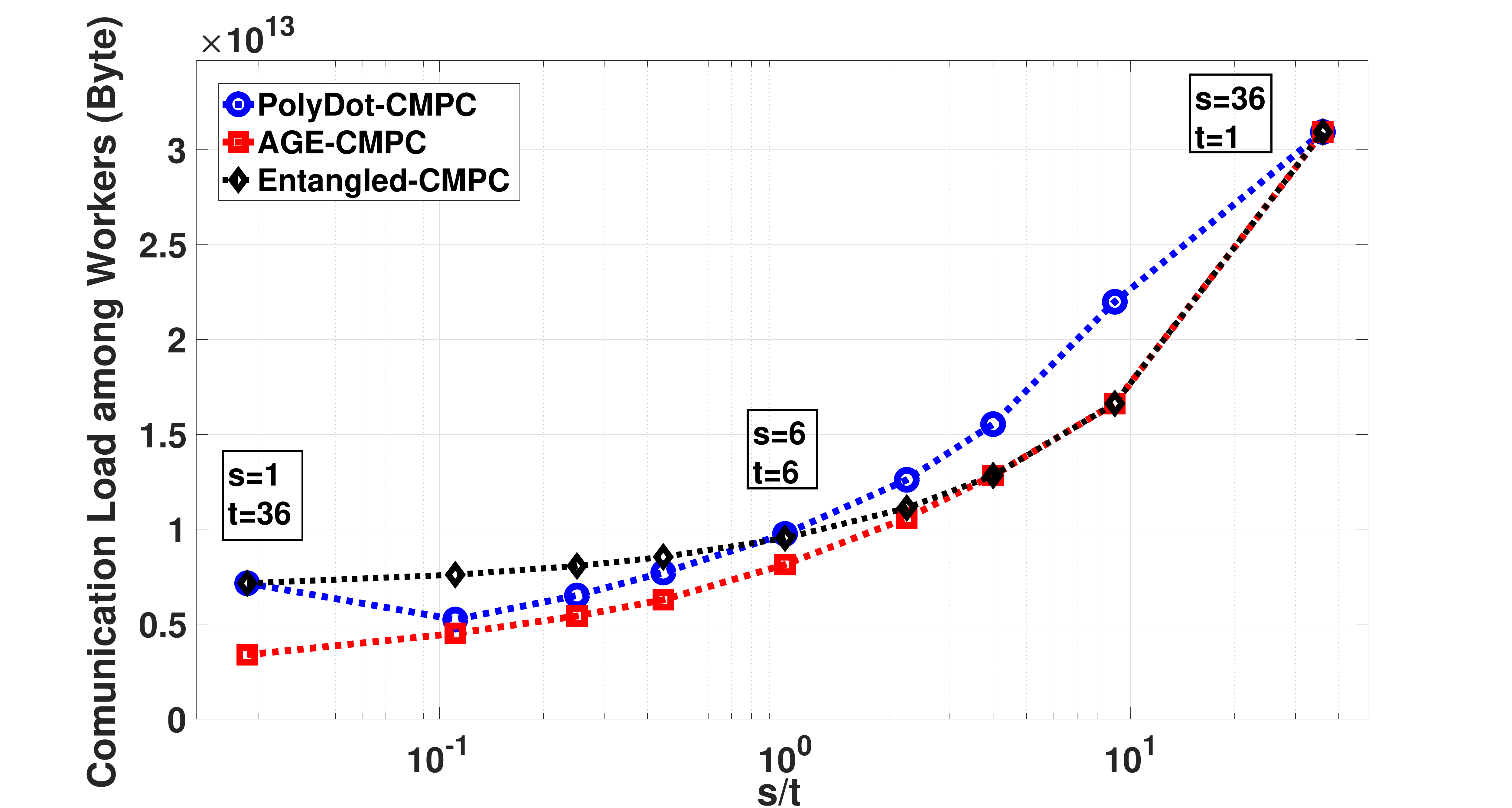}} } 
\caption{(a) Computation,  (b) storage , and (c) communication loads.}
\vspace{-15pt}
\label{fig:storage-comm-comp}
\end{figure*}

In Fig.~\ref{fig:N-AGE} and Fig.~\ref{fig:storage-comm-comp} the system parameters are considered as follows: the size of each matrix $A$ and $B$ is $m\times m=36000 \times 36000$, the number of colluding workers is $z=42$, and the number of partitions of matrices $A$ and $B$ is $st=36$.

Fig.~\ref{fig:N-AGE} shows the required number of workers needed to compute the multiplication of $Y=A^TB$ versus $s/t$, the ratio of number of row partitions over the number of column partitions. As seen, the required number of workers by AGE-CMPC is always less than or equal to the other baselines. 
{Moreover, 
PolyDot-CMPC requires less number of workers than the other baseline methods for $(s,t) \in \{(2,18),(3,12),(4,9)\}$, since in these scenarios we have $42=z>ts=36$, and thus $p=\min\{\floor{\frac{z-1}{ts-t}},t-1\}$ is equal to $2$, $1$ and $1$, for $t=18$, $12$, $9$, respectively
. Thus, $p<\frac{t-1}{s}$ and condition 1 in Lemmas \ref{lemma: regions where N_polydot<N_entangled}, \ref{lemma: regions where N_polydot<N_ssmm}, and \ref{lemma: regions where N_polydot<N_gcsana} are satisfied. For $(s,t) \in \{(1,36),(6,6),(9,4),(12,3),(18,2),(36,1)\}$, PolyDot-CMPC requires equal number of workers or larger compared with the other baselines. These observations are aligned with Lemmas \ref{lemma: regions where N_polydot<N_entangled}, \ref{lemma: regions where N_polydot<N_ssmm}, and \ref{lemma: regions where N_polydot<N_gcsana}}.
%
%

Fig.~\ref{fig:storage-comm-comp}(a) shows the computation load per worker {defined in Section \ref{sec:recoverythreshold,comp,com,S analysis-polydot}.A} versus $s/t$. {AGE-CMPC reduces the computation load per worker. The reason is that based on (\ref{eq:polydot_computation_cost}), computation load grows linearly with $N$, the required number of workers. Since the required number of workers by AGE-CMPC is less than the other methods, the computation load required by AGE-CMPC is also less than the other methods.} 
Also, as seen in Fig.~\ref{fig:storage-comm-comp}(a), computation load per worker does not have a monotonic behavior by increasing $s/t$. {The reason is that according to the equation (\ref{eq:polydot_computation_cost}), computation load per worker has a direct relationship with the required number of workers, which as shown in Fig.~\ref{fig:N-AGE} decreases by increasing $s/t$. On the other hand, it has an inverse relationship with $t$ (for fixed $m$ and $st$, where $st=36$). Therefore, these two parameters have apposing effects on the computation load per worker. In other words, as seen in the figure, for fixed $st=36$, if we decrease $t$ from $36$ to $9$, the effect of decreasing $N$ dominates the effect of decreasing $t$ and thus the computation load per worker will decrease. If we decrease $t$ from $9$ to $1$, the effect of decreasing $t$ dominates the effect of decreasing $N$ and thus the computation load per worker will increase. 
}

Fig.~\ref{fig:storage-comm-comp}(b) shows the storage load per worker, {defined in Section \ref{sec:recoverythreshold,comp,com,S analysis-polydot}.B}, where the size of each stored scalar is 1 Byte, versus $s/t$. AGE-CMPC reduces the storage load per worker as compared to baselines. 
The reason is that {based on (\ref{eq:polydot_storage})} there is a direct relationship between storage load per worker in CMPC setup and the required number of workers. 
Therefore, the smaller number of workers required by AGE-CMPC results in the smaller storage load per worker as compared to PolyDot-CMPC and Entangled-CMPC. 

Fig.~\ref{fig:storage-comm-comp}(c) shows the communication load, {defined in Section \ref{sec:recoverythreshold,comp,com,S analysis-polydot}.C}, versus $s/t$. 
We assume that each scalar that is transmitted among workers is 1 Byte. 
{Based on (\ref{eq:polydot_comm})} the communication load among workers has a direct relationship with the required number of workers, \ie larger number of workers results in larger communication load among workers. Therefore, the communication load among workers of AGE-CMPC is less than or equal to the other methods. 
\vspace{-5pt}
\section{Conclusion} \label{sec:conc}

We have investigated coded privacy-preserving computation using Shamir's secret sharing. We have proposed a new coded privacy-preserving computation mechanism; PolyDot-CMPC, which is designed by employing PolyDot codes and using ``garbage terms'' that naturally arise when polynomials are constructed in the design of PolyDot codes. Motivated by this observation, we have designed a novel coded computation method; AGE codes that is customized for coded privacy-preserving computations to create the optimum number of ``garbage terms''. We also designed coded privacy-preserving computation mechanisms; AGE coded MPC (AGE-CMPC) 
by employing AGE codes. 
Also, we have analyzed AGE-CMPC and PloyDot-CMPC in terms of the required number of workers as well as its computation, storage, and communication overhead. We showed that PolyDot-CMPC outperforms the other state of the art methods for a range of colluding workers. We also showed that AGE-CMPC provides significant improvement {and always performs better than the other methods}. 

\vspace{-5pt}
\begingroup
\def\bibfont{\small}
\bibliographystyle{IEEEtran}
\bibliography{refs}

\begin{thebibliography}{10}
\providecommand{\url}[1]{#1}
\csname url@samestyle\endcsname
\providecommand{\newblock}{\relax}
\providecommand{\bibinfo}[2]{#2}
\providecommand{\BIBentrySTDinterwordspacing}{\spaceskip=0pt\relax}
\providecommand{\BIBentryALTinterwordstretchfactor}{4}
\providecommand{\BIBentryALTinterwordspacing}{\spaceskip=\fontdimen2\font plus
\BIBentryALTinterwordstretchfactor\fontdimen3\font minus
  \fontdimen4\font\relax}
\providecommand{\BIBforeignlanguage}[2]{{%
\expandafter\ifx\csname l@#1\endcsname\relax
\typeout{** WARNING: IEEEtran.bst: No hyphenation pattern has been}%
\typeout{** loaded for the language `#1'. Using the pattern for}%
\typeout{** the default language instead.}%
\else
\language=\csname l@#1\endcsname
\fi
#2}}
\providecommand{\BIBdecl}{\relax}
\BIBdecl

\bibitem{AGE-CMPC}
E.~Vedadi, Y.~Keshtkarjahromi, and H.~Seferoglu, ``Adaptive gap entangled
  polynomial coding for multi-party computation at the edge,'' in \emph{2022
  IEEE International Symposium on Information Theory (ISIT)}, 2022, pp.
  1217--1222.

\bibitem{PolyDot-CMPC}
------, ``Polydot coded privacy preserving multi-party computation at the
  edge,'' in \emph{2022 IEEE 23rd International Workshop on Signal Processing
  Advances in Wireless Communication (SPAWC)}, 2022, pp. 1--5.

\bibitem{IDCReport}
R.~Swearingen, ``Idc report 2020: Iot growth demands rethink of long-term
  storage strategies, says idc,'' 2020.

\bibitem{DemocratizingNetworkEdge}
\BIBentryALTinterwordspacing
L.~Peterson, T.~Anderson, S.~Katti, N.~McKeown, G.~Parulkar, J.~Rexford,
  M.~Satyanarayanan, O.~Sunay, and A.~Vahdat, ``Democratizing the network
  edge,'' \emph{SIGCOMM Comput. Commun. Rev.}, vol.~49, no.~2, pp. 31--36, May
  2019. [Online]. Available: \url{http://doi.acm.org/10.1145/3336937.3336942}
\BIBentrySTDinterwordspacing

\bibitem{EdgeComputingVideo}
\BIBentryALTinterwordspacing
P.~Levine and A.~Horowitz, ``Return to the edge and the end of cloud
  computing,'' 2017. [Online]. Available:
  \url{https://www.youtube.com/watch?v=-QRXQTSZxdQ}
\BIBentrySTDinterwordspacing

\bibitem{EdgeEatCloud}
G.~M. Research, ``The edge will eat the cloud,'' 2017.

\bibitem{scalableMPC}
J.~Saia and M.~Zamani, ``Recent results in scalable multi-party computation,''
  in \emph{SOFSEM 2015: Theory and Practice of Computer Science}, G.~F.
  Italiano, T.~Margaria-Steffen, J.~Pokorn{\'y}, J.-J. Quisquater, and
  R.~Wattenhofer, Eds.\hskip 1em plus 0.5em minus 0.4em\relax Berlin,
  Heidelberg: Springer Berlin Heidelberg, 2015, pp. 24--44.

\bibitem{Yao}
A.~C.-C. Yao, ``How to generate and exchange secrets,'' in \emph{27th Annual
  Symposium on Foundations of Computer Science (sfcs 1986)}, 1986, pp.
  162--167.

\bibitem{GMW}
S.~M. O.~Goldreich and A.~Wigderson, ``How to play any mental game,'' in
  \emph{Proc. of the 19th STOC}, 1987, pp. 218--229.

\bibitem{BGW}
M.~Ben-Or, S.~Goldwasser, and A.~Wigderson, ``Completeness theorems for
  non-cryptographic fault-tolerant distributed computation,'' in
  \emph{Providing Sound Foundations for Cryptography: On the Work of Shafi
  Goldwasser and Silvio Micali}, 2019, pp. 351--371.

\bibitem{ShamirSS}
A.~Shamir, ``How to share a secret,'' \emph{Communications of the ACM},
  vol.~22, no.~11, pp. 612--613, 1979.

\bibitem{10.1007/3-540-48405-1_4}
U.~Maurer, ``Information-theoretic cryptography,'' in \emph{Advances in
  Cryptology --- CRYPTO' 99}, M.~Wiener, Ed.\hskip 1em plus 0.5em minus
  0.4em\relax Berlin, Heidelberg: Springer Berlin Heidelberg, 1999, pp. 47--65.

\bibitem{SpeedUp-journal}
K.~{Lee}, M.~{Lam}, R.~{Pedarsani}, D.~{Papailiopoulos}, and K.~{Ramchandran},
  ``Speeding up distributed machine learning using codes,'' \emph{IEEE
  Transactions on Information Theory}, vol.~64, no.~3, March 2018.

\bibitem{Tradeoff-journal}
S.~{Li}, M.~A. {Maddah-Ali}, Q.~{Yu}, and A.~S. {Avestimehr}, ``A fundamental
  tradeoff between computation and communication in distributed computing,''
  \emph{IEEE Transactions on Information Theory}, vol.~64, no.~1, pp. 109--128,
  Jan 2018.

\bibitem{8613446}
H.~A. Nodehi, S.~R.~H. Najarkolaei, and M.~A. Maddah-Ali, ``Entangled
  polynomial coding in limited-sharing multi-party computation,'' in \emph{2018
  IEEE Information Theory Workshop (ITW)}, 2018, pp. 1--5.

\bibitem{Zhu2021ImprovedCF}
J.~Zhu, Q.~Yan, and X.~Tang, ``Improved constructions for secure multi-party
  batch matrix multiplication,'' \emph{IEEE Transactions on Communications},
  vol.~69, pp. 7673--7690, 2021.

\bibitem{9333639}
Z.~Chen, Z.~Jia, Z.~Wang, and S.~A. Jafar, ``Gcsa codes with noise alignment
  for secure coded multi-party batch matrix multiplication,'' \emph{IEEE
  Journal on Selected Areas in Information Theory}, vol.~2, no.~1, pp.
  306--316, 2021.

\bibitem{SpeedingUp}
K.~Lee, M.~Lam, R.~Pedarsani, D.~Papailiopoulos, and K.~Ramchandran, ``Speeding
  up distributed machine learning using codes,'' \emph{IEEE Transactions on
  Information Theory}, vol.~64, no.~3, pp. 1514--1529, 2018.

\bibitem{burges2005learning}
C.~Burges, T.~Shaked, E.~Renshaw, A.~Lazier, M.~Deeds, N.~Hamilton, and
  G.~Hullender, ``Learning to rank using gradient descent,'' in
  \emph{Proceedings of the 22nd international conference on Machine learning},
  2005, pp. 89--96.

\bibitem{zhang2004solving}
T.~Zhang, ``Solving large scale linear prediction problems using stochastic
  gradient descent algorithms,'' in \emph{Proceedings of the twenty-first
  international conference on Machine learning}, 2004, p. 116.

\bibitem{bottou2010large}
L.~Bottou, ``Large-scale machine learning with stochastic gradient descent,''
  in \emph{Proceedings of COMPSTAT'2010}.\hskip 1em plus 0.5em minus
  0.4em\relax Springer, 2010, pp. 177--186.

\bibitem{YuFundamentalLimits2018}
Q.~Yu, M.~A. Maddah-Ali, and A.~S. Avestimehr, ``Straggler mitigation in
  distributed matrix multiplication: Fundamental limits and optimal coding,''
  \emph{IEEE Transactions on Information Theory}, vol.~66, no.~3, pp.
  1920--1933, 2020.

\bibitem{FerdinandAnytime}
N.~S. Ferdinand and S.~C. Draper, ``Anytime coding for distributed
  computation,'' in \emph{2016 54th Annual Allerton Conference on
  Communication, Control, and Computing (Allerton)}, 2016, pp. 954--960.

\bibitem{YuPolynomial2017}
Q.~Yu, M.~A. Maddah-Ali, and S.~Avestimehr, ``Polynomial codes: an optimal
  design for high-dimensional coded matrix multiplication,'' in \emph{NIPS},
  2017, pp. 4406--4416.

\bibitem{LeeHighDim2017}
K.~Lee, C.~Suh, and K.~Ramchandran, ``High-dimensional coded matrix
  multiplication,'' in \emph{2017 IEEE International Symposium on Information
  Theory (ISIT)}, 2017, pp. 2418--2422.

\bibitem{PolyDotMatDot}
M.~Fahim, H.~Jeong, F.~Haddadpour, S.~Dutta, V.~Cadambe, and P.~Grover, ``On
  the optimal recovery threshold of coded matrix multiplication,'' in
  \emph{2017 55th Annual Allerton Conference on Communication, Control, and
  Computing (Allerton)}.\hskip 1em plus 0.5em minus 0.4em\relax IEEE, 2017, pp.
  1264--1270.

\bibitem{ShortDot}
S.~Dutta, V.~Cadambe, and P.~Grover, ``“short-dot”: Computing large linear
  transforms distributedly using coded short dot products,'' \emph{IEEE
  Transactions on Information Theory}, vol.~65, no.~10, pp. 6171--6193, 2019.

\bibitem{DuttaCodedConvolution2017}
------, ``Coded convolution for parallel and distributed computing within a
  deadline,'' in \emph{2017 IEEE International Symposium on Information Theory
  (ISIT)}, 2017, pp. 2403--2407.

\bibitem{TandonGradientCoding2017}
\BIBentryALTinterwordspacing
R.~Tandon, Q.~Lei, A.~G. Dimakis, and N.~Karampatziakis, ``Gradient coding:
  Avoiding stragglers in distributed learning,'' in \emph{Proceedings of the
  34th International Conference on Machine Learning}, ser. Proceedings of
  Machine Learning Research, D.~Precup and Y.~W. Teh, Eds., vol.~70.\hskip 1em
  plus 0.5em minus 0.4em\relax PMLR, 06--11 Aug 2017, pp. 3368--3376. [Online].
  Available: \url{http://proceedings.mlr.press/v70/tandon17a.html}
\BIBentrySTDinterwordspacing

\bibitem{HalbawiImprovingDistGradient2018}
W.~Halbawi, N.~Azizan, F.~Salehi, and B.~Hassibi, ``Improving distributed
  gradient descent using reed-solomon codes,'' in \emph{2018 IEEE International
  Symposium on Information Theory (ISIT)}, 2018, pp. 2027--2031.

\bibitem{RavivGradientCyclic2018}
N.~Raviv, I.~Tamo, R.~Tandon, and A.~G. Dimakis, ``Gradient coding from cyclic
  mds codes and expander graphs,'' \emph{IEEE Transactions on Information
  Theory}, vol.~66, no.~12, pp. 7475--7489, 2020.

\bibitem{KarakusRedundancy2019}
\BIBentryALTinterwordspacing
C.~Karakus, Y.~Sun, S.~Diggavi, and W.~Yin, ``Redundancy techniques for
  straggler mitigation in distributed optimization and learning,''
  \emph{Journal of Machine Learning Research}, vol.~20, no.~72, pp. 1--47,
  2019. [Online]. Available: \url{http://jmlr.org/papers/v20/18-148.html}
\BIBentrySTDinterwordspacing

\bibitem{YuCodedFourier2017}
Q.~Yu, M.~A. Maddah-Ali, and A.~S. Avestimehr, ``Coded fourier transform,'' in
  \emph{2017 55th Annual Allerton Conference on Communication, Control, and
  Computing (Allerton)}, 2017, pp. 494--501.

\bibitem{YangComputeLinear2017}
Y.~Yang, P.~Grover, and S.~Kar, ``Computing linear transformations with
  unreliable components,'' \emph{IEEE Transactions on Information Theory},
  vol.~63, no.~6, pp. 3729--3756, 2017.

\bibitem{8382305}
H.~Yang and J.~Lee, ``Secure distributed computing with straggling servers
  using polynomial codes,'' \emph{IEEE Transactions on Information Forensics
  and Security}, vol.~14, no.~1, pp. 141--150, Jan 2019.

\bibitem{SecureCoded}
J.~Kakar, S.~Ebadifar, and A.~Sezgin, ``On the capacity and
  straggler-robustness of distributed secure matrix multiplication,''
  \emph{IEEE Access}, vol.~7, pp. 45\,783--45\,799, 2019.

\bibitem{GASP}
R.~G.~L. D’Oliveira, S.~El~Rouayheb, and D.~Karpuk, ``Gasp codes for secure
  distributed matrix multiplication,'' \emph{IEEE Transactions on Information
  Theory}, vol.~66, no.~7, pp. 4038--4050, 2020.

\bibitem{bitar_trans_PRAC}
R.~Bitar, Y.~Xing, Y.~Keshtkarjahromi, V.~Dasari, S.~El~Rouayheb, and
  H.~Seferoglu, ``Private and rateless adaptive coded matrix-vector
  multiplication,'' \emph{EURASIP Journal on Wireless Communications and
  Networking}, 2021.

\bibitem{BPR17}
R.~Bitar, P.~Parag, and S.~El~Rouayheb, ``Minimizing latency for secure
  distributed computing,'' in \emph{Information Theory (ISIT), 2017 IEEE
  International Symposium on}.\hskip 1em plus 0.5em minus 0.4em\relax IEEE,
  2017, pp. 2900--2904.

\bibitem{LCC1}
Q.~Yu, N.~Raviv, J.~So, and A.~S. Avestimehr, ``Lagrange coded computing:
  Optimal design for resiliency, security and privacy,'' \emph{arXiv preprint,
  arXiv:1806.00939}, 2018.

\bibitem{LCC2}
Q.~Yu, N.~Raviv, and A.~S. Avestimehr, ``Coding for private and secure
  multiparty computing,'' in \emph{2018 IEEE Information Theory Workshop
  (ITW)}, 2018, pp. 1--5.

\bibitem{PolynomCMPC}
H.~Akbari-Nodehi and M.~A. Maddah-Ali, ``Secure coded multi-party computation
  for massive matrix operations,'' \emph{IEEE Transactions on Information
  Theory}, vol.~67, no.~4, pp. 2379--2398, 2021.

\bibitem{Dutta2018AUC}
S.~Dutta, Z.~Bai, H.~Jeong, T.~M. Low, and P.~Grover, ``A unified coded deep
  neural network training strategy based on generalized polydot codes,''
  \emph{2018 IEEE International Symposium on Information Theory (ISIT)}, pp.
  1585--1589, 2018.

\end{thebibliography}
\endgroup

\vspace{-45pt}
\begin{IEEEbiography}
[{\includegraphics[width=1in
,clip
]{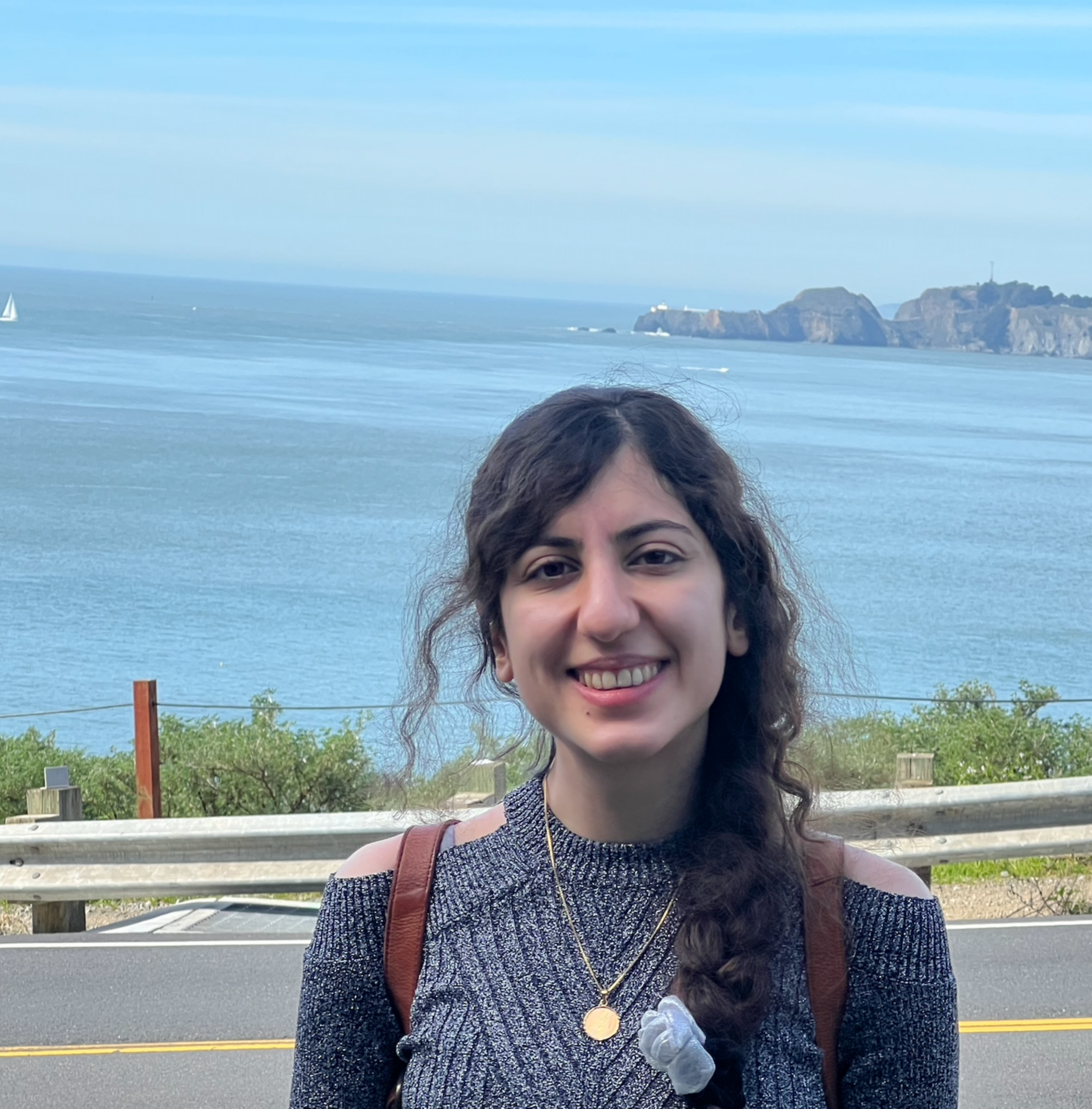}}]
{Elahe Vedadi} is a Ph.D. student in the Electrical and Computer Engineering Department of University of Illinois at Chicago under supervision of Prof. Hulya Seferoglu. She received her B.S. degree in Electrical Engineering from Sharif University of Technology, Iran, in 2018. She worked as a research intern at Google in spring 2023, and Seagate Technology in both summer 2021 and 2022.
\end{IEEEbiography}

\vspace{-50pt}
\begin{IEEEbiography}
[{\includegraphics[width=1in
]{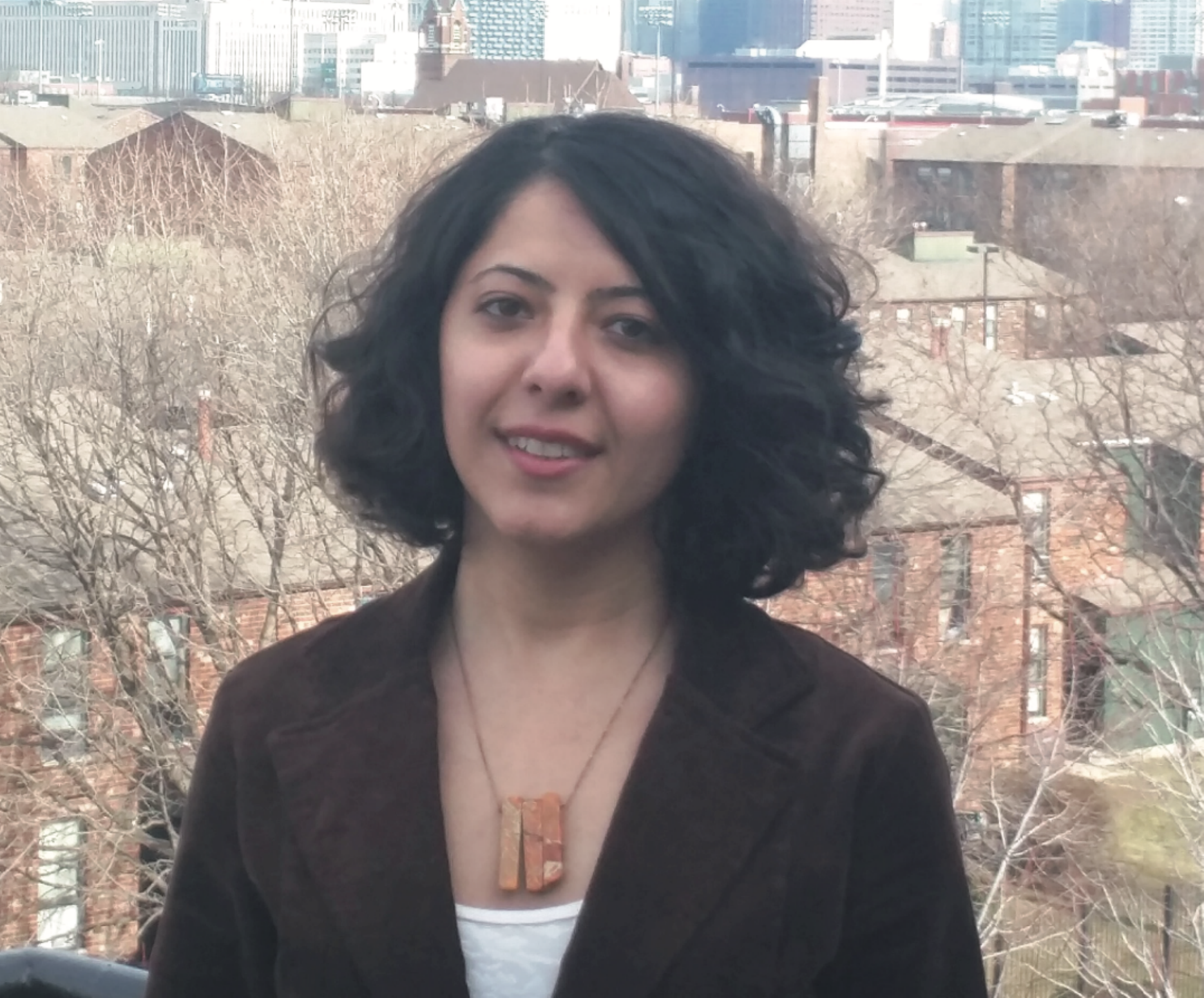}}]
{Yasaman Keshtkarjahromi} is an Engineering Manager at Seagate Technology, Research Group. Before joining Seagate, she was an ORAU Postdoc Fellow during 2018-2019. She received the B.S. degree in Electrical and Computer Engineering from Shiraz University, Iran, M.S. degree in Electrical and Computer Engineering from Tehran University, and Ph.D. degree in Electrical and Computer Engineering from the University of Illinois at Chicago. 
She worked as a summer intern at Huawei R\&D and Alcatel-Lucent Bell Labs in 2016, and 2015, respectively.
\end{IEEEbiography}
\vspace{-40pt}

\begin{IEEEbiography}[{\includegraphics[width=1in]{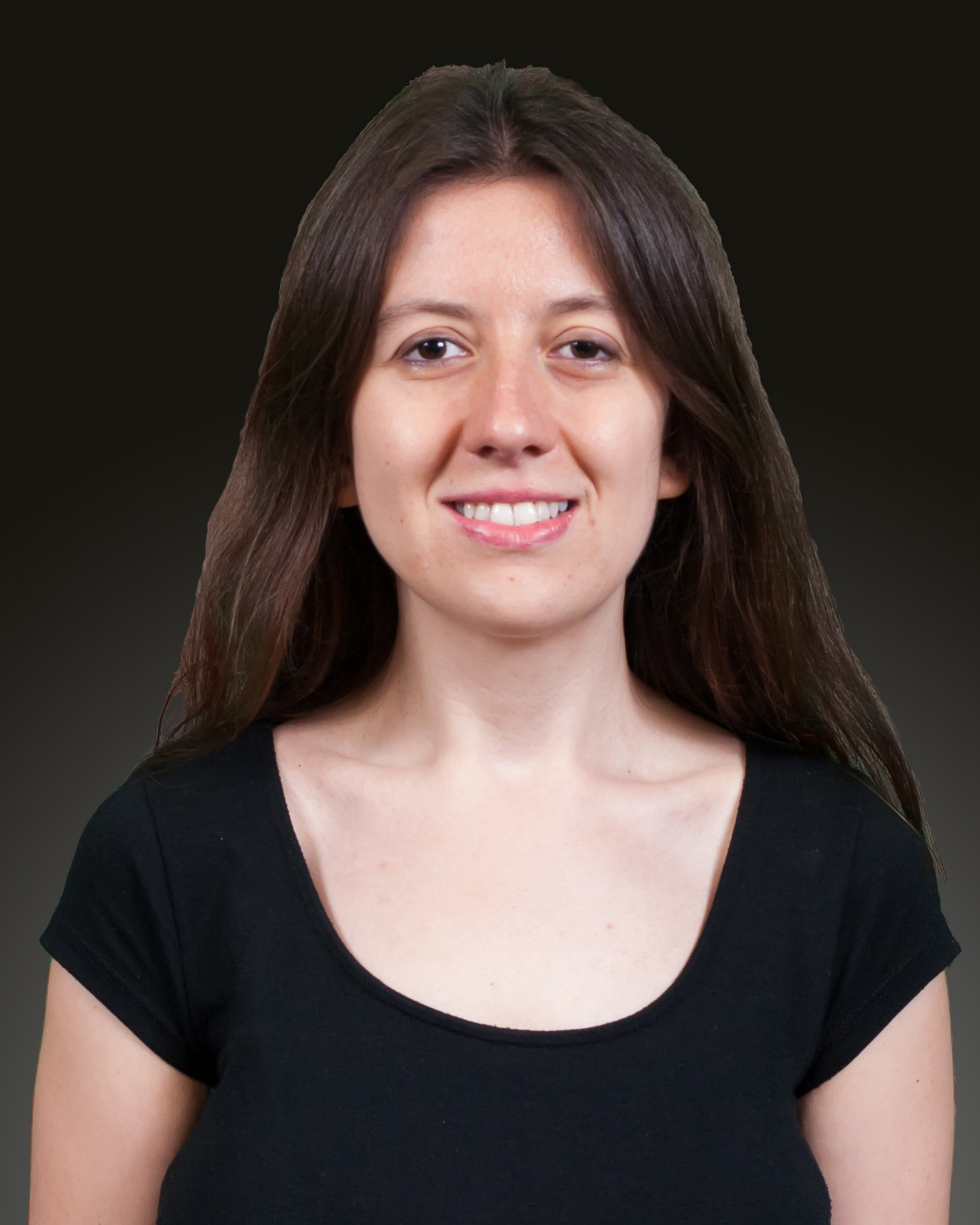}}]
{Hulya Seferoglu} 
is an Associate Professor in the Electrical and Computer Engineering Department of University of Illinois at Chicago. 
Before joining University of Illinois at Chicago, she was a Postdoctoral Associate at Massachusetts Institute of Technology. 
She received her Ph.D. degree in Electrical and Computer Engineering from University of California, Irvine, M.S. degree in Electrical Engineering and Computer Science from Sabanci University, and B.S. degree in Electrical Engineering from Istanbul University. 
She has served as an associate editor for IEEE Transactions on Mobile Computing and IEEE/ACM Transactions on Networking. 
She received the NSF CAREER award in 2020. 
\end{IEEEbiography} 


\newpage
\section{\textbf{Supplemental Materials}}
\section*{Appendix A: Proof of Theorem \ref{th:S_i(x)andC_i(x)-strategy-polydot}} 
We first 
determine $\mathbf{P}(S_A(x))$ and $\mathbf{P}(S_B(x))$ and then derive $F_{A}(x)$ and $F_{B}(x)$, accordingly.

Based on our strategy for determining $\mathbf{P}(S_A(x))$ and $\mathbf{P}(S_B(x))$, we: (i) first find all elements of $\mathbf{P}(S_{A}(x))$, starting from the minimum possible element, satisfying C1 in (\ref{eq:non_eq-polydot-thrm4}), (ii) then fix $\mathbf{P}(S_{A}(x))$, containing the $z$ smallest elements, in C2 of (\ref{eq:non_eq-polydot-thrm4}), and find all elements of the subset of $\mathbf{P}(S_B(x))$, starting from the minimum possible element, that satisfies C2; we call this subset as $\mathbf{P'}(S_B(x))$, (iii) find all elements of the subset of $\mathbf{P}(S_B(x))$, starting from the minimum possible element, that satisfies C3 in (\ref{eq:non_eq-polydot-thrm4}); we call this subset as $\mathbf{P''}(S_B(x))$, and (iv) finally, find the intersection of $\mathbf{P'}(S_B(x))$ and $\mathbf{P''}(S_B(x))$ to form $\mathbf{P}(S_B(x))$. Next, we explain these steps in details. 

\emph{(i) Find all elements of $\mathbf{P}(S_A(x))$ satisfying C1 in (\ref{eq:non_eq-polydot-thrm4}).} 

For this step, using (\ref{eq:polydot-p(CB)-th}) and C1 in (\ref{eq:non_eq-polydot-thrm4}), we have: 
\begin{align}\label{non_eq1-polydot'}
    &i+t(s-1)+tl(2s-1) \not\in \{t(s-1)-tq+tl'(2s-1)\} \nonumber \\
    & +\mathbf{P}(S_A(x)),  0 \leq q \leq s-1,\; 0 \leq i, l, l' \leq t-1,\; s, \nonumber \\
    & t \in \mathbb{N}
\end{align}
which is equivalent to:
\begin{align}\label{eq:non_eq1-polydot_2}
    \beta+\theta' l'' \not\in \mathbf{P}(S_A(x)),
\end{align}
 for $l''=(l-l')$, $\theta' = t(2s-1)$ and $\beta = i+tq$. From (\ref{non_eq1-polydot'}), the range of the variables $\beta$ and $l''$ are derived as $\beta \in \{0,\ldots,ts-1\}$ and $l'' \in \{-(t-1),\ldots,(t-1)\}$. However, knowing the fact that all powers in $\mathbf{P}(S_A(x))$ are from $\mathbb{N}$, we consider only $l'' \in \{0,\ldots,t-1\}$.\footnote{The reason is that for the largest value of $\beta$, \ie $\beta = ts-1$ and largest value of $l'' \in \{-(t-1),\ldots,-1\}$, \ie $l''=-1$, $\beta+\theta'(l'')$ is equal to $ts-1+(2ts-t)(-1) = t(1-s)-1$, which is negative for $s,t \in \mathbb{N}$. Therefore, for all $l'' \in \{-(t-1),\ldots,-1\}$ in (\ref{eq:non_eq1-polydot_2}), $\beta+\theta'l''$ is negative.} 
Considering different values of $l''$ from the interval $l''\in\{0,...,t-1\}$ in (\ref{eq:non_eq1-polydot_2}), we have: 
\begin{align}
   &\mathbf{P}(S_A(x)) \notin \{0,\ldots,ts-1\}, \nonumber \\
   &\mathbf{P}(S_A(x)) \notin \{\theta',\ldots,ts-1+\theta'\}, \nonumber \\
   &\mathbf{P}(S_A(x)) \notin \{2\theta',\ldots,ts-1+2\theta'\}, \nonumber \\
& \ldots\nonumber \\
   & \mathbf{P}(S_A(x)) \notin \{(t-1)\theta',\ldots,ts-1+(t-1)\theta'\}.
\end{align}
Using the complement of the above intervals, the intervals that $\mathbf{P}(S_A(x))$ can be selected from, is derived as follows: 
\begin{align}\label{eq:P(RA)_set_representation}
    \mathbf{P}(S_A(x)) \in & \{ts,\ldots,\theta'-1\} \cup \{ts+\theta',\ldots,2\theta'-1\} \cup \ldots \nonumber \\
    & \cup \{ts+(t-1)\theta',\ldots,+\infty\}, s,t > 1
\end{align}
\begin{align}\label{eq:P(RA)_set_representation_s=1}
    \mathbf{P}(S_A(x)) \in 
    \{t^2,\ldots,+\infty\}, s=1
\end{align}
\begin{align}\label{eq:P(RA)_set_representation_t=1}
    \mathbf{P}(S_A(x)) \in 
    \{s,\ldots,+\infty\}, t=1
\end{align}
Note that the required number of powers with non-zero coefficients for the secret term $S_A(x)$ is $z$, \ie
\begin{equation}
    |\mathbf{P}(S_A(x))| = z.
\end{equation}
Since our goal is to make the degree of polynomial $F_A(x)$ as small as possible, we choose the $z$ smallest powers from the sets in (\ref{eq:P(RA)_set_representation}) to form $\mathbf{P}(S_A(x))$. 
Note that in (\ref{eq:P(RA)_set_representation}), there are $t-1$ finite sets and one infinite set, where each finite set contains $\theta'-ts$ elements. 
Therefore, based on the value of $z$, we use the first interval and as many remaining intervals as required for $z > \theta'-ts$, and the first interval only for $ z \leq \theta'-ts$. 

\begin{lemma}\label{lem:P(SA)-z large}
If $z > \theta'-ts$ and $s,t \neq 1$, the subsets of all powers of polynomial $S_A(x)$ with non-zero coefficients is defined as the following:
\begin{align}\label{eq:finite_P(RA)_set_representation-z large}
    \mathbf{P}(S_A(x)) = &\Big(\bigcup\limits_{l=0}^{p-1} \{ts+\theta' l,\ldots,(l+1)\theta'-1\}\Big) \nonumber \\ & \cup \{ts+p\theta',\ldots,ts+p\theta'+z-1-p(\theta'-ts)\}\\
    =&\{ts+\theta'l+w, l\in\Omega_0^{p-1}, w\in \Omega_0^{t(s-1)-1}\} \nonumber \\  &\cup \{ts+\theta'p+u, u\in\Omega_0^{z-1-pt(s-1)}\}\label{eq:psa12}.
\end{align}
\end{lemma}
{\em Proof:}
For the case of $z > \theta'-ts$ and $s,t \neq 1$, the number of elements in the first interval of (\ref{eq:P(RA)_set_representation}), which is equal to $\theta'-ts$, is not sufficient for selecting $z$ powers. Therefore, more than one interval is used; we show the number of selected intervals with $p+1$, where $p \ge 1$ is defined as $p=\min\{\floor{\frac{z-1}{\theta'-ts}},t-1\}$. With this definition, the first $p$ selected intervals are selected in full, in other words, in total we select $p(\theta'-ts)$ elements to form the first $p$ intervals in (\ref{eq:finite_P(RA)_set_representation-z large}). The remaining $z-p(\theta'-ts)$ elements are selected from the $(p+1)^\text{st}$ interval of (\ref{eq:P(RA)_set_representation}) as shown as the last interval of (\ref{eq:finite_P(RA)_set_representation-z large}). (\ref{eq:psa12}) can be derived from (\ref{eq:finite_P(RA)_set_representation-z large}) by replacing $\theta'$ with its equivalence, $2ts-t$.
\hfill $\Box$

\begin{lemma}\label{lem:P(SA)-z small}
If $z \leq \theta'-ts$ and $s,t \neq 1$, the subsets 
of all powers of polynomial $S_A(x)$ with non-zero coefficients is defined as the following:
\begin{align}\label{eq:finiteP(SA)-polydot-second-scenario}
\mathbf{P}(S_A(x)) = \{ts,\dots,ts+z-1\}
= \{ts+u, u\in \Omega_0^{z-1}\}.
\end{align}
\end{lemma}
{\em Proof:}
In this scenario for $z \leq \theta'-ts$, the first interval of (\ref{eq:P(RA)_set_representation}) is sufficient to select all $z$ elements of $\mathbf{P}(S_A(x))$, therefore, $z$ elements are selected from the first interval of (\ref{eq:P(RA)_set_representation}), as shown in (\ref{eq:finiteP(SA)-polydot-second-scenario}). 
\hfill $\Box$

\begin{lemma}\label{lem:P(SA)-s=1t=1}
If $s=1$, the subsets of all powers of polynomial $S_A(x)$ with non-zero coefficients is defined as the following:
\begin{align}\label{eq:psafors=1}
    \mathbf{P}(S_A(x)) = \{t^2,\dots,t^2+z-1\}
= \{t^2+u, u\in \Omega_0^{z-1}\},
\end{align}
and if $t=1$, it is defined as:
\begin{align}\label{eq:psafort=1}
    \mathbf{P}(S_A(x)) = \{s,\dots,s+z-1\}
= \{s+u, u\in \Omega_0^{z-1}\}.
\end{align}
\end{lemma}
{\em Proof:}
If $s=1$, $z$ smallest elements are selected from (\ref{eq:P(RA)_set_representation_s=1}), as shown in (\ref{eq:psafors=1}) and if $t=1$, $z$ smallest elements are selected from (\ref{eq:P(RA)_set_representation_t=1}), as shown in (\ref{eq:psafort=1}). 
\hfill $\Box$

\emph{(ii) Fix $\mathbf{P}(S_{A}(x))$ in C2 of (\ref{eq:non_eq-polydot-thrm4}), and find the subset of $\mathbf{P}(S_B(x))$ that satisfies C2; we call this subset as $\mathbf{P'}(S_B(x))$.}

In this step, we consider the four cases of $s=1$, $t=1$, $z>\theta'-ts,\;s,t \neq 1$, and $z \leq \theta'-ts,\; s,t \neq 1$ and derive $\mathbf{P'}(S_B(x))$ as summarized in Lemmas \ref{lem:P'(SB)-s=1}, \ref{lem:P'(SB)-t=1}, \ref{lem:P'(SB)-z large} and \ref{lem:P'(SB)-z small}, respectively.

\begin{lemma}\label{lem:P'(SB)-s=1}
If $s=1$, $\mathbf{P'}(S_B(x))$ is defined as the following:
\begin{align}\label{eq:P'(RB)_set_s=1}
    \mathbf{P'}(S_B(x)) =  \{0,\ldots,+\infty\}.
\end{align}
\end{lemma}
{\em Proof:}
In this scenario, we use (\ref{eq:psafors=1}) defined for $\mathbf{P}(S_A(x))$. By replacing $\mathbf{P}(S_A(x))$ in C2 we have the following: 
\begin{align}\label{eq:non_eq2-polydot''-s=1}
   & \text{C2: } i+tl \not\in \{t^2,\dots,t^2+z-1\}+\mathbf{P'}(S_B(x)),
\end{align}
which can be equivalently written as:
\begin{align}\label{eq:non_eq3-polydot''-s=1}
   & \text{C2: } \{0,\dots,t^2-1\} \not\in \{t^2,\dots,t^2+z-1\}+\mathbf{P'}(S_B(x)).
\end{align}
From the above equation, any non-negative elements for $\mathbf{P'}(S_B(x))$ satisfies this constraint.  
This completes the proof.\hfill $\Box$

\begin{lemma}\label{lem:P'(SB)-t=1}
If $t=1$, $\mathbf{P'}(S_B(x))$ is defined as the following:
\begin{align}\label{eq:P'(RB)_set_t=1}
    \mathbf{P'}(S_B(x)) =  \{0,\ldots,+\infty\}.
\end{align}
\end{lemma}
{\em Proof:}
In this scenario, we use (\ref{eq:psafort=1}) defined for $\mathbf{P}(S_A(x))$. By replacing $\mathbf{P}(S_A(x))$ in C2 we have the following: 
\begin{align}\label{eq:non_eq2-polydot''-t=1}
   & \text{C2: } s-1 \not\in \{s,\dots,s+z-1\}+\mathbf{P'}(S_B(x)).
\end{align}
From the above equation, any non-negative elements for $\mathbf{P'}(S_B(x))$ satisfies this constraint. 
This completes the proof.\hfill $\Box$

   \begin{lemma}\label{lem:P'(SB)-z large}
If $z > \theta'-ts$ and $s,t \neq 1$, $\mathbf{P'}(S_B(x))$ is defined as the following:
\begin{align}\label{eq:P'(RB)_set_representation-zlarge}
    \mathbf{P'}(S_B(x)) = 
    &\Big(\bigcup\limits_{l'=0}^{t-2} \{\theta' l',\ldots,(l'+1)\theta'-ts\}\Big) \\ 
    &\cup \{(t-1)\theta',\ldots,+\infty\}.
\end{align}
\end{lemma}
{\em Proof:}
In this scenario, we use (\ref{eq:finite_P(RA)_set_representation-z large}) defined for $\mathbf{P}(S_A(x))$ when $z > \theta'-ts$, which can be equivalently written as:
\begin{align}
 \mathbf{P}(S_A(x)) = \Bigg\{ \begin{array}{cc}
   ts+\theta' l''+w,  &  l''\in \Omega_{0}^{p-1},\; w \in \Omega_{0}^{\theta'-ts-1} \\
   ts+\theta' l''+u,  &  l''= p,\; u\in \Omega_{0}^{z-1-p(\theta'-ts)}
\end{array}
\end{align}
and then replace $\mathbf{P}(S_A(x))$ in C2 using the above equation:
\begin{align}\label{eq:non_eq2-polydot''}
   & \text{C2: } i+t(s-1)+\theta' l \not\in \nonumber \\
   & \Bigg\{ \begin{array}{cc}
   ts+\theta' l''+w+\mathbf{P'}(S_B(x)),  &  l''\in \Omega_{0}^{p-1},\; w \in \Omega_{0}^{\theta'-ts-1}  \\
   ts+\theta' l''+u+\mathbf{P'}(S_B(x)),  &  l''=p,\; u\in \Omega_{0}^{z-1-p(\theta'-ts)}
\end{array}
\end{align}
Equivalently:
\begin{align}\label{eq:non_eq2-polydot'''-}
  & \mathbf{P'}(S_B(x)) \not \in \nonumber \\
  & \Bigg\{ \begin{array}{cc}
   i-t-w+\theta'(l-l''), &  l''\in \Omega_{0}^{p-1},\; w \in \Omega_{0}^{\theta'-ts-1}  \\
   i-t-u+\theta'(l-p),  &  u \in \Omega_{0}^{z-1-p(\theta'-ts)} 
\end{array}
\end{align}
By simplifying the above equation, we have:
\begin{align}\label{eq:non_eq2prime-polydot'''}
  & \mathbf{P'}(S_B(x)) \not \in \nonumber \\
  & \Bigg\{ \begin{array}{cc}
   \hat{i}-w+\theta'\hat{l}, &  
   \hat{i}\in \Omega_{-t}^{-1},\;
   \hat{l}\in \Omega_{-(p-1)}^{t-1},\; w \in \Omega_{0}^{\theta'-ts-1}  \\
   \hat{i}-u+\theta'\Tilde{l},  & 
   \hat{i}\in \Omega_{-t}^{-1},\; \tilde{l}\in \Omega_{-p}^{t-1-p},\; u \in \Omega_{0}^{z-1-p(\theta'-ts)} 
\end{array}
\end{align}
Knowing the fact that all powers in $\mathbf{P'}(S_B(x))$ are in $ \mathbb{N}$, we consider only $\hat{l}, \tilde{l} \ge 1$ as $\hat{l}, \tilde{l} < 1$ results in negative powers of $\mathbf{P'}(S_B(x))$\footnote{The reason is that $i'-w$ and $i'-u$ are always negative. If $\hat{l}, \tilde{l}$ are also negative or equal to zero, $i'-w+\theta'\hat{l}$ and $i'-u+\theta'\tilde{l}$ are negative.}. 
This results in:
\begin{align}\label{eq:non_eq2-polydot'''}
& \mathbf{P'}(S_B(x)) \not \in 
\Bigg\{ \begin{array}{cc}
   \mathbf{V}_1\\
   \mathbf{V}_2,
\end{array} \\
& =\Bigg\{ \begin{array}{cc}
   \hat{i}-w+\theta'\hat{l}, &  
   \hat{i}\in \Omega_{-t}^{-1},\;
   \hat{l}\in \Omega_1^{t-1},\; w \in \Omega_{0}^{\theta'-ts-1}\\
   \hat{i}-u+\theta'\Tilde{l},  & 
   \hat{i}\in \Omega_{-t}^{-1},\; \tilde{l}\in \Omega_1^{t-1-p},\; u \in \Omega_{0}^{z-1-p(\theta'-ts)}
\end{array}
\end{align} 
\begin{lemma} \label{lemma:D1_D2}
$\mathbf{V}_2$ defined in (\ref{eq:non_eq2-polydot'''}) is a subset of $\mathbf{V}_1$: $V_2 \subset V_1$. 
\end{lemma}
{\em Proof:}
To prove this lemma, we consider two cases of\footnote{Note that from the definition of $p=\min\{\floor{\frac{z-1}{\theta'-ts}},t-1\}$, $p$ is less than or equal to $t-1$.} (i) $p=t-1$ and (ii) $p<t-1$. For the first case of $p=t-1$, $V_2$ is an empty set as the upper bound of $\tilde{l}$, \ie $t-1-p$, becomes less than its lower bound, \ie $1$. Thus $\mathbf{V}_2 \subset \mathbf{V}_1$ for $p=t-1$. In the following, we consider the second case of $p<t-1$ and prove that $V_2 \subset V_1$. 
\begin{align}\label{eq:p}
    & p = \min\{\floor{\frac{z-1}{\theta'-ts}},t-1\}, \;\;\; p<t-1 \nonumber \\
    \Rightarrow & p = \floor{\frac{z-1}{\theta'-ts}} \nonumber \\
    \Rightarrow & p+1 > \frac{z-1}{\theta'-ts} \nonumber \\
    \Rightarrow & \theta'-ts > z-1-p(\theta'-ts) \nonumber \\
    \Rightarrow & \theta'-ts \geq z-p(\theta'-ts).
\end{align}

Using (\ref{eq:p}), $u \subset w$ in (\ref{eq:non_eq2-polydot'''}). In addition, $\Tilde{l} \subset \hat{l}$, as $p\ge 0$. Therefore, $\mathbf{V}_2$ is a subset of $\mathbf{V}_1$ for the second case of $p<t-1$, as well.  
This completes the proof. \hfill $\Box$

Using Lemma \ref{lemma:D1_D2}, we can reduce (\ref{eq:non_eq2-polydot'''}) to:
\begin{align}
    \mathbf{P'}(S_B(x)) \not\in \hat{i}-w+\theta' \hat{l}, \hat{i}\in \Omega_{-t}^{-1},\;
   \hat{l}\in \Omega_1^{t-1},\; w \in \Omega_{0}^{\theta'-ts-1}
\end{align}
By replacing $\theta'$ with its equivalence $t(2s-1)$, the range of variation for $\hat{i}-w$ is $\hat{i}-w \in \{-ts+1, \ldots, -1\}$. Therefore, by considering different values of $\hat{l}$, the above equation is expanded as:
\begin{align}
%
   & \mathbf{P'}(S_B(x)) \not\in \{\theta'-ts+1,\ldots,\theta'-1\},  \nonumber \\
   & \mathbf{P'}(S_B(x)) \not\in \{2\theta'-ts+1,\ldots,2\theta'-1\}, \nonumber \\ 
& \ldots \nonumber \\ 
   & \mathbf{P'}(S_B(x)) \not\in \{(t-1)\theta'-ts+1,\ldots,(t-1)\theta'-1\}.
\end{align}
Using the complement of the above intervals, the intervals that $\mathbf{P'}(S_B(x))$ can be selected from, is derived as follows:
\begin{align}
     \mathbf{P'}(S_B(x)) & \in \{0,\ldots,\theta'-ts\} \cup \{\theta',\ldots,2\theta'-ts\} \cup \ldots \nonumber \\ & \cup \{(t-1)\theta',\ldots,+\infty\}.
\end{align}
This completes the proof of Lemma \ref{lem:P'(SB)-z large}.
\hfill $\Box$\\
  \begin{lemma}\label{lem:P'(SB)-z small}
If $z \leq \theta'-ts$ and $s,t \neq 1$, $\mathbf{P'}(S_B(x))$ is defined as the following:
\begin{align}\label{eq:P'(RB)_set_representation-z small}
    \mathbf{P'}(S_B(x)) = 
    &\Big(\bigcup\limits_{l'=0}^{t-2} \{\theta' l',\ldots,(l'+1)\theta'-z-t\}\Big) \\ 
    &\cup \{(t-1)\theta',\ldots,+\infty\}.
\end{align} 
\end{lemma}
{\em Proof:}
To determine $\mathbf{P'}(S_B(x))$, we need to find a subset of $\mathbf{P}(S_B(x))$ that satisfies C2. By replacing $\mathbf{P}(S_A(x))$ from Lemma \ref{lem:P(SA)-z small} in C2, we have:
\begin{align}\label{eq:satisfyC2-polydot-secondscenario}
   & \text{C2: } i+t(s-1)+\theta' l \not\in ts+r+\mathbf{P'}(S_B(x)).
   \end{align}
   Equivalently:
   \begin{align}\label{eq:satisfyC2-polydot-secondscenario'}
       \mathbf{P}(S_B(x)) \not\in i-r-t+\theta' l,
   \end{align}
   where $i,l \in \{0,\dots,t-1\}$ and $r \in \{0,\dots,z-1\}$. By expanding the above equation we have:
\begin{align}
   & \mathbf{P'}(S_B(x)) \not\in \{-z-t+1,\ldots,-1\}, \nonumber \\
   & \mathbf{P'}(S_B(x)) \not\in \{\theta'-z-t+1,\ldots,\theta'-1\},  \nonumber \\
   & \mathbf{P'}(S_B(x)) \not\in \{2\theta'-z-t+1,\ldots,2\theta'-1\}, \nonumber \\ 
& \ldots \nonumber \\ 
   & \mathbf{P'}(S_B(x)) \not\in \{(t-1)\theta'-z-t+1,\ldots,(t-1)\theta'-1\}. 
\end{align}
Using the complement of the above intervals, the intervals that $\mathbf{P'}(S_B(x))$ can be selected from, is derived as follows:
\begin{align}
    & \mathbf{P'}(S_B(x)) = \{0,\ldots,\theta'-z-t\} \cup \{\theta',\ldots,2\theta'-z-t\} \cup \nonumber \\ &\ldots \cup
     \{(t-1)\theta',\ldots,+\infty\}.
\end{align} 
This completes the proof.
\hfill $\Box$\\ \\ \\
\emph{(iii) Find the subset of $\mathbf{P}(S_B(x))$ that satisfies C3 in (\ref{eq:non_eq-polydot-thrm4}); we call this subset as $\mathbf{P''}(S_B(x))$}.

In this step, we consider the three cases of $s=1$, $t=1$ and $s,t \geq 2$, and derive $\mathbf{P''}(S_B(x))$ as summarized in Lemmas \ref{lem:p''SB-s=1}, \ref{lem:p''SB-t=1} and \ref{lem:p''SB-s,t not 1}.

\begin{lemma}\label{lem:p''SB-s=1}
If $s=1$, $\mathbf{P''}(S_B(x))$ is defined as the following:
\begin{align}\label{eq:P''(SB)_set_s=1}
    \mathbf{P''}(S_B(x))=\{t^2,\dots,+\infty\}.
\end{align}    
\end{lemma}
{\em Proof:}
By replacing $\mathbf{P}(C_A(x))$ from (\ref{eq:polydot-p(CA)-th}) in C3, we have
\begin{align}\label{eq:C3-polydot-eq1-s=1}
     \text{C3: }& i+tl \not\in  
     \{0,\ldots,t-1\}+\mathbf{P''}(S_B(x)),
\end{align}
which can be equivalently written as:
\begin{align}
      \{0,\dots,t^2-1\} \not\in  
     \{0,\ldots,t-1\}+\mathbf{P''}(S_B(x)) \nonumber \\ 
     \Rightarrow  \{-t+1,\dots,t^2-1\} \not\in  
     \mathbf{P''}(S_B(x))\label{eq:C3-polydot-eq3-s=1}
\end{align}
From the above equation, the elements of $\mathbf{P''}(S_B(x))$ can be selected from any positive integer greater than $t^2-1$. This completes the proof. \hfill $\Box$

\begin{lemma}\label{lem:p''SB-t=1}
If $t=1$, $\mathbf{P''}(S_B(x))$ is defined as the following:
\begin{align}\label{eq:P''(SB)_set_t=1}
    \mathbf{P''}(S_B(x))=\{s,\dots,+\infty\}.
\end{align}    
\end{lemma}
{\em Proof:}
By replacing $\mathbf{P}(C_A(x))$ from (\ref{eq:polydot-p(CA)-th}) in C3, we have
\begin{align}\label{eq:C3-polydot-eq1-t=1}
     \text{C3: }& s-1 \not\in  
     \{0,\ldots,s-1\}+\mathbf{P''}(S_B(x)),
\end{align}
which can be equivalently written as:
\begin{align}\label{eq:C3-polydot-eq2-t=1}
     & \{0,\dots,s-1\} \not\in \mathbf{P''}(S_B(x)).
\end{align}
From the above equation, the elements of $\mathbf{P''}(S_B(x))$ can be selected from any positive integer greater than $s$
This completes the proof. \hfill $\Box$

\begin{lemma}\label{lem:p''SB-s,t not 1}
For any $s,t \geq 2$ and $z \in \mathbb{N}$, $\mathbf{P''}(S_B(x))$ is defined as the following:
\begin{align}\label{eq:P''(RB)_set_representation}
    \mathbf{P''}(S_B(x)) & = 
    \Big(\bigcup\limits_{l''=0}^{t-2} \{ts+\theta' l'',\ldots,(l''+1)\theta'-t\}\Big) \nonumber \\ & \cup \{ts+(t-1)\theta',\ldots,+\infty\}.
\end{align}.
\end{lemma}

{\em Proof:}
By replacing $\mathbf{P}(C_A(x))$ from (\ref{eq:polydot-p(CA)-th}) in C3, we have
\begin{align}\label{eq:non_eq3-polydot'}
     & i+t(s-1)+\theta' l \not\in  
     \{0,\ldots,ts-1\}+\mathbf{P''}(S_B(x)).
\end{align}
Equivalently,
\begin{align}
    \mathbf{P''}(S_B(x)) \not\in 
    \{-t+1,\ldots,ts-1\}+\theta' l.
\end{align}
By expanding the above equation for different values of $l$, we have:
\begin{align}
   & \mathbf{P''}(S_B(x)) \not\in \{-t+1,\ldots,ts-1\},  \nonumber \\
& \mathbf{P''}(S_B(x)) \not\in \{-t+1+\theta',\ldots,ts-1+\theta'\},   \nonumber \\
  &  \mathbf{P''}(S_B(x)) \not\in \{-t+1+2\theta',\ldots,ts-1+2\theta'\},   \nonumber \\
& \ldots   \nonumber \\
  &  \mathbf{P''}(S_B(x)) \not\in \{-t+1+(t-1)\theta',\ldots,ts-1+(t-1)\theta'\}. \nonumber  
\end{align}
We define $\mathbf{P''}(S_B(x))$ as the complement of the above intervals:
\begin{align}\label{eq:P''(RB)_set_representation-1}
    \mathbf{P''}(S_B(x))&= 
    \Big(\bigcup\limits_{l''=0}^{t-2} \{ts+\theta' l'',\ldots,(l''+1)\theta'-t\}\Big) \nonumber \\  & \cup \{ts+(t-1)\theta',\ldots,+\infty\}.
\end{align}
\hfill $\Box$

\emph{(iv) 
Find the intersection of $\mathbf{P'}(S_B(x))$ and $\mathbf{P''}(S_B(x))$ to form $\mathbf{P}(S_B(x))$.} 

In this step, we consider four regions for the range of variable $z$, (a) $z > \theta'-ts, s,t \neq 1$, (b) $\theta'-ts-t < z \leq \theta'-ts, s,t \neq 1$, (c) $\frac{\theta'-ts-t+1}{2} < z \leq \theta'-ts-t, s,t \neq 1$, and (d) $z \leq \frac{\theta'-ts-t+1}{2}, s,t \neq 1$, as well as the special cases of (e) $s=1$ and (f) $t=1$, and calculate $\mathbf{P}(S_B(x)$ for each case, as summarized in Lemmas \ref{lem:P(SB)-z large}, \ref{lem:P(SB)-z medium}, \ref{lem:P(SB)-z small}, \ref{lem:P(SB)-z very small}, \ref{lem:P(SB)-s=1}, and \ref{lem:P(SB)-t=1}, respectively.

\begin{lemma}\label{lem:P(SB)-z large}
If $z > \theta'-ts$ and $s,t \neq 1$, the subsets of all powers of polynomials $S_B(x)$ with non-zero coefficients is defined as the following
\begin{align}\label{eq:P(RB)_set_representation-z large}
      \mathbf{P}(S_B(x)) 
        = 
        \{ts+(t-1)\theta'+r, 
        \; r \in \Omega_0^{z-1},\; \theta'=t(2s-1) 
    \}.
\end{align}
\end{lemma}
{\em Proof:}
For this region, we use $\mathbf{P'}(S_B(x))$ defined in Lemma \ref{lem:P'(SB)-z large} and 
$\mathbf{P''}(S_B(x))$ defined in (\ref{eq:P''(RB)_set_representation}):
\begin{align}\label{P'(SB) set representation-M'}
   & \mathbf{P'}(S_B(x)) = \mathbf{M}_1' \cup \mathbf{M}_2', \nonumber \\
    & \mathbf{P''}(S_B(x)) = \mathbf{M}_1'' \cup \mathbf{M}_2'', 
\end{align} where,
\begin{align}\label{eq:def-M1'_M1''_M2'_M2''}
    & \mathbf{M}_1'= \bigcup\limits_{l'=0}^{t-2} \{\theta' l',\ldots,(l'+1)\theta'-ts\}, \nonumber \\
    & \mathbf{M}_1'' = \bigcup\limits_{l''=0}^{t-2} \{ts+\theta' l'',\ldots,(l''+1)\theta'-t\}, \nonumber \\
    & \mathbf{M}_2' = \{(t-1)\theta',\ldots,+\infty\}, \nonumber \\
   & \mathbf{M}_2'' = \{ts+(t-1)\theta',\ldots,+\infty\}.
\end{align}

The intersection of $\mathbf{P'}(S_B(x))$ and $\mathbf{P''}(S_B(x))$ is calculated as:
\begin{align}\label{P(SB)=P'(SB)capP''(SB)}
      \mathbf{P'}(S_B(x)) \cap \mathbf{P''}(S_B(x))
      = &\big(\mathbf{M}_1' \cup \mathbf{M}_2' \big) \cap \big(\mathbf{M}_1'' \cup \mathbf{M}_2'' \big) \nonumber \\
      = & \big(\mathbf{M}_1' \cap \mathbf{M}_1''\big) \cup \big(\mathbf{M}_2' \cap \mathbf{M}_1''\big) \nonumber \\ & \cup  \big(\mathbf{M}_1' \cap \mathbf{M}_2''\big) \cup \big(\mathbf{M}_2' \cap \mathbf{M}_2''\big).
\end{align}
In the following, we calculate $\big(\mathbf{M}_1' \cap \mathbf{M}_1''\big)$, $\big(\mathbf{M}_2' \cap \mathbf{M}_1''\big)$, $\big(\mathbf{M}_1' \cap \mathbf{M}_2''\big)$, and $\big(\mathbf{M}_2' \cap \mathbf{M}_2''\big)$, separately.
\begin{itemize}
    \item Calculating $\big(\mathbf{M}_1' \cap \mathbf{M}_1''\big)$

To calculate $\big(\mathbf{M}_1' \cap \mathbf{M}_1''\big)$, we consider each subset of $\mathbf{M}_1'$, \ie $\{\theta' l',\ldots,(l'+1)\theta'-ts\}$ and show that this subset does not have any overlap with any of the subsets of $\mathbf{M}_1''$, \ie $\{ts+\theta' l'',\ldots,(l''+1)\theta'-t\},  0\leq l''< t-2$; This results in $\big(\mathbf{M}_1' \cap \mathbf{M}_1''\big)=\emptyset$. For this purpose, (i) first we consider the subsets of $\mathbf{M}_1''$, for which $l''< l'$ and show that $\{\theta' l',\ldots,(l'+1)\theta'-ts\}$ falls to the right side of all intervals $\{ts+\theta' l'',\ldots,(l''+1)\theta'-t\}, 0\leq l''< l'$, and (ii) second we consider the subsets of $\mathbf{M}_1''$, for which $l''\ge l'$ and show that $\{\theta' l',\ldots,(l'+1)\theta'-ts\}$ falls to the left  side of all intervals $\{ts+\theta' l'',\ldots,(l''+1)\theta'-t\}, l' \leq l'' \leq t-2 $. 

\begin{figure*}[ht]
		\centering
		\scalebox{1.1}{
		\includegraphics[width=15cm]{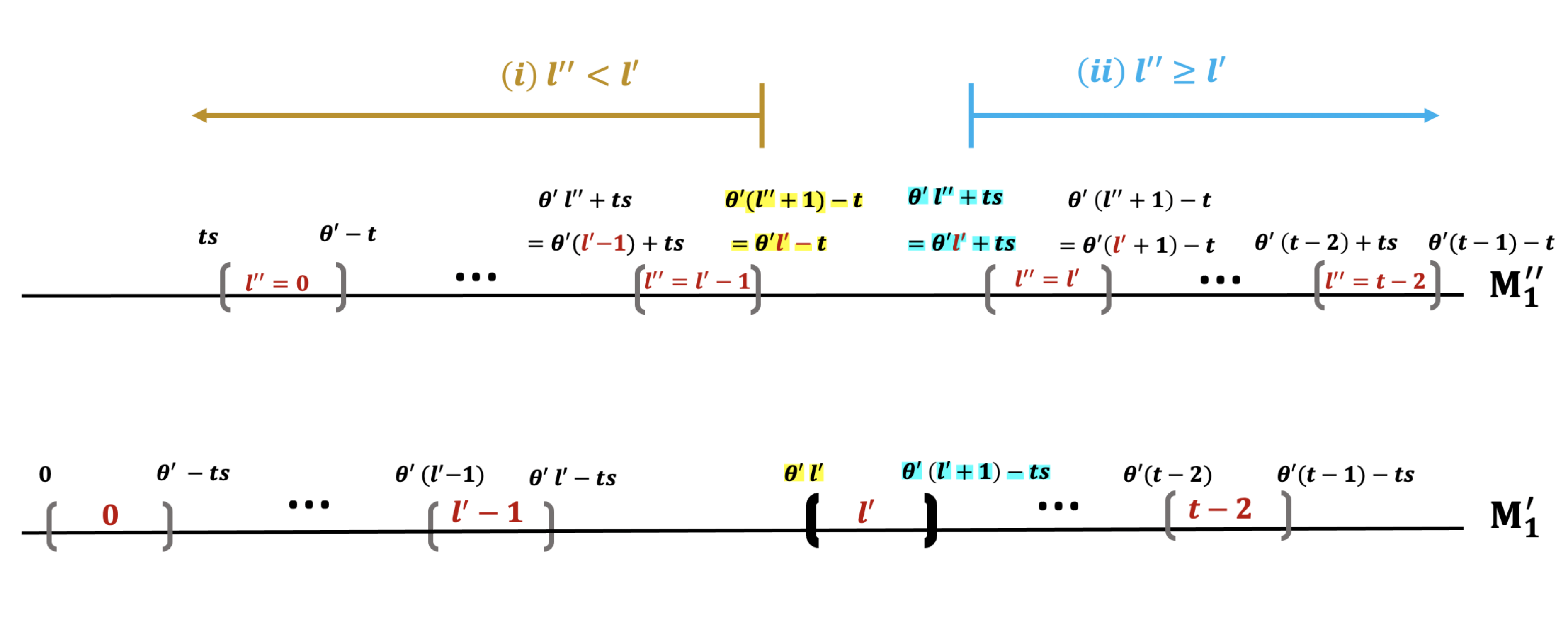}}
	\caption{An illustration showing that $\mathbf{M}_1' \cap \mathbf{M}_1''=\emptyset$ holds in Lemma \ref{lem:P(SB)-z large}. 
	}
\label{Fig_union-M_i'andM_j''=empty}
\vspace{-5pt}
\end{figure*}

(i) 
$l''<l'$: In this case, the largest element of all subsets of $\mathbf{M}_1''$, \ie $\theta' (l''+1)-t$ is less than the smallest element of $\{\theta' l',\ldots,(l'+1)\theta'-ts\}$, as shown in Fig. \ref{Fig_union-M_i'andM_j''=empty}. The reason is that:
    \begin{align}\label{eq:M'1capM''1-l''<l'}
      l'' <l' \Rightarrow & l''+1 \leq l', \nonumber \\
      \Rightarrow &  \theta'(l''+1) \leq \theta' l', \nonumber \\
      \Rightarrow & \theta'(l''+1)-t < \theta' l'.
\end{align}
    
(ii) 
$l'' \ge l'$. In this case, the smallest element of all subsets of $\mathbf{M}_1''$, \ie $\theta' l''+ts$, is greater than the largest element of $\{\theta' l',\ldots,(l'+1)\theta'-ts\}$, as shown in Fig. \ref{Fig_union-M_i'andM_j''=empty}. The reason is that:
    \begin{align}
          l' \leq l''\Rightarrow & \theta' l' \leq \theta' l'', \nonumber \\
          \Rightarrow & \theta' l' -t < \theta' l'', \nonumber \\
          \Rightarrow & \theta' l' -t+ts < \theta' l''+ts, \nonumber \\ 
          \Rightarrow & \theta' l' -t+2ts-ts < \theta' l''+ts, \nonumber \\
          \Rightarrow & \theta' l' +\theta'-ts < \theta' l''+ts, \nonumber \\
          \Rightarrow & \theta' (l'+1)-ts < \theta' l''+ts.
   \end{align}
From (i) and (ii) discussed in the above, we conclude that:
\begin{equation}\label{eq:m'1capm''1} \mathbf{M}_1' \cap \mathbf{M}_1''=\emptyset \end{equation}

\item Calculating 
$\big(\mathbf{M}_2' \cap \mathbf{M}_1''\big)$

The largest element of $\mathbf{M}_1''$, $(t-1)\theta' -t$, is always less than $(t-1)\theta'$, which is the smallest element of $\mathbf{M}_2'$. This results in: \begin{equation}\label{eq:m'2capm''1} \mathbf{M}_2' \cap \mathbf{M}_1'' = \emptyset\end{equation}

\item Calculating 
$\big(\mathbf{M}_1' \cap \mathbf{M}_2''\big)$

The largest element of $\mathbf{M}_1'$, \ie $(t-1)\theta' -ts$ is always less than $(t-1)\theta'+ts$, which is the smallest element of $\mathbf{M}_2''$. This results in: \begin{equation}\label{eq:m'1capm''2}\mathbf{M}_1' \cap \mathbf{M}_2'' = \emptyset \end{equation}
\item Calculating $\big(\mathbf{M}_2' \cap \mathbf{M}_2''\big)$

\begin{align}\label{eq:m'2capm''2}
      \mathbf{M}_2' \cap \nonumber \mathbf{M}_2'' = &\{(t-1)\theta',\ldots,+\infty\} \cap \\ &\{ts+(t-1)\theta',\ldots,+\infty\} \nonumber \\
      = & \{ts+(t-1)\theta',\ldots,+\infty\}.
\end{align}
\end{itemize}
From (\ref{P(SB)=P'(SB)capP''(SB)}), (\ref{eq:m'1capm''1}), (\ref{eq:m'2capm''1}), (\ref{eq:m'1capm''2}), and (\ref{eq:m'2capm''2}), we have:

\begin{align}\label{eq:P(RB)_set_representation}
      \mathbf{P'}(S_B(x)) \cap \mathbf{P''}(S_B(x))
      = & \{ts+(t-1)\theta',\ldots,+\infty\},
\end{align}
from which the elements of $\mathbf{P}(S_B(x))$ can be selected. As there are $z$ colluding workers, the size of $\mathbf{P}(S_B(x))$ should be $z$
, \ie $|\mathbf{P}(S_B(x))|=z$. On the other hand, since our goal is to reduce the degree of $F_B(x)$ as much as possible, we select the $z$ smallest elements of the set shown in (\ref{eq:P(RB)_set_representation}) to form  
$\mathbf{P}(S_B(x))$: 
\begin{align}
    \mathbf{P}(S_B(x)) = &  \{ts+(t-1)\theta',\ldots,ts+(t-1)\theta'+z-1\} \nonumber \\
\end{align}
This completes the proof of Lemma \ref{lem:P(SB)-z large}.
\hfill $\Box$\\

\begin{lemma}\label{lem:P(SB)-z medium}
If $\theta'-ts-t < z \leq \theta'-ts$ and $s,t \neq 1$, the subsets of all powers of polynomials $S_B(x)$ with non-zero coefficients is defined as the following:
\begin{align}\label{eq:P(RB)_set_representation-z medum}
      \mathbf{P}(S_B(x)) 
       = 
       \{ts+(t-1)\theta'+r,
        \; &0 \leq r \leq z-1,\; \nonumber \\
    & \theta'=t(2s-1)
    \}.
\end{align}
\end{lemma}
{\em Proof:}
For this region, we use $\mathbf{P'}(S_B(x))$ defined in Lemma \ref{lem:P'(SB)-z small} and 
$\mathbf{P''}(S_B(x))$ defined in (\ref{eq:P''(RB)_set_representation}):
\begin{align}\label{P'(SB) set representation-M'--}
   & \mathbf{P'}(S_B(x)) = \mathbf{M}_1' \cup \mathbf{M}_2', \nonumber \\
    & \mathbf{P''}(S_B(x)) = \mathbf{M}_1'' \cup \mathbf{M}_2'', 
\end{align} where,

\begin{align}\label{eq:def-M1'_M1''_M2'_M2''case2}
    & \mathbf{M}_1'= \bigcup\limits_{l'=0}^{t-2} \{\theta' l',\ldots,(l'+1)\theta'-z-t\}, \nonumber \\
    & \mathbf{M}_1'' = \bigcup\limits_{l''=0}^{t-2} \{ts+\theta' l'',\ldots,(l''+1)\theta'-t\}, \nonumber \\
    & \mathbf{M}_2' = \{(t-1)\theta',\ldots,+\infty\}, \nonumber \\
   & \mathbf{M}_2'' = \{ts+(t-1)\theta',\ldots,+\infty\}.
\end{align}
Similar to the proof of Lemma \ref{lem:P(SB)-z large}, we find $\mathbf{P'}(S_B(x)) \cap \mathbf{P''}(S_B(x))$ by calculating
$\big(\mathbf{M}_1' \cap \mathbf{M}_1''\big) \cup \big(\mathbf{M}_2' \cap \mathbf{M}_1''\big) \cup \big(\mathbf{M}_1' \cap \mathbf{M}_2''\big) \cup \big(\mathbf{M}_2' \cap \mathbf{M}_2''\big)$ with the only difference that the definition of $\mathbf{M}_1'$ in (\ref{eq:def-M1'_M1''_M2'_M2''}) is different from the definition of $\mathbf{M}_1'$ in (\ref{eq:def-M1'_M1''_M2'_M2''case2}).

\begin{itemize}
    \item Calculating $\big(\mathbf{M}_1' \cap \mathbf{M}_1''\big)$
    
    We show that each subset of $\mathbf{M}_1'$, \ie $\{\theta' l',\ldots,(l'+1)\theta'-z-t\}$ does not have any overlap with any of the subsets of $\mathbf{M}_1''$, \ie $\{ts+\theta' l'',\ldots,(l''+1)\theta'-t\}, 0\leq l''< t-2$. Similar to the proof of Lemma \ref{lem:P(SB)-z large}, we consider two cases of $l''<l'$ and $l'' \ge l'$.
    
    (i) $l''<l'$: As shown in (\ref{eq:M'1capM''1-l''<l'}), all subsets of $\mathbf{M}_1''$ falls to the left of the subset of $\mathbf{M}_1'$.
    
    (ii) $l'' \ge l'$: In this case, the smallest element of all subsets of $\mathbf{M}_1''$, \ie $\theta' l''+ts$, is greater than the largest element of $\{\theta' l',\ldots,(l'+1)\theta'-z-t\}$. The reason is that:
    \begin{align}\label{eq:M'1capM''1-scenario2-p1}
          l' \leq l''\Rightarrow & \theta' l' \leq \theta' l'', \nonumber \\
          \Rightarrow & \theta' l'+ts < \theta' l''+ts.
   \end{align}
On the other hand we have:
\begin{align}\label{eq:M'1capM''1-scenario2-p2}
& \theta'-ts-t < z \nonumber \\
& \Rightarrow \theta' l'-z < \theta' l'-\theta'+ts+t \nonumber \\
& \Rightarrow (l'+1)\theta'-z-t < \theta' l'+ts.
\end{align}
Therefore, from (\ref{eq:M'1capM''1-scenario2-p1}) and (\ref{eq:M'1capM''1-scenario2-p2}) we have:
\begin{align}
& (l'+1)\theta'-z-t < \theta' l''+ts.
\end{align}
From (i) and (ii) discussed in the above, we conclude that:
\begin{equation}\label{eq:m'1capm''12} \mathbf{M}_1' \cap \mathbf{M}_1''=\emptyset \end{equation}

\item Calculating $\big(\mathbf{M}_2' \cap \mathbf{M}_1''\big)$

The largest element of $\mathbf{M}_1''$, $(t-1)\theta' -t$, is always less than $(t-1)\theta'$, which is the smallest element of $\mathbf{M}_2'$. This results in: \begin{equation}\label{eq:m'2capm''12} \mathbf{M}_2' \cap \mathbf{M}_1'' = \emptyset\end{equation}

\item Calculating $\big(\mathbf{M}_1' \cap \mathbf{M}_2''\big)$

The largest element of $\mathbf{M}_1'$, \ie $(t-1)\theta'-z -t$ is always less than $(t-1)\theta'+ts$, which is the smallest element of $\mathbf{M}_2''$. This results in: \begin{equation}\label{eq:m'1capm''22}\mathbf{M}_1' \cap \mathbf{M}_2'' = \emptyset \end{equation}

\item Calculating $\big(\mathbf{M}_2' \cap \mathbf{M}_2''\big)$

\begin{align}\label{eq:m'2capm''22}
      \mathbf{M}_2' \cap \nonumber \mathbf{M}_2'' = &\{(t-1)\theta',\ldots,+\infty\} \cap \\ &\{ts+(t-1)\theta',\ldots,+\infty\} \nonumber \\
      = & \{ts+(t-1)\theta',\ldots,+\infty\}.
\end{align}
\end{itemize}
From (\ref{P(SB)=P'(SB)capP''(SB)}), (\ref{eq:m'1capm''12}), (\ref{eq:m'2capm''12}), (\ref{eq:m'1capm''22}), and (\ref{eq:m'2capm''22}), we have:
\begin{align}\label{eq:P(RB)_set_representation2}
      \mathbf{P'}(S_B(x)) \cap \mathbf{P''}(S_B(x))
      = & \{ts+(t-1)\theta',\ldots,+\infty\}.
\end{align}
$\mathbf{P}(S_B(x))$ is formed by selecting the $z$ smallest elements of the set shown in (\ref{eq:P(RB)_set_representation2}):
\begin{align}
    \mathbf{P}(S_B(x)) = \{ts+(t-1)\theta',\ldots,ts+(t-1)\theta'+z-1\}
\end{align}
This completes the proof.
\hfill $\Box$

\begin{lemma}\label{lem:P(SB)-z small}
If $\frac{\theta'-ts-t+1}{2} <z \leq \theta'-ts-t$ and $s,t \neq 1$, the subsets of all powers of polynomials $S_B(x)$ with non-zero coefficients is defined as the following:
\begin{align}\label{eq:P(RB)_set_representation-zsmall}
    \mathbf{P}(S_B(x)) = & \mathbf{P'}(S_B(x)) \cap \mathbf{P''}(S_B(x)) \nonumber \\
    = & \Big(\bigcup\limits_{l''=0}^{p'-1}\{ts+\theta'l'', \ldots, (l''+1)\theta'-z-t\}\Big) \nonumber \\ & \cup \{ts+p'\theta',\dots, \nonumber \\ & ts+p'\theta'+z-1-p'(\theta'-t-ts-z+1)\}
    \end{align}
    \begin{align}
    =&\{ts+\theta'l'+d, d\in\Omega_0^{\theta'-t-ts-z}, l'\in \Omega_0^{p'-1}\} \nonumber \\
    &\cup \{ts+\theta'p'+v, v\in\Omega_0^{z-1-p'(\theta'-ts-t-z+1)}\}\label{eq:psb3}.
\end{align}
\end{lemma}
{\em Proof:}
In this scenario, $\mathbf{P'}(S_B(x))$ and $\mathbf{P''}(S_B(x))$ are equal to the previous case, as shown in (\ref{P'(SB) set representation-M'--}) and (\ref{eq:def-M1'_M1''_M2'_M2''case2}). The difference between this case and the previous case is that $\big(\mathbf{M}_1' \cap \mathbf{M}_1''\big)$ is no longer an empty set. The reason is that as we can see in Fig. \ref{fig:M1'M1''overlap}, each $l^{\text{th}}$ subset of $\mathbf{M}_1'$, \ie $\{\theta' l',\ldots,(l'+1)\theta'-z-t\}, l'=l-1$ has overlap with each $l^{\text{th}}$ subset of $\mathbf{M}_1''$, \ie $\{ts+\theta' l'',\ldots,(l''+1)\theta'-t\}, l''=l-1$: 

\begin{figure}[t]
		\centering
		\includegraphics[width=9.2cm]{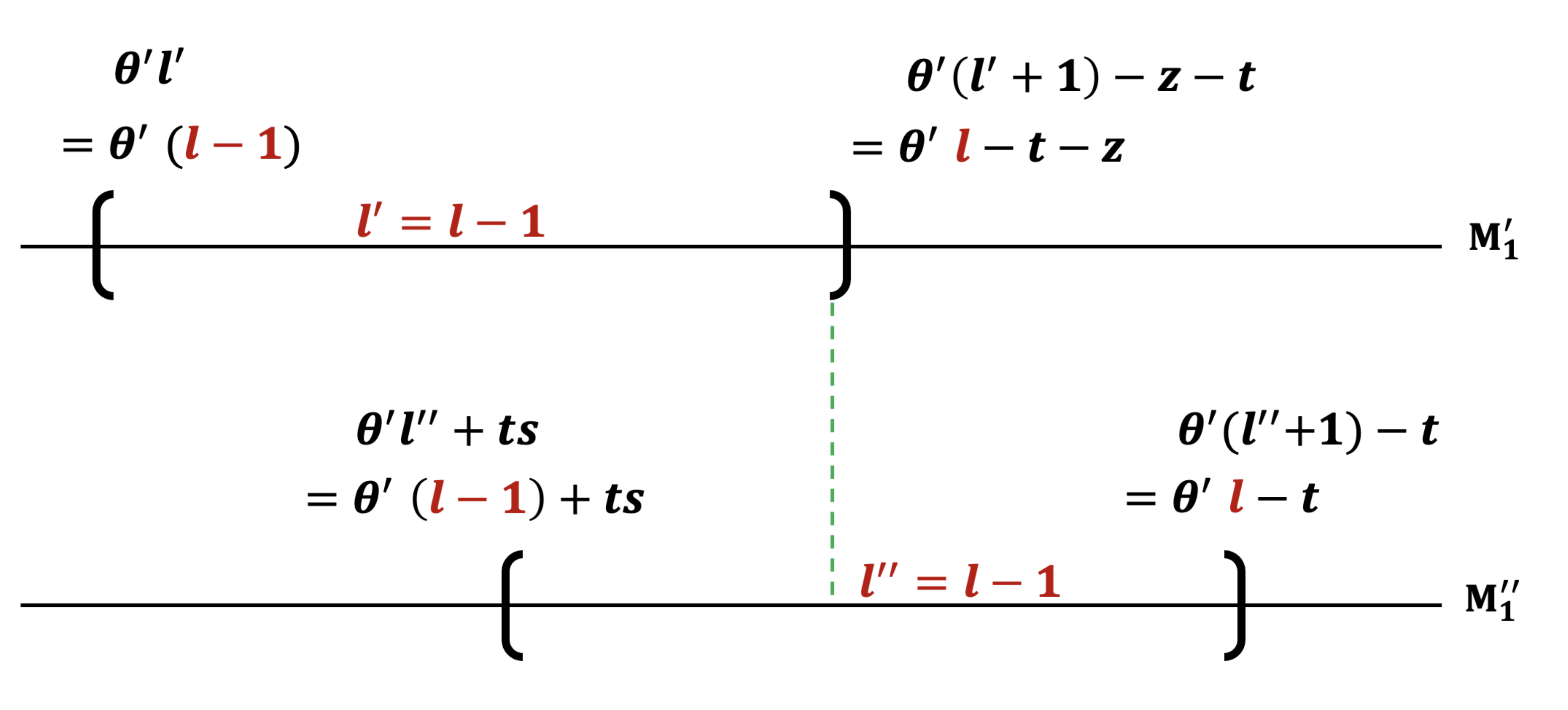}
	\caption{Illustration of the overlap between $\mathbf{M}_1'$ and $\mathbf{M}_1''$ in Lemma \ref{lem:P(SB)-z small}.
	}
\label{fig:M1'M1''overlap}
\vspace{5pt}
\end{figure}

\begin{align}
     & z \leq \theta'-ts-t \nonumber \\
    & \Rightarrow -z \geq -\theta'+ts+t \nonumber \\ 
    & \Rightarrow \theta' l -t-z \geq \theta' (l-1)+ts \nonumber \\ 
    & \Rightarrow \theta' l -t-z > \theta' (l-1)+ts \nonumber \\
    & \Rightarrow \theta'(l-1) < ts+\theta'(l-1) < l\theta'-z-t < l\theta'-t.
\end{align}
Therefore, we have:
\begin{equation}\label{eq:m'1capm''13}
    \mathbf{M}_1' \cap \mathbf{M}_1'' = \bigcup\limits_{l''=0}^{t-2}\{ts+\theta'l'', \ldots, (l''+1)\theta'-z-t\}
\end{equation}

$(\mathbf{M}_2' \cap \mathbf{M}_1''), (\mathbf{M}_1' \cap \mathbf{M}_2'')$, and $(\mathbf{M}_2' \cap \mathbf{M}_2'')$ can be calculated the same way as they are calculated in the previous case. Therefore, from (\ref{P(SB)=P'(SB)capP''(SB)}), (\ref{eq:m'1capm''13}), (\ref{eq:m'2capm''12}), (\ref{eq:m'1capm''22}), and (\ref{eq:m'2capm''22}), we have:
\begin{align}\label{eq:P(RB)_set_representation when z less than alpha-ts-t}
    &\mathbf{P'}(S_B(x)) \cap \mathbf{P''}(S_B(x)) \nonumber \\
    = &\bigcup\limits_{l''=0}^{t-2}\{ts+\theta'l'', \ldots, (l''+1)\theta'-z-t\} \nonumber \\ &\cup \{ts+(t-1)\theta',\dots,\infty\}.
\end{align}
$\mathbf{P}(S_B(x))$ is formed by selecting the $z$ smallest elements of the set shown in (\ref{eq:P(RB)_set_representation when z less than alpha-ts-t}).
This set consists of $t-1$ finite sets and one infinite set, where each finite set contains $(\theta'-ts-t-z+1)=(ts-2t-z+1)$\footnote{$\theta'$ is defined as $\theta'=t(2s-1)$.} elements. For the case of 
$\frac{\theta'-ts-t+1}{2} < z \leq \theta'-ts-t$, or equivalently $\frac{t(s-2)+1}{2} < z \leq t(s-2)$, $z$ is greater than $ts-2t-z+1$ and thus more than one finite set of (\ref{eq:P(RB)_set_representation when z less than alpha-ts-t}) is required to form $\mathbf{P}(S_B(x))$. Therefore we select $p'+1 \ge 2$ sets, where $p'$ is defined as $p'=
\min\{\floor{\frac{z-1}{ts-2t-z+1}},t-1\}$. With this definition, the first $p$ selected intervals are selected in full, in other words, we select $p'(ts-2t-z+1)$ elements to form the first $p'$ intervals of $\mathbf{P}(S_B(x))$. The remaining $z-p'(ts-2t-z+1)=z-p'(\theta'-t-ts-z+1)$ elements are selected from the $(p'+1)^{\text{st}}$ interval of (\ref{eq:P(RB)_set_representation when z less than alpha-ts-t}). This results in:

\begin{align} 
    \mathbf{P}(S_B(x)) 
    = &\{ts,\dots,\theta'-t-z\} \nonumber \\ & \cup \{ts+\theta',\dots,2\theta'-t-z\} \cup \dots \nonumber \\ & \cup \{ts+p'\theta',\dots, \nonumber \\ & ts+p'\theta'+z-1-p'(\theta'-t-ts-z+1)\}. \nonumber
\end{align}
This completes the proof.
\hfill $\Box$
\begin{lemma}\label{lem:P(SB)-z very small}
If $z \leq \frac{\theta'-ts-t+1}{2}$ and $s,t \neq 1$, the subsets of all powers of polynomial $S_B(x)$ with non-zero coefficients is defined as the following:
\begin{align}\label{eq:P(RB)_set_representation-z very small}
\mathbf{P}(S_B(x)) & = \{ts,\dots,ts+z-1\}\nonumber \\
&=\{ts+v, v\in \Omega_0^{z-1}\}.
\end{align}
\end{lemma}
{\em Proof:}
This case is similar to the previous case, where $\frac{\theta'-ts-t+1}{2} < z \leq \theta'-ts-t$, with the difference that the first subset of (\ref{eq:P(RB)_set_representation when z less than alpha-ts-t}) is sufficient to form $\mathbf{P}(S_B(x))$. The reason is that:
\begin{align}
     & z \leq \frac{\theta'-ts-t+1}{2} \nonumber \\
    & \Rightarrow z \leq \theta'-ts-t-z+1,
    & \Rightarrow z \leq ts-2t-z+1,
\end{align}
and thus the first subset with $ts-2t-z+1$ elements is sufficient to form $z$ elements of $\mathbf{P}(S_B(x))$ as shown in (\ref{eq:P(RB)_set_representation-z very small}). This completes the proof. \hfill $\Box$

\begin{lemma}\label{lem:P(SB)-s=1}
If $s=1$, the set of all powers of polynomial $S_B(x)$ with non-zero coefficients is defined as the following:
\begin{align}\label{eq:finiteP(BA)-s=1}
\mathbf{P}(S_B(x)) = \{t^2,\dots,t^2+z-1\}, \nonumber\\
= \{t^2+r, r\in \Omega_0^{z-1}\}.
\end{align}
\end{lemma}
{\em Proof:}
In this scenario, from lemma \ref{lem:P'(SB)-s=1}, we have $\mathbf{P'}(S_B(x))=\{0,\dots,+\infty\}$, and from Lemma \ref{lem:p''SB-s=1} we have $\mathbf{P''}(S_B(x))=\{t^2,\dots,+\infty\}$. Therefor, in this scenario the intersection of $\mathbf{P'}(S_B(x))$ and $\mathbf{P''}(S_B(x))$ is equal to $\{t^2,\dots,+\infty\}$, and $\mathbf{P}(S_B(x))$ is formed by selecting the $z$ smallest elements of $\{t^2,\dots,+\infty\}$, as shown in (\ref{eq:finiteP(BA)-s=1}). This completes the proof. \hfill $\Box$

\begin{lemma}\label{lem:P(SB)-t=1}
If $t=1$, the set of all powers of polynomial $S_B(x)$ with non-zero coefficients is defined as the following:
\begin{align}\label{eq:finiteP(BA)-t=1}
\mathbf{P}(S_B(x)) = \{s,\dots,s+z-1\}, \nonumber\\
= \{s+r, r\in \Omega_0^{z-1}\}.
\end{align}
\end{lemma}
{\em Proof:}
In this scenario, from lemma \ref{lem:P'(SB)-t=1}, we have $\mathbf{P'}(S_B(x))=\{0,\dots,+\infty\}$, and from Lemma \ref{lem:p''SB-t=1} we have $\mathbf{P''}(S_B(x))=\{s,\dots,+\infty\}$. Therefor, in this scenario the intersection of $\mathbf{P'}(S_B(x))$ and $\mathbf{P''}(S_B(x))$ is equal to $\{s,\dots,+\infty\}$, and $\mathbf{P}(S_B(x))$ is formed by selecting the $z$ smallest elements of $\{s,\dots,+\infty\}$, as shown in (\ref{eq:finiteP(BA)-t=1}). This completes the proof. \hfill $\Box$ 
%

$S_A(x)$ in (\ref{eq:FA1andFA2PolyDotCMPC}) can be directly derived from Lemmas \ref{lem:P(SA)-z large}, \ref{lem:P(SA)-z small}, and \ref{lem:P(SA)-s=1t=1}. Note that (i) when $z\leq \theta'-ts$, we have $p=0$ by definition and thus $ts+\theta' p+u$ in (\ref{eq:FA2PolyDotCMPC}) is equal to $ts+u$ in (\ref{eq:finiteP(SA)-polydot-second-scenario}), (ii) when $s=1$, we have $p=t-1$ and $\theta'=t$ by definition and thus $ts+\theta' p+u$ in (\ref{eq:FA2PolyDotCMPC}) is equal to $t^2+u$ in (\ref{eq:psafors=1}), and (iii) when $t=1$, we have $p=0$ by definition and thus $ts+\theta' p+u$ in (\ref{eq:FA2PolyDotCMPC}) is equal to $s+u$ in (\ref{eq:psafort=1}). 
%
%
Next we explain how to derive (\ref{eq:FBPolyDotCMPC}).

$S_B(x)$ in (\ref{eq:FBPolyDotCMPC}) can be directly derived from Lemmas \ref{lem:P(SB)-z large}, \ref{lem:P(SB)-z medium}, \ref{lem:P(SB)-z small}, \ref{lem:P(SB)-z very small}, \ref{lem:P(SB)-s=1}, and \ref{lem:P(SB)-t=1}. Note that (i) when $z > \theta'-ts, s,t \neq 1$ or $\theta'-ts-t < z \leq \theta'-ts, s,t \neq 1$, $\mathbf{P}(S_B(x))$ in (\ref{eq:P(RB)_set_representation-z large}) and (\ref{eq:P(RB)_set_representation-z medum}) is equal to the powers of $S_B(x)$ in (\ref{eq:FB1PolyDotCMPC}), (ii) when $\frac{\tau+1}{2}=\frac{\theta'-ts-t+1}{2} <z \leq \theta'-ts-t=\tau, s,t \neq 1$, $\mathbf{P}(S_B(x))$ in (\ref{eq:psb3}) is equal to the powers of $S_B(x)$ in (\ref{eq:FB2PolyDotCMPC}), (iii) when $z \leq \frac{\theta'-ts-t+1}{2}, s,t \neq 1$, $\mathbf{P}(S_B(x))$ in (\ref{eq:P(RB)_set_representation-z very small}) is equal to the powers of $S_B(x)$ in (\ref{eq:FB3PolyDotCMPC}), (iv) when $s=1$, we have $\theta' = t$ by definition, and thus $ts+\theta' (t-1)+r$ in (\ref{eq:FB1PolyDotCMPC}) is equal to $t^2+r$ in (\ref{eq:finiteP(BA)-s=1}), and (v) when $t=1$, $ts+\theta' (t-1)+r$ in (\ref{eq:FB1PolyDotCMPC}) is equal to $s+r$ in (\ref{eq:finiteP(BA)-t=1}). 

This completes the derivation of (\ref{eq:FA1andFA2PolyDotCMPC}) and (\ref{eq:FBPolyDotCMPC}).
\hfill $\Box$

\section*{Appendix B: Proof of Theorem \ref{th:N_PolyDot}}
To prove this theorem, we first consider the two cases of $t=1$ and $s=1$ separately and in the rest of this appendix, we consider $s, t \neq 1$.

\begin{lemma}\label{lemma:psit=1}
For $t=1$, $N_{\text{PolyDot-CMPC}}=2s+2z-1=(p+2)ts+\theta'(t-1)+2z-1=\psi_1$.\end{lemma}

{\em Proof:} For $t=1$, $p=0$ by definition. From (\ref{eq:FA2PolyDotCMPC}) and (\ref{eq:FB1PolyDotCMPC}) and by replacing $p$ with $0$, $F_A(x)$ and $F_B(x)$ are calculated as the following:
\begin{align}\label{eq:FA PolyDot-CMPC-t=1}
    F_{A}(x) = & \sum_{j=0}^{s-1} A_{j}x^{j}
    + \sum_{u=0}^{z-1}\bar{A}_{u}x^{s+u},
\end{align}
\begin{align}\label{eq:FB PolyDot-CMPC-t=1}
    F_{B}(x) = & \sum_{k=0}^{s-1} B_{k}x^{s-1-k}
    + \sum_{r=0}^{z-1}\bar{B}_{r}x^{s+r},
\end{align}
which are equal to the secret shares of Entangled-CMPC \cite{8613446}, for $t=1$. Thus, in this case PolyDot-CMPC and Entangled-CMPC are equivalent and as a result we have $N_{\text{PolyDot-CMPC}}=N_{\text{Entangled-CMPC}}=2s+2z-1$ \cite{8613446}, where by replacing $p=0$, we have $2s+2z-1=(p+2)ts+\theta'(t-1)+2z-1=\psi_1$. This completes the proof. \hfill $\Box$

\begin{lemma}\label{lemma:psis=1}
For $s=1$, 
\begin{align}
    N_{\text{PolyDot-CMPC}}
    =\begin{cases}
    2t^2+2z-1=\psi_1 & z>t \\
    t^2+2t+tz-1=\psi_6 & z \leq t
     \end{cases}
\end{align}
\end{lemma}
{\em Proof:}
For $s=1$, $\theta'=t$ and $p=t-1$ by definition. From (\ref{eq:FA2PolyDotCMPC}) and (\ref{eq:FB1PolyDotCMPC}) and by replacing $\theta'$ and $p$ with $t$ and $t-1$, respectively, $F_A(x)$ and $F_B(x)$ are calculated as the following:
\begin{align}\label{eq:FA PolyDot-CMPC-s=1}
    F_{A}(x) = & \sum_{i=0}^{t-1} A_{i}x^{i}
    + \sum_{u=0}^{z-1}\bar{A}_{u}x^{t^2+u},
\end{align}
\begin{align}\label{eq:FB PolyDot-CMPC-s=1}
    F_{B}(x) = & \sum_{l=0}^{t-1} B_{l}x^{tl}
    + \sum_{r=0}^{z-1}\bar{B}_{r}x^{t^2+r},
\end{align}
which are equal to the secret shares of Entangled-CMPC \cite{8613446}, for $s=1$. Thus, in this case PolyDot-CMPC and Entangled-CMPC are equivalent and as a result, we have:
\begin{align}
    &N_{\text{PolyDot-CMPC}}=\nonumber\\
    &N_{\text{Entangled-CMPC}}
    =\begin{cases}
    2t^2+2z-1 & z>t \\
    t^2+2t+tz-1 & z \leq t,
     \end{cases}
\end{align}
where by replacing $p=t-1$ and $\theta'=t$, we have $\psi_1=(p+2)ts+\theta'(t-1)+2z-1=2t^2+2z-1$ and $\psi_6=t^2+2t+tz-1$. This completes the proof. \hfill $\Box$

Now, we consider $s,t \neq 1$. The required number of workers is equal to the number of terms in $H(x)=F_A(x)F_B(x)$ with non-zero coefficients. The set of all powers in polynomial $H(x)$ with non-zero coefficients, shown by $\mathbf{P}({H}(x))$, is equal to:
\begin{align}\label{eq:PHx}
 \mathbf{P}({H}(x)) = \mathbf{D}_1 \cup  \mathbf{D}_2\cup \mathbf{D}_3 \cup \mathbf{D}_4,
 \end{align}
where
\begin{align}\label{eq:d1definition}
     & \mathbf{D}_1 = \mathbf{P}(C_A(x))+\mathbf{P}(C_B(x))
 \end{align}
\begin{align}\label{eq:d2definition}
    & \mathbf{D}_2  =\mathbf{P}(C_A(x))+\mathbf{P}(S_B(x))
\end{align}
\begin{align}\label{eq:d3definition}
    & \mathbf{D}_3=\mathbf{P}(S_A(x))+\mathbf{P}(C_B(x))
\end{align}
\begin{align}\label{eq:d4definition}
    & \mathbf{D}_4=\mathbf{P}(S_A(x))+\mathbf{P}(S_B(x))
\end{align}
%
 
Using (\ref{eq:polydot-p(CA)-th}) and (\ref{eq:polydot-p(CB)-th}), $\mathbf{D}_1$ is calculated as:
 \begin{align}\label{eq:d1}
      \mathbf{D}_1 = & \mathbf{P}(C_{A}(x))+\mathbf{P}(C_B(x)) \nonumber \\
     = &  \{i'+tj
     : 0 \leq i' \leq t-1,\; 0 \leq j \leq s-1,\} \nonumber \\ 
     &+ \{tq'+\theta' l' 
     : 0 \leq l' \leq t-1,\; 0 \leq q' \leq s-1\} \nonumber \\
     = &  \{i'+t(j+q')+\theta' l':0 \leq i',l' \leq t-1,\; \nonumber \\
     & 0 \leq j, q' \leq s-1,\}\nonumber \\
     = & \{i'+tj'+\theta' l':0 \leq i',l' \leq t-1,\; 0 \leq j' \leq 2s-2 \}\nonumber \\
     = & \{0,\ldots,t(2s-1)-1\} + \{\theta'l' : 0\leq l'\leq t-1\} \nonumber\\
     = & \{0,\ldots,\theta'-1\} + \{\theta'l': 0\leq l'\leq t-1\} \nonumber\\
     = & \{0,\ldots,t\theta'-1\}.
      \end{align}
In the following, we consider different regions for the value of $z$ and calculate $|\mathbf{P}({H}(x))|$ through calculation of $\mathbf{D}_2$, $\mathbf{D}_3$, and $\mathbf{D}_4$ for each region. In addition, we use the following lemma, which in some cases helps us to calculate $\mathbf{P}({H}(x))$ without requiring to calculate all of the terms $\mathbf{D}_2$, $\mathbf{D}_3$, and $\mathbf{D}_4$.
\begin{lemma}\label{lemma:UpperBoundPHx}
\begin{align}\label{eq:UpperBoundPHx}
    |\mathbf{P}({H}(x))|\leq& \deg(S_A(x))+\max\{\deg(S_B(x)), \deg(C_B(x))\}\nonumber\\&+1.
\end{align}
\end{lemma}
{\em Proof:} 
$|\mathbf{P}({H}(x))|$ which is equal to the number of terms in $H(x)$ with non-zero coefficients is less than or equal to the number of all terms, which is equal to $\deg(H(x))+1$:
\begin{align}\label{eq:phxUpperBound}
    |\mathbf{P}({H}(x))|\leq &\deg(H(x))+1 \nonumber \\ =&\deg((C_A(x)+S_A(x))(C_B(x)+S_B(x)))+1 \nonumber \\
    =&\max\{\deg(C_A(x)),\deg(S_A(x))\} \nonumber \\
    &+\max\{\deg(S_B(x), \deg(C_B(x))\}+1.
\end{align}
From (\ref{eq:polydot-p(CA)-th}), $\deg(C_A(x))=ts-1$. On the other hand, from (\ref{eq:psa12}) and (\ref{eq:finiteP(SA)-polydot-second-scenario}), $\deg(S_A(x))\ge ts$. Therefore, $\max\{\deg(C_A(x)),\deg(S_A(x))\} = \deg(S_A(x))$, which results in (\ref{eq:UpperBoundPHx}). This completes the proof. \hfill $\Box$

%
\begin{lemma}\label{lemma:non-zero-coeff-polydot}
For $z > ts$ or $t = 1$:
\begin{equation}
    |\mathbf{P}({H}(x))|= \psi_1=(p+2)ts+\theta'(t-1)+2z-1
\end{equation}
 \end{lemma}
 {\em Proof:} 
 To prove this lemma, we first calculate $\mathbf{D}_2$ from (\ref{eq:polydot-p(CA)-th}) and (\ref{eq:P(RB)_set_representation-z large}):
 \begin{align}\label{eq:d2}
    \mathbf{D}_2 = & \mathbf{P}(C_{A}(x))+\mathbf{P}(S_B(x)) \nonumber \\
    = & \{0,\ldots,ts-1\} + \{ts+(t-1)\theta',\ldots,ts+(t-1)\theta'+ \nonumber \\
    & z-1\} \nonumber \\
    = & \{ts+(t-1)\theta',\ldots,ts-1+ts+(t-1)\theta'+z-1\} \nonumber \\
    = & \{t\theta'-t(s-1),\ldots,t\theta'+t+z-2\}.
\end{align}
 
From (\ref{eq:d1}) and (\ref{eq:d2}), we can calculate $\mathbf{D}_{1} \cup \mathbf{D}_2$ as:
\begin{align}\label{eq:d12}
    \mathbf{D}_{12} = & \mathbf{D}_1 \cup \mathbf{D}_2 \nonumber \\
    = & \{0,\ldots,t\theta'-1\} \nonumber \\
    &\cup \{t\theta'-t(s-1),\ldots,t\theta'+t+z-2\} \nonumber \\
    = & \{0,\ldots,t\theta'+t+z-2\},
\end{align}
where the last equality comes from the fact that $t(s-1)\ge 0$ and thus $(t\theta'-1)+1\ge t\theta'-t(s-1)$. Next, we calculate $\mathbf{D}_4$ and its union with $\mathbf{D}_{12}$.

From (\ref{eq:psa12}) and (\ref{eq:P(RB)_set_representation-z large}), we have:
\begin{align}
     \mathbf{D}_4= &\mathbf{P}(S_A(x))+\mathbf{P}(S_B(x))  \nonumber \\
     =&\{ts+\theta'l+w, l\in\Omega_0^{p-1}, w\in \Omega_0^{t(s-1)-1}\} \nonumber \\
    &\cup \{ts+\theta'p+u, u\in\Omega_0^{z-1-pt(s-1)}\} \nonumber\\
    & + \{ts+(t-1)\theta'+r, \; 0 \leq r \leq z-1\} \nonumber \\
    = & \bigcup\limits_{l=0}^{p-1} \{2ts+(t-1+l)\theta',\ldots,2ts+(t-1+l)\theta'+ \nonumber \\
    & \quad \quad t(s-1)-1+z-1\}
     \nonumber \\
     & \cup \{2ts+(t-1+p)\theta',\ldots,2ts+\theta'p+(t-1)\theta'+z \nonumber \\
     & \quad -1-pt(s-1)+z-1\}\nonumber \\
    = & \bigcup\limits_{l=0}^{p-1} \{2ts+(t-1+l)\theta',\ldots,(t+l)\theta'+ts+z-2\}
     \nonumber \\
     & \cup \{2ts+(t-1+p)\theta',\ldots,(p+2)ts+\theta'(t-1)+ \nonumber \\
     & \quad \quad 2z-2\} \label{eq:d4subsets}\\
     = &\{2ts+(t-1)\theta',\ldots,(p+2)ts+\theta'(t-1)+2z-2\}\label{eq:d42},
 \end{align}
 where the last equality comes from the fact that there is no gap between each two consecutive subsets of (\ref{eq:d4subsets}). The reason is that:
 \begin{align}
    ts < z  
    \Rightarrow &  ts \leq z-1 \nonumber \\
    \Rightarrow & 2ts \leq ts+z-1 \nonumber \\
    \Rightarrow & 2ts+(t+l)\theta' \leq ((t+l)\theta'+ts+z-2)+1.
\end{align}
\begin{figure*}
		\centering
		\includegraphics[width=14cm]{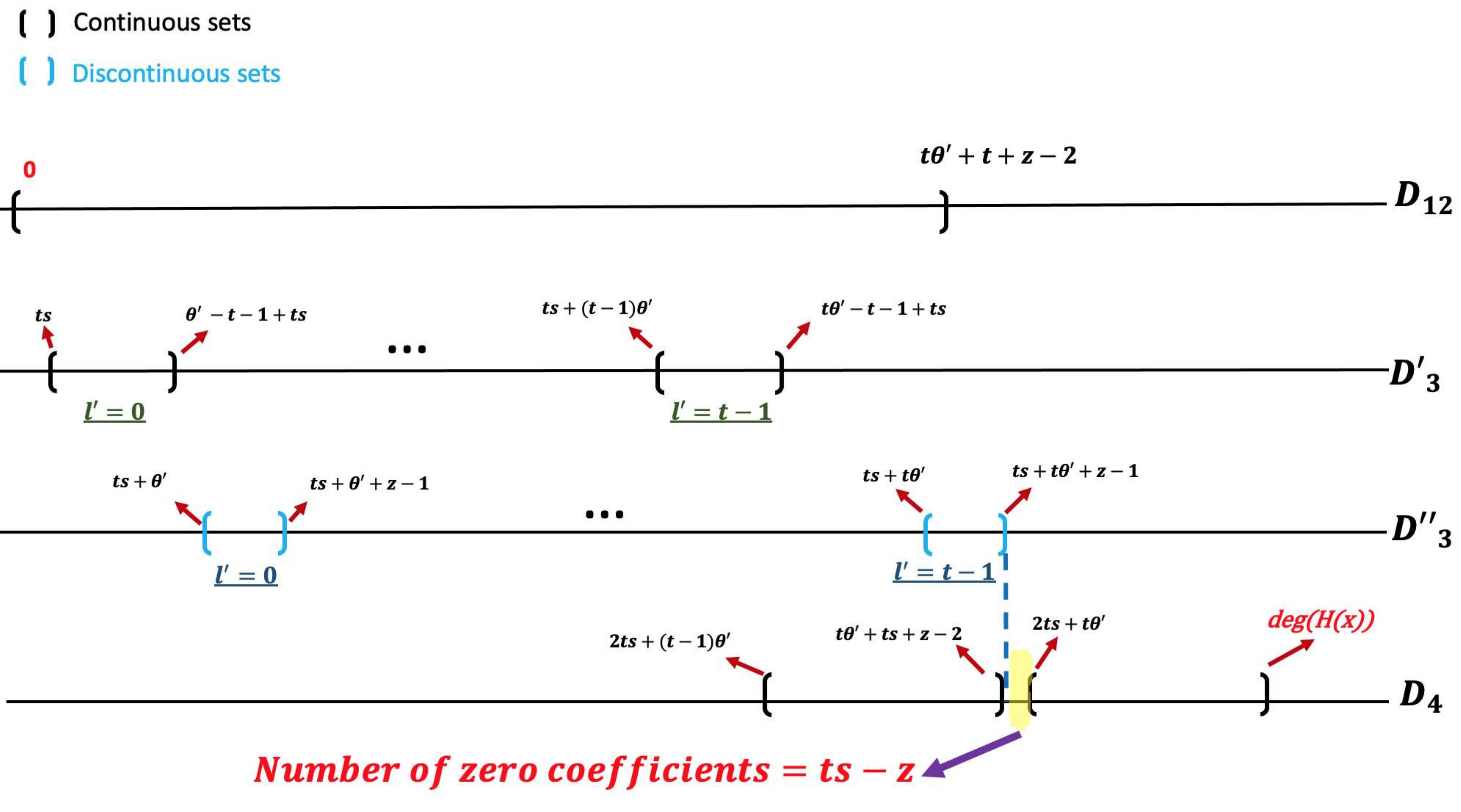}
	\caption{Illustration of $\mathbf{D}_{12} \cup \mathbf{D}_{3} \cup \mathbf{D}_{4}$ for $\theta'-ts<z \leq ts$.
	}
\label{Numberof zerocoeffslemma17-revised}
\vspace{-15pt}
\end{figure*}
Now, we calculate $\mathbf{D}_{12} \cup \mathbf{D}_4$. From (\ref{eq:d12}) and (\ref{eq:d42}), we have:
\begin{align}\label{eq:d124--}
    \mathbf{D}_{1} \cup& \mathbf{D}_2 \cup  \mathbf{D}_4 =\mathbf{D}_{12} \cup \mathbf{D}_4 = \{0,\ldots,t\theta'+t+z-2\}\cup \nonumber \\ & \{2ts+(t-1)\theta',\ldots,(p+2)ts+\theta'(t-1)+2z-2\}\nonumber \\
    &=\{0,\ldots,(p+2)ts+\theta'(t-1)+2z-2\},
\end{align} 
where the last equality comes from the fact that $\mathbf{D}_{12}$ has overlap with $\mathbf{D}_4$ and the upper bound of $\mathbf{D}_4$ is larger than the upper bound of $\mathbf{D}_{12}$. The reason is that:
\begin{align}
0 \leq z-2 
\Rightarrow & 2ts-2ts+t \leq t+z-2 \nonumber \\
\Rightarrow & 2ts-\theta' \leq t+z-2\nonumber \\
\Rightarrow & 2ts + (t-1)\theta' \leq t\theta'+t+z-2,
\end{align}
and
\begin{align}
0 < pts+&z 
\Rightarrow t < pts+t+z \nonumber \\
\Rightarrow & t < (p+2)ts-t(2s-1)+z\nonumber \\
\Rightarrow & t\theta'+t+z-2 < (p+2)ts+\theta'(t-1)+2z-2.
\end{align}
On the other hand, from (\ref{eq:UpperBoundPHx}), (\ref{eq:finite_P(RA)_set_representation-z large}), (\ref{eq:P(RB)_set_representation-z large}), and (\ref{eq:polydot-p(CB)-th}), $|\mathbf{P}({H}(x))|$ is upper bounded by: 
\begin{align}\label{eq:phxUpper}
    |\mathbf{P}({H}(x))|&\leq \deg(S_A(x))+\max\{\deg(S_B(x), \deg(C_B(x))\}\nonumber\\
    &+1\nonumber\\
    &=ts+p\theta'+z-1-p(\theta'-ts)+\nonumber\\
    &\max\{ts+(t-1)\theta'+z-1, t(s-1)+\theta'(t-1)\}\nonumber\\
    &+1\nonumber\\
    &=ts+p\theta'+z-1-p(\theta'-ts)+\nonumber\\
    &ts+(t-1)\theta'+z-1+1,\nonumber\\
    &=(p+2)ts+\theta'(t-1)+2z-1.
\end{align}
From (\ref{eq:PHx}) and (\ref{eq:d124--}), $|\mathbf{P}({H}(x))|$ is lower bounded by:
\begin{align}\label{eq:phxLower}
    |\mathbf{P}({H}(x))|&\ge |\mathbf{D}_1\cup\mathbf{D}_2\cup\mathbf{D}_4|\nonumber\\
    &=(p+2)ts+\theta'(t-1)+2z-1.
\end{align}
From (\ref{eq:phxUpper}) and (\ref{eq:phxLower}), $|\mathbf{P}({H}(x))|=(p+2)ts+\theta'(t-1)+2z-1$. This completes the proof. \hfill $\Box$

\begin{lemma}\label{lemma:non-zero-coeff-polydot-p=1&ts-t<z<ts}
For $\theta'-ts < z \leq ts$ and $s,t \neq 1$:
\begin{equation}
    |\mathbf{P}({H}(x))|= \psi_2=2ts+\theta'(t-1)+3z-1
\end{equation}
\end{lemma}
{\em Proof:}
For $\theta'-ts < z \leq ts$, $\mathbf{D}_1$ and $\mathbf{D}_2$ are calculated as (\ref{eq:d1}) and (\ref{eq:d2}) and thus from (\ref{eq:d12}), $\mathbf{D}_{12}$ is equal to:
\begin{align}\label{eq:d122}
    \mathbf{D}_{12} = \mathbf{D}_1 \cup \mathbf{D}_2=\{0,\ldots,t\theta'+t+z-2\},
\end{align}
Next, we calculate $\mathbf{D}_4$ and $\mathbf{D}_3$. We note that $p$ is equal to 1. The reason is that for this region of $z$, we have:
\begin{align}
\theta'-ts < z \leq ts 
\Rightarrow &\theta'-ts \leq z-1 < ts  \nonumber \\
\Rightarrow &\theta'-ts \leq z-1 < ts+t(s-2) \nonumber \\
\Rightarrow & \theta'-ts \leq z-1 < 2ts-2t\nonumber \\
\Rightarrow & \theta'-ts \leq z-1 < 2\theta'-ts \nonumber \\
\Rightarrow & p = \min\{\floor{\frac{z-1}{\theta'-ts}},t-1\} = 1.
\end{align}
By replacing $p$ with 1 in (\ref{eq:finite_P(RA)_set_representation-z large}) and using (\ref{eq:P(RB)_set_representation-z large}), $\mathbf{D}_4$ is equal to:
\begin{align}
    \mathbf{D}_4 =&\mathbf{P}(S_A(x))+\mathbf{P}(S_B(x))\nonumber \\ 
    =&\{ts,\ldots,\theta'-1\}\cup \{ts+\theta',\ldots,2ts+z-1\}\nonumber\\
    &+\{ts+(t-1)\theta',\ldots,ts+(t-1)\theta'+z-1\}\nonumber\\
    =&\{2ts+(t-1)\theta',\ldots,t\theta'+ts+z-2\}\nonumber\\
    &\cup \{2ts+t\theta',\ldots,3ts+\theta'(t-1)+2z-2\}.
\end{align}
Using (\ref{eq:finite_P(RA)_set_representation-z large}) with $p=1$ and (\ref{eq:polydot-p(CB)-th}), $\mathbf{D}_3$ is equal to:
\begin{align}\label{eq:d32}
    \mathbf{D}_3
    =&\mathbf{P}(S_A(x))+\mathbf{P}(C_B(x))\nonumber \\ 
    =&\{ts,\ldots,\theta'-1\}\cup \{ts+\theta',\ldots,2ts+z-1\}\nonumber\\
    &+\{tq'+\theta'l', 0\leq l' \leq t-1, 0\leq q' \leq s-1\}\nonumber\\
    =&\mathbf{D}'_3 \cup \mathbf{D}''_3,
\end{align}
where $\mathbf{D}'_3$ and $\mathbf{D}''_3$ are defined as follows.
\begin{align}\label{eq:d'3}
    \mathbf{D}'_3=&\{ts,\ldots,\theta'-1\}\nonumber\\
    &+\{tq'+\theta'l', 0\leq l' \leq t-1, 0\leq q' \leq s-1\}\nonumber\\
    =&\bigcup\limits_{l'=0}^{t-1}\bigcup\limits_{q'=0}^{s-1} \{ts+tq'+\theta'l',\ldots,\theta'-1+tq'+\theta'l'\}\nonumber\\
    =&\bigcup\limits_{l'=0}^{t-1}
    \{ts+\theta'l',\ldots,\theta'-1+t(s-1)+\theta'l'\},
\end{align}
where the last equality comes from the fact that there is no gap between each two consecutive subsets of $\bigcup\limits_{q'=0}^{s-1} \{ts+tq'+\theta'l',\ldots,\theta'-1+tq'+\theta'l'\}$. The reason is that:
\begin{align}
s \ge 2 
\Rightarrow &st \ge 2t  \nonumber \\
\Rightarrow &t(2s-1) \ge ts+t \nonumber \\
\Rightarrow & \theta' \ge ts+t\nonumber \\
\Rightarrow & (\theta'-1+tq'+\theta'l')+1 \ge ts+t(q'+1)+\theta'l'.
\end{align}
$\mathbf{D}''_3$ is defined and calculated as:
\begin{align}\label{eq:d''3}
    \mathbf{D}''_3=&\{ts+\theta',\ldots,2ts+z-1\}\nonumber\\
    &+\{tq'+\theta'l', 0\leq l' \leq t-1, 0\leq q' \leq s-1\}\nonumber\\
    =&\bigcup\limits_{l'=0}^{t-1}\bigcup\limits_{q'=0}^{s-1} \{ts+\theta'+tq'+\theta'l',\ldots,\nonumber\\
    &\quad \quad \quad \quad 2ts+z-1+tq'+\theta'l'\}.
\end{align}
To calculate $\mathbf{D}_1 \cup \mathbf{D}_2 \cup \mathbf{D}_3 \cup \mathbf{D}_4$, we first calculate $\mathbf{D}_{12} \cup \mathbf{D}'_3$ using (\ref{eq:d122}) and (\ref{eq:d'3}):
\begin{align}\label{eq:d12Capd'3}
    \mathbf{D}_{12} \cup \mathbf{D}'_3 =& \{0,\ldots,t\theta'+t+z-2\} \nonumber \\ &\cup \bigcup\limits_{l'=0}^{t-1}
    \{ts+\theta'l',\ldots,\theta'-1+t(s-1)+\theta'l'\}\nonumber\\
    =&\mathbf{D}_{12},
\end{align}
where the last equality comes from the fact that the largest element of $\mathbf{D}'_3$, \ie $\theta'-1+t(s-1)+\theta'(t-1)$ is smaller than the largest element of $\mathbf{D}_{12}$, \ie $t\theta'+t+z-2$, as illustrated in Fig. \ref{Numberof zerocoeffslemma17-revised} and shown below:
\begin{align}
z > \theta'-ts 
&\Rightarrow z > ts-t  \nonumber \\
\Rightarrow &z > ts-t-(t-1) \nonumber \\
\Rightarrow &t+z-2 > ts-t-1 \nonumber \\
\Rightarrow &t\theta'+t+z-2 > \theta'-1+t(s-1)+\theta'(t-1).
\end{align}
Next, we calculate $\mathbf{D}_{12} \cup \mathbf{D}_4$ as demonstrated in Fig. \ref{Numberof zerocoeffslemma17-revised}:
\begin{align}\label{eq:d12Cupd4}
    &\mathbf{D}_{12}\cup \mathbf{D}_4\nonumber \\
    =&\{0,\ldots,t\theta'+t+z-2\} \nonumber \\
    & \cup \{2ts+(t-1)\theta',\ldots,t\theta'+ts+z-2\}\nonumber\\
    &\cup \{2ts+t\theta',\ldots,3ts+\theta'(t-1)+2z-2\}\nonumber\\
    =&\{0,\ldots,3ts+\theta'(t-1)+2z-2\} \nonumber \\
    & - \{t\theta'+ts+z-1, \dots, 2ts+t\theta'-1\}.
\end{align}
$z\leq ts$ results in the non-empty set of $\{t\theta'+ts+z-1, \dots, 2ts+t\theta'-1\}$ in the above equation. Now we calculate $\mathbf{D}_{12} \cup \mathbf{D}_4 \cup \mathbf{D}''_3$ using (\ref{eq:d12Cupd4}) and (\ref{eq:d''3}):
\begin{align}\label{eq:d12Capd4Capd''3}
    &\mathbf{D}_{12}\cup \mathbf{D}_4 \cup \mathbf{D}''_3\nonumber \\
    =&(\{0,\ldots,3ts+\theta'(t-1)+2z-2\} \nonumber \\
    & - \{t\theta'+ts+z-1, \dots, 2ts+t\theta'-1\})\nonumber\\
    &\cup \mathbf{D}''_3\nonumber\\
    =&\{0,\ldots,3ts+\theta'(t-1)+2z-2\} \nonumber \\
    & - \{t\theta'+ts+z, \dots, 2ts+t\theta'-1\},
\end{align}
where the last equality comes from the fact that $\mathbf{D}''_3 \subset \{0,\ldots,3ts+\theta'(t-1)+2z-2\}$\footnote{ The reason is that the largest element of $\mathbf{D}''_3$, \ie $t\theta'+ts+z-1$ is smaller than the largest element of $\{0,\ldots,3ts+\theta'(t-1)+2z-2\}$.} and $\mathbf{D}''_3 \cap (\{t\theta'+ts+z-1, \dots, 2ts+t\theta'-1\}) = \{t\theta'+ts+z-1\}$. From (\ref{eq:d122}), (\ref{eq:d32}), (\ref{eq:d12Capd'3}), and (\ref{eq:d12Capd4Capd''3}), we have:
\begin{align}
    \mathbf{D}_1 \cup \mathbf{D}_2 \cup \mathbf{D}_3 \cup \mathbf{D}_4 =&\{0,\ldots,3ts+\theta'(t-1)+2z-2\} \nonumber \\
    &- \{t\theta'+ts+z, \dots, 2ts+t\theta'-1\},
\end{align}
and thus from (\ref{eq:PHx}):
\begin{align}
 |\mathbf{P}({H}(x))|=&(3ts+\theta'(t-1)+2z-2)+1\nonumber\\
 &-(2ts+t\theta'-1-(t\theta'+ts+z)+1)\nonumber\\
 =&2ts+\theta'(t-1)+3z-1.
\end{align}
This completes the proof.
\hfill $\Box$ 

\begin{lemma}\label{lemma:non-zero-coeff-polydot-zless than alpha minus ts1}
For $\theta'-ts-t < z \leq \theta'-ts$ and $s,t \neq 1$:
\begin{equation}
    |\mathbf{P}({H}(x))|= \psi_3= 2ts+\theta'(t-1)+2z-1
\end{equation}
\end{lemma}
{\em Proof:} For $\theta'-ts-t < z \leq \theta'-ts$, $\mathbf{P}(S_B(x))$ is derived from (\ref{eq:P(RB)_set_representation-z medum}), which is equal to $\mathbf{P}(S_B(x))$ used in (\ref{eq:d2}). Therefore, $\mathbf{D}_{2}$ is equal to:
\begin{equation}
    \mathbf{D}_2=\{t\theta'-t(s-1),\ldots,t\theta'+t+z-2\},
\end{equation}
and thus using (\ref{eq:d12}), we have:
\begin{equation}
    \mathbf{D}_1 \cup \mathbf{D}_2 = \{0,\ldots,t\theta'+t+z-2\}.
\end{equation}
From (\ref{eq:finiteP(SA)-polydot-second-scenario}) and (\ref{eq:P(RB)_set_representation-z medum}), $\mathbf{D}_4$ is calculated as:
\begin{align}
    \mathbf{D}_4 =& \mathbf{P}(S_A(x))+\mathbf{P}(S_B(x)) \nonumber\\
    =& \{2ts+\theta'(t-1),\ldots,2ts+\theta'(t-1)+2z-2\}.
\end{align}
Now, from the above two equations, we calculate $\mathbf{D}_1 \cup \mathbf{D}_2 \cup \mathbf{D}_4$:
\begin{align}
    \mathbf{D}_1 \cup \mathbf{D}_2 \cup \mathbf{D}_4 = \{0,\ldots,2ts+\theta'(t-1)+2z-2\}, 
\end{align}
where the equality comes from the fact that:
\begin{align}
z \ge 1 
\Rightarrow &t+z-2+1 \ge t  \nonumber \\
\Rightarrow &(t\theta'+t+z-2)+1 \ge 2ts+\theta'(t-1),
\end{align}
and
\begin{align}
t < t+z 
&\Rightarrow t\theta'+t+z-2 < 2ts+\theta'(t-1)+2z-2.
\end{align}
Therefore, $|\mathbf{P}({H}(x))| \ge |\mathbf{D}_1 \cup \mathbf{D}_2 \cup \mathbf{D}_4|=(2ts+\theta'(t-1)+2z-2)+1$. On the other hand, from (\ref{eq:UpperBoundPHx}), (\ref{eq:finiteP(SA)-polydot-second-scenario}), and (\ref{eq:P(RB)_set_representation-z medum}), we have:
\begin{align}
    |\mathbf{P}({H}(x))|\leq& \deg(S_A(x))+\max\{\deg(S_B(x), \deg(C_B(x))\}\nonumber\\&+1\nonumber\\
    =&(ts+z-1)+\max\{ts+(t-1)\theta'+z-1,\nonumber\\
    &\quad \quad \quad \quad \quad \quad \quad \quad \quad  t(s-1)+\theta'(t-1)\}+1\nonumber\\
    =&2ts+\theta'(t-1)+2z-1.
\end{align}
This results in $|\mathbf{P}({H}(x))|=2ts+\theta'(t-1)+2z-1$, which completes the proof.
\hfill $\Box$ 

\begin{lemma}\label{lemma:non-zero-coeff-polydot-zless than alpha minus ts}
For $\frac{\theta'-ts-t+1}{2}< z \leq \theta'-ts-t$:
\begin{equation}\label{eq:equation_in_lemma30}
    |\mathbf{P}({H}(x))|= \max\{\theta' t+z,(p'+2)ts+p'(z+t-1)+2z-1\}
\end{equation}
\end{lemma}
{\em Proof:}
For $\frac{\theta'-ts-t+1}{2}< z \leq \theta'-ts-t$, $\mathbf{D}_2$ is calculated using (\ref{eq:polydot-p(CA)-th}) and (\ref{eq:P(RB)_set_representation-zsmall}):
\begin{align}\label{eq:d24}
    \mathbf{D}_2 = & \mathbf{P}(C_{A}(x))+\mathbf{P}(S_B(x)) \nonumber \\
    = & \{0,\ldots,ts-1\} + \nonumber \\
    \Big(&\bigcup\limits_{l''=0}^{p'-1}\{ts+\theta'l'', \ldots, (l''+1)\theta'-z-t\}\Big) \nonumber \\ 
    &\cup \{ts+p'\theta',\dots, \nonumber \\ 
    & \quad \quad ts+p'\theta'+z-1-p'(\theta'-t-ts-z+1)\} \nonumber \\
    =\Big(&\bigcup\limits_{l''=0}^{p'-1}\{ts+\theta'l'', \ldots, (l''+1)\theta'-z-t+ts-1\}\Big) \nonumber \\
    &\cup \{ts+p'\theta',\dots, \nonumber \\ 
    & ts+p'\theta'+z-1-p'(\theta'-t-ts-z+1)+ts-1\}\nonumber
    \end{align}
    \begin{align}
    =\Big(&\bigcup\limits_{l''=0}^{p'-1}\{ts+\theta'l'', \ldots, (l''+1)\theta'-z-t+ts-1\}\Big) \nonumber \\
    &\cup \{ts+p'\theta',\dots,2ts+p'(t+ts+z-1)+z-2\}
\end{align}
From (\ref{eq:d1}) and (\ref{eq:d24}), $\mathbf{D}_1 \cup \mathbf{D}_2$ is equal to:
\begin{align}\label{eq:---d124}
    &\mathbf{D}_{12} = \mathbf{D}_1 \cup \mathbf{D}_2 \nonumber \\
    = & \{0,\ldots,t\theta'-1\} \cup \nonumber \\
    \Big(&\bigcup\limits_{l''=0}^{p'-1}\{ts+\theta'l'', \ldots, (l''+1)\theta'-z-t+ts-1\}\Big) \nonumber \\
    &\cup \{ts+p'\theta',\dots,2ts+p'(t+ts+z-1)+z-2\}\nonumber\\
    =& \{0,\dots,\max\{2ts+p'(ts+z+t-1)+z-2 , t\theta'-1\}\},
\end{align}
where the last equality comes from the fact that $\mathbf{D}_1$ has overlap with the last subset of $\mathbf{D}_2$, as shown below:
\begin{align}
    &p' \leq t-1 \nonumber \\
    &\Rightarrow p'\theta' \leq (t-1)\theta' \nonumber \\
    &\Rightarrow p'\theta'+ts \leq t\theta'-ts+t < t\theta'-1 \nonumber \\
    &\Rightarrow p'\theta'+ts < t\theta'-1.
\end{align}
From (\ref{eq:d3definition}), (\ref{eq:finiteP(SA)-polydot-second-scenario}) and (\ref{eq:polydot-p(CB)-th}), $\mathbf{D}_3$ is calculated as:
\begin{align}\label{eq:d34}
    \mathbf{D}_3 = \bigcup\limits_{l'=0}^{t-1}\bigcup\limits_{q'=0}^{s-1} \{ts+tq'+\theta'l',\ldots,ts+z-1+tq'+\theta'l'\}.
\end{align}
From (\ref{eq:d4definition}), (\ref{eq:finiteP(SA)-polydot-second-scenario}), and (\ref{eq:P(RB)_set_representation-zsmall}), $\mathbf{D}_4$ is calculated as:
\begin{align}\label{eq:d441}
    \mathbf{D}_4 = &\Big(\bigcup\limits_{l''=0}^{p'-1}\{2ts+\theta'l'', \ldots, ts-1+(l''+1)\theta'-t\}\Big) \nonumber \\
    &\cup \{2ts+p'\theta',\dots,2ts+p'(t+ts+z-1)+2z-2\}.
\end{align}
To calculate $\mathbf{D}_1 \cup \mathbf{D}_2 \cup \mathbf{D}_3 \cup \mathbf{D}_4$, we consider two cases of (i) $2ts+p'(ts+z+t-1)+z-2 \ge t\theta'-1$ and (ii) $2ts+p'(ts+z+t-1)+z-2 < t\theta'-1$.

(i) $2ts+p'(ts+z+t-1)+z-2 \ge t\theta'-1$: For this case, from (\ref{eq:---d124}), $\mathbf{D}_{12}$ is equal to:
\begin{align}\label{eq:d1242}
    \mathbf{D}_1 \cup \mathbf{D}_2=\{0,\dots,2ts+p'(ts+z+t-1)+z-2\}
\end{align}
From (\ref{eq:d441}) and (\ref{eq:d1242}), we have:
\begin{align}
    \mathbf{D}_1 \cup &\mathbf{D}_2 \cup \mathbf{D}_4 = \{0,\dots,2ts+p'(ts+z+t-1)+z-2\} \nonumber \\
    & \cup \{2ts+p'\theta',\dots,2ts+p'(t+ts+z-1)+2z-2\}\label{eq:Cupd4LastSubset} \\
    & = \{0,\dots,2ts+p'(t+ts+z-1)+2z-2\}\label{eq:d12Cupd441},
\end{align}
where (\ref{eq:Cupd4LastSubset}) and (\ref{eq:d12Cupd441}) come from the fact that each subset of $S_B(x)$ in (\ref{eq:P(RB)_set_representation-zsmall}) is designed to be non-empty:
\begin{align}
&ts+p'\theta' \leq ts+p'\theta'+z-1-p'(\theta'-t-ts-z+1)\nonumber\\
&\Rightarrow 2ts+p'\theta' \leq (2ts+p'(ts+z+t-1)+z-2)+1,
\end{align}
and $2ts+p'(ts+z+t-1)+z-2 < 2ts+p'(ts+z+t-1)+2z-2$. On the other hand,  from the condition considered in (i), the largest element of $\mathbf{D}_3$, \ie $ts+z-1+t(s-1)+\theta'(t-1)=z-1+\theta't$ is less than or equal to $(2ts+p'(ts+z+t-1)+z-2)+z=2ts+p'(t+ts+z-1)+2z-2$, and thus $\mathbf{D}_3 \subset \{0,\dots,2ts+p'(t+ts+z-1)+2z-2\}$:
\begin{align}\label{eq:d1d2d3d4i}
    \mathbf{D}_1 \cup &\mathbf{D}_2 \cup \mathbf{D}_3 \cup \mathbf{D}_4 =\nonumber \\ &\{0,\dots,2ts+p'(t+ts+z-1)+2z-2\},\nonumber \\
    & \text{for  } (2ts+p'(ts+z+t-1)+z-2) \ge t\theta'-1
\end{align}

(ii) $2ts+p'(ts+z+t-1)+z-2 < t\theta'-1$: For this case, from (\ref{eq:---d124}), $\mathbf{D}_{12}$ is equal to:
\begin{align}\label{eq:d1243}
    \mathbf{D}_1 \cup \mathbf{D}_2=\{0,\dots,t\theta'-1\}
\end{align} 
From (\ref{eq:d34}) and (\ref{eq:d1242}), we have:
\begin{align}
    \mathbf{D}_1 \cup \mathbf{D}_2 \cup \mathbf{D}_3 &= \{0,\dots,t\theta'-1\} \nonumber \\
    & \quad \cup \{ts+t(t-1)+\theta'(t-1),\ldots,\nonumber\\
    & \quad \quad ts+z-1+t(s-1)+\theta'(t-1)\} \nonumber \\
    &= \{0,\dots,t\theta'-1\} \cup \{t\theta',\ldots,t\theta'+z-1\} \nonumber \\
    & = \{0,\dots,t\theta'+z-1\}\label{eq:d12Cupd441--},
\end{align}
where the first equality comes from the fact that $\{0,\ldots,t\theta'-1\}$ has overlap with all subsets of $\mathbf{D}_3$ in (\ref{eq:d34}) except for the last subset. On the other hand, from the condition considered in (ii), the largest element of $\mathbf{D}_4$, \ie $2ts+p'(t+ts+z-1)+2z-2$ is less than $t\theta'+z-1$, and thus $\mathbf{D}_4 \subset \{0,\ldots,t\theta'+z-1\}$:
\begin{align}\label{eq:d1d2d3d4ii}
    \mathbf{D}_1 \cup &\mathbf{D}_2 \cup \mathbf{D}_3 \cup \mathbf{D}_4 =\{0,\dots,t\theta'+z-1\} \nonumber\\
    &\text{for  } (2ts+p'(ts+z+t-1)+z-2) < t\theta'-1
\end{align}

From (\ref{eq:d1d2d3d4i}) and (\ref{eq:d1d2d3d4ii}), we have:
\begin{align}
    &|\mathbf{P}({H}(x))|= | \mathbf{D}_1 \cup \mathbf{D}_2 \cup \mathbf{D}_3 \cup \mathbf{D}_4| \nonumber \\
    &= \max\{\theta' t+z,(p'+2)ts+p'(z+t-1)+2z-1\}
\end{align}
This completes the proof.
\hfill $\Box$
\begin{lemma}\label{lemma:psi4-clarification}
For $\frac{\theta'-ts-t+1}{2}<z \leq ts-2t-s+2$ and $s,t \neq 1$:
\begin{align}\label{eq:lemma31FirPart}
    |\mathbf{P}({H}(x))| = t\theta'+z
\end{align}
and for $\max\{st-2t-s+2,\frac{\theta'-ts-t+1}{2}\} < z \leq \theta'-ts-t$:
\begin{align}\label{eq:lemma31SecPart}
    &|\mathbf{P}({H}(x))| = \psi_4= (t+1)ts+(t-1)(z+t-1)+2z-1
\end{align} 
\end{lemma}
{\em Proof:}
To prove this lemma, first, we determine the condition for which $p'=t-1$ and the condition that $p'<t-1$:
\begin{align}\label{eq:p'value determination}
    p'=&\min\{\floor{\frac{z-1}{\theta'-ts-t-z+1}},t-1\}\nonumber\\
    &\begin{cases}
   =t-1 & z>st-2t-s+2\\
   <t-1 & z \leq st-2t-s+2,
\end{cases}
\end{align}
The above equation comes from the following:
\begin{align}
    & z \leq st-2t-s+2\nonumber\\
    \Rightarrow & z-1 < st-2t-s+2 \nonumber \\
    \Rightarrow & t(z-1) < t(s-2)(t-1) \nonumber \\
    \Rightarrow & z-1 < t(s-2)(t-1)-(t-1)(z-1) \nonumber \\
    \Rightarrow & z-1 < (ts-2t-z+1)(t-1) \nonumber \\
    \Rightarrow & \frac{z-1}{\theta'-ts-t-z+1} < t-1 \nonumber \\
    \Rightarrow & \floor{\frac{z-1}{\theta'-ts-t-z+1}} < t-1
\end{align}
Next, we decompose (\ref{eq:equation_in_lemma30}) to determine in which region $|\mathbf{P}({H}(x))| =\psi'_4= t\theta'+z$ and in which region $|\mathbf{P}({H}(x))| =\psi''_4= (t+1)ts+(t-1)(z+t-1)+2z-1$ when $\frac{\theta'-ts-t+1}{2}<z\leq \theta'-ts-t$. For this purpose, we calculate $\psi'_4 - \psi''_4$ as follows:
\begin{align}\label{eq:psi'4-psi''4}
    & \psi'_4-\psi''_4 \nonumber \\
   & = \theta't+z-(p'+2)ts-p'(z+t-1)-2z+1 \nonumber \\
   & = 2st^2-t^2+z-(p'+2)ts-p'(t-1)-z(p'+2)+1 \nonumber \\ 
   & = ts(2t-p'-2)-t(t+p')+p'+1-z(p'+1) \nonumber \\
   & = (p'+1)(ts(\frac{2t-p'-1-1}{p'+1})-t(\frac{p'+1+t-1}{p'+1})+1-z) \nonumber \\
   & = (p'+1)(ts(\frac{2t-2+1}{p'+1}-1)-t(\frac{t-1}{p'+1}+1)+1-z) \nonumber 
   \end{align}
   \begin{align}
   & = (p'+1)(ts(\frac{2t-1}{p'+1})-t(\frac{t-1}{p'+1})-(ts+t)+1-z) \nonumber \\
   & = (p'+1)(2ts(\frac{t-1/2}{p'+1})-t(\frac{t-1}{p'+1})-(ts+t)+1-z) \nonumber \\
   & = (p'+1)((\frac{t-1}{p'+1})(2ts-t)+\frac{ts}{p'+1}-(ts+t)+1-z) \nonumber \\
   & = (p'+1)(y-z),
\end{align}
Next, we consider the two cases of (i) $\max\{st-2t-s+2,\frac{\theta'-ts-t+1}{2}\} < z \leq \theta'-ts-t$ and (ii) $\frac{\theta'-ts-t+1}{2}<z \leq ts-2t-s+2$ and calculate $\psi'_4 - \psi''_4$ through comparison of $y$ and $z$.

(i) $\max\{st-2t-s+2,\frac{\theta'-ts-t+1}{2}\} < z \leq \theta'-ts-t$: For this case, from (\ref{eq:p'value determination}), $p'=t-1$ and from (\ref{eq:psi'4-psi''4}), $\psi'_4 - \psi''_4$ is calculated as:
\begin{align}
    \psi'_4 - \psi''_4 & =t(y-z) \nonumber\\
    & = (t-1)(2ts-t)+ts-t(ts+t)+t-tz \nonumber \\
    & = t(-2t-s+2+ts-z)\nonumber\\
    & < 0,
\end{align}
where the last inequality comes from the condition of (i). Therefore, for $\max\{st-2t-s+2,\frac{\theta'-ts-t+1}{2}\} < z \leq \theta'-ts-t$, we have $\max\{\psi'_4, \psi''_4\}=\psi''_4= (t+1)ts+(t-1)(z+t-1)+2z-1$. Since the condition of (i) is a subset of the condition considered in Lemma \ref{lemma:non-zero-coeff-polydot-zless than alpha minus ts}, \ie $\frac{\theta'-ts-t+1}{2} < z \leq \theta'-ts-t$, from (\ref{eq:equation_in_lemma30}), we have $|\mathbf{P}({H}(x))|= \max\{\theta' t+z,(p'+2)ts+p'(z+t-1)+2z-1\} = (t+1)ts+(t-1)(z+t-1)+2z-1$. This proves (\ref{eq:lemma31SecPart}).

(ii) $\frac{\theta'-ts-t+1}{2}<z \leq ts-2t-s+2$: For this case, from (\ref{eq:p'value determination}), $p'<t-1$ and from (\ref{eq:psi'4-psi''4}), $\psi'_4 - \psi''_4$ is calculated as:
\begin{align}
    \psi'_4 - \psi''_4 & =(p'+1)(y-z) \nonumber \\
    & > (p'+1)(\frac{t-1}{t}(2ts-t)+\frac{ts}{t}-(ts+t)+1-z) \nonumber \\
    & = (p'+1)((t-1)(2s-1)+s-(ts+t)+1-z)\nonumber\\
    & = (p'+1)(-s-2t+2+ts-z)\nonumber\\
    & \ge 0,
\end{align}
where the last inequality comes from the condition of (ii). Therefore, for $\frac{\theta'-ts-t+1}{2}<z \leq ts-2t-s+2$, we have $\max\{\psi'_4, \psi''_4\}=\psi'_4= t\theta'+z$. Since the condition of (ii) is a subset of the condition considered in Lemma \ref{lemma:non-zero-coeff-polydot-zless than alpha minus ts}, \ie $\frac{\theta'-ts-t+1}{2} < z \leq \theta'-ts-t$\footnote{This comes from the fact that $0\ge 2-s$ and thus $\theta'-ts-t=ts-2t \ge ts-2t-s+2$.}, from (\ref{eq:equation_in_lemma30}), we have $|\mathbf{P}({H}(x))|= \max\{\theta' t+z,(p'+2)ts+p'(z+t-1)+2z-1\} = \theta' t+z$. This proves (\ref{eq:lemma31FirPart}).

This completes the proof.  \hfill $\Box$

\begin{lemma}\label{lemma:non-zero-coeff-polydot-zless than (alpha minus ts minus t)/2}
For $z \leq \frac{\theta'-ts-t+1}{2}$:
\begin{equation}
    |\mathbf{P}({H}(x))|= t\theta'+z
\end{equation}
\end{lemma}
{\em Proof:}
For $z \leq \frac{\theta'-ts-t+1}{2}$, $\mathbf{P}(S_A(x))$ and $\mathbf{P}(S_B(x))$ are calculated from (\ref{eq:finiteP(SA)-polydot-second-scenario}) and (\ref{eq:P(RB)_set_representation-z very small}). Therefore, using (\ref{eq:polydot-p(CA)-th}) and (\ref{eq:polydot-p(CB)-th}), $\mathbf{D}_2, \mathbf{D}_3$, and $\mathbf{D}_4$ are equal to:
\begin{align}
    \mathbf{D}_2=&\mathbf{P}(C_A(x))+\mathbf{P}(S_B(x))=\{ts,\ldots,2ts+z-2\}\nonumber\\
    \mathbf{D}_3=&\mathbf{P}(S_A(x))+\mathbf{P}(C_B(x))\nonumber \\
    =&\bigcup\limits_{l'=0}^{t-1}\bigcup\limits_{q'=0}^{s-1} \{ts+tq'+\theta'l',\ldots,ts+z-1+tq'+\theta'l'\}\nonumber\\
    \mathbf{D}_3=&\mathbf{P}(S_A(x))+\mathbf{P}(S_B(x))=\{2ts,\ldots,2ts+2z-2\}
\end{align}
From (\ref{eq:d1}) and the above equations, we calculate $\mathbf{D}_1 \cup \mathbf{D}_2 \cup \mathbf{D}_3 \cup \mathbf{D}_4$ as follows:
\begin{align}
    &\mathbf{D}_1 \cup \mathbf{D}_2 \cup \mathbf{D}_3 \cup \mathbf{D}_4 =\{0,\dots,t\theta'-1\} \cup \nonumber \\
    &\{ts,\dots,2ts+z-2\} \cup \nonumber \\
    &\bigcup\limits_{l'=0}^{t-1}\bigcup\limits_{q'=0}^{s-1} \{ts+tq'+\theta'l',\ldots,ts+z-1+tq'+\theta'l'\}\nonumber\\
    &\cup \{2ts,\ldots,2ts+2z-2\} \nonumber
        \end{align}
    \begin{align}
    &=\{0,\dots,t\theta'-1\} \cup \{ts,\dots,2ts+z-2\} \cup \nonumber \\
    &\{ts+t(s-1)+\theta'(t-1),\ldots,\nonumber\\
    &ts+z-1+t(s-1)+\theta'(t-1)\}\nonumber\\
    &\cup \{2ts,\ldots,2ts+2z-2\}\label{eq:d1d2d3d46}\\
    &=\{0,\dots,t\theta'-1\} \cup \{ts,\dots,2ts+2z-2\} \cup \nonumber \\
    &\quad \quad \{\theta't,\ldots,\theta't+z-1\}\nonumber\\
    &=\{0,\ldots,\theta't+z-1\}\cup\{ts,\ldots,2ts+2z-2\}\nonumber\\
    &=\{0,\ldots,t\theta'+z-1\}\label{eq:d1d2d3d48},
\end{align}
where (\ref{eq:d1d2d3d46}) comes from the fact that all subsets of $\mathbf{D}_3$ except for the last one is subsets of $\{0,\ldots,t\theta'-1\}$ and (\ref{eq:d1d2d3d48}) comes from the fact that $2ts+2z-2<t\theta'+z-1$. The reason is that:
\begin{align}
    2ts+2z-2 \leq & 2ts+(\theta'-ts-t+1)-2 \nonumber \\
    = & 2\theta'-ts-1 \nonumber \\
    \leq & t\theta'-ts-1\nonumber \\
    <& t\theta'\nonumber\\
    \leq& t\theta'+z-1.
\end{align}
From (\ref{eq:d1d2d3d48}) we have:
\begin{align}
    &|\mathbf{P}({H}(x))|= | \mathbf{D}_1 \cup \mathbf{D}_2 \cup \mathbf{D}_3 \cup \mathbf{D}_4|= t\theta'+z
\end{align}
This completes the proof.
\hfill $\Box$
\begin{lemma}\label{lemma:psi5-clarification}
For $ z \leq \max\{st-2t-s+2,\frac{\theta'-ts-t+1}{2}\}$ and $s,t \neq 1$:
\begin{align}
    &|\mathbf{P}({H}(x))| = \psi_5= t\theta'+z
\end{align} 
\end{lemma}
{\em Proof:} To prove this lemma we consider two scenarios:

(i) $\frac{\theta'-ts-t+1}{2}<st-2t-s+2$: From Lemma \ref{lemma:psi4-clarification}, for $\frac{\theta'-ts-t+1}{2} < z \leq st-2t-s+2$, we have $|\mathbf{P}({H}(x))| = t\theta'+z$. On the other hand, from Lemma \ref{lemma:non-zero-coeff-polydot-zless than (alpha minus ts minus t)/2}, for $z\leq \frac{\theta'-ts-t+1}{2}$, we have $|\mathbf{P}({H}(x))| = t\theta'+z$. Therefore, we conclude that for $z \leq st-2t-s+2=$, we have $|\mathbf{P}({H}(x))| = t\theta'+z$.

(ii) $st-2t-s+2 \leq \frac{\theta'-ts-t+1}{2}$: 
From Lemma \ref{lemma:non-zero-coeff-polydot-zless than (alpha minus ts minus t)/2}, for $z\leq \frac{\theta'-ts-t+1}{2}$, we have $|\mathbf{P}({H}(x))| = t\theta'+z$.

From (i) and (ii), for $ z \leq \max\{st-2t-s+2,\frac{\theta'-ts-t+1}{2}\}$, $|\mathbf{P}({H}(x))| = t\theta'+z$. This completes the proof. \hfill $\Box$

The required number of workers, $N_{\text{PolyDot-CMPC}}$, is equal to $|\mathbf{P}({H}(x))|$. Therefore, from Lemmas (\ref{lemma:psit=1}), (\ref{lemma:psis=1}), (\ref{lemma:non-zero-coeff-polydot}), (\ref{lemma:non-zero-coeff-polydot-p=1&ts-t<z<ts}), (\ref{lemma:non-zero-coeff-polydot-zless than alpha minus ts1}), (\ref{lemma:psi4-clarification}), and (\ref{lemma:psi5-clarification}), Theorem \ref{th:N_PolyDot} is proved.

\section*{Appendix C: Proof of Lemmas \ref{lemma: regions where N_polydot<N_entangled}, \ref{lemma: regions where N_polydot<N_ssmm}, and \ref{lemma: regions where N_polydot<N_gcsana}}\label{app:appendixC}
\subsection{Proof of Lemma \ref{lemma: regions where N_polydot<N_entangled} (PolyDot-CMPC Versus Entangled-CMPC)}
To prove this lemma, we consider different regions for the value of $z$ and compare the required number of workers for PolyDot-CMPC, $N_{\text{PolyDot-CMPC}}$, with Entangled-CMPC, $N_{\text{Entangled-CMPC}}$, in each region. From \cite{8613446}, $N_{\text{Entangled-CMPC}}$ is equal to:
\begin{align}\label{eq:N Entang-CMPC}
N_{\text{Entangled-CMPC}}=\begin{cases}
   2st^2+2z-1,&z>ts-s\\
   st^2+3st-2s+tz-t+1,&z \leq ts-s,
\end{cases} 
\end{align}
and we use (\ref{eq:N-PolyDot-DMPC}) for $N_{\text{PolyDot-CMPC}}$ in each region.

(i) $ts<z \text{ or } t=1$: From (\ref{eq:N-PolyDot-DMPC}), $N_{\text{PolyDot-CMPC}} = \psi_1 = (p+2)ts+\theta'(t-1)+2z-1$ and from (\ref{eq:N Entang-CMPC}), $N_{\text{Entangled-CMPC}}=2st^2+2z-1$, thus we have:
\begin{align}\label{eq:compare psi-1 with Entang}
 & N_{\text{PolyDot-CMPC}} - N_{\text{Entangled-CMPC}} \nonumber \\
 = & (p+2)ts+\theta'(t-1)+2z-1-(2st^2+2z-1) \nonumber \\
 = & pts+2ts+(2ts-t)(t-1)+2z-1-2st^2-2z+1 \nonumber \\
 = & t(ps-t+1).
 \end{align} 
From the above equation, if $p<\frac{t-1}{s}$ and $t\neq 1$, we have $N_{\text{PolyDot-CMPC}}<N_{\text{Entangled-CMPC}}$, otherwise, $N_{\text{PolyDot-CMPC}} \ge N_{\text{Entangled-CMPC}}$\footnote{Note that for $t=1$, $N_{\text{PolyDot-CMPC}} = N_{\text{Entangled-CMPC}}$.}. This along with the condition of (i), provides condition 1 for $N_{\text{PolyDot-CMPC}}<N_{\text{Entangled-CMPC}}$ in Lemma \ref{lemma: regions where N_polydot<N_entangled}.

(ii) $ts-t < z \leq ts$ and $s,t \neq 1$: From (\ref{eq:N-PolyDot-DMPC}), $N_{\text{PolyDot-CMPC}} = \psi_2=2ts+\theta'(t-1)+3z-1$ and from (\ref{eq:N Entang-CMPC}), $N_{\text{Entangled-CMPC}}=2st^2+2z-1$ for $z>ts-s$ and $N_{\text{Entangled-CMPC}}=st^2+3st-2s+t(z-1)+1$ for $z\leq ts-s$, thus we have:

(a) $z>ts-s$ and $t-1>s$: For this case, we have:
\begin{align}
 & N_{\text{PolyDot-CMPC}} - N_{\text{Entangled-CMPC}} \nonumber \\
 = & 2ts+\theta'(t-1)+3z-1-(2st^2+2z-1) \nonumber \\
 = & 2ts+(2ts-t)(t-1)+3z-1-2st^2-2z+1 \nonumber \\
 = & z-t(t-1)\nonumber\\
 < & z-ts\label{eq:temppppps}\\
 \leq & 0,
\end{align}
where (\ref{eq:temppppps}) comes from the condition of (a), $t-1>s$ and the last inequality comes from the condition of (ii), $z\leq ts$. Therefore, for the combination of conditions (ii) and (a), \ie $ts-s<z\leq ts$ and $t-1>s$, we have $ N_{\text{PolyDot-CMPC}} < N_{\text{Entangled-CMPC}}$. This provides condition 2 for $N_{\text{PolyDot-CMPC}}<N_{\text{Entangled-CMPC}}$ in Lemma \ref{lemma: regions where N_polydot<N_entangled}.

(b) $z>ts-s$ and $s= t-1$: For this case, we have:
\begin{align}\label{eq:compare psi-2 with Entang-(ii-a)}
 & N_{\text{PolyDot-CMPC}} - N_{\text{Entangled-CMPC}} \nonumber \\
 = & 2ts+\theta'(t-1)+3z-1-(2st^2+2z-1) \nonumber \\
 = & 2ts+(2ts-t)(t-1)+3z-1-2st^2-2z+1 \nonumber \\
 = & z-(t^2-t)\nonumber\\
 \leq & 0,
\end{align} 
where the last inequality comes from the condition of (ii), $z\leq ts=t(t-1)$. From the above equation, for $z<t^2-t$, we have $N_{\text{PolyDot-CMPC}}<N_{\text{Entangled-CMPC}}$, otherwise, $N_{\text{PolyDot-CMPC}}=N_{\text{Entangled-CMPC}}$. By replacing $s$ with $t-1$ and combining the conditions of (ii), (b), and $z<t^2-t$, \ie $t^2-2t+1<z<t^2-t, s=t-1$, condition 3 for $N_{\text{PolyDot-CMPC}}<N_{\text{Entangled-CMPC}}$ in Lemma \ref{lemma: regions where N_polydot<N_entangled} is derived.

(c) $z>ts-s$ and $s> t-1$: For this case, we have:
\begin{align}
 & N_{\text{PolyDot-CMPC}} - N_{\text{Entangled-CMPC}} \nonumber \\
 = & 2ts+\theta'(t-1)+3z-1-(2st^2+2z-1) \nonumber \\
 = & 2ts+(2ts-t)(t-1)+3z-1-2st^2-2z+1 \nonumber \\
 = & z-t(t-1)\nonumber\\
 \ge & z-t(s-1)\label{eq:tempp54}\\
 > & 0,
\end{align} 
where (\ref{eq:tempp54}) comes from the condition of (c), $s>t-1$ and the last inequality comes from the condition of (ii), $z>ts-t$.

(d) $z\leq ts-s, t>3$: For this case, we have:
\begin{align}\label{eq:compare psi-2 with Entang-(ii-d)}
 & N_{\text{PolyDot-CMPC}} - N_{\text{Entangled-CMPC}} \nonumber \\
 = & 2ts+\theta'(t-1)+3z-1\nonumber\\
 &\quad \quad -(st^2+3st-2s+t(z-1)+1) \nonumber \\
 = & 2ts+(2ts-t)(t-1)+3z-1\nonumber\\
 &\quad \quad -st^2-3st+2s-tz+t-1 \nonumber \\
 = & st^2-t^2+2t-3st+2s-2-z(t-3) \nonumber \\
 = & st^2-3st-t^2+3t-t+3+2s-5-z(t-3) \nonumber \\
 = & st(t-3)-t(t-3)-(t-3)+2s-5-z(t-3) \nonumber \\
 = &(t-3)(st-t-1+\frac{2s-5}{t-3})-(t-3)z.
\end{align} 
From the above equation, if $z>(st-t-1+\frac{2s-5}{t-3})$, we have $ N_{\text{PolyDot-CMPC}} < N_{\text{Entangled-CMPC}}$\footnote{Note that in this case $t\ge 3$.}, otherwise $ N_{\text{PolyDot-CMPC}} \ge N_{\text{Entangled-CMPC}}$. By combining the conditions of (ii), (d), and $z>(st-t-1+\frac{2s-5}{t-3})$, \ie $ts-t-\min\{0,1-\frac{2s-5}{t-3}\}<z\leq ts-s, t>3$, condition 4 for $N_{\text{PolyDot-CMPC}}<N_{\text{Entangled-CMPC}}$ in Lemma \ref{lemma: regions where N_polydot<N_entangled} is derived.

(e) $z\leq ts-s, t=3$: For this case, we have:
\begin{align}\label{eq:compare psi-2 with Entang-(ii-e)}
 & N_{\text{PolyDot-CMPC}} - N_{\text{Entangled-CMPC}} \nonumber \\
 = & 2ts+\theta'(t-1)+3z-1\nonumber\\
 &\quad \quad -(st^2+3st-2s+t(z-1)+1) \nonumber \\
 = & 2ts+(2ts-t)(t-1)+3z-1\nonumber\\
 &\quad \quad -st^2-3st+2s-tz+t-1 \nonumber \\
 =&2s-5.
\end{align} 
From the above equation, if $s=2$, $N_{\text{PolyDot-CMPC}} < N_{\text{Entangled-CMPC}}$, otherwise $N_{\text{PolyDot-CMPC}} > N_{\text{Entangled-CMPC}}$. By combining the conditions of (ii), (e), and $s=2$, \ie $s=2, t=3,z=4$, condition 5 for $N_{\text{PolyDot-CMPC}}<N_{\text{Entangled-CMPC}}$ in Lemma \ref{lemma: regions where N_polydot<N_entangled} is derived.

(f) $z\leq ts-s, t=2$: This condition is not possible, because $s\ge 2$ and thus $2s-2 \ge s$. Therefore, there is no overlap between the condition of (ii), $z>ts-t=2s-2$ and the condition of (f), $z\leq ts-s=s$. 

(iii) $ts-2t < z \leq ts-t$ and $s,t \neq 1$: From (\ref{eq:N-PolyDot-DMPC}), $N_{\text{PolyDot-CMPC}} = \psi_3=2ts+\theta'(t-1)+2z-1$ and from (\ref{eq:N Entang-CMPC}), $N_{\text{Entangled-CMPC}}=2st^2+2z-1$ for $z>ts-s$ and $N_{\text{Entangled-CMPC}}=st^2+3st-2s+t(z-1)+1$ for $z\leq ts-s$, thus we have:

(a) $t\ge s$: For this case, we have:
\begin{align}\label{eq:compare psi-3 with Entang-iii-a}
 & N_{\text{PolyDot-CMPC}} - N_{\text{Entangled-CMPC}} \nonumber \\
 = & 2ts+\theta'(t-1)+2z-1 \nonumber\\
 &\quad \quad - (st^2+3st-2s+t(z-1)+1) \nonumber \\
 = & 2ts+(2ts-t)(t-1)+2z-1 \nonumber\\
 &\quad \quad -st^2-3st+2s-tz+t-1 \nonumber \\
 = & st^2-2st-st-t^2+2t+2s-2-z(t-2) \nonumber \\
 = & st(t-2)-t(t-2)-s(t-2)-2-z(t-2) \nonumber \\
 = & (t-2)(st-t-s-\frac{2}{t-2})-z(t-2).
\end{align} 
From the above equation, if $t=2$, $N_{\text{PolyDot-CMPC}} < N_{\text{Entangled-CMPC}}$. By replacing $t=2$ in the conditions of (iii) and (a), \ie $t=2, s=2, z=1,2$, condition 6 for $N_{\text{PolyDot-CMPC}}<N_{\text{Entangled-CMPC}}$ in Lemma \ref{lemma: regions where N_polydot<N_entangled} is derived. In addition, if $t>2$ and $z>st-t-s-\frac{2}{t-2}$, $N_{\text{PolyDot-CMPC}} < N_{\text{Entangled-CMPC}}$, otherwise, $N_{\text{PolyDot-CMPC}} \ge N_{\text{Entangled-CMPC}}$. By combining the conditions of (iii), (a), and $t>2, z>st-t-s-\frac{2}{t-2}$, \ie $\max\{st-t-s-\frac{2}{t-2}, ts-2t\} < z \leq ts-t, t>2, t\ge s$, condition 7 for $N_{\text{PolyDot-CMPC}}<N_{\text{Entangled-CMPC}}$ in Lemma \ref{lemma: regions where N_polydot<N_entangled} is derived.

(b) $2t \ge s>t, z>ts-s$: For this case, we have:
\begin{align}\label{eq:compare psi-3 with Entang-iii-b}
 & N_{\text{PolyDot-CMPC}} - N_{\text{Entangled-CMPC}} \nonumber \\
 = & 2ts+\theta'(t-1)+2z-1-(2st^2+2z-1) \nonumber \\
 = & 2ts+(2ts-t)(t-1)+2z-1-2st^2-2z+1 \nonumber \\
 = & -t(t-1)\nonumber\\
 <&0,
\end{align}
From the above equation, for this case, $N_{\text{PolyDot-CMPC}} < N_{\text{Entangled-CMPC}}$. By combining the conditions of (iii) and (b), \ie $t<s\leq 2t, ts-s<z\leq ts-t$, condition 8 for $N_{\text{PolyDot-CMPC}}<N_{\text{Entangled-CMPC}}$ in Lemma \ref{lemma: regions where N_polydot<N_entangled} is derived.

(c) $2t \ge s>t, z\leq ts-s$: For this case, we have:
\begin{align}\label{eq:compare psi-3 with Entang-iii-c}
 & N_{\text{PolyDot-CMPC}} - N_{\text{Entangled-CMPC}} \nonumber \\
 = & 2ts+\theta'(t-1)+2z-1 \nonumber \\
 -& (st^2+3st-2s+t(z-1)+1) \nonumber \\
 = & 2ts+(2ts-t)(t-1)+2z-1\nonumber\\
 &\quad \quad -st^2-3st+2s-tz+t-1 \nonumber 
\end{align}
\begin{align}
 = & st(t-2)-t(t-2)-s(t-2)-2-z(t-2) \nonumber \\
 = & (t-2)(st-t-s-\frac{2}{t-2})-z(t-2).
\end{align} 
From the above equation, if $t=2$, $N_{\text{PolyDot-CMPC}} < N_{\text{Entangled-CMPC}}$. By replacing $t=2$ in the conditions of (iii) and (c), \ie $t=2, 3\leq s \leq 4, 2(s-2)<z\leq 2(s-1)$, condition 9 for $N_{\text{PolyDot-CMPC}}<N_{\text{Entangled-CMPC}}$ in Lemma \ref{lemma: regions where N_polydot<N_entangled} is derived. In addition, if $t>2$ and $z>st-t-s-\frac{2}{t-2}$, $N_{\text{PolyDot-CMPC}} < N_{\text{Entangled-CMPC}}$, otherwise, $N_{\text{PolyDot-CMPC}} \ge N_{\text{Entangled-CMPC}}$. By combining the conditions of (iii), (c), and $t>2, z>st-t-s-\frac{2}{t-2}$, \ie $st-2t < z \leq ts-s, t>2, t< s\leq 2t$, condition 10 for $N_{\text{PolyDot-CMPC}}<N_{\text{Entangled-CMPC}}$ in Lemma \ref{lemma: regions where N_polydot<N_entangled} is derived.

(d) $s>2t$: For this case, we have:
\begin{align}
 & N_{\text{PolyDot-CMPC}} - N_{\text{Entangled-CMPC}} \nonumber \\
 = & 2ts+\theta'(t-1)+2z-1-(2st^2+2z-1) \nonumber \\
 = & 2ts+(2ts-t)(t-1)+2z-1-2st^2-2z+1 \nonumber \\
 = & -t(t-1)\nonumber\\
 <&0,
\end{align}
From the above equation, for this case, $N_{\text{PolyDot-CMPC}} < N_{\text{Entangled-CMPC}}$. By combining the conditions of (iii) and (d), \ie $s>2t, ts-2t<z\leq ts-t$, condition 11 for $N_{\text{PolyDot-CMPC}}<N_{\text{Entangled-CMPC}}$ in Lemma \ref{lemma: regions where N_polydot<N_entangled} is derived.

(iv) $\max\{ts-2t-s+2, \frac{ts-2t+1}{2}\} < z \leq st-2t$ and $s,t \ne 1$: From (\ref{eq:N-PolyDot-DMPC}), $N_{\text{PolyDot-CMPC}} = \psi_4=(t+1)ts+(t-1)(z+t-1)+2z-1$ and from (\ref{eq:N Entang-CMPC}), $N_{\text{Entangled-CMPC}}=2st^2+2z-1$ for $z>ts-s$ and $N_{\text{Entangled-CMPC}}=st^2+3st-2s+t(z-1)+1$ for $z\leq ts-s$, thus we have:

(a) $2t\ge s$: For this case, we have:
\begin{align}\label{eq:compare psi-4 with Entang-iv-a}
 & N_{\text{PolyDot-CMPC}} - N_{\text{Entangled-CMPC}} \nonumber \\
 = & (t+1)ts+(t-1)(z+t-1)+2z-1 \nonumber \\ -&(st^2+3st-2s+t(z-1)+1) \nonumber \\
 = & z-(2ts-t^2+t-2s+1).
\end{align} 
From the above equation, if $z<(2ts-t^2+t-2s+1)$, we have $N_{\text{PolyDot-CMPC}} < N_{\text{Entangled-CMPC}}$, otherwise, $N_{\text{PolyDot-CMPC}} \ge N_{\text{Entangled-CMPC}}$. By combining the conditions of (iv), (a), and $z<(2ts-t^2+t-2s+1)$, \ie $2t\ge s, \max\{ts-2t-s+2, \frac{ts-2t+1}{2}\} < z \leq \min\{st-2t, 2ts-t^2+t-2s+1\}$, condition 12 for $N_{\text{PolyDot-CMPC}} < N_{\text{Entangled-CMPC}}$ in Lemma \ref{lemma: regions where N_polydot<N_entangled} is derived.

(b) $2t<s, ts-s<z\leq st-2t, t\neq 2$: For this case, $\max\{ts-2t-s+2, \frac{ts-2t+1}{2}\}=ts-2t-s+2<ts-s$. The reason is summarized as follows:
\begin{align}\label{eq:temmp}
    & s>2t \nonumber\\
    \Rightarrow & s(t-2)>2t\nonumber\\
    \Rightarrow & s(t-2)+3>2t\nonumber\\
    \Rightarrow & ts-2t-2s+4>1\nonumber\\
    \Rightarrow & ts-2t-s+2>\frac{ts-2t+1}{2}
\end{align}
For this case, we have:
\begin{align}
 & N_{\text{PolyDot-CMPC}} - N_{\text{Entangled-CMPC}} \nonumber \\
 = & (t+1)ts+(t-1)(z+t-1)+2z-1 \nonumber \\ 
 -&(2st^2+2z-1) \nonumber \\
 = & (t-1)(z-1+t-ts)\nonumber\\
 <&0,
\end{align} 
where the last inequality comes from the condition of (b), $z\leq st-2t$, as $st-2t<st-t+1$ and thus $z< st-t+1$. By combining the conditions of (iv) and (b) \ie $s>2t, ts-s<z\leq ts-2t, t\neq 2$, condition 13 for $N_{\text{PolyDot-CMPC}}<N_{\text{Entangled-CMPC}}$ in Lemma \ref{lemma: regions where N_polydot<N_entangled} is derived.

(c) $2t<s, ts-s<z\leq st-2t, t= 2$: By replacing $t=2$ in conditions of (iv) and (c), we have $4<s<z<2s-4$. Therefore, for this case, we have:
\begin{align}
 & N_{\text{PolyDot-CMPC}} - N_{\text{Entangled-CMPC}} \nonumber \\
 = & (t+1)ts+(t-1)(z+t-1)+2z-1 \nonumber \\ 
 -&(2st^2+2z-1) \nonumber \\
 = & (t-1)(z-1+t-ts)\nonumber\\
 = & z-1+2-2s\nonumber\\
 <&-3 < 0.
\end{align}
The condition of this case, \ie $4<s<z<2s-4, t=2$, provides condition 14 for $N_{\text{PolyDot-CMPC}}<N_{\text{Entangled-CMPC}}$ in Lemma \ref{lemma: regions where N_polydot<N_entangled}.

(d) $2t<s, ts-2t-s+2<z\leq ts-s$: For this case, we have:
\begin{align}
 & N_{\text{PolyDot-CMPC}} - N_{\text{Entangled-CMPC}} \nonumber \\
 = & (t+1)ts+(t-1)(z+t-1)+2z-1 \nonumber \\ -&(st^2+3st-2s+t(z-1)+1) \nonumber \\
 = & z-(2ts-t^2+t-2s+1).
\end{align} 
From the above equation, if $z<2ts-t^2+t-2s+1$, we have $ N_{\text{PolyDot-CMPC}} < N_{\text{Entangled-CMPC}}$, otherwise, $N_{\text{PolyDot-CMPC}} \ge N_{\text{Entangled-CMPC}}$.
On the other hand, $\max\{ts-2t-s+2, \frac{ts-2t+1}{2}\}=ts-2t-s+2<ts-s$, which is derived from (\ref{eq:temmp}) for $t\neq 2$. For $t=2$, $\max\{ts-2t-s+2, \frac{ts-2t+1}{2}\}=\max\{s-2, s-2+1/2\}=s-1.5$, however, we consider $s-2=ts-2t-s+2$ as $s$ and $z$ are integers and $z>s-1.5$ is equivalent to $z>s-2$. Therefore, by combining the conditions of (iv) and (d), \ie 
$ts-2t-s+2<z<ts-s, 2t<s$,
condition 15 for $N_{\text{PolyDot-CMPC}}<N_{\text{Entangled-CMPC}}$ in Lemma \ref{lemma: regions where N_polydot<N_entangled} is derived. The reason for this combination is that:
\begin{align}
    & 2ts-t^2+t-2s+1 \nonumber \\
    &= 2ts-2s-t(t-1)+1 \nonumber \\
    &= ts-s+s(t-1)-t(t-1)+1 \nonumber \\
    &= ts-s+(t-1)(s-t)+1 \nonumber \\
    &> ts-s+(t-1)(2t-t)+1 \text{\;since $s>2t$} \nonumber \\
    &= ts-s+t(t-1)+1 \nonumber \\
   &>ts-s \nonumber \\
   \Rightarrow & \min\{ts-s,2ts-t^2+t-2s+1\} = ts-s.
\end{align}
%

(v) $z \leq \max\{ts-2t-s+2, \frac{ts-2t+1}{2}\}$ and $s,t \neq 1$: For this case, we have, $z\leq ts-s$. The reason is that $ts-s>ts-s-2t+2$ and $ts-s>\frac{ts-2t+1}{2}$\footnote{This can be directly derived from the fact that $s(t-2)\ge 0 >-2t+1$.}, therefore, from (\ref{eq:N Entang-CMPC}), 
$N_{\text{Entangled-CMPC}}=st^2+3st-2s+t(z-1)+1$ and from (\ref{eq:N-PolyDot-DMPC}), $N_{\text{PolyDot-CMPC}} = \psi_5=\theta' t+z$, thus we have:
%
\begin{align}
 & N_{\text{PolyDot-CMPC}} - N_{\text{Entangled-CMPC}} \nonumber \\
 = & \theta't+z-(st^2+3st-2s+t(z-1)+1) \nonumber \\
 = & 2st^2-t^2+z-st^2-3st+2s-tz+t-1 \nonumber \\
 = & (t-1)(st-2s-t-\frac{1}{t-1})-z(t-1).
\end{align}
From the above equation, if $z>st-2s-t-\frac{1}{t-1}$, we have $N_{\text{PolyDot-CMPC}} < N_{\text{Entangled-CMPC}}$, otherwise, $N_{\text{PolyDot-CMPC}} \ge N_{\text{Entangled-CMPC}}$. By combining (v), (a), and $z>st-2s-t-\frac{1}{t-1}$, \ie $st-2s-t-\frac{1}{t-1}<z\leq \max\{ts-2t-s+2, \frac{ts-2t+1}{2}\}$, condition 16 for $N_{\text{PolyDot-CMPC}}<N_{\text{Entangled-CMPC}}$ in Lemma \ref{lemma: regions where N_polydot<N_entangled} is derived.

(vi) $s=1 \text{ and } t \ge z$: From (\ref{eq:N-PolyDot-DMPC}), $N_{\text{PolyDot-CMPC}} = \psi_6=t^2+2t+tz-1$ and from (\ref{eq:N Entang-CMPC}), $N_{\text{Entangled-CMPC}}=2t^2+2z-1$ for $z>t-1$ and $N_{\text{Entangled-CMPC}}=t^2+3t-2+t(z-1)+1$ for $z\leq t-1$, thus we have:

(a) $z=t$: For this case, we have:
\begin{align}
 & N_{\text{PolyDot-CMPC}} - N_{\text{Entangled-CMPC}} \nonumber \\
 = & t^2+2t+tz-1-(2t^2+2z-1) \nonumber \\
 = & (z-t)(t-2) \nonumber \\
 =&0.
\end{align}
From the above equation, for this condition, $N_{\text{PolyDot-CMPC}} = N_{\text{Entangled-CMPC}}$.

(b) $z\leq t-1$: For this case, we have:
\begin{align}
 & N_{\text{PolyDot-CMPC}} - N_{\text{Entangled-CMPC}} \nonumber \\
 = & t^2+2t+tz-1-(t^2+3t-2+t(z-1)+1) \nonumber \\
 =&0.
\end{align}
From the above equation, for this condition, $N_{\text{PolyDot-CMPC}} = N_{\text{Entangled-CMPC}}$.
\subsection{Proof of Lemma \ref{lemma: regions where N_polydot<N_ssmm} (PolyDot-CMPC Versus SSMM)}
To prove this lemma, we consider different regions for the value of $z$ and compare the required number of workers for PolyDot-CMPC, $N_{\text{PolyDot-CMPC}}$, with SSMM, $N_{\text{SSMM}}$, in each region. From \cite{Zhu2021ImprovedCF}, $N_{\text{SSMM}} = (t+1)(ts+z)-1$ and we use (\ref{eq:N-PolyDot-DMPC}) for $N_{\text{PolyDot-CMPC}}$ in each region.

(i) $ts<z \text{ or } t=1$: From (\ref{eq:N-PolyDot-DMPC}), $N_{\text{PolyDot-CMPC}} = \psi_1 = (p+2)ts+\theta'(t-1)+2z-1$ and thus we have:
\begin{align}\label{eq:compare psi-1 with N-SSMM}
    & N_{\text{PolyDot-CMPC}} - N_{\text{SSMM}} \nonumber \\
    = & (p+2)ts+\theta'(t-1)+2z-1-(t+1)(ts+z)+1 \nonumber \\
    = & pts+2ts+2t^2s-2ts-t^2+t+2z-t^2s-ts-(t+1)z \nonumber \\
    = & pts+(t-1)ts-t(t-1)-(t-1)z. 
\end{align}
From the above equation, if $z>\frac{pts}{t-1}+ts-t$ and $t\neq 1$, we have $N_{\text{PolyDot-CMPC}} < N_{\text{SSMM}}$, otherwise $N_{\text{PolyDot-CMPC}} \geq N_{\text{SSMM}}$\footnote{Note that for $t=1$, $N_{\text{PolyDot-CMPC}} = N_{\text{SSMM}}$.}. Therefore, from the condition of (i), we have $N_{\text{PolyDot-CMPC}} < N_{\text{SSMM}}$ only if $z > \max\{ts,ts-t+\frac{pts}{t-1}\}, t\neq 1$. This provides one of the conditions that $N_{\text{PolyDot-CMPC}} < N_{\text{SSMM}}$ in Lemma \ref{lemma: regions where N_polydot<N_ssmm}.

(ii) $ts-t < z \leq ts $: From (\ref{eq:N-PolyDot-DMPC}), $N_{\text{PolyDot-CMPC}} = \psi_2=2ts+\theta'(t-1)+3z-1$ and thus we have:
\begin{align}\label{eq:compare psi-2 with N-SSMM}
 & N_{\text{PolyDot-CMPC}} - N_{\text{SSMM}} \nonumber \\
 = & 2ts+\theta'(t-1)+3z-1 - (t+1)(ts+z)+1 \nonumber \\
 = & 2ts+2t^2s-2ts-t^2+t\nonumber \\
 &+3z-1-t^2s-ts-(t+1)z+1 \nonumber \\
 = & st^2-st-t^2+t-(t-2)z \nonumber \\
 = & st(t-1)-t(t-1)-(t-2)z \nonumber \\
 = & (t-1)(st-t)-(t-2)z.
\end{align}
From the above equation, if $z>\frac{(st-t)(t-1)}{t-2}$, we have $N_{\text{PolyDot-CMPC}}<N_{\text{SSMM}}$ otherwise,  $N_{\text{PolyDot-CMPC}} \geq N_{\text{SSMM}}$. Therefore, from the condition of (ii), we have $N_{\text{PolyDot-CMPC}}<N_{\text{SSMM}}$ only if $\frac{t-1}{t-2}(st-t)<z \leq ts$. This provides the other condition that $N_{\text{PolyDot-CMPC}}<N_{\text{SSMM}}$ in Lemma \ref{lemma: regions where N_polydot<N_ssmm}.

(iii) $ts-2t < z \leq ts-t$: From (\ref{eq:N-PolyDot-DMPC}), $N_{\text{PolyDot-CMPC}} = \psi_3=2ts+\theta'(t-1)+2z-1$ and thus we have:
\begin{align}\label{eq:compare psi-3 with N-SSMM}
 & N_{\text{PolyDot-CMPC}} - N_{\text{SSMM}} \nonumber \\
 &= 2ts+\theta'(t-1)+2z-1 - (t+1)(ts+z)+1 \nonumber \\
 &= 2ts+2t^2s-2ts-t^2+t-st^2-st-(t-1)z \nonumber \\
 &=-t^2+t+st^2-st-(t-1)z \nonumber \\
 &= (ts-t)(t-1)-(t-1)z.
\end{align}
From the above equation and the condition of (iii), $N_{\text{PolyDot-CMPC}} \geq N_{\text{SSMM}}$ for $ts-2t <z \leq ts-t$. 

(iv) $\max\{ts-2t-s+2, \frac{ts-2t+1}{2}\} < z \leq st-2t$: From (\ref{eq:N-PolyDot-DMPC}), $N_{\text{PolyDot-CMPC}} = \psi_4=(t+1)ts+(t-1)(z+t-1)+2z-1$ and thus we have:
\begin{align}\label{eq:compare psi-4 with N-SSMM}
 & N_{\text{PolyDot-CMPC}} - N_{\text{SSMM}} \nonumber \\
 = & (t+1)ts+(t-1)(z+t-1)\nonumber \\
 &+2z-1 -(t+1)(ts+z)+1
\nonumber \\
= &  (t+1)ts+(t+1)z+(t-1)^2-(t+1)(ts+z) \nonumber \\
= & (t-1)^2 > 0.
\end{align}
From the above equation, $N_{\text{PolyDot-CMPC}} > N_{\text{SSMM}}$ for $\max\{ts-2t-s+2, \frac{ts-2t+1}{2}\} < z \leq st-2t$.

(v) $z \leq \max\{ts-2t-s+2, \frac{ts-2t+1}{2}\}$: From (\ref{eq:N-PolyDot-DMPC}), $N_{\text{PolyDot-CMPC}} = \psi_5=\theta' t+z$ and thus we have:
\begin{align}
 & N_{\text{PolyDot-CMPC}} - N_{\text{SSMM}} \nonumber \\
 &= \theta't+z - (t+1)(ts+z)+1 \nonumber \\
 &= 2t^2s-t^2+z-t^2s-ts-(t+1)z+1 \nonumber \\
 &= t^2s-t^2-ts+1-tz \nonumber \\
 &= t(ts-t-s+\frac{1}{t}-z)\nonumber\\
 & \ge t(\max\{ts-2t-s+2, \frac{ts-2t+1}{2}\}-z) \label{eq:compare1v1}\\
 & \geq 0 \label{eq:compare psi-5 with N-SSMM},
 \end{align}
where, (\ref{eq:compare1v1}) comes from:
\begin{align}
  & ts-t-s+\frac{1}{t} - (ts-2t-s+2) \nonumber \\
  = & ts-t-s+\frac{1}{t}-ts+2t+s-2\nonumber \\
  = & t+\frac{1}{t}-2>0,
\end{align}
and
\begin{align}
  & ts-t-s+\frac{1}{t} - (\frac{ts-2t+1}{2}) \nonumber \\
  = & \frac{s(t-2)+2/t-1}{2} \ge 0,
\end{align}
and (\ref{eq:compare psi-5 with N-SSMM}) comes from the condition of the (v), \ie $z \leq ts-2t-s+1$. Therefore, $N_{\text{PolyDot-CMPC}} \ge N_{\text{SSMM}}$ for $z \leq ts-2t-s+1$.

(vi) $s=1 \text{ and } t \ge z$: From (\ref{eq:N-PolyDot-DMPC}), $N_{\text{PolyDot-CMPC}} = \psi_6=t^2+2t+tz-1$ and thus we have:
\begin{align}
 & N_{\text{PolyDot-CMPC}} - N_{\text{SSMM}} \nonumber \\
 &= t^2+2t+tz-1 - (t+1)(ts+z)+1 \nonumber \\
 &= t^2+2t-t^2s-ts-z \nonumber \\
 &= t^2+2t-t^2-t-z \nonumber \\
 &= t-z\nonumber\\
 & \geq 0 \label{eq:compare psi-6 with N-SSMM},
 \end{align}
From (i), (ii), (iii), (iv), (v) and (vi), the only conditions that $N_{\text{PolyDot-CMPC}} < N_{\text{SSMM}}$, are $z > \max\{ts,ts-t+\frac{pts}{t-1}\}, t\neq 1$ and $\frac{t-1}{t-2}(st-t)<z \leq ts$. In all other conditions, we have $N_{\text{PolyDot-CMPC}} \ge N_{\text{SSMM}}$.
This completes the proof. \hfill $\Box$

\subsection{Proof of Lemma \ref{lemma: regions where N_polydot<N_gcsana} (PolyDot-CMPC Versus GCSA-NA)}\label{subsec:polydot-cmpc vs gcsa-na}
To prove this lemma, we consider different regions for the value of $z$ and compare the required number of workers for PolyDot-CMPC, $N_{\text{PolyDot-CMPC}}$, with GCSA-NA, $N_{\text{GCSA-NA}}$, in each region. From \cite{9333639}, $N_{\text{GCSA-NA}}$ for one matrix multiplication (the number of batch is one) is equal to $N_{\text{GCSA-NA}} = 2st^2+2z-1$ and we use (\ref{eq:N-PolyDot-DMPC}) for $N_{\text{PolyDot-CMPC}}$ in each region.

(i) $ts<z \text{ or } t=1$: From (\ref{eq:N-PolyDot-DMPC}), $N_{\text{PolyDot-CMPC}} = \psi_1 = (p+2)ts+\theta'(t-1)+2z-1$ and thus we have:
\begin{align}\label{eq:compare psi-1 with GCSA-NA}
 & N_{\text{PolyDot-CMPC}} - N_{\text{GCSA-NA}} \nonumber \\
 = & (p+2)ts+\theta'(t-1)+2z-1-(2st^2+2z-1) \nonumber \\
 = & pts+2ts+(2ts-t)(t-1)+2z-1-2st^2-2z+1 \nonumber \\
 = & t(ps-t+1).
 \end{align} 
From the above equation, if $p<\frac{t-1}{s}$ and $t\neq 1$, we have $N_{\text{PolyDot-CMPC}}<N_{\text{GCSA-NA}}$, otherwise, $N_{\text{PolyDot-CMPC}} \ge N_{\text{GCSA-NA}}$\footnote{Note that for $t=1$, $N_{\text{PolyDot-CMPC}} = N_{\text{GCSA-NA}}$.}. This along with the condition of (i), provides one of the conditions that $N_{\text{PolyDot-CMPC}}<N_{\text{GCSA-NA}}$ in Lemma \ref{lemma: regions where N_polydot<N_gcsana}.

(ii) $ts-t < z \leq ts$: From (\ref{eq:N-PolyDot-DMPC}), $N_{\text{PolyDot-CMPC}} = \psi_2=2ts+\theta'(t-1)+3z-1$ and thus we have:
\begin{align}\label{eq:compare psi-2 with GCSA-NA}
 & N_{\text{PolyDot-CMPC}} - N_{\text{GCSA-NA}} \nonumber \\
 = & 2ts+\theta'(t-1)+3z-1-(2st^2+2z-1) \nonumber \\
 = & 2ts+(2ts-t)(t-1)+3z-1-2st^2-2z+1 \nonumber \\
 = & z-(t^2-t).
\end{align} 
From the above equation, if $z<t(t-1)$, we have $N_{\text{polyDot-CMPC}}<N_{\text{GCSA-NA}}$, otherwise, $N_{\text{polyDot-CMPC}}\ge N_{\text{GCSA-NA}}$. From the condition of (ii), $ts-t<z\leq ts$. Therefore, $N_{\text{polyDot-CMPC}}<N_{\text{GCSA-NA}}$ only if $ts-t<z \leq \min\{ts, t(t-1)-1\}$, which also requires that $s<t$. This is another condition that $N_{\text{PolyDot-CMPC}}<N_{\text{GCSA-NA}}$in Lemma \ref{lemma: regions where N_polydot<N_gcsana}.

(iii) $ts-2t < z \leq ts-t$: From (\ref{eq:N-PolyDot-DMPC}), $N_{\text{PolyDot-CMPC}} = \psi_3=2ts+\theta'(t-1)+2z-1$ and thus we have:
\begin{align}\label{eq:compare psi-3 with GCSA-NA}
 & N_{\text{PolyDot-CMPC}} - N_{\text{GCSA-NA}} \nonumber \\
 = & 2ts+\theta'(t-1)+2z-1-(2st^2+2z-1) \nonumber \\
 = & 2ts+(2ts-t)(t-1)+2z-1-2st^2-2z+1 \nonumber \\
 = & t(1-t)\nonumber\\
 <&0.
\end{align}
From the above equation, for $ts-2t < z \leq ts-t$, we have $N_{\text{PolyDot-CMPC}} < N_{\text{GCSA-NA}}$. This provides part of the third condition that $N_{\text{PolyDot-CMPC}}<N_{\text{GCSA-NA}}$ in Lemma \ref{lemma: regions where N_polydot<N_gcsana}.

(iv) $\max\{ts-2t-s+2, \frac{ts-2t+1}{2}\} < z \leq st-2t$: From (\ref{eq:N-PolyDot-DMPC}), $N_{\text{PolyDot-CMPC}} = \psi_4=(t+1)ts+(t-1)(z+t-1)+2z-1$ and thus we have:
\begin{align}\label{eq:compare psi-4 with GCSA-NA}
 & N_{\text{PolyDot-CMPC}} - N_{\text{GCSA-NA}} \nonumber \\
 = & (t+1)ts+(t-1)(z+t-1)+2z-1-(2st^2+2z-1) \nonumber \\
 = & t^2s+ts+(t-1)(z+t-1)-2st^2 \nonumber \\
 = & (t-1)(z-(st-t+1)).
\end{align} 
From the above equation, if $z<st-t+1$, we have $N_{\text{polyDot-CMPC}}<N_{\text{GCSA-NA}}$. This condition is satisfied for the condition of (iv), $\max\{ts-2t-s+2, \frac{ts-2t+1}{2}\} < z \leq st-2t$, as $st-t-t<st-t+1$. Therefore, for $\max\{ts-2t-s+2, \frac{ts-2t+1}{2}\} < z \leq st-2t$, we have $N_{\text{PolyDot-CMPC}} < N_{\text{GCSA-NA}}$. This provides part of the third condition that $N_{\text{PolyDot-CMPC}}<N_{\text{GCSA-NA}}$ in Lemma \ref{lemma: regions where N_polydot<N_gcsana}.

(v) $z \leq \max\{ts-2t-s+2, \frac{ts-2t+1}{2}\}$: From (\ref{eq:N-PolyDot-DMPC}), $N_{\text{PolyDot-CMPC}} = \psi_5=\theta' t+z$ and thus we have:
\begin{align}\label{eq:compare psi-5 with GCSA-NA}
 & N_{\text{PolyDot-CMPC}} - N_{\text{GCSA-NA}} \nonumber \\
 = & \theta't+z-(2st^2+2z-1) \nonumber \\
 = & 2st^2-t^2+z-2st^2-2z+1 \nonumber \\
 = & -t^2-z+1\nonumber\\
 <&0.
\end{align} 
From the above equation, for $z \leq \max\{ts-2t-s+2, \frac{ts-2t+1}{2}\}$, we have $N_{\text{PolyDot-CMPC}} < N_{\text{GCSA-NA}}$. This provides part of the third condition that $N_{\text{PolyDot-CMPC}}<N_{\text{GCSA-NA}}$ in Lemma \ref{lemma: regions where N_polydot<N_gcsana}.

(vi) $s=1 \text{ and } t \ge z$: From (\ref{eq:N-PolyDot-DMPC}), $N_{\text{PolyDot-CMPC}} = \psi_6=t^2+2t+tz-1$ and thus we have:
\begin{align}\label{eq:compare psi-6 with GCSA-NA}
 & N_{\text{PolyDot-CMPC}} - N_{\text{GCSA-NA}} \nonumber \\
 = & t^2+2t+tz-1-(2st^2+2z-1) \nonumber \\
 = & 2t+tz-t^2-2z \nonumber \\
 = & (2-t)(t-z)\nonumber\\
 \leq &0.
\end{align}
From the above equation and the condition of (vi), if $s=1, t>z$ and $t\neq 2$, we have $N_{\text{polyDot-CMPC}}<N_{\text{GCSA-NA}}$. This provides the last condition that $N_{\text{PolyDot-CMPC}}<N_{\text{GCSA-NA}}$ in Lemma \ref{lemma: regions where N_polydot<N_gcsana}, and completes the proof. \hfill $\Box$
\section*{Appendix D: Proof of Theorem \ref{th:decodabilityofAGEcodes}} 
For AGE codes with $\{\alpha, \beta,\theta\}=\{1,s,ts+\lambda\}$, (\ref{eq:generalEntangled}) is reduced to:
\begin{align}\label{eq:AGECodes}
    C_A(x) = & \sum\limits_{i=0}^{t-1}\sum\limits_{j=0}^{s-1}A_{i,j}x^{j+is}, \nonumber \\
    C_B(x) = & \sum\limits_{k=0}^{s-1}\sum\limits_{l=0}^{t-1}B_{k,l}x^{(s-1-k)+(ts+\lambda)l},
\end{align}

To prove the decodability of AGE codes, we need to prove that the polynomial $C_Y(x)=C_A(x)C_B(x)=\sum\limits_{i=0}^{t-1}\sum\limits_{j=0}^{s-1} \sum\limits_{k=0}^{s-1}\sum\limits_{l=0}^{t-1} A_{i,j} B_{k,l}x^{j+is+(s-1-k)+(ts+\lambda)l}$ consists of $t^2$ distinct terms with coefficients $Y_{i,l} = \sum\limits_{j=0}^{s-1}A_{i,j}B_{j,l}, 0\leq i, l \leq t-1$; {which are the important coefficients that are required for decoding}. For this purpose, we define two sets of (i) $\mathbf{P}_1=\{s-1+is+(ts+\lambda)l, 0 \leq i,l \leq t-1\}$, representing the potential set of powers of the terms in $C_Y(x)$ with coefficients $Y_{i,l}$ (resulting from $j=k$)
, and (ii) $\mathbf{P}_2=\{j+is+(s-1-k)+(ts+\lambda)l, 0 \leq i,l \leq t-1, 0\leq k,j \leq s-1, j \neq k\}$, the set of powers of the remaining terms in $C_Y(x)$. Then, we prove that (i) $\mathbf{P}_1$ consists of $t^2$ distinct elements, and (ii) $\mathbf{P}_1$ and $\mathbf{P}_2$ do not have any overlap.

{\em{(i) Proving that $\mathbf{P}_1$ consists of $t^2$ distinct elements}}: From the definition of $\mathbf{P}_1$, it is equal to:
\begin{align}
    \mathbf{P}_1=\bigcup\limits_{l=0}^{t-1} \bigcup\limits_{i=0}^{t-1} \{s-1+is+(ts+\lambda)l\}.
\end{align}
For a given $l$, each subset of $\bigcup\limits_{i=0}^{t-1} \{s-1+is+(ts+\lambda)l\}$ consists of $t$ distinct elements. In addition, for two different values of $l=l_1$ and $l=l_2$ ($l_1 \neq l_2$), there is no overlap between $\bigcup\limits_{i=0}^{t-1} \{s-1+is+(ts+\lambda)l_1\}$ and $\bigcup\limits_{i=0}^{t-1} \{s-1+is+(ts+\lambda)l_2\}$. The reason is that for $0 \leq l_1 < l_2 \leq t-1$\footnote{Note that the assumption of $l_1<l_2$ does not result in loss of generality.}, the largest element of $\bigcup\limits_{i=0}^{t-1} \{s-1+is+(ts+\lambda)l_1\}$, \ie $ts-1+(ts+\lambda)l_1$ is less than the smallest element of $\bigcup\limits_{i=0}^{t-1} \{s-1+is+(ts+\lambda)l_2\}$, \ie $s-1+(ts+\lambda)l_2$:
\begin{align}
0 < s+\lambda 
\Rightarrow &ts-1 < s-1+ts+\lambda  \nonumber \\
\Rightarrow &ts-1+(ts+\lambda)l_1 < s-1+(ts+\lambda)(l_1+1) \nonumber \\
\Rightarrow &ts-1+(ts+\lambda)l_1 < s-1+(ts+\lambda)l_2.
\end{align}
Therefore, $\mathbf{P}_1$ consists of $t^2$ distinct elements.

{\em (ii) Proving that $\mathbf{P}_1$ and $\mathbf{P}_2$ have no overlap:} From the definition of $\mathbf{P}_1$ and $\mathbf{P}_2$, we have:
\begin{align}
    \mathbf{P}_1=&\bigcup\limits_{l_1=0}^{t-1} \bigcup\limits_{i_1=0}^{t-1} \mathbf{P}_1(l_1,i_1)\nonumber\\
    =&\bigcup\limits_{l_1=0}^{t-1} \bigcup\limits_{i_1=0}^{t-1} \{s-1+i_1s+(ts+\lambda)l_1\}
\end{align}
and
\begin{align}
    \mathbf{P}_2=&\bigcup\limits_{l_2=0}^{t-1} \bigcup\limits_{i_2=0}^{t-1}\mathbf{P}_2(l_2,i_2)\nonumber\\ =&\bigcup\limits_{l_2=0}^{t-1} \bigcup\limits_{i_2=0}^{t-1}\bigcup\limits_{\substack{j'=-(s-1)\\j'\neq 0}}^{s-1}\{j'+i_2s+s-1+(ts+\lambda)l_2\}.
\end{align}
To prove $\mathbf{P}_1 \cap \mathbf{P}_2 = \emptyset$, we consider the following five cases; (a) $l_1=l_2,i_1=i_2$, (b) $l_1=l_2,i_1<i_2$, (c) $l_1=l_2,i_1>i_2$, (d) $l_1>l_2$, (e) $l_1<l_2$. We prove that $\mathbf{P}_1(i_1,l_1) \cap \mathbf{P}_2(i_2,l_2)=\emptyset$ holds for each case.

(a) $l_1=l_2,i_1=i_2$: 
For this case, $\mathbf{P}_1(l_1,i_1)$ consists of the only element of $s-1+i_1s+(ts+\lambda)l_1$ which is not a member of $\mathbf{P}_2(l_2,i_2)$ as  $j'\neq0$. Therefore, $\mathbf{P}_1(i_1,l_1) \cap \mathbf{P}_2(i_2,l_2)=\emptyset$ for this case.

(b) $l_1=l_2,i_1<i_2$:
For this case, the smallest element of $\mathbf{P}_2(l_2,i_2)$ is always greater than $\mathbf{P}_1(l_1,i_1)=s-1+i_1s+(ts+\lambda)l_1$, as shown below:
\begin{align}
s-1+i_1s+(ts+&\lambda)l_1 < (i_1+1)s+(ts+\lambda)l_1\nonumber\\
&\leq i_2s+(ts+\lambda)l_2 \nonumber \\
&=-(s-1)+i_2s+(s-1)+(ts+\lambda)l_2
\end{align}
Therefore, $\mathbf{P}_1(i_1,l_1) \cap \mathbf{P}_2(i_2,l_2)=\emptyset$ holds for this case.

(c) $l_1=l_2,i_1>i_2$:
For this case, the largest element of $\mathbf{P}_2(l_2,i_2)$ is always less than $\mathbf{P}_1(l_1,i_1)=s-1+i_1s+(ts+\lambda)l_1$, as shown below:
\begin{align}
s-1+i_1s+(ts+\lambda)l_1 &> s-2+i_1s+(ts+\lambda)l_1\nonumber\\
&\geq s-2+(i_2+1)s+(ts+\lambda)l_2 \nonumber \\
&=2s-2+i_2s+(ts+\lambda)l_2
\end{align}
Therefore, $\mathbf{P}_1(i_1,l_1) \cap \mathbf{P}_2(i_2,l_2)=\emptyset$ holds for this case.

(d) $l_1>l_2$:
For this case, the smallest element of $\mathbf{P}_1(l_1,i_1)$, \ie $\mathbf{P}_1(l_1,0)$ is always greater than the largest element of $\mathbf{P}_2(l_2,i_2)$ \ie $\mathbf{P}_2(l_2, t-1), j'=s-1$, as shown below:
\begin{align}
s-1+(ts+\lambda)l_1 &\ge s-1+(ts+\lambda)(l_2+1)\nonumber\\
&= s-1+ts+\lambda+(ts+\lambda)l_2 \nonumber \\
&>s-1+(ts+\lambda)l_2
\end{align}
Therefore, $\mathbf{P}_1(i_1,l_1) \cap \mathbf{P}_2(i_2,l_2)=\emptyset$ holds for this case.

(e) $l_1<l_2$: 
For this case, the largest element of $\mathbf{P}_1(l_1,i_1)$, \ie $\mathbf{P}_1(l_1,t-1)$ is always less than the smallest element of $\mathbf{P}_2(l_2,i_2)$ \ie $\mathbf{P}_2(l_2, 0), j'=-(s-1)$, as shown below:
\begin{align}
(t-1)s+(ts+\lambda)l_1 &< (t-1)s+(ts+\lambda)(l_2-1)\nonumber\\
&= (t-1)s-ts-\lambda+(ts+\lambda)l_2 \nonumber \\
&=-s-\lambda+(ts+\lambda)l_2\nonumber\\
&< (ts+\lambda)l_2
\end{align}
Therefore, $\mathbf{P}_1(i_1,l_1) \cap \mathbf{P}_2(i_2,l_2)=\emptyset$ holds for this case.

This completes the proof of Theorem \ref{th:decodabilityofAGEcodes}.
\section*{Appendix E: Proof of Theorem \ref{th:FA-FB-AGE-thrm}} 
We first show that $\mathbf{P}(S_B(x))=\{ts+(ts+\lambda)(t-1),\ldots,ts+(ts+\lambda)(t-1)+z-1\}$ in (\ref{eq:S-B}) satisfies C4 in (\ref{eq:non_eq-AGE-conditions}). Then, we fix $\mathbf{P}(S_{B}(x))$ in C6 of (\ref{eq:non_eq-AGE-conditions}), and find $\mathbf{P}(S_A(x))$ that satisfies C5 and C6. Next, we explain these steps in details.

\emph{Showing that $\mathbf{P}(S_B(x))=\{ts+(ts+\lambda)(t-1),\ldots,ts+(ts+\lambda)(t-1)+z-1\}$ in (\ref{eq:S-B}) satisfies C4 in (\ref{eq:non_eq-AGE-conditions}).}
%
The largest element of the left side of C4 is equal to $(s-1)+(t-1)s+(ts+\lambda)(t-1)=ts+(ts+\lambda)(t-1)-1$ and the smallest element of the right side of C4 is equal to the smallest element of $\mathbf{P}(S_B(x))$, \ie $ts+(ts+\lambda)(t-1)$ plus the smallest element of $\mathbf{P}(C_A(x))$, \ie $0$. As $ts+(ts+\lambda)(t-1)-1$ is less than $ts+(ts+\lambda)(t-1)$, C4 is satisfied.

\emph{Fixing $\mathbf{P}(S_{B}(x))$ in C6 of (\ref{eq:non_eq-AGE-conditions}), and find $\mathbf{P}(S_A(x))$ that satisfies C5 and C6.}
C6 is satisfied for any choice of $\mathbf{P}(S_A(x))$ with non-negative elements. The reason is that the largest element of the left side of C6 is less than the smallest element of $\mathbf{P}(S_B(x))$. Next, we find $\mathbf{P}(S_A(x))$ with the smallest elements that satisfies C5, so (\ref{eq:non_eq-AGE-conditions}) is equal to
\begin{align}\label{eq:non_eq1-AGE_1}
    & (s-1)+si+(ts+\lambda)l \not\in \mathbf{P}(S_A(x)) \nonumber \\
    & +\{(s-1-k)+l'(ts+\lambda)\},
\end{align}
where $0\leq k \leq s-1,\; 0 \leq i,l,l' \leq t-1, 0 \leq \lambda \leq z$. The above equation is equivalent to:
\begin{align}\label{eq:non_eq1-AGE_2}
    \beta'+\theta l'' \not\in \mathbf{P}(S_A(x)),
\end{align}
for $l''=(l-l')$, $\theta = ts+\lambda$ and $\beta' = si+k$. The range of variable $\beta'$ is $\{si+k, 0\leq i \leq t-1, 0 \leq k \leq s-1\} = \bigcup\limits_{i=0}^{t-1} \{si,\ldots,si+s-1\} = \{0,\ldots,ts-1\}$. Therefore, we have
\begin{align}\label{eq:non_eq1-AGE_3}
    \mathbf{P}(S_A(x)) \not\in \bigcup\limits_{l=-(t-1)}^{t-1} \{\theta l,\ldots,\theta l+ts-1\},
\end{align}
Using the complement of the above intervals and the fact that the elements of $\mathbf{P}(S_A(x))$ is non-negative, we have
\begin{align}\label{eq:non_eq1-AGE_4}
    \mathbf{P}(S_A(x)) \in &\bigcup\limits_{l=0}^{t-2} \{ts+\theta l,\ldots,(l+1)\theta-1\}\nonumber\\
    &\cup \{ts+\theta (t-1), \ldots, +\infty\}, t>1
\end{align}
\begin{align}\label{eq:non_eq1-AGE_4t=1}
    \mathbf{P}(S_A(x)) \in 
    & \{s, \ldots, +\infty\}, t=1
\end{align}
Note that the required number of powers with non-zero coefficients for the secret term $S_A(x)$ is $z$, \ie
\begin{equation}
    |\mathbf{P}(S_A(x))| = z.
\end{equation}
Since our goal is to make the degree of polynomial $F_A(x)$ as small as possible, we choose the $z$ smallest powers from the sets in (\ref{eq:non_eq1-AGE_4}) to form $\mathbf{P}(S_A(x))$. 
Note that in (\ref{eq:non_eq1-AGE_4}), there are $t-1$ finite sets and one infinite set, where each finite set contains $\lambda=\theta-ts$ elements. Therefore, based on the value of $z$, we use the first interval and as many remaining intervals as required for $z > \lambda$, and the first interval only for $z = \lambda$ (Note that $0\leq \lambda \leq z$).
\begin{lemma}\label{lem:P(SA)-z large-AGE}
If $z > \lambda$ and $t \neq 1$, the set of all powers of polynomial $S_A(x)$ with non-zero coefficients is defined as 
\begin{align}\label{eq:finite_P(SA)_set_representation-z large-AGE}
    \mathbf{P}(S_A(x)) = &\Big(\bigcup\limits_{l=0}^{q-1} \{ts+\theta l,\ldots,(l+1)\theta-1\}\Big) \nonumber\\
    & \cup \{ts+q\theta,\ldots,ts+q\theta+z-1-q(\theta-ts)\}\\
    =&\{ts+\theta l+w, l\in\Omega_0^{q-1}, w\in \Omega_0^{\lambda-1}\} \nonumber \\
    &\cup \{ts+\theta q+u, u\in\Omega_0^{z-1-q\lambda}\}\label{eq:psa12-AGE}.
\end{align}
\end{lemma}
{\em Proof:}
For the case of $z > \lambda$, the number of elements in the first interval of (\ref{eq:non_eq1-AGE_4}), which is equal to $\lambda$, is not sufficient for selecting $z$ powers. Therefore, more than one interval is used. We show the number of selected intervals with $q+1$, where $q \ge 1$ is defined as $q=\min\{\floor{\frac{z-1}{\lambda}},t-1\}$. With this definition, the first $q$ intervals of (\ref{eq:non_eq1-AGE_4}) are selected in full. In other words, in total, we select $q\lambda$ elements to form the first $q$ intervals in (\ref{eq:finite_P(SA)_set_representation-z large-AGE}). The remaining $z-q\lambda$ elements are selected from the $(q+1)^\text{st}$ interval of (\ref{eq:non_eq1-AGE_4}) to form the last interval of   (\ref{eq:finite_P(SA)_set_representation-z large-AGE}). We can derive (\ref{eq:psa12-AGE}) from  (\ref{eq:finite_P(SA)_set_representation-z large-AGE}) by replacing $\theta$ with its equivalent value, $ts+\lambda$.
\hfill $\Box$

\begin{lemma}\label{lem:P(SA)-z small-AGE}
If {$z \leq \lambda$} and $t \ne 1$, the set of all powers of polynomial $S_A(x)$ with non-zero coefficients is defined as the following:
\begin{align}\label{eq:finiteP(SA)-second-scenario-AGE}
\mathbf{P}(S_A(x)) = \{ts,\dots,ts+z-1\}, \nonumber\\
= \{ts+u, u\in \Omega_0^{z-1}\}.
\end{align}
\end{lemma}
{\em Proof:}
In this scenario since {$z \leq \lambda$}, the first interval of (\ref{eq:non_eq1-AGE_4}) is sufficient to select all $z$ elements of $\mathbf{P}(S_A(x))$. Therefore, $z$ elements are selected from the first interval of (\ref{eq:non_eq1-AGE_4}), as shown in (\ref{eq:finiteP(SA)-second-scenario-AGE}).\hfill $\Box$

\begin{lemma}\label{lem:P(SA)-t=1}
If $t=1$ , the set of all powers of polynomial $S_A(x)$ with non-zero coefficients is defined as the following:
\begin{align}\label{eq:finiteP(SA)-third-scenario-AGE}
\mathbf{P}(S_A(x)) = \{s,\dots,s+z-1\}, \nonumber\\
= \{s+u, u\in \Omega_0^{z-1}\}.
\end{align}
\end{lemma}
{\em Proof:}
In this scenario, z smallest elements are selected from (\ref{eq:non_eq1-AGE_4t=1}) 
as shown in (\ref{eq:finiteP(SA)-third-scenario-AGE}).\hfill $\Box$

This completes the proof of Theorem \ref{th:FA-FB-AGE-thrm}.
\section*{Appendix F: Proof of Theorem \ref{th:N_AGE}}
To prove this theorem, we first consider the case that $t=1$. Then, we consider that case thats $t\neq 1$.
\begin{lemma}\label{lemma:t=1NAGECMPC} $N_{\text{AGE-CMPC}}=2s+2z-1$ when $t=1$. s
\end{lemma}
{\em Proof}: $F_A(x)$ and $F_B(x)$ are expressed as in the following for $t=1$ using 
(\ref{eq:AGE-p(CA)-th}), (\ref{eq:AGE-p(CB)-th}), (\ref{eq:S-A}) and (\ref{eq:S-B}). 
\begin{align}\label{eq:FA AGE-CMPC-t=1}
    F_{A}(x) = & \sum_{j=0}^{s-1} A_{j}x^{j}
    + \sum_{u=0}^{z-1}\bar{A}_{u}x^{s+u},
\end{align}
\begin{align}\label{eq:FB AGE-CMPC-t=1}
    F_{B}(x) = & \sum_{k=0}^{s-1} B_{k}x^{s-1-k}
    + \sum_{r=0}^{z-1}\bar{B}_{r}x^{s+r}. 
\end{align}
$F_A(x)$ and $F_B(x)$ are equal to the secret shares of Entangled-CMPC \cite{8613446}, for $t=1$. Thus, in this case, AGE-CMPC and Entangled-CMPC are equivalent, so we have $N_{\text{AGE-CMPC}}=N_{\text{Entangled-CMPC}}=2s+2z-1$ \cite{8613446}. 
This completes the proof. \hfill $\Box$
%
%
%

Now, we consider $t\neq 1$. The required number of workers is equal to the number of terms in $H(x)=F_A(x)F_B(x)$ with non-zero coefficients. The set of all powers of polynomial $H(x)$ with non-zero coefficients, shown by $\mathbf{P}({H}(x))$, is expressed as 
 \begin{align}\label{eq:PHx-AGE}
 \mathbf{P}({H}(x)) = \mathbf{D}_1 \cup  \mathbf{D}_2\cup \mathbf{D}_3 \cup \mathbf{D}_4,
 \end{align}
 where
  \begin{align}\label{eq:D1age}
     & \mathbf{D}_1 = \mathbf{P}(C_A(x))+\mathbf{P}(C_B(x))
 \end{align}
\begin{align}
    & \mathbf{D}_2  =\mathbf{P}(C_A(x))+\mathbf{P}(S_B(x))
 \end{align}
   \begin{align}\label{eq:d3definition-AGE}
    & \mathbf{D}_3=\mathbf{P}(S_A(x))+\mathbf{P}(C_B(x))
 \end{align}
   \begin{align}\label{eq:d4definition-AGE}
    & \mathbf{D}_4=\mathbf{P}(S_A(x))+\mathbf{P}(S_B(x))
 \end{align}
 
Using (\ref{eq:AGE-p(CA)-th}) and (\ref{eq:AGE-p(CB)-th}), $\mathbf{D}_1$ is calculated as:
 \begin{align}
      \mathbf{D}_1 = & \mathbf{P}(C_{A}(x))+\mathbf{P}(C_B(x)) \nonumber \\
     = &  \{j+si
     : 0 \leq i \leq t-1,\; 0 \leq j \leq s-1,\} \nonumber \\ 
     &+ \{s-1-k+\theta l 
     : 0 \leq l \leq t-1,\; 0 \leq k \leq s-1\} \nonumber \\
     = &  \{j+si+s-1-k+\theta l:0 \leq i,l \leq t-1,\; \nonumber \\
     & 0 \leq j,k \leq s-1,\}\nonumber \\
     = & \bigcup\limits_{i=0}^{t-1}\{is,\ldots,(i+2)s-2\} + \{\theta l : 0\leq l\leq t-1\} \nonumber\\
     = & \{0,\ldots,ts+s-2\} + \{\theta l : 0\leq l\leq t-1\} \label{eq:d1AGE1}\\
     = & \bigcup\limits_{l=0}^{t-1}\{\theta l,\ldots,ts+s-2+\theta l\}, \label{eq:d1-AGE}
      \end{align}
where (\ref{eq:d1AGE1}) comes from the fact that the largest element of each $i^{\text{th}}$ subset of $\bigcup\limits_{i=0}^{t-1}\{is,\ldots,(i+2)s-2\}$ plus one, \ie $(i+2)s-2+1$ is greater than or equal to the smallest element of the $(i+1)^{\text{st}}$ subset, \ie $(i+1)s$ as $s\ge 1$. 
Using (\ref{eq:AGE-p(CA)-th}) and (\ref{eq:S-B}), $\mathbf{D}_2$ is calculated as:
 \begin{align}\label{eq:d2-AGE}
      \mathbf{D}_2 = & \mathbf{P}(C_{A}(x))+\mathbf{P}(S_B(x)) \nonumber \\
     = &  \{j+si
     : 0 \leq i \leq t-1,\; 0 \leq j \leq s-1,\} \nonumber \\ 
     &+ \{ts+\theta(t-1)+r: 
      0 \leq r \leq z-1\} \nonumber \\
     = &  \{j+si+ts+\theta(t-1)+r:0 \leq i \leq t-1,\; \nonumber \\
     & 0 \leq j\leq s-1,\; 0 \leq r \leq z-1\}\nonumber \\
     = & \bigcup\limits_{i=0}^{t-1} \{is, \ldots, (i+1)s+z-2\} +ts+\theta(t-1)\nonumber \\
     = & \{ts+\theta(t-1),\ldots,2ts+\theta(t-1)+z-2\},
      \end{align}       
where the last equality comes from the fact that there is no gap between the subsets of $\bigcup\limits_{i=0}^{t-1} \{is, \ldots, (i+1)s+z-2\}$. The reason is that the largest element of the $i^{\text{th}}$ subset, \ie $(i+1)s+z-2$ plus one is larger than or equal to the smallest element of the $(i+1)^{\text{st}}$ subset, \ie $(i+1)s$ as $z\ge 1$.

In the following, we consider different regions for the values of $z$ and $\lambda$ and calculate $|\mathbf{P}({H}(x))|$ through calculation of $\mathbf{D}_3$ and $\mathbf{D}_4$. In addition, we use the following lemma, whichhelps us to calculate $\mathbf{P}({H}(x))$ without requiring to consider all of the terms of $\mathbf{D}_3$  in some cases.
\begin{lemma}\label{lemma:UpperBoundPHx-AGE} The following inequality holds. 
\begin{align}\label{eq:UpperBoundPHx-AGE}
    |\mathbf{P}({H}(x))|\leq& \deg(S_A(x))+\deg(S_B(x))+1\nonumber\\
    &=\max\{\mathbf{D}_4\}+1
\end{align}
\end{lemma}
{\em Proof:} 
$|\mathbf{P}({H}(x))|$, which is equal to the number of terms in $H(x)$ with non-zero coefficients, is less than or equal to the number of all terms, which is equal to $\deg(H(x))+1$. Thus, 
\begin{align}\label{eq:phxUpperBound-AGE}
    |\mathbf{P}({H}(x))|\leq &\deg(H(x))+1 \nonumber \\ =&\deg((C_A(x)+S_A(x))(C_B(x)+S_B(x)))+1 \nonumber \\
    =&\max\{\deg(C_A(x)),\deg(S_A(x)) \} \nonumber \\
    &+\max\{\deg(S_B(x)), \deg(C_B(x))\}+1.
\end{align}
From (\ref{eq:AGE-p(CA)-th}), $\deg(C_A(x))=ts-1$. On the other hand, from (\ref{eq:S-A}), $\deg(S_A(x))\ge ts$. Therefore, $\max\{\deg(C_A(x)),\deg(S_A(x))\} = \deg(S_A(x))$. Moreover, From (\ref{eq:AGE-p(CB)-th}), $\deg(C_B(x))=s-1+\theta(t-1)$, and from (\ref{eq:S-B}), $\deg(S_B(x))\ge ts+\theta(t-1)$. Therefore, $\max\{\deg(C_B(x)),\deg(S_B(x))\} = \deg(S_B(x))$,
which results in the first inequality of (\ref{eq:UpperBoundPHx-AGE}).

On the other hand, from (\ref{eq:d4definition-AGE}), $\max\{\mathbf{D}_4\}=\max\{\mathbf{P}(S_A(x))\}+\max\{\mathbf{P}(S_B(x))\}=\deg(S_A(x))+\deg(S_B(x))$.

This completes the proof. \hfill $\Box$

\begin{lemma}\label{lemma:non-zero-coeff-AGE-lambda=0 and z greater ts-t}
For $z > ts-s, t\neq 1$ and $\lambda=0$, we have
\begin{align}\label{eq:non-zero-coef-AGE-lambda=0 z greater ts-s}
    |\mathbf{P}({H}(x))|= \Upsilon_1(0)= 2st^2+2z-1
\end{align}
\end{lemma}
{\em Proof:} By replacing $\lambda$ with $0$ in AGE-CMPC formulations, the scheme is equivalent to Entangled-CMPC in \cite{8613446}. Therefor, the proof of this lemma can be derived directly from the proof of Theorem 1 in \cite{8613446}.\hfill $\Box$

\begin{lemma}\label{lemma:non-zero-coeff-AGE-lambda=0 and z less ts-t}
For $z \leq ts-s, t\neq 1$ and $\lambda=0$, we have
\begin{align}
    |\mathbf{P}({H}(x))|&=\Upsilon_2(0)= st^2+3st-2s+t(z-1)+1
\end{align}
\end{lemma}
{\em Proof:} For this case, AGE-CMPC is equivalent to Entangled-CMPC. Therefore, the proof of this lemma can be derived directly from the proof of Theorem 1 in \cite{8613446}.\hfill $\Box$

\begin{lemma}\label{lemma:non-zero-coeff-AGE-lambda=z}
 For $\lambda=z, t\neq 1$, we have
\begin{align}
    |\mathbf{P}({H}(x))|&=\Upsilon_3(z)= 2ts+\theta(t-1)+2z-1\nonumber\\
    &=(ts+z)(1+t)-1
\end{align}
\end{lemma}
 
 {\em Proof:} 
 To prove this lemma, we first calculate $\mathbf{D}_3$ from (\ref{eq:AGE-p(CB)-th}) and (\ref{eq:S-A}):
 \begin{align}\label{eq:d3-AGE- lambda=z}
      \mathbf{D}_3 = & \mathbf{P}(S_{A}(x))+\mathbf{P}(C_B(x)) \nonumber \\
      = & \{ts+u: 0 \leq u \leq z-1\} \nonumber \\
      + & \{s-1-k+\theta l:0 \leq l \leq t-1,\; 0 \leq k \leq s-1,\}\nonumber \\
     = & \{ts,\dots,ts+z+s-2\}+\{\theta l: 0 \leq l \leq t-1\} \nonumber \\
     = & \bigcup_{l=0}^{t-1} \{\theta l+ts,\dots,\theta l +ts+z+s-2\}
     .
\end{align}
From (\ref{eq:d1-AGE}) and (\ref{eq:d3-AGE- lambda=z}), we can calculate $\mathbf{D}_{13}=\mathbf{D}_1 \cup \mathbf{D}_3$ as:
\begin{align}
    \mathbf{D}_{13}=&\mathbf{D}_1 \cup \mathbf{D}_3\nonumber\\
    =&\bigcup\limits_{l=0}^{t-1}\{\theta l,\ldots,ts+s-2+\theta l\}\nonumber\\
    &\cup\bigcup_{l=0}^{t-1} \{\theta l+ts,\dots,\theta l +ts+z+s-2\}\nonumber\\
    =&\bigcup_{l=0}^{t-1} \{\theta l,\ldots,\theta l +ts+z+s-2\}\label{eq:d13AGE11}\\
    =&\{0,\ldots,\theta (t-1) +ts+z+s-2\}\label{eq:d13AGE12},
\end{align}
where (\ref{eq:d13AGE11}) comes from the fact that $\theta l < \theta l+ts \leq ts+s-2+\theta l < \theta l +ts+z+s-2$ and (\ref{eq:d13AGE12}) comes from the fact that there is no gap between each two consecutive subsets of $\bigcup_{l=0}^{t-1} \{\theta l,\ldots,\theta l +ts+z+s-2\}$ as $\theta l+ts+z+s-2+1 = \theta l+\theta+s-1 \ge \theta (l+1)$. 
Next, we calculate $\mathbf{D}_{123}=\mathbf{D}_{13} \cup \mathbf{D}_{2}$ from (\ref{eq:d13AGE12}) and (\ref{eq:d2-AGE})
\begin{align}\label{eq:d123AGE1}
    \mathbf{D}_{123}=&\mathbf{D}_{1} \cup \mathbf{D}_{3} \cup \mathbf{D}_{2}\nonumber\\
    =&\{0,\ldots,\theta (t-1) +ts+z+s-2\} \nonumber\\
    &\cup\{ts+\theta(t-1),\ldots,2ts+\theta(t-1)+z-2\}\nonumber\\
    =&\{0,\ldots,2ts+\theta(t-1) +z-2\},
\end{align}
where the last equality 
comes from the fact that $0 < ts+\theta(t-1) \leq \theta(t-1)+ts+z+s-2 < 2ts+\theta(t-1)+z-2$. Next, we first calculate $\mathbf{D}_4$, and then its union with $\mathbf{D}_{123}$.
From (\ref{eq:S-A}) and (\ref{eq:S-B}), we have
 \begin{align}\label{eq:d4-AGE- lambda=z}
      \mathbf{D}_4 = & \mathbf{P}(S_{A}(x))+\mathbf{P}(S_B(x)) \nonumber \\
      = & \{ts,\dots,ts+z-1\} \nonumber \\
      + & \{ts+\theta(t-1),\dots,ts+\theta(t-1)+z-1\} \nonumber \\
      = & \{2ts+\theta(t-1),\dots,2ts+\theta(t-1)+2z-2\}.
\end{align}
From (\ref{eq:PHx-AGE}), (\ref{eq:d123AGE1}) and (\ref{eq:d4-AGE- lambda=z}), we have
\begin{align}
    \mathbf{P}({H}(x)) = & \mathbf{D}_{123} \cup \mathbf{D}_4 \nonumber\\
    &\{0,\ldots,2ts+\theta(t-1) +z-2\} \cup\nonumber\\ &\{2ts+\theta(t-1),\dots,2ts+\theta(t-1)+2z-2\}\nonumber\\
    =&\{0,\ldots,2ts+\theta(t-1)+2z-2\}.
\end{align}
Therefore, $|\mathbf{P}({H}(x))|=2ts+\theta(t-1)+2z-2+1$. 
This completes the proof. \hfill $\Box$ 


For the remaining regions of the values of $z$ and $\lambda$, where $\lambda < z$, we use the following lemma to calculate $\mathbf{P}({H}(x))$. 

\begin{lemma}
For $\lambda < z$, we have
\begin{align}\label{eq:d1cupd2cupd3AGE4}
    \mathbf{D}_1 \cup  \mathbf{D}_2\cup \mathbf{D}_3 = \mathbf{\widehat{D}}_{123'} \cup \mathbf{\widetilde{D}}'_3 \cup \mathbf{\widetilde{D}}''_3,
\end{align}
\end{lemma}
where $\mathbf{\widehat{D}}_{123'} =\{0,\dots,2ts+\theta(t-1)+z-2\}$, $\mathbf{\widetilde{D}}'_3=\bigcup_{l'=t-1}^{t+q-2} \{ts+\theta l',\dots,\theta(l'+1)+s-2\}$ and $\mathbf{\widetilde{D}}''_3=\{ts+(q+t-1)\theta,\dots,(q+1)ts+(t-1)\theta+s+z-2\}$.

{\em Proof:}
To prove this lemma, we first calculate and decompose $\mathbf{\widehat{D}}_{3}$ using (\ref{eq:S-A}) and (\ref{eq:AGE-p(CB)-th}):
\begin{align}
      \mathbf{D}_3 = & \mathbf{P}(S_{A}(x))+\mathbf{P}(C_B(x)) \nonumber \\
      = &(\{ts+\theta l+w, 0 \leq l \leq q-1, 0\leq w \leq \lambda-1\}\cup\nonumber \\
      &\{ts+\theta q+u, 0\leq u \leq z-1-q\lambda\})\nonumber\\
      +&\{s-1-k+l(ts+\lambda), 0 \leq k \leq s-1, 0 \leq l \leq t-1\}\nonumber\\
      =&\{ts+\theta l'+w', 0 \leq l' \leq t+q-2, \nonumber\\
      & \quad \quad \quad \quad \quad \quad 0 \leq w' \leq \lambda+s-2\}\nonumber\\
      &\cup \{ts+\theta l''+u', q \leq l'' \leq t+q-1, \nonumber\\
      & \quad \quad \quad \quad \quad \quad \quad 0 \leq u' \leq s+z-q\lambda-2\}\nonumber\\
      =&\{ts+\theta l'+w', 0 \leq l' \leq t+q-2, \nonumber\\
      & \quad \quad \quad \quad \quad \quad 0 \leq w' \leq \lambda+s-2\}\nonumber\\
      &\cup \{ts+\theta l''+u', q \leq l'' \leq t+q-2, \nonumber\\
      & \quad \quad \quad \quad \quad \quad \quad 0 \leq u' \leq s+z-q\lambda-2\}\nonumber\\
      &\cup \{ts+\theta l''+u', l'' = t+q-1, \nonumber\\
      & \quad \quad \quad \quad \quad \quad \quad 0 \leq u' \leq s+z-q\lambda-2\}\nonumber\\
      =&\{ts+\theta l'+w', 0 \leq l' \leq t+q-2, \nonumber\\
      & \quad \quad \quad \quad \quad \quad 0 \leq w' \leq \lambda+s-2\} \label{d3''hatSubsetd3'}\\
      &\cup \{ts+\theta (t+q-1)+u', 0 \leq u' \leq s+z-q\lambda-2\}\nonumber\\
      =&\{ts+\theta l'+w', 0 \leq l' \leq t-2, \nonumber\\
      & \quad \quad \quad \quad \quad \quad 0 \leq w' \leq \lambda+s-2\} \nonumber\\
      &\cup\{ts+\theta l'+w', t-1 \leq l' \leq t+q-2, \nonumber\\
      & \quad \quad \quad \quad \quad \quad 0 \leq w' \leq \lambda+s-2\} \nonumber\\
      &\cup \{ts+\theta (t+q-1)+u', 0 \leq u' \leq s+z-q\lambda-2\}\nonumber\\
      =&\mathbf{\widehat{D}}'_3\cup\mathbf{\widetilde{D}}'_3 \cup \mathbf{\widetilde{D}}''_3, \label{eq:d3-AGE-z grater than ts-1}
\end{align}
where
\begin{align}\label{eq:d'3hatAGE}
    \mathbf{\widehat{D}}'_3=\bigcup_{l'=0}^{t-2}\{ts+\theta l',\ldots,ts+\theta l'+\lambda+s-2\},
\end{align}
\begin{align}\label{eq:d'3tildeAGE}
    \mathbf{\widetilde{D}}'_3=\bigcup_{l'=t-1}^{t+q-2}\{ts+\theta l',\ldots,ts+\theta l'+\lambda+s-2\},
\end{align}
\begin{align}\label{eq:d''3tildeAGE}
    &\mathbf{\widetilde{D}}''_3=\nonumber\\
    &\{ts+\theta(t+q-1),\ldots,ts+\theta(t+q-1)+s+z-q\lambda-2\}\nonumber\\
    &=\nonumber\\
    &\{ts+\theta(t+q-1),\ldots,(q+1)ts+(t-1)\theta+s+z-2\},
\end{align}
and (\ref{d3''hatSubsetd3'}) comes from the fact that
\begin{align}
    z > \lambda &\Rightarrow z-1 \ge \lambda \nonumber\\
    & \Rightarrow q=\min\{\floor{\frac{z-1}{\lambda}},t-1\}=\floor{\frac{z-1}{\lambda}}\nonumber\\
    & \quad \quad  \& \quad q+1 > \frac{z-1}{\lambda} \nonumber\\
    & \Rightarrow \lambda+s-2 > s+z-q\lambda -3\nonumber \\
    & \Rightarrow \lambda+s-2 \ge s+z-q\lambda -2\nonumber
        \end{align}
    \begin{align}
    & \Rightarrow \{ts+\theta l''+u', q \leq l'' \leq t+q-2, \nonumber\\
      & \quad \quad \quad \quad \quad \quad \quad 0 \leq u' \leq s+z-q\lambda-2\}\nonumber\\
      &\subset \{ts+\theta l'+w', 0 \leq l' \leq t+q-2, \nonumber\\
      & \quad \quad \quad \quad \quad \quad 0 \leq w' \leq \lambda+s-2\}\nonumber\\
      .
\end{align}
Next, we calculate $\mathbf{\widehat{D}}_{123'} = \mathbf{D}_1 \cup \mathbf{\widehat{D}}'_3 \cup \mathbf{D}_2$ using (\ref{eq:d1-AGE}), (\ref{eq:d'3hatAGE}), and (\ref{eq:d2-AGE}):
\begin{align}
    \mathbf{\widehat{D}}_{123'} =& \mathbf{D}_1 \cup \mathbf{\widehat{D}}'_3 \cup \mathbf{D}_2\nonumber\\
    =&\bigcup\limits_{l=0}^{t-1}\{\theta l,\ldots,ts+s-2+\theta l\}\cup\nonumber\\
    &\bigcup_{l'=0}^{t-2}\{ts+\theta l',\ldots,ts+\theta l'+\lambda+s-2\}\cup\nonumber\\
    &\{ts+\theta(t-1),\ldots,2ts+\theta(t-1)+z-2\}\nonumber\\
    =&\bigcup\limits_{l=0}^{t-2}\{\theta l,\ldots,ts+s-2+\theta l\}\cup\nonumber\\
    &\{\theta (t-1),\ldots,ts+s-2+\theta (t-1)\}\cup\nonumber\\
    &\bigcup_{l'=0}^{t-2}\{ts+\theta l',\ldots,ts+\theta l'+\lambda+s-2\}\cup\nonumber\\
    &\{ts+\theta(t-1),\ldots,2ts+\theta(t-1)+z-2\}\nonumber\\
    =&\bigcup\limits_{l=0}^{t-2}\{\theta l,\ldots,ts+s-2+\theta l+\lambda\}\cup\nonumber\\
    &\{\theta (t-1),\ldots,ts+s-2+\theta (t-1)\}\cup\nonumber\\
    &\{ts+\theta(t-1),\ldots,2ts+\theta(t-1)+z-2\}\label{eq:dhat123'AGE1}\\
    =&\bigcup\limits_{l=0}^{t-2}\{\theta l,\ldots,ts+s-2+\theta l+\lambda\}\cup\nonumber\\
    &\{\theta (t-1),\ldots,2ts+\theta(t-1)+z-2\}\nonumber\\
    =&\{0,\ldots,ts+s-2+\theta(t-2)+\lambda\}\cup\nonumber\\
    &\{\theta (t-1),\ldots,2ts+\theta(t-1)+z-2\}\label{eq:dhat123'AGE2}\\
    =&\{0,\ldots,2ts+\theta(t-1)+z-2\}\label{eq:dhat123'AGEfinal},
\end{align}
where (\ref{eq:dhat123'AGE1}) comes from the fact that $s<ts+z$. Thus, $ts+s-2+\theta (t-1)< 2ts+\theta(t-1)+z-2$. We obtain (\ref{eq:dhat123'AGE2}) from the fact that $ts+s-2+\theta l+\lambda+1=s-1+\theta(l+1)\ge \theta(l+1)$ and the last equality comes from the fact that:
\begin{align}
    &\Rightarrow 0 \leq s-1 <2ts+z-2 \nonumber\\
    & \Rightarrow \theta(t-1) \leq s-1+\theta(t-1) < 2ts+z-2+\theta(t-1) \nonumber
    \end{align}
    \begin{align}
    & \Rightarrow \theta(t-1) \leq ts+s-1+\theta(t-2)+\lambda \nonumber\\
    &\quad \quad \quad \quad \quad \quad \quad \quad \quad \quad \quad< 2ts+\theta(t-1)+z-2.
\end{align}
We can derive (\ref{eq:d1cupd2cupd3AGE4}) from (\ref{eq:d3-AGE-z grater than ts-1}), (\ref{eq:d'3tildeAGE}), (\ref{eq:d''3tildeAGE}), and (\ref{eq:dhat123'AGEfinal}). 
This completes the proof. \hfill $\Box$ 

From (\ref{eq:S-A}) and (\ref{eq:S-B}), $\mathbf{D}_4$ for $z>\lambda$ is calculated as
 \begin{align}\label{eq:d4-AGE}
   \mathbf{D}_4 =& \mathbf{P}(S_{A}(x))+\mathbf{P}(S_B(x)) \nonumber \\
   = & \bigcup_{l=0}^{q-1} \{2ts+\theta(l+t-1),\dots,\nonumber\\
   &\quad \quad \quad \quad 2ts+\theta(l+t-1)+z-1+\lambda-1\} \nonumber \\
    &\cup  \{2ts+\theta(q+t-1),\dots, \nonumber \\
    & \quad \quad \quad 2ts+\theta(q+t-1)+2z-2-q(\theta-ts)\} \nonumber \\
   = & \bigcup_{l=0}^{q-1} \{2ts+\theta(l+t-1),\dots,ts+\theta(l+t)+z-2\} \nonumber \\
   &\cup \{2ts+\theta(q+t-1),\dots,\nonumber\\
   & \quad \quad \quad (q+2)ts+\theta(t-1)+2z-2\},
 \end{align}
where for $z>ts$. The above equation is a continuous set as there exist no gaps between each of its two consecutive subsets
. The reason is that, for $z>ts$, the greatest element of each subset plus one, \ie $ts+\theta(l+t)+z-1$, is greater than or equal to the smallest element of it's consecutive subset, \ie $2ts+\theta(l+t)$ for $l=\{0,\dots,q-1\}$. This is shown as
 \begin{align}
     & z > ts \Rightarrow ts+z-1 \geq 2ts \nonumber \\
     &  \Rightarrow ts+z-1+\theta(l+t) \geq 2ts+\theta(l+t).
 \end{align}
Therefore, for $z>\max\{ts,\lambda\}$, $\mathbf{D}_4$ is equal to:
\begin{align}\label{eq:d4-z greater ts-AGE}
    \mathbf{D}_4 = & 
    \{2ts+\theta(t-1),\dots,(q+2)ts+\theta(t-1)+2z-2\}.
\end{align}

\begin{lemma}\label{lemma:non-zero-coeff-AGE-z greater than ts}
For $z > ts, t\neq 1$ and $0 < \lambda < z$:
\begin{equation}
    |\mathbf{P}({H}(x))|= \Upsilon_4(\lambda)= (q+2)ts+\theta(t-1)+2z-1
\end{equation}
\end{lemma}
{\em Proof:} 
To prove this lemma, we calculate $\mathbf{P}({H}(x)) = \mathbf{D}_1 \cup  \mathbf{D}_2\cup \mathbf{D}_3 \cup \mathbf{D}_4$ using (\ref{eq:d1cupd2cupd3AGE4}) and (\ref{eq:d4-z greater ts-AGE}):
\begin{align}
    &\mathbf{P}({H}(x)) = \mathbf{D}_1 \cup  \mathbf{D}_2\cup \mathbf{D}_3 \cup \mathbf{D}_4\nonumber\\
    =&\mathbf{\widehat{D}}_{123'} \cup \mathbf{\widetilde{D}}'_3 \cup \mathbf{\widetilde{D}}''_3\cup\mathbf{D}_4\nonumber\\
    =&\{0,\dots,2ts+\theta(t-1)+z-2\}\nonumber\\
    &\cup\{2ts+\theta(t-1),\dots,(q+2)ts+\theta(t-1)+2z-2\}\nonumber\\
    &\cup \mathbf{\widetilde{D}}'_3 \cup \mathbf{\widetilde{D}}''_3\nonumber\\
    =&\{0,\dots,(q+2)ts+\theta(t-1)+2z-2\}\nonumber\\
    &\cup \mathbf{\widetilde{D}}'_3 \cup \mathbf{\widetilde{D}}''_3
\end{align}
From the above equation, $|\mathbf{P}({H}(x))| \ge (q+2)ts+\theta(t-1)+2z-1$.
On the other hand, from (\ref{eq:UpperBoundPHx-AGE}), $|\mathbf{P}({H}(x))| \leq \max\{\mathbf{D}_4\}+1=(q+2)ts+\theta(t-1)+2z-2+1$. Therefore, $|\mathbf{P}({H}(x))| = (q+2)ts+\theta(t-1)+2z-1$.  
This completes the proof.
\hfill $\Box$

\begin{lemma}\label{lemma:non-zero-coeff-AGE-z less than ts: 1}
For $z\leq ts<\lambda+s-1, t\neq 1$ and $0< \lambda<z$, we have
\begin{equation}
    |\mathbf{P}({H}(x))|= \Upsilon_5(\lambda)= 3ts+\theta(t-1)+2z-1
\end{equation}
\end{lemma}
{\em Proof:} 
For the conditions of this lemma, \ie $ ts-s+2 \leq \lambda \leq z-1$ and $z-1 \leq ts-1$, the range of variation of $\frac{z-1}{\lambda}$ and thus the value of $q$ is calculated as follows:
\begin{align}
     &1 \leq \frac{z-1}{\lambda} \leq \frac{ts-1}{ts-s+2}\nonumber \\
      \Rightarrow &1 \leq \frac{z-1}{\lambda} \leq \frac{ts-1}{ts-s+2} < 2 \label{eq:qforlemma40} \\
     \Rightarrow &q = \floor{\frac{z-1}{\lambda}} = 1,
\end{align}
where (\ref{eq:qforlemma40}) comes from the fact that $s(t-2)\ge 0$ and thus $s(t-2)+5>0$. By replacing $q=1$ in (\ref{eq:d1cupd2cupd3AGE4}) and (\ref{eq:d4-AGE}), we calculate $\mathbf{P}({H}(x)) = \mathbf{D}_1 \cup  \mathbf{D}_2\cup \mathbf{D}_3 \cup \mathbf{D}_4$ as
\begin{align}
    \mathbf{P}({H}(x)) =& \mathbf{D}_1 \cup  \mathbf{D}_2\cup \mathbf{D}_3 \cup \mathbf{D}_4\nonumber\\
    =&\mathbf{\widehat{D}}_{123'} \cup \mathbf{\widetilde{D}}'_3 \cup \mathbf{\widetilde{D}}''_3\cup\mathbf{D}_4\nonumber\\
    =&\{0,\dots,2ts+\theta(t-1)+z-2\}\nonumber\\
    &\cup \{ts+\theta(t-1),\dots,\theta t+s-2\} \nonumber\\
    &\cup \{ts+t\theta,\dots,2ts+(t-1)\theta+s+z-2\}\nonumber\\
    &\cup \{2ts+\theta(t-1),\dots,ts+\theta t+z-2\} \nonumber \\
   &\cup \{2ts+\theta t,\dots,3ts+\theta(t-1)+2z-2\}\nonumber\\
   =&\{0,\dots,2ts+\theta(t-1)+z-2\}\nonumber\\
   &\cup \{ts+t\theta,\dots,2ts+(t-1)\theta+s+z-2\}\nonumber\\
    &\cup \{2ts+\theta(t-1),\dots,ts+\theta t+z-2\} \nonumber \\
   &\cup \{2ts+\theta t,\dots,3ts+\theta(t-1)+2z-2\}\label{eq:d1cupd2cupd3cup4AGE39a}\\
   =&\{0,\dots,2ts+(t-1)\theta+s+z-2\}\nonumber\\
    &\cup \{2ts+\theta(t-1),\dots,ts+\theta t+z-2\} \nonumber \\
   &\cup \{2ts+\theta t,\dots,3ts+\theta(t-1)+2z-2\}\label{eq:d1cupd2cupd3cup4AGE39b}\\
   =&\{0,\dots,ts+\theta t+z-2\}\nonumber\\
    &\cup \{2ts+\theta(t-1),\ldots,3ts+\theta(t-1)+2z-2\}\label{eq:d1cupd2cupd3cup4AGE39c} \\
    =&\{0,\dots,ts+\theta (t-1)+\theta+z-2\}\nonumber\\ 
    &\cup \{2ts+\theta(t-1),\ldots,3ts+\theta(t-1)+2z-2\}\nonumber\\
     =&\{0,\dots,2ts+\theta (t-1)+\lambda+z-2\}\nonumber\\ 
    &\cup \{2ts+\theta(t-1),\ldots,3ts+\theta(t-1)+2z-2\}\nonumber\\
    =&\{0,\dots,3ts+\theta(t-1)+2z-2\}\label{eq:d1cupd2cupd3cup4AGE39final}
\end{align}
where (\ref{eq:d1cupd2cupd3cup4AGE39a}) comes from 
\begin{align}
    & \lambda < z \nonumber \\
    \Rightarrow & s(1-t)<0 < z-\lambda \nonumber \\
    \Rightarrow & s-2 < ts-\lambda z+z-2 \nonumber \\
    \Rightarrow & \theta t+s-2 < 2ts+\theta(t-1)+z-2\nonumber\\
    \Rightarrow & \{ts+\theta(t-1),\dots,\theta t+s-2\} \nonumber\\
    &\subset \{0,\dots,2ts+\theta(t-1)+z-2\},
\end{align} and
(\ref{eq:d1cupd2cupd3cup4AGE39b}) comes from 
\begin{align}
    & \lambda < z \nonumber \\
    \Rightarrow & \lambda \leq z-1 \nonumber \\
    \Rightarrow & ts+t\theta \leq 2ts+\theta(t-1)+z-2+1,
\end{align} and
(\ref{eq:d1cupd2cupd3cup4AGE39c}) comes from 
\begin{align}
    & ts-s+1 < \lambda \nonumber \\
    \Rightarrow & s(t-1)+1 < \lambda \nonumber \\
    \Rightarrow & s(2-1)+1 \leq s(t-1)+1 < \lambda \nonumber \\
    \Rightarrow & s < \lambda \nonumber \\
    \Rightarrow & 2ts+(t-1)\theta+s+z-2 < ts+\theta t +z-2,
\end{align}
and (\ref{eq:d1cupd2cupd3cup4AGE39final}) comes from the fact that $\lambda+z-2\ge 0$ (because $\lambda \ge 1$ and $z \ge 1$) and $\lambda < z$ (and thus $\lambda < z+ts$ and $2ts+\theta(t-1)+\lambda+z-2<3ts+\theta(t-1)+2z-2$). 
From (\ref{eq:d1cupd2cupd3cup4AGE39final}), $|\mathbf{P}({H}(x))|=3ts+\theta(t-1)+2z-1$. This completes the proof. \hfill $\Box$

In order to calculate $\mathbf{P}({H}(x))=\mathbf{\widehat{D}}_{123'} \cup \mathbf{\widetilde{D}}'_3 \cup \mathbf{\widetilde{D}}''_3\cup\mathbf{D}_4$ for the remaining regions of the values of $z$ and $\lambda$, \ie $z>\lambda>0$, we first calculate $\mathbf{\widehat{D}}_{123'}\cup\mathbf{D}_4$ using (\ref{eq:dhat123'AGEfinal}) and (\ref{eq:d4-AGE}) as follows
\begin{align}
    &\mathbf{P}({H}(x)) = \mathbf{D}_1 \cup  \mathbf{D}_2\cup \mathbf{D}_3 \cup \mathbf{D}_4\nonumber\\
    =&\mathbf{\widehat{D}}_{123'} \cup \mathbf{\widetilde{D}}'_3 \cup \mathbf{\widetilde{D}}''_3\cup\mathbf{D}_4\nonumber\\
    =&\{0,\ldots,2ts+\theta(t-1)+z-2\}\nonumber\\
    &\cup\bigcup_{l=0}^{q-1} \{2ts+\theta(l+t-1),\dots,ts+\theta(l+t)+z-2\} \nonumber \\
   &\cup \{2ts+\theta(q+t-1),\dots ,(q+2)ts+\theta(t-1)+2z-2\}\nonumber\\
    &\cup \mathbf{\widetilde{D}}'_3 \cup \mathbf{\widetilde{D}}''_3\nonumber\\
    =&\{0,\ldots,2ts+\theta(t-1)+z-2\}\nonumber\\
    &\cup\{2ts+\theta(t-1),\dots,ts+\theta t+z-2\}\nonumber\\
    &\cup\bigcup_{l=1}^{q-1} \{2ts+\theta(l+t-1),\dots,ts+\theta(l+t)+z-2\} \nonumber \\
   &\cup \{2ts+\theta(q+t-1),\dots ,(q+2)ts+\theta(t-1)+2z-2\}\nonumber\\
    &\cup \mathbf{\widetilde{D}}'_3 \cup \mathbf{\widetilde{D}}''_3\nonumber\\
    =&\{0,\ldots,ts+\theta t+z-2\}\nonumber\\
    &\cup\bigcup_{l=1}^{q-1} \{2ts+\theta(l+t-1),\dots,ts+\theta(l+t)+z-2\} \nonumber \\
   &\cup \{2ts+\theta(q+t-1),\dots ,(q+2)ts+\theta(t-1)+2z-2\}\nonumber\\
    &\cup \mathbf{\widetilde{D}}'_3 \cup \mathbf{\widetilde{D}}''_3 \label{eq:d1cupd2cupd3cupd3AGELastFour1}\\
    =&(\mathbf{\widehat{D}}_{123'4}\cup\mathbf{\widetilde{D}}'_3)\cup(\mathbf{\widehat{D}}_{123'4}\cup\mathbf{\widetilde{D}}''_3)\quad \quad \text{for    } 0<\lambda<z \label{eq:phxforthe4lastcases},
\end{align}
where (\ref{eq:d1cupd2cupd3cupd3AGELastFour1}) comes from the fact that $\lambda>0$ and thus $ts+\theta t+z-2 > 2ts+\theta(t-1)+z-2$ and $\mathbf{\widehat{D}}_{123'4}$ is equal to
\begin{align}\label{eq:d1234}
    &\mathbf{\widehat{D}}_{123'4} =\{0,\ldots,ts+\theta t+z-2\}\nonumber\\
    &\cup\bigcup_{l=1}^{q-1} \{2ts+\theta(l+t-1),\dots,ts+\theta(l+t)+z-2\} \nonumber \\
   &\cup \{2ts+\theta(q+t-1),\dots ,(q+2)ts+\theta(t-1)+2z-2\}.
\end{align}
Next, we calculate $\mathbf{\widehat{D}}_{123'4}\cup\mathbf{\widetilde{D}}'_3$ and $\mathbf{\widehat{D}}_{123'4}\cup\mathbf{\widetilde{D}}''_3$ for different regions of values of $z$ and $\lambda$.

\begin{lemma}\label{lemma:d1234cupd'3AGE}
\begin{align}
\mathbf{\widehat{D}}_{123'4}\cup\mathbf{\widetilde{D}}'_3 =\begin{cases}
\mathbf{D}_{123'4(a)}, & z > \lambda+s-1 \\
   \mathbf{D}_{123'4(b)}, & z \leq \lambda+s-1,
\end{cases}
\end{align}
\begin{align}
    &\mathbf{D}_{123'4(a)} = \{0,\ldots,ts+\theta t+z-2\}\nonumber\\
    &\cup\bigcup_{l=t}^{t+q-2} \{2ts+\theta l,\dots,ts+\theta(l+1)+z-2\} \nonumber \\
   &\cup \{2ts+\theta(q+t-1),\dots ,(q+2)ts+\theta(t-1)+2z-2\},
\end{align}
\begin{align}
    &\mathbf{D}_{123'4(b)}= \{0,\ldots,\theta(t+1)+s-2\}\nonumber\\
    &\cup\bigcup_{l=t}^{t+q-3} \{2ts+\theta  l,\dots,\theta(l+2)+s-2\} \nonumber\\
    &\cup \{2ts+\theta (t+q-2),\dots,ts+\theta(t+q-1)+z-2\} \nonumber \\
   &\cup \{2ts+\theta(q+t-1),\dots ,(q+2)ts+\theta(t-1)+2z-2\}
\end{align}
\end{lemma}
{\em Proof:}
From (\ref{eq:d1234}) and (\ref{eq:d1cupd2cupd3AGE4}), we have:
\begin{align}
&\mathbf{\widehat{D}}_{123'4}\cup\mathbf{\widetilde{D}}'_3 = \{0,\ldots,ts+\theta t+z-2\}\nonumber\\
    &\cup\bigcup_{l=1}^{q-1} \{2ts+\theta(l+t-1),\dots,ts+\theta(l+t)+z-2\} \nonumber \\
   &\cup \{2ts+\theta(q+t-1),\dots ,(q+2)ts+\theta(t-1)+2z-2\}\nonumber\\
   &\cup \bigcup_{l'=t-1}^{t+q-2}\{ts+\theta l',\ldots,\theta (l'+1)+s-2\}\nonumber\\
   = &\{0,\ldots,ts+\theta t+z-2\}\nonumber
         \end{align}
   \begin{align}
    &\cup\bigcup_{l=1}^{q-1} \{2ts+\theta(l+t-1),\dots,ts+\theta(l+t)+z-2\} \nonumber \\
   &\cup \{2ts+\theta(q+t-1),\dots ,(q+2)ts+\theta(t-1)+2z-2\}\nonumber\\
   &\cup \{ts+\theta (t-1),\ldots,\theta t+s-2\}\nonumber\\
   &\cup \bigcup_{l'=t}^{t+q-2}\{ts+\theta l',\ldots,\theta (l'+1)+s-2\}\nonumber\\
   = &\{0,\ldots,ts+\theta t+z-2\}\nonumber\\
    &\cup\bigcup_{l=1}^{q-1} \{2ts+\theta(l+t-1),\dots,ts+\theta(l+t)+z-2\} \nonumber
       \end{align}
   \begin{align}
   &\cup \{2ts+\theta(q+t-1),\dots ,(q+2)ts+\theta(t-1)+2z-2\}\nonumber\\
   &\cup \bigcup_{l'=t}^{t+q-2}\{ts+\theta l',\ldots,\theta (l'+1)+s-2\}\label{eq:d1cupd2cupd3cupd4cupdthilde3a}\\
   = &\{0,\ldots,ts+\theta t+z-2\}\nonumber
      \end{align}
   \begin{align}
    &\cup\bigcup_{l=t}^{t+q-2} \{2ts+\theta l,\dots,ts+\theta(l+1)+z-2\} \nonumber
    \end{align}
    \begin{align}
   &\cup \{2ts+\theta(q+t-1),\dots ,(q+2)ts+\theta(t-1)+2z-2\}\nonumber\\
   &\cup \bigcup_{l'=t}^{t+q-2}\{ts+\theta l',\ldots,\theta (l'+1)+s-2\}\nonumber\\
   = &\{0,\ldots,ts+\theta t+z-2\}\nonumber\\
    &\cup\bigcup_{l=t}^{t+q-2} \{2ts+\theta l,\dots,ts+\theta(l+1)+z-2\} \nonumber \\
   &\cup \{2ts+\theta(q+t-1),\dots ,(q+2)ts+\theta(t-1)+2z-2\}\nonumber\\
   &\cup \{ts+\theta t,\ldots,\theta (t+1)+s-2\}\nonumber\\
   &\cup \bigcup_{l'=t}^{t+q-3}\{ts+\theta (l'+1),\ldots,\theta (l'+2)+s-2\}
   \label{eq:d1cupd2cupd3cupd4cupdthilde3b}
\end{align}
where (\ref{eq:d1cupd2cupd3cupd4cupdthilde3a}) comes from the fact that $s<ts+z$ and thus $\theta t+s-2<ts+\theta t+z-2$; this results in $\{ts+\theta (t-1),\ldots,\theta t+s-2\} \subset \{0,\ldots,ts+\theta t+z-2\}$. 

Now, we consider the two cases;  Case 1: $z > \lambda+s-1$, and Case 2: $z \leq \lambda+s-1$, and simplify (\ref{eq:d1cupd2cupd3cupd4cupdthilde3b}) for each case. 

\begin{figure*}
		\centering
		\includegraphics[width=14cm]{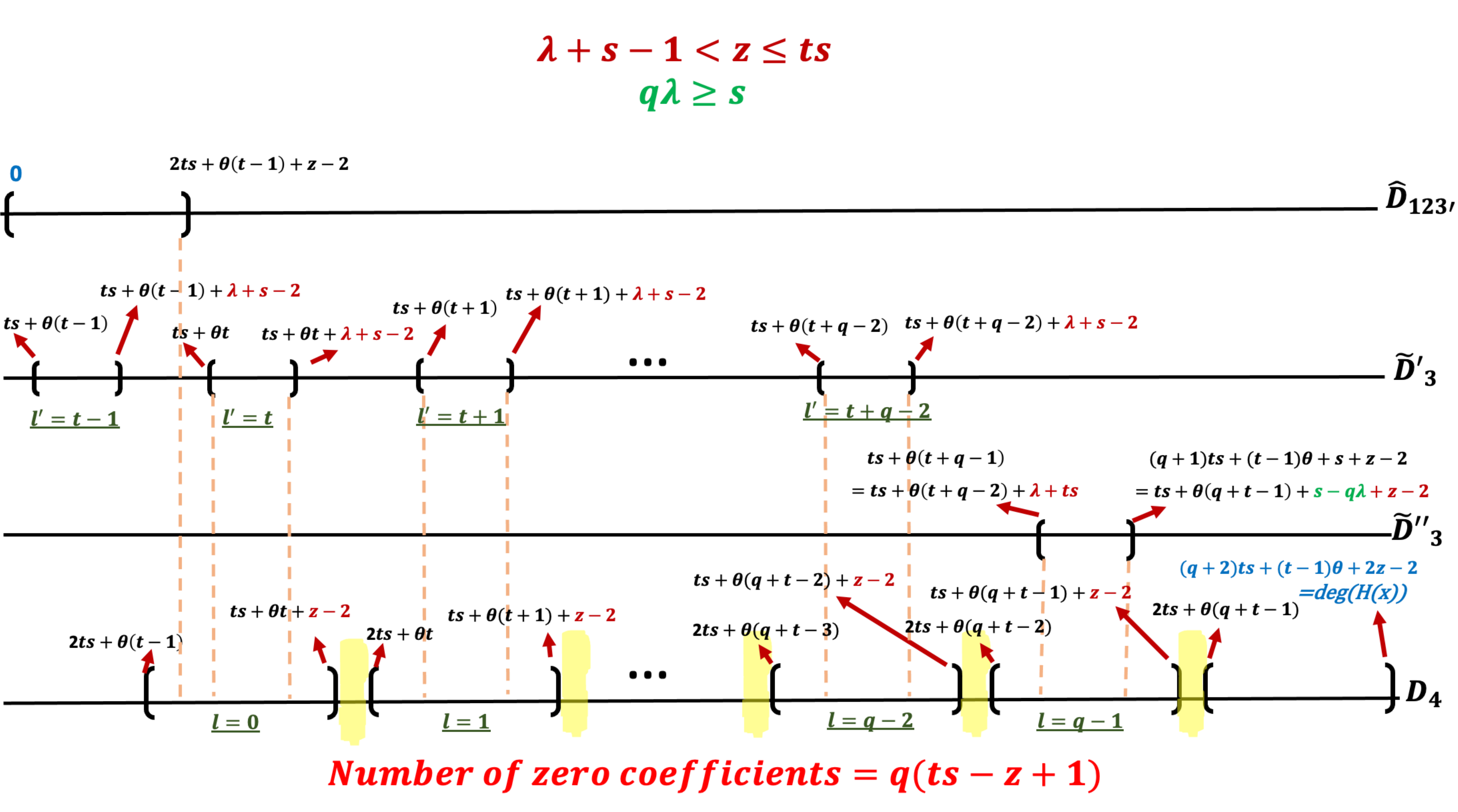}
	\caption{Illustration of $ \mathbf{\widehat{D}}_{123'} \cup \mathbf{\widetilde{D}}'_3 \cup \mathbf{\widetilde{D}}''_3 \cup \mathbf{D}_4$ for $\lambda+s-1 < z \leq ts$ and $q\lambda \geq s$.
	}
\label{fig:non-zero-coeff-AGE-z less than ts: 2}
\vspace{-10pt}
\end{figure*}

\begin{figure*}
		\centering
		\includegraphics[width=14cm]{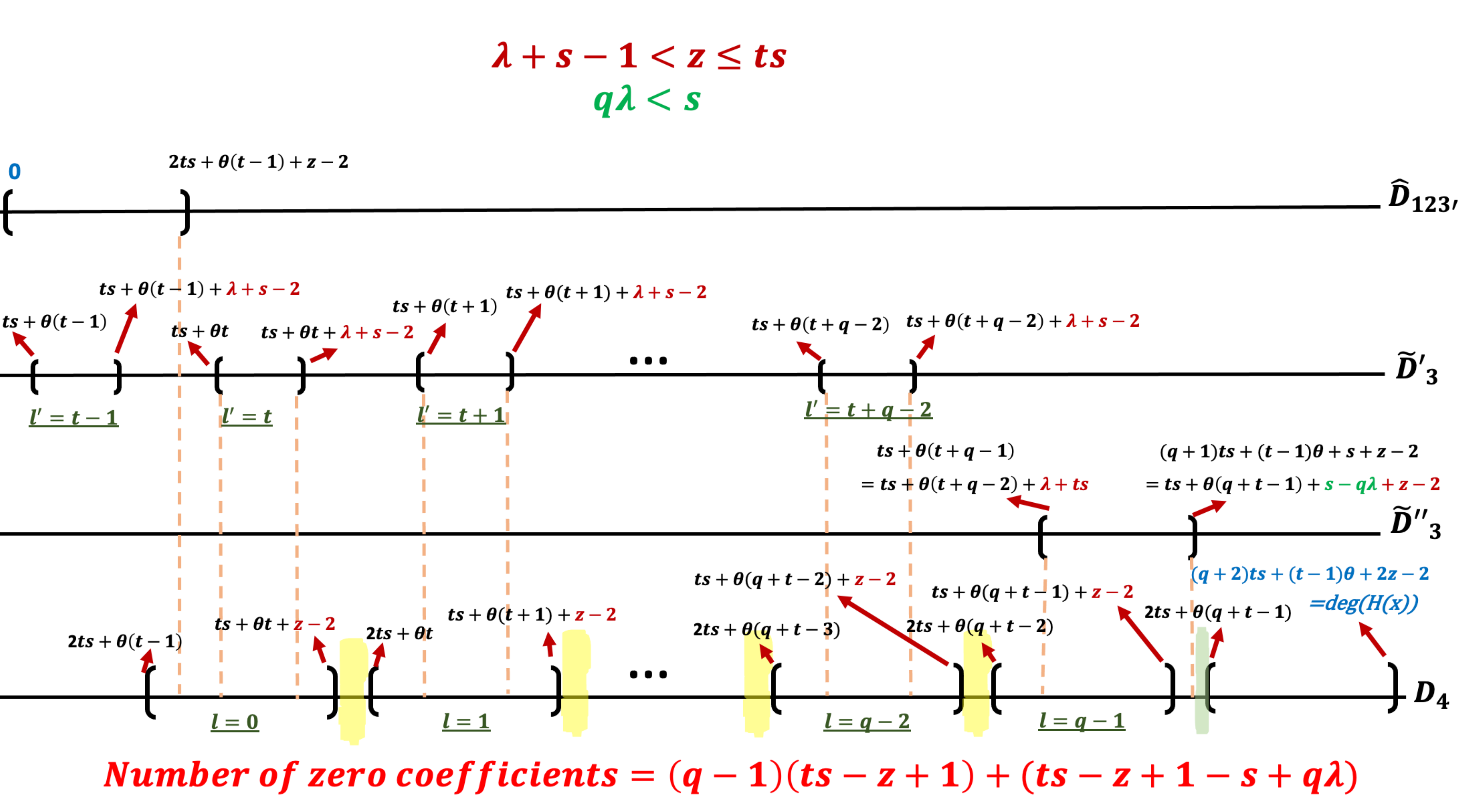}
	\caption{Illustration of $ \mathbf{\widehat{D}}_{123'} \cup \mathbf{\widetilde{D}}'_3 \cup \mathbf{\widetilde{D}}''_3 \cup \mathbf{D}_4$ for $\lambda+s-1 < z \leq ts$ and $q\lambda < s$.
	}
\label{fig:non-zero-coeff-AGE-z less than ts: 3}
\vspace{-10pt}
\end{figure*}

Case 1: $z > \lambda+s-1$. For this case, $\bigcup_{l'=t}^{t+q-3} \{ts+\theta (l'+1),\dots,\theta(l'+2)+s-2\}$ is a subset of $\bigcup_{l=t}^{t+q-2} \{2ts+\theta l,\dots,ts+\theta(l+1)+z-2\}$. This is formulated in the following and demonstrated in Fig. \ref{fig:non-zero-coeff-AGE-z less than ts: 2} and \ref{fig:non-zero-coeff-AGE-z less than ts: 3}.
\begin{align}\label{eq:subsetPrrofb}
    & 0 < \lambda \nonumber \\
    \Rightarrow & ts < \theta \nonumber \\
    \Rightarrow & 2ts+\theta l < ts+\theta(l+1),
\end{align}
and 
\begin{align}\label{eq:subsetPrrofc}
    & \lambda+s-1 < z \nonumber \\
    \Rightarrow & \lambda+s \leq z \nonumber \\
    \Rightarrow & \theta+s-2 \leq ts+z-2\nonumber\\
    \Rightarrow & \theta(l+2)+s-2 \leq ts+\theta(l+1)+z-2.
\end{align}

On the other hand, $\{ts+\theta t,\ldots,\theta (t+1)+s-2\}$ is a subset of $\{0,\ldots,ts+\theta t+z-2\}$. This is expressed in the following and demonstrated in Fig. \ref{fig:non-zero-coeff-AGE-z less than ts: 2} and \ref{fig:non-zero-coeff-AGE-z less than ts: 3}.
\begin{align}\label{eq:subsetProofa}
    & \lambda+s-1 < z \nonumber \\
    \Rightarrow & s+\lambda \leq z \nonumber \\
    \Rightarrow & ts+z \leq \theta+s\nonumber\\
    \Rightarrow & \theta (t+1)+s-2 \leq ts+\theta t+z-2,
\end{align}
Therefore, for the case of $z>\lambda+s-1$, (\ref{eq:d1cupd2cupd3cupd4cupdthilde3b}) is simplified as
\begin{align}
    &\mathbf{\widehat{D}}_{123'4}\cup\mathbf{\widetilde{D}}'_3 = \{0,\ldots,ts+\theta t+z-2\}\nonumber\\
    &\cup\bigcup_{l=t}^{t+q-2} \{2ts+\theta l,\dots,ts+\theta(l+1)+z-2\} \nonumber \\
   &\cup \{2ts+\theta(q+t-1),\dots ,(q+2)ts+\theta(t-1)+2z-2\}
\end{align}

\begin{figure*}
		\centering
		\includegraphics[width=14cm]{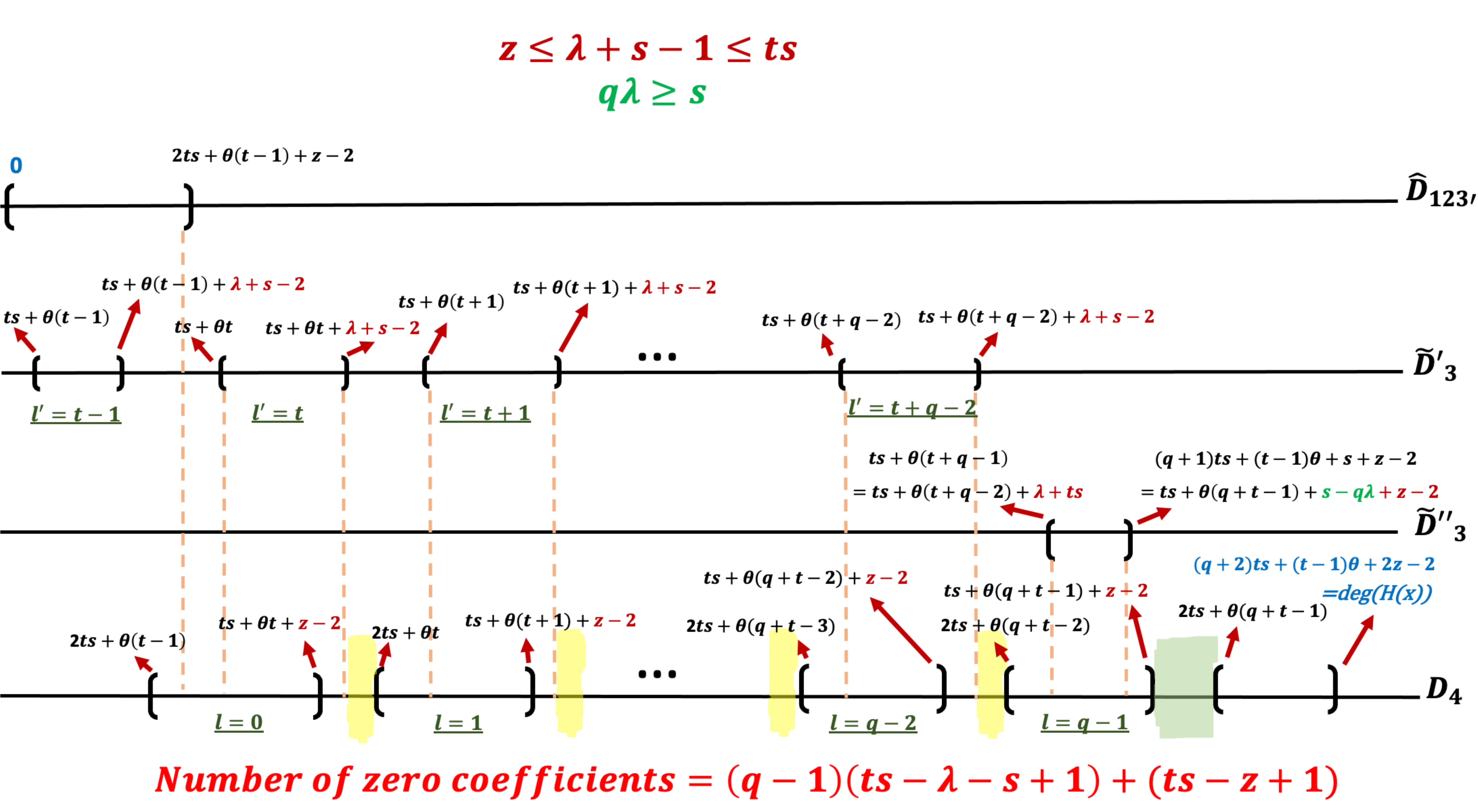}
	\caption{Illustration of $ \mathbf{\widehat{D}}_{123'} \cup \mathbf{\widetilde{D}}'_3 \cup \mathbf{\widetilde{D}}''_3 \cup \mathbf{D}_4$ for $z \leq \lambda+s-1 \leq ts$ and $q\lambda \geq s$.
	}
\label{fig:non-zero-coeff-AGE-z less than ts: 4}
\vspace{-10pt}
\end{figure*}

\begin{figure*}
		\centering
		\includegraphics[width=14cm]{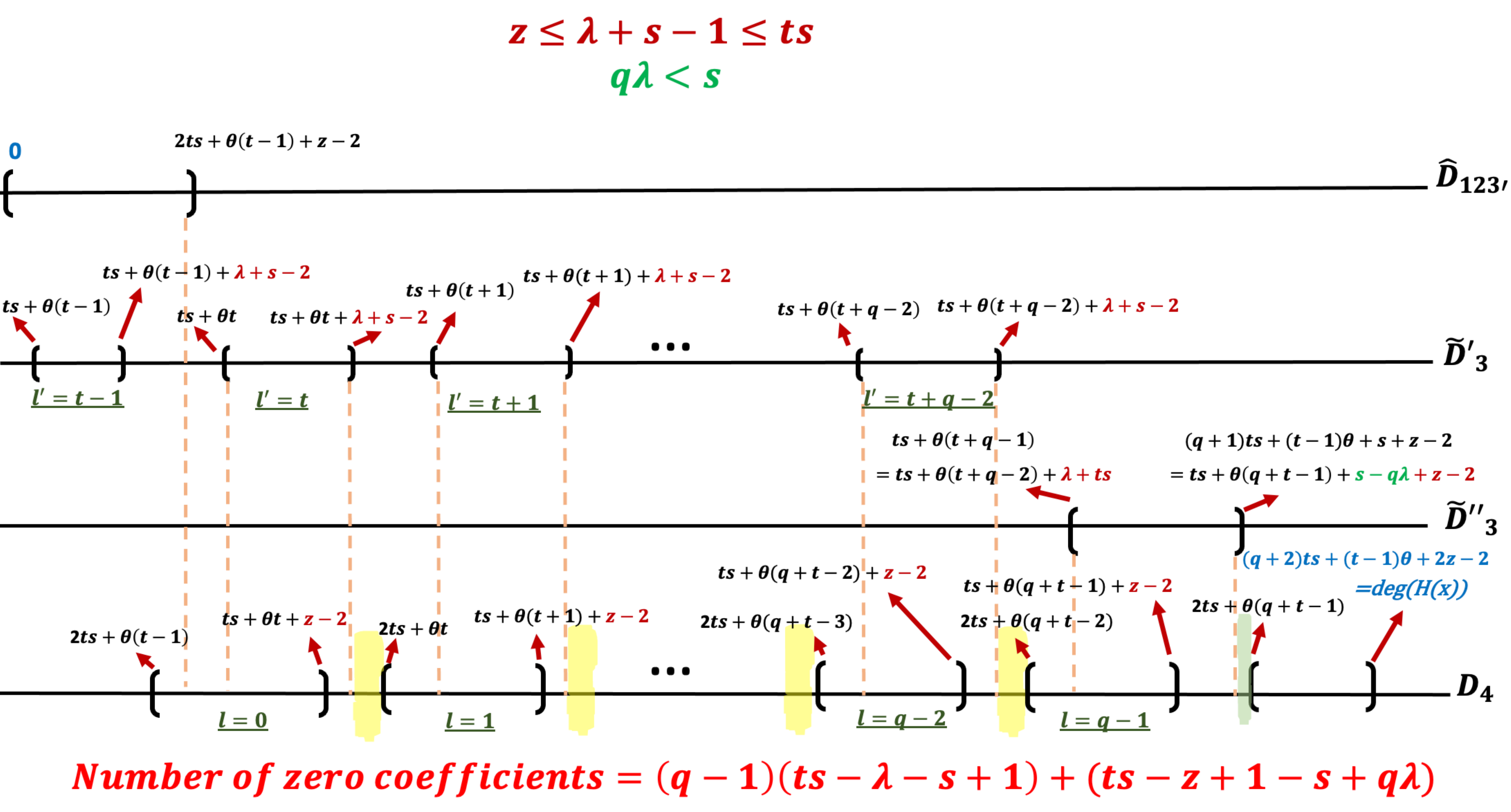}
	\caption{Illustration of $ \mathbf{\widehat{D}}_{123'} \cup \mathbf{\widetilde{D}}'_3 \cup \mathbf{\widetilde{D}}''_3 \cup \mathbf{D}_4$ for $z \leq \lambda+s-1 \leq ts$ and $q\lambda < s$.
	}
\label{fig::non-zero-coeff-AGE-z less than ts: 5}
\end{figure*}

Case 2: $z \leq \lambda+s-1$.
For this case, the union of $\{ts+\theta t,\ldots,\theta (t+1)+s-2\}$ and $\{0,\ldots,ts+\theta t+z-2\}$ is equal to $\{0,\ldots,\theta (t+1)+s-2\}$. This can be derived from (\ref{eq:subsetProofa}) and demonstrated in Fig. \ref{fig:non-zero-coeff-AGE-z less than ts: 4} and \ref{fig::non-zero-coeff-AGE-z less than ts: 5}.
On the other hand, the union of $\bigcup_{l'=t}^{t+q-3} \{ts+\theta (l'+1),\dots,\theta(l'+2)+s-2\}$ and $\bigcup_{l=t}^{t+q-2} \{2ts+\theta l,\dots,ts+\theta(l+1)+z-2\}$ is equal to $\bigcup_{l=t}^{t+q-3} \{2ts+\theta  l,\dots,\theta(l+2)+s-2\} \cup \{2ts+\theta (t+q-2),\dots,ts+\theta(t+q-1)+z-2\}$. This can be derived  from (\ref{eq:subsetPrrofb}) and (\ref{eq:subsetPrrofc}) and demonstrated in Fig. \ref{fig:non-zero-coeff-AGE-z less than ts: 4} and \ref{fig::non-zero-coeff-AGE-z less than ts: 5}.
Therefore, for the case of $z\leq \lambda+s-1$, (\ref{eq:d1cupd2cupd3cupd4cupdthilde3b}) is simplified as
\begin{align}
    &\mathbf{\widehat{D}}_{123'4}\cup\mathbf{\widetilde{D}}'_3 = \{0,\ldots,\theta(t+1)+s-2\}\nonumber\\
    &\cup\bigcup_{l=t}^{t+q-3} \{2ts+\theta  l,\dots,\theta(l+2)+s-2\} \nonumber\\
    &\cup \{2ts+\theta (t+q-2),\dots,ts+\theta(t+q-1)+z-2\} \nonumber \\
   &\cup \{2ts+\theta(q+t-1),\dots ,(q+2)ts+\theta(t-1)+2z-2\}
\end{align}
This completes the proof. \hfill $\Box$

\begin{lemma}\label{lemma:d1234cupd''3AGE} The following equalities hold.
\begin{align}
\mathbf{\widehat{D}}_{123'4}\cup\mathbf{\widetilde{D}}''_3 =\begin{cases}
\mathbf{D}_{123''4(a)}, & q\lambda \ge s \\
   \mathbf{D}_{123''4(b)}, & q\lambda < s,
\end{cases}
\end{align}
\begin{align}
&\mathbf{D}_{123''4(a)}= \{0,\ldots,ts+\theta t+z-2\}\nonumber\\
    &\cup\bigcup_{l=1}^{q-2} \{2ts+\theta(l+t-1),\dots,ts+\theta(l+t)+z-2\} \nonumber \\
    &\cup \{2ts+\theta(q+t-2),\dots,ts+\theta(q+t-1)+z-2\}\nonumber\\
   &\cup \{2ts+\theta(q+t-1),\dots ,(q+2)ts+\theta(t-1)+2z-2\},
\end{align}
\begin{align}
&\mathbf{D}_{123''4(b)}=\{0,\ldots,ts+\theta t+z-2\}\nonumber\\
    &\cup\bigcup_{l=1}^{q-2} \{2ts+\theta(l+t-1),\dots,ts+\theta(l+t)+z-2\} \nonumber \\
    &\cup \{2ts+\theta(q+t-2),\dots,(q+1)ts+(t-1)\theta+s+z-2\}\nonumber\\
   &\cup \{2ts+\theta(q+t-1),\dots ,(q+2)ts+\theta(t-1)+2z-2\}
\end{align}
\end{lemma}
{\em Proof:}
From (\ref{eq:d1234}) and (\ref{eq:d1cupd2cupd3AGE4}), we have
\begin{align}\label{eq:d1cupd2cupd3''cupd4AGE}
&\mathbf{\widehat{D}}_{123'4}\cup\mathbf{\widetilde{D}}''_3 = \{0,\ldots,ts+\theta t+z-2\}\nonumber\\
    &\cup\bigcup_{l=1}^{q-1} \{2ts+\theta(l+t-1),\dots,ts+\theta(l+t)+z-2\} \nonumber\\
   &\cup \{2ts+\theta(q+t-1),\dots ,(q+2)ts+\theta(t-1)+2z-2\}\nonumber\\
   &\cup \{ts+(q+t-1)\theta,\dots,(q+1)ts+(t-1)\theta+s+z-2\}\nonumber\\
   =& \{0,\ldots,ts+\theta t+z-2\}\nonumber\\
    &\cup\bigcup_{l=1}^{q-2} \{2ts+\theta(l+t-1),\dots,ts+\theta(l+t)+z-2\} \nonumber \\
    &\cup \{2ts+\theta(q+t-2),\dots,ts+\theta(q+t-1)+z-2\}\nonumber\\
   &\cup \{2ts+\theta(q+t-1),\dots ,(q+2)ts+\theta(t-1)+2z-2\}\nonumber\\
   &\cup \{ts+(q+t-1)\theta,\dots,(q+1)ts+(t-1)\theta+s+z-2\}
\end{align}
To simplify the above equation, we consider the two cases; Case 1: $q\lambda \ge s$, and Case 2: $q\lambda < s$. 

Case 1: $q\lambda \ge s$. For this case, $\{ts+(q+t-1)\theta,\dots,(q+1)ts+(t-1)\theta+s+z-2\}$ is a subset of $\{2ts+\theta(q+t-2),\dots,ts+\theta(q+t-1)+z-2\}$. This is shown mathematically in the following and demonstrated in Fig. \ref{fig:non-zero-coeff-AGE-z less than ts: 2} and \ref{fig:non-zero-coeff-AGE-z less than ts: 4}:
\begin{align}\label{eq:subsetProofd}
    & q\lambda \ge s \nonumber \\
    \Rightarrow & \theta q+z-2 \ge qts+s+z-2 \nonumber \\
    \Rightarrow & ts+\theta(q+t-1)+z-2 \ge\nonumber\\
    &\quad \quad \quad (q+1)ts+(t+1)\theta+s+z-2,
\end{align}
and
\begin{align}\label{eq:subsetProofe}
    & ts < ts+\lambda \nonumber \\
    \Rightarrow & 2ts+\theta(q+t-2) < ts+(q+t-1)\theta.
\end{align}
Therefore, for the case of $q\lambda \ge s$, (\ref{eq:d1cupd2cupd3''cupd4AGE}) is simplified as:
\begin{align}
&\mathbf{\widehat{D}}_{123'4}\cup\mathbf{\widetilde{D}}''_3 = \{0,\ldots,ts+\theta t+z-2\}\nonumber\\
    &\cup\bigcup_{l=1}^{q-2} \{2ts+\theta(l+t-1),\dots,ts+\theta(l+t)+z-2\} \nonumber \\
    &\cup \{2ts+\theta(q+t-2),\dots,ts+\theta(q+t-1)+z-2\}\nonumber\\
   &\cup \{2ts+\theta(q+t-1),\dots ,(q+2)ts+\theta(t-1)+2z-2\}
\end{align}

Case 2: $q\lambda < s$. For this case, the union of $\{ts+(q+t-1)\theta,\dots,(q+1)ts+(t-1)\theta+s+z-2\}$ and $\{2ts+\theta(q+t-2),\dots,ts+\theta(q+t-1)+z-2\}$ is equal to $\{2ts+\theta(q+t-2),\dots,(q+1)ts+(t-1)\theta+s+z-2\}$. This can be derived mathematically from (\ref{eq:subsetProofd}) and (\ref{eq:subsetProofe}) and demonstrated in Fig. \ref{fig:non-zero-coeff-AGE-z less than ts: 3} and \ref{fig::non-zero-coeff-AGE-z less than ts: 5}. Therefore, for the case of $q\lambda < s$, (\ref{eq:d1cupd2cupd3''cupd4AGE}) is simplified as
\begin{align}
&\mathbf{\widehat{D}}_{123'4}\cup\mathbf{\widetilde{D}}''_3 = \{0,\ldots,ts+\theta t+z-2\}\nonumber\\
    &\cup\bigcup_{l=1}^{q-2} \{2ts+\theta(l+t-1),\dots,ts+\theta(l+t)+z-2\} \nonumber \\
    &\cup \{2ts+\theta(q+t-2),\dots,(q+1)ts+(t-1)\theta+s+z-2\}\nonumber\\
   &\cup \{2ts+\theta(q+t-1),\dots ,(q+2)ts+\theta(t-1)+2z-2\}
\end{align}
This completes the proof. \hfill $\Box$

\begin{lemma}\label{lemma:non-zero-coeff-AGE-z less than ts: 2}
For $\lambda+s-1 < z \leq ts, 0<\lambda <z, t\neq 1$ and $q\lambda \geq s$, we have
\begin{equation}
    |\mathbf{P}({H}(x))|=\Upsilon_6(\lambda)= 2ts+\theta(t-1)+(q+2)z-q-1
\end{equation}
\end{lemma}
{\em Proof:} 
From (\ref{eq:phxforthe4lastcases}) and Lemmas \ref{lemma:d1234cupd'3AGE} and {\ref{lemma:d1234cupd''3AGE}}, we have
\begin{align}
    &\mathbf{P}({H}(x))=(\mathbf{\widehat{D}}_{123'4}\cup\mathbf{\widetilde{D}}'_3)\cup(\mathbf{\widehat{D}}_{123'4}\cup\mathbf{\widetilde{D}}''_3)\nonumber\\
    =&\{0,\ldots,ts+\theta t+z-2\}\nonumber\\
    &\cup\bigcup_{l=t}^{t+q-2} \{2ts+\theta l,\dots,ts+\theta(l+1)+z-2\} \nonumber \\
   &\cup \{2ts+\theta(q+t-1),\dots ,(q+2)ts+\theta(t-1)+2z-2\}\nonumber\\
   &\cup\{0,\ldots,ts+\theta t+z-2\}\nonumber\\
    &\cup\bigcup_{l=1}^{q-2} \{2ts+\theta(l+t-1),\dots,ts+\theta(l+t)+z-2\} \nonumber \\
    &\cup \{2ts+\theta(q+t-2),\dots,ts+\theta(q+t-1)+z-2\}\nonumber\\
   &\cup \{2ts+\theta(q+t-1),\dots ,(q+2)ts+\theta(t-1)+2z-2\}\nonumber
    \end{align}
    \begin{align}
   =&\{0,\ldots,ts+\theta t+z-2\}\nonumber\\
    &\cup\bigcup_{l=t}^{t+q-2} \{2ts+\theta l,\dots,ts+\theta(l+1)+z-2\} \nonumber \\
   &\cup \{2ts+\theta(q+t-1),\dots ,(q+2)ts+\theta(t-1)+2z-2\}\nonumber\\
    &\cup \{2ts+\theta(q+t-2),\dots,ts+\theta(q+t-1)+z-2\}\nonumber\\
   =&\{0,\ldots,ts+\theta t+z-2\}\nonumber\\
    &\cup\bigcup_{l=t}^{t+q-2} \{2ts+\theta l,\dots,ts+\theta(l+1)+z-2\} \nonumber \\
   &\cup \{2ts+\theta(q+t-1),\dots ,(q+2)ts+\theta(t-1)+2z-2\}\label{eq:phxAGEtheLastFoura}
\end{align}
Next, we show that the subsets shown in (\ref{eq:phxAGEtheLastFoura}) do not have overlap. 
\begin{align}
    & z \leq ts \nonumber \\
    \Rightarrow & z-2 < ts \nonumber \\
    \Rightarrow & ts+\theta t +z-2 < 2ts+\theta t\nonumber\\
    & \& \quad ts+\theta(l+1)+z-2 < 2ts+\theta(l+1).
\end{align}
Therefore, by calculating the size of each subset, we can calculate the number of elements of $\mathbf{P}({H}(x))$. The size of $\{0,\ldots,ts+\theta t+z-2\}$ is equal to $ts+\theta t+z-1$. The size of $\bigcup_{l=t}^{t+q-2} \{2ts+\theta l,\dots,ts+\theta(l+1)+z-2\}$ is equal to $(q-1)(\lambda +z-1)$. The size of $\{2ts+\theta(q+t-1),\dots ,(q+2)ts+\theta(t-1)+2z-2\}$ is equal to $-\lambda q+2z-1$. Therefore, $\mathbf{P}({H}(x))$ is equal to the sum of all these sizes, \ie $\mathbf{P}({H}(x))=ts+\theta t+z-1+(q-1)(\lambda +z-1)-\lambda q+2z-1=2ts+\theta(t-1)+(q+2)z-q-1$. This completes the proof. \hfill $\Box$
 \begin{lemma}\label{lemma:non-zero-coeff-AGE-z less than ts: 3}
For $\lambda+s-1 < z \leq ts, 0<\lambda <z, t\neq 1$ and $q\lambda < s$, we have
\begin{align}
    |\mathbf{P}({H}(x))|=\Upsilon_7(\lambda) 
    =& \theta(t+1)+q(z-1)-2\lambda +z+ts\nonumber\\
    &+\min\{0, z+s(1-t)-\lambda q-1\}
\end{align}
\end{lemma}
{\em Proof:} 
From (\ref{eq:phxforthe4lastcases}) and Lemmas \ref{lemma:d1234cupd'3AGE} and {\ref{lemma:d1234cupd''3AGE}}, we have
\begin{align}
    &\mathbf{P}({H}(x))=(\mathbf{\widehat{D}}_{123'4}\cup\mathbf{\widetilde{D}}'_3)\cup(\mathbf{\widehat{D}}_{123'4}\cup\mathbf{\widetilde{D}}''_3)\nonumber\\
    =& \{0,\ldots,ts+\theta t+z-2\}\nonumber
            \end{align}
    \begin{align}
    &\cup\bigcup_{l=t}^{t+q-2} \{2ts+\theta l,\dots,ts+\theta(l+1)+z-2\} \nonumber \\
   &\cup \{2ts+\theta(q+t-1),\dots ,(q+2)ts+\theta(t-1)+2z-2\}\nonumber\\
   &\cup \{0,\ldots,ts+\theta t+z-2\}\nonumber\\
    &\cup\bigcup_{l=1}^{q-2} \{2ts+\theta(l+t-1),\dots,ts+\theta(l+t)+z-2\} \nonumber \\
    \cup & \{2ts+\theta(q+t-2),\dots,(q+1)ts+(t-1)\theta+s+z-2\}\nonumber\\
   &\cup \{2ts+\theta(q+t-1),\dots ,(q+2)ts+\theta(t-1)+2z-2\}\nonumber\\
   =& \{0,\ldots,ts+\theta t+z-2\}\nonumber\\
    &\cup\bigcup_{l=t}^{t+q-2} \{2ts+\theta l,\dots,ts+\theta(l+1)+z-2\} \nonumber 
    \end{align}
    \begin{align}
   &\cup \{2ts+\theta(q+t-1),\dots ,(q+2)ts+\theta(t-1)+2z-2\}\nonumber\\
    &\cup\bigcup_{l=t}^{t+q-3} \{2ts+\theta l,\dots,ts+\theta(l+1)+z-2\} \nonumber \\
    \cup & \{2ts+\theta(q+t-2),\dots,(q+1)ts+(t-1)\theta+s+z-2\}\nonumber\\
      =& \{0,\ldots,ts+\theta t+z-2\}\nonumber\\
    &\cup\bigcup_{l=t}^{t+q-2} \{2ts+\theta l,\dots,ts+\theta(l+1)+z-2\} \nonumber \\
    \cup & \{2ts+\theta(q+t-2),\dots,(q+1)ts+(t-1)\theta+s+z-2\}\nonumber\\
   &\cup \{2ts+\theta(q+t-1),\dots ,(q+2)ts+\theta(t-1)+2z-2\}\nonumber\\
   =& \{0,\ldots,ts+\theta t+z-2\}\nonumber\\
    &\cup\bigcup_{l=t}^{t+q-3} \{2ts+\theta l,\dots,ts+\theta(l+1)+z-2\} \nonumber \\
    \cup & \{2ts+\theta(q+t-2),\dots,(q+1)ts+(t-1)\theta+s+z-2\}\nonumber\\
   &\cup \{2ts+\theta(q+t-1),\dots ,(q+2)ts+\theta(t-1)+2z-2\}\label{eq:phxAGEtheLastFourb}
\end{align}
Next, we show that the subsets shown in (\ref{eq:phxAGEtheLastFourb}), do not have overlap except for the last two subsets of $\{2ts+\theta(q+t-2),\dots,(q+1)ts+(t-1)\theta+s+z-2\}$ and $\{2ts+\theta(q+t-1),\dots ,(q+2)ts+\theta(t-1)+2z-2\}$
\begin{align}
    & z \leq ts \nonumber \\
    \Rightarrow & z-2 < ts \nonumber \\
    \Rightarrow & ts+\theta(l+1)+z-2 < 2ts+\theta(l+1)
\end{align}
Therefore, by calculating the size of each subset, we can calculate the number of elements of $\mathbf{P}({H}(x))$. 
The size of $\{0,\ldots,ts+\theta t+z-2\}$ is equal to $ts+\theta t+z-1$.
The size of $\bigcup_{l=t}^{t+q-3} \{2ts+\theta l,\dots,ts+\theta(l+1)+z-2\}$ is equal to $(q-2)(\lambda+z-1)$. 
The size of $\{2ts+\theta(q+t-2),\dots,(q+1)ts+(t-1)\theta+s+z-2\}\cup \{2ts+\theta(q+t-1),\dots ,(q+2)ts+\theta(t-1)+2z-2\}$ is equal to $\min\{\lambda(1-2q)+3z+s-2, \theta -\lambda q+2z-1\}$. Therefore, $|\mathbf{P}({H}(x))|$ is equal to the sum of all these sizes, \ie $|\mathbf{P}({H}(x))|=\theta(t+1)+q(z-1)-2\lambda +z+ts+\min\{0, z+s(1-t)-\lambda q-1\}$. This completes the proof.\hfill $\Box$ 

\begin{lemma}\label{lemma:non-zero-coeff-AGE-z less than ts: 4}
For $z \leq \lambda+s-1 \leq ts, 0<\lambda <z, t\neq 1$ and $q\lambda \geq s$, we have
\begin{equation}
    |\mathbf{P}({H}(x))|=\Upsilon_8(\lambda)= 2ts+\theta(t-1)+3z+(\lambda+s-1)q-\lambda-s-1
\end{equation}
\end{lemma}
{\em Proof:} 
From (\ref{eq:phxforthe4lastcases}) and Lemmas \ref{lemma:d1234cupd'3AGE} and {\ref{lemma:d1234cupd''3AGE}}, we have
\begin{align}
    &\mathbf{P}({H}(x))=(\mathbf{\widehat{D}}_{123'4}\cup\mathbf{\widetilde{D}}'_3)\cup(\mathbf{\widehat{D}}_{123'4}\cup\mathbf{\widetilde{D}}''_3)\nonumber\\
    &=\{0,\ldots,\theta(t+1)+s-2\}\nonumber\\
    &\cup\bigcup_{l=t}^{t+q-3} \{2ts+\theta  l,\dots,\theta(l+2)+s-2\} \nonumber\\
    &\cup \{2ts+\theta (t+q-2),\dots,ts+\theta(t+q-1)+z-2\} \nonumber \\
   &\cup \{2ts+\theta(q+t-1),\dots ,(q+2)ts+\theta(t-1)+2z-2\}\nonumber\\
   &\cup \{0,\ldots,ts+\theta t+z-2\}\nonumber\\
    &\cup\bigcup_{l=1}^{q-2} \{2ts+\theta(l+t-1),\dots,ts+\theta(l+t)+z-2\} \nonumber \\
    &\cup \{2ts+\theta(q+t-2),\dots,ts+\theta(q+t-1)+z-2\}\nonumber\\
   &\cup \{2ts+\theta(q+t-1),\dots ,(q+2)ts+\theta(t-1)+2z-2\}\nonumber  \\
   &=\{0,\ldots,\theta(t+1)+s-2\}\nonumber\\
    &\cup\bigcup_{l=t}^{t+q-3} \{2ts+\theta  l,\dots,\theta(l+2)+s-2\} \nonumber\\
    &\cup \{2ts+\theta (t+q-2),\dots,ts+\theta(t+q-1)+z-2\} \nonumber \\
   &\cup \{2ts+\theta(q+t-1),\dots ,(q+2)ts+\theta(t-1)+2z-2\}\nonumber\\
    &\cup\bigcup_{l=t}^{t+q-3} \{2ts+\theta l,\dots,ts+\theta(l+1)+z-2\} \nonumber 
    \end{align}
    \begin{align}
   &=\{0,\ldots,\theta(t+1)+s-2\}\nonumber\\
    &\cup\bigcup_{l=t}^{t+q-3} \{2ts+\theta  l,\dots,\theta(l+2)+s-2\} \nonumber\\
    &\cup \{2ts+\theta (t+q-2),\dots,ts+\theta(t+q-1)+z-2\} \nonumber \\
   &\cup \{2ts+\theta(q+t-1),\dots ,(q+2)ts+\theta(t-1)+2z-2\}\label{eq:phxAGEtheLastFourc}
\end{align}
Next, we show that the subsets shown in (\ref{eq:phxAGEtheLastFourc}) do not have overlap
\begin{align}
    & \lambda+s-1 \leq ts \nonumber \\
    \Rightarrow & \lambda+s-2 < ts \nonumber \\
    \Rightarrow & \theta(l+1)+s-2 < 2ts+\theta l
\end{align}
and
\begin{align}
    & z \leq ts \nonumber \\
    \Rightarrow & z-2 < ts \nonumber \\
    \Rightarrow & ts+\theta(t+q-1)+z-2 < 2ts+\theta(q+t-1)
\end{align}
Therefore, by calculating the size of each subset, we can calculate the number of elements of $\mathbf{P}({H}(x))$. 
The size of $\{0,\ldots,\theta(t+1)+s-2\}$ is equal to $\theta(t+1)+s-1$. 
The size of $\bigcup_{l=t}^{t+q-3} \{2ts+\theta  l,\dots,\theta(l+2)+s-2\}$ is equal to $(q-2)(2\lambda +s-1)$. 
The size of $\{2ts+\theta (t+q-2),\dots,ts+\theta(t+q-1)+z-2\}$ is equal to $z+\lambda-1$. 
The size of $\{2ts+\theta(q+t-1),\dots ,(q+2)ts+\theta(t-1)+2z-2\}$ is equal to $2z-q\lambda -1$. 
Therefore, $|\mathbf{P}({H}(x))|$ is equal to the sum of all these sizes, \ie $|\mathbf{P}({H}(x))|=2ts+\theta(t-1)+3z+(\lambda+s-1)q-\lambda-s-1$. This completes the proof.\hfill $\Box$

\begin{lemma}\label{lemma:non-zero-coeff-AGE-z less than ts: 5}
For $z \leq \lambda+s-1 \leq ts, 0<\lambda <z, t\neq 1$ and $q\lambda < s$:
\begin{align}
    |\mathbf{P}({H}(x))|=\Upsilon_9(\lambda)=&\theta(t+1)+q(s-1)-3\lambda+3z-1\nonumber\\
    &+\min\{0,ts-z+1+\lambda q-s\}
\end{align}
\end{lemma}
{\em Proof:} 
From (\ref{eq:phxforthe4lastcases}) and Lemmas \ref{lemma:d1234cupd'3AGE} and {\ref{lemma:d1234cupd''3AGE}}, we have
\begin{align}
    &\mathbf{P}({H}(x))=(\mathbf{\widehat{D}}_{123'4}\cup\mathbf{\widetilde{D}}'_3)\cup(\mathbf{\widehat{D}}_{123'4}\cup\mathbf{\widetilde{D}}''_3)\nonumber
    \end{align}
    \begin{align}
    &=\{0,\ldots,\theta(t+1)+s-2\}\nonumber\\
    &\cup\bigcup_{l=t}^{t+q-3} \{2ts+\theta  l,\dots,\theta(l+2)+s-2\} \nonumber\\
    &\cup \{2ts+\theta (t+q-2),\dots,ts+\theta(t+q-1)+z-2\} \nonumber \\
   &\cup \{2ts+\theta(q+t-1),\dots ,(q+2)ts+\theta(t-1)+2z-2\}\nonumber\\
   &\cup\{0,\ldots,ts+\theta t+z-2\}\nonumber\\
    &\cup\bigcup_{l=1}^{q-2} \{2ts+\theta(l+t-1),\dots,ts+\theta(l+t)+z-2\} \nonumber\\
    &\cup \{2ts+\theta(q+t-2),\dots,(q+1)ts+(t-1)\theta+s+z-2\}
             \end{align}
    \begin{align}
   &\cup \{2ts+\theta(q+t-1),\dots ,(q+2)ts+\theta(t-1)+2z-2\}\nonumber\\
   &=\{0,\ldots,\theta(t+1)+s-2\}\nonumber\\
    &\cup\bigcup_{l=t}^{t+q-3} \{2ts+\theta  l,\dots,\theta(l+2)+s-2\} \nonumber
            \end{align}
    \begin{align}
   &\cup \{2ts+\theta(q+t-1),\dots ,(q+2)ts+\theta(t-1)+2z-2\}\nonumber\\
    &\cup\bigcup_{l=t}^{t+q-3} \{2ts+\theta l,\dots,ts+\theta(l+1)+z-2\} \nonumber \\
    &\cup \{2ts+\theta(q+t-2),\dots,(q+1)ts+(t-1)\theta+s+z-2\}\nonumber\\
   &=\{0,\ldots,\theta(t+1)+s-2\}\nonumber\\
    &\cup\bigcup_{l=t}^{t+q-3} \{2ts+\theta  l,\dots,\theta(l+2)+s-2\} \nonumber\\
    &\cup \{2ts+\theta(q+t-2),\dots,(q+1)ts+(t-1)\theta+s+z-2\}\nonumber\\
    &\cup \{2ts+\theta(q+t-1),\dots ,(q+2)ts+\theta(t-1)+2z-2\}\label{eq:phxAGEtheLastFourd}
\end{align}

Next, we show that the subsets shown in (\ref{eq:phxAGEtheLastFourd}), do not have overlap except for the last two subsets of $\{2ts+\theta(q+t-2),\dots,(q+1)ts+(t-1)\theta+s+z-2\}$ and $\{2ts+\theta(q+t-1),\dots ,(q+2)ts+\theta(t-1)+2z-2\}$
\begin{align}
    & \lambda+s-1 \leq ts \nonumber \\
    \Rightarrow & \lambda+s-2 < ts \nonumber \\
    \Rightarrow & \theta(l+1)+s-2 < 2ts+\theta l
\end{align}
Therefore, by calculating the size of each subset, we can calculate the number of elements of $\mathbf{P}({H}(x))$.
The size of $\{0,\ldots,\theta(t+1)+s-2\}$ is equal to $\theta(t+1)+s-1$.
The size of $\bigcup_{l=t}^{t+q-3} \{2ts+\theta  l,\dots,\theta(l+2)+s-2\}$ is equal to $(q-2)(2\lambda+s-1)$.
The size of $\{2ts+\theta(q+t-2),\dots,(q+1)ts+(t-1)\theta+s+z-2\}\cup \{2ts+\theta(q+t-1),\dots ,(q+2)ts+\theta(t-1)+2z-2\}$ is equal to $\min\{\lambda(1-2q)+3z+s-2, \theta -\lambda q+2z-1\}$. Therefore, $|\mathbf{P}({H}(x))|$ is equal to the sum of all these sizes, \ie $|\mathbf{P}({H}(x))|=\theta(t+1)+q(s-1)-3\lambda+3z-1+\min\{0,ts-z+1+\lambda q-s\}$. This completes the proof.\hfill $\Box$

From Lemmas \ref{lemma:t=1NAGECMPC}, \ref{lemma:non-zero-coeff-AGE-lambda=0 and z greater ts-t}, \ref{lemma:non-zero-coeff-AGE-lambda=0 and z less ts-t}, \ref{lemma:non-zero-coeff-AGE-lambda=z}, \ref{lemma:non-zero-coeff-AGE-z greater than ts}, \ref{lemma:non-zero-coeff-AGE-z less than ts: 1}, \ref{lemma:non-zero-coeff-AGE-z less than ts: 2}, \ref{lemma:non-zero-coeff-AGE-z less than ts: 3}, \ref{lemma:non-zero-coeff-AGE-z less than ts: 4} and \ref{lemma:non-zero-coeff-AGE-z less than ts: 5}, Theorem \ref{th:N_AGE} is proved.

\section*{Appendix G: Proof of Lemma \ref{corol:AGE-Vs-Ent-SSMM-GC-Poly}} 
\setcounter{subsection}{0}
\subsection{AGE-CMPC Versus Entangled-CMPC} 
$N_{\text{AGE-CMPC}}=2s+2z-1$ when $t=1$ using (\ref{eq:eq:N-AGE-CMPC-optimization}). On the other hand,  $N_{\text{Entangled-CMPC}}=2s+2z-1$ from \cite{8613446}. Thus, $N_{\text{AGE-CMPC}}=N_{\text{Entangled-CMPC}}$ when $t=1$.


$N_{\text{AGE-CMPC}}$ is expressed as in the following when $t \neq 1$ using (\ref{eq:eq:N-AGE-CMPC-optimization}) and (\ref{eq:N-AGE-CMPC}). 
\begin{align}
    N_{\text{AGE-CMPC}}&=\displaystyle\min_{\lambda} \Gamma\nonumber\\
    \leq& \Gamma \text{ for } \lambda=0\nonumber\\
    =&\begin{cases}
   2st^2+2z-1, & z>ts-s,\\\
   st^2+3st-2s+t(z-1)+1, & z \leq ts-s,\
   \end{cases}\nonumber\\
   =&N_{\text{Entangled-CMPC}},
\end{align}
where the last equality comes from \cite{8613446}.

From the above discussion, we conclude that $N_{\text{AGE-CMPC}}<N_{\text{Entangled-CMPC}}$ when $0 < \lambda^* \leq z$. For the case of $\lambda^*=0$, $N_{\text{AGE-CMPC}}=N_{\text{Entangled-CMPC}}$. This completes the {comparison between $N_{\text{AGE-CMPC}}$ and $N_{\text{Entangled-CMPC}}$}.

\subsection{AGE-CMPC Versus SSMM}\label{subsec-AGE-VS-SSMM}
$N_{\text{AGE-CMPC}}=2s+2z-1$ when $t=1$ using (\ref{eq:eq:N-AGE-CMPC-optimization}). On the other hand, $N_{\text{SSMM}}=2s+2z-1$ from \cite{Zhu2021ImprovedCF}. Thus, $N_{\text{AGE-CMPC}}=N_{\text{SSMM}}$ when $t=1$. Next, we consider the case of $t\neq 1$ and compare $N_{\text{AGE-CMPC}}$ with $N_{\text{SSMM}}$.

$N_{\text{AGE-CMPC}}$ is expressed as the following when $t \neq 1$ using (\ref{eq:eq:N-AGE-CMPC-optimization}) and (\ref{eq:N-AGE-CMPC}).
\begin{align}
    N_{\text{AGE-CMPC}}=&\displaystyle\min_{\lambda} \Gamma\nonumber\\
    \leq& \Gamma \text{ for } \lambda=z\nonumber\\
    =&2ts+(ts+z)(t-1)+2z-1\nonumber\\
    =&(t+1)(ts+z)-1\nonumber\\
    =&N_{\text{SSMM}},
\end{align}
where the last equality comes from Theorem 1 in \cite{Zhu2021ImprovedCF}. 

From the above discussion, we conclude that $N_{\text{AGE-CMPC}}<N_{\text{SSMM}}$ when $0 \leq \lambda^* < z$. For the case of $\lambda^*=z$, $N_{\text{AGE-CMPC}}=N_{\text{SSMM}}$. This completes the {comparison between $N_{\text{AGE-CMPC}}$ and $N_{\text{SSMM}}$}.

\subsection{AGE-CMPC Versus GCSA-NA}
$N_{\text{AGE-CMPC}}=2s+2z-1$ when $t=1$ using (\ref{eq:eq:N-AGE-CMPC-optimization}). On the other hand, $N_{\text{GCSA-NA}}=2s+2z-1$ from \cite{9333639}. Thus, $N_{\text{AGE-CMPC}}=N_{\text{GCSA-NA}}$ when $t=1$. Next, we consider the case of $t\neq 1$ and compare $N_{\text{AGE-CMPC}}$ with $N_{\text{GCSA-NA}}$.

$N_{\text{AGE-CMPC}}$ is expressed as the following when $t \neq 1$ using (\ref{eq:eq:N-AGE-CMPC-optimization}) and (\ref{eq:N-AGE-CMPC}).
\begin{align}
    N_{\text{AGE-CMPC}}&=\displaystyle\min_{\lambda} \Gamma\nonumber\\
    \leq& \Gamma \text{ for } \lambda=0\nonumber
        \end{align}
    \begin{align}
    =&\begin{cases}
   2st^2+2z-1, & z>ts-s,\\\
   st^2+3st-2s+t(z-1)+1, & z \leq ts-s,\
   \end{cases}\nonumber\\
   &\begin{cases}
   =2st^2+2z-1, & z>ts-s,\\\
   \leq 2st^2+2z-1
   , & z \leq ts-s,\
   \end{cases}\label{eq:AGEvsGCSANA1}\\
       &\begin{cases}
   =N_{\text{GCSA-NA}}, & z>ts-s\\\
   \leq N_{\text{GCSA-NA}}
   , & z \leq ts-s\label{eq:AGEvsGCSANA2}\
   \end{cases},
\end{align}
where (\ref{eq:AGEvsGCSANA1}) comes from the condition of $z\leq ts-s$ as described in the following:
\begin{align}
    st^2+&3st-2s+t(z-1)+1\nonumber\\
    &=st^2+3st-2s+tz-t-2z+2z+1\nonumber\\
    &= st^2+3st-2s+(t-2)(z)-t+2z+1\nonumber\\
    &\leq st^2+3st-2s+(t-2)(ts-s)-t+2z+1\nonumber\\
    &=2st^2+2z-1-t+2\nonumber\\
    &\leq 2st^2+2z-1
\end{align}
and (\ref{eq:AGEvsGCSANA2}) comes from Theorem 1 in \cite{9333639}.

From the above discussion, we conclude that $N_{\text{AGE-CMPC}}<N_{\text{GCSA-NA}}$ when $0 < \lambda^* \leq z$. For the case of $\lambda^*=0$, $N_{\text{AGE-CMPC}}\leq N_{\text{GCSA-NA}}$. This completes the {comparison between $N_{\text{AGE-CMPC}}$ and $N_{\text{GCSA-NA}}$}. 

\subsection{AGE-CMPC Versus PolyDot-CMPC}
To prove this lemma, we consider different regions for the value of $z$, and prove that in all of the regions, the inequality of $N_{\text{AGE-CMPC}} \leq N_{\text{PolyDot-CMPC}}$ is valid.

(i) $z>ts, t\neq 1$: For this region, We consider the two cases of (a) $s\neq 1$ and (b) $s=1$.

(a) $s\neq 1$: From (\ref{eq:eq:N-AGE-CMPC-optimization}) and (\ref{eq:N-AGE-CMPC}), we have
\begin{align}
    N_{\text{AGE-CMPC}}=&\displaystyle\min_{\lambda} \Gamma\nonumber\\
    \leq& \Gamma \text{ for } 0<\lambda=ts-t<z\nonumber
        \end{align}
    \begin{align}
    =&(q+2)ts+\theta(t-1)+2z-1\text{ for } \lambda=ts-t\nonumber\\
    =&(\min\{\floor{\frac{z-1}{2ts-t-ts}},t-1\}+2)ts+\nonumber\\
    &\quad \quad \quad \quad (2ts-t)(t-1)+2z-1\nonumber\\
    =&N_{\text{PolyDot-CMPC}},
\end{align}
where the last equality comes from  (\ref{eq:N-PolyDot-DMPC}).

(b) $s=1$: From (\ref{eq:eq:N-AGE-CMPC-optimization}) and (\ref{eq:N-AGE-CMPC}), we have
\begin{align}
    N_{\text{AGE-CMPC}}=&\displaystyle\min_{\lambda} \Gamma\nonumber\\
    \leq& \Gamma \text{ for } \lambda=0\nonumber\\
    =&2t^2+2z-1\nonumber\\
    =&N_{\text{PolyDot-CMPC}},
\end{align}
where the last equality comes from $N_{\text{PolyDot-CMPC}}$ defined in (\ref{eq:N-PolyDot-DMPC}) for $s=1$ and $z>ts$.

(ii) $\frac{t-1}{t-2}(ts-t)<z\leq ts, t\neq 1$: This condition exists only if the constraint of $\frac{t-1}{t-2}(ts-t)< ts$ is satisfied. This constraint is satisfied when
\begin{equation}\label{eq:temp1}
    s+1<t.
\end{equation} 

Next, we show that, for $0<\lambda=ts-t<z$, $\Gamma$ is equal to one of $\Upsilon_i(\lambda)$'s where $i=6,7,8,9$. Then, we show that each $\Upsilon_i(\lambda), i=6,7,8,9$ for $\lambda=ts-t$ is less than $N_{\text{PolyDot-CMPC}}$. For this purpose, we first assert that the conditions for this case, \ie $\lambda=ts-t<\frac{t-1}{t-2}(ts-t)<z\leq ts, t\neq 1$, do not satisfy the conditions for $\Upsilon_i(\lambda), i=1,2,3,4,5$.
The reason is that $0<t(s-1)=\lambda$ does not satisfy the condition for $\Upsilon_i(\lambda), i=1,2,3$. On the other hand, $z\leq ts$ does not satisfy the condition for $\Upsilon_4(\lambda)$. In addition, from (\ref{eq:temp1}), $s-1<t$ and thus $ts-s+1>ts-t=\lambda$, which does not satisfy the condition for $\Upsilon_5(\lambda)$.

We consider the following cases;  $q=0$ and $q=1$.

(a) $q=0$: For this case, based on the definition of $q$, we should have either (1) $\frac{z-1}{\lambda}=\frac{z-1}{ts-t}<1$, which is not possible as this contradicts the condition of (ii) that requires $ts-t< \frac{t-1}{t-2}ts-t<z$, so $ts-t\leq z-1$, or (2) $t=1$, which is not possible as (\ref{eq:temp1}) results in $s<0$, which is not a valid inequality. 

(b) $q=1$: From (\ref{eq:N-AGE-CMPC}), this falls under the condition of $\Upsilon_6(\lambda)$ and $\Upsilon_8(\lambda)$ as $q\lambda=\lambda=s(t-1)\ge s$. For this case, either the condition of $\lambda+s-1<z$ (condition of $\Upsilon_6(\lambda)$) or $z\leq \lambda+s-1$ (condition of $\Upsilon_8(\lambda)$) is satisfied. Both $\Upsilon_6(\lambda)$ and $\Upsilon_8(\lambda)$ are less than $N_{\text{PolyDot-CMPC}}$ as shown below.

For $s\neq 1$, we have 
\begin{align}\label{eq:gamma6vspolydotsneq1}
   \Upsilon_6(\lambda) &= 2ts+\theta(t-1)+(q+2)z-q-1 \nonumber \\
  & =  2ts+(ts+ts-t)(t-1)+3z-2 \nonumber\\
   & <  2ts+(2ts-t)(t-1)+3z-1\nonumber\\
   &= N_{\text{PolyDot-CMPC}},
\end{align}
where the last equality comes from $N_{\text{PolyDot-CMPC}}$ defined in (\ref{eq:N-PolyDot-DMPC}) for $ts-t<z\leq ts, s\neq 1$. Next, we consider the case  $s=1$.

For $s=1$, from (\ref{eq:temp1}), we have $t>2$ and from the condition of (ii), we have $z\leq t$. Therefore, we have
\begin{align}
   \Upsilon_6(\lambda) &= 2ts+\theta(t-1)+(q+2)z-q-1 \nonumber \\
  & = 2t+(2t-t)(t-1)+3z-2 \nonumber\\
  & = t^2+t+3z-2\nonumber\\
   & < t^2+t+2z+z-1\nonumber\\
   & < t^2+t+tz+t-1\label{eq:tempp2}\\
   & = t^2+2t+tz-1\nonumber\\
   &= N_{\text{PolyDot-CMPC}}, 
   \label{eq:gamma6vspolydots=1}
\end{align}
where (\ref{eq:tempp2}) comes from $t>2$ and $z\leq t$. The last equality comes from $N_{\text{PolyDot-CMPC}}$ defined in (\ref{eq:N-PolyDot-DMPC}) for $s=1, z\leq t$.

From (\ref{eq:gamma6vspolydotsneq1}) and (\ref{eq:gamma6vspolydots=1}), we conclude $\Upsilon_6(\lambda)<N_{\text{PolyDot-CMPC}}$.

For $s\neq 1$, we have
\begin{align}\label{eq:gamma8vspolydotsneq1}
   \Upsilon_8(\lambda) &= 2ts+\theta(t-1)+3z+(\lambda+s-1)q-\lambda-s-1 \nonumber \\
  & =  2ts+(2ts-t)(t-1)+3z \nonumber \\
  &+\lambda+s-1-\lambda-s-1 \nonumber \\
  & = 2ts+(2ts-t)(t-1)+3z-2 \nonumber \\
  & <  2ts+(2ts-t)(t-1)+3z-1\nonumber\\
   &= N_{\text{PolyDot-CMPC}},
\end{align}
where the last equality comes from $N_{\text{PolyDot-CMPC}}$ defined in (\ref{eq:N-PolyDot-DMPC}) for $ts-t<z\leq ts, s\neq 1$. Next, we consider the case of $s=1$.

For $s=1$, from (\ref{eq:temp1}), we have $t>2$ and from the condition of (ii), we have $z\leq t$. Therefore, similar to (\ref{eq:tempp2}), we have
\begin{align}\label{eq:gamma8vspolydots=1}
   \Upsilon_8(\lambda) &= 2ts+\theta(t-1)+3z+(\lambda+s-1)q-\lambda-s-1 \nonumber \\
  & =  2t+(2t-t)(t-1)+3z+\lambda+1-1-\lambda-1-1 \nonumber \\
  & = t^2+t+3z-2 \nonumber \\
  & <  t^2+2t+tz-1\nonumber\\
   &= N_{\text{PolyDot-CMPC}},
\end{align}
where the last inequality comes from $t>2$ and $z\leq t$ and the last equality comes from $N_{\text{PolyDot-CMPC}}$ defined in (\ref{eq:N-PolyDot-DMPC}) for $s=1, z\leq t$.

From (\ref{eq:gamma8vspolydotsneq1}) and (\ref{eq:gamma8vspolydots=1}), we conclude $\Upsilon_8(\lambda)<N_{\text{PolyDot-CMPC}}$.

From the above discussion, $\Gamma$ for $\lambda=ts-t$ is less than $N_{\text{PolyDot-CMPC}}$ for the condition of (ii). Therefore, we have:
\begin{align}
    N_{\text{AGE-CMPC}}=&\displaystyle\min_{\lambda} \Gamma\nonumber\\
    \leq& \Gamma \text{ for } 0<\lambda=ts-t<z\nonumber\\
    <&N_{\text{PolyDot-CMPC}},
\end{align}

(iii) $z \leq \frac{t-1}{t-2}(ts-t), s,t \neq 1$\footnote{Note that for this case, we have $s \neq 1$ as $z\leq \frac{t-1}{t-2}t(s-1)$.}: It is shown in the proof of Lemma \ref{lemma: regions where N_polydot<N_ssmm} in Appendix C, that for the condition of (iii), $N_{\text{SSMM}} \leq N_{\text{PolyDot-CMPC}}$. On the other hand, from {the comparison of $N_{\text{AGE-CMPC}}$ and $N_{\text{SSMM}}$ in Section~\ref{subsec-AGE-VS-SSMM} of this appendix}, $N_{\text{AGE-CMPC}} \leq N_{\text{SSMM}}$. Therefore, for this region, $N_{\text{AGE-CMPC}} \leq N_{\text{PolyDot-CMPC}}$.
 
(iv) $t=1$: For this region, from (\ref{eq:eq:N-AGE-CMPC-optimization}), we have:
\begin{align}
 N_{\text{AGE-CMPC}}&=2s+2z-1\nonumber\\
 &=N_{\text{PolyDot-CMPC}},
\end{align}
where the last equality comes from $N_{\text{PolyDot-CMPC}}$ defined in (\ref{eq:N-PolyDot-DMPC}) for $t=1$.

From (i), (ii), (iii), and (iv), the number of workers required by AGE-CMPC method is always less than or equal to the number of workers required by PolyDot-CMPC.
This completes the {comparison between $N_{\text{AGE-CMPC}}$ and $N_{\text{PolyDot-CMPC}}$}.

 \section*{Appendix H: Proof of requiring more number of workers for $\lambda>z$ than $\lambda=z$ 
\footnote{Note that according to (\ref{eq:eq:N-AGE-CMPC-optimization}) for the case of $t=1$, the required number of workers is independent of $\lambda$, therefore in this appendix we just consider the case of $t \neq 1$.}}\label{appendix:H-lambda range} 
{
Intuitively, for $\lambda>z$, the created gaps in the powers of $C_B(x)$ in (\ref{eq:generalEntangled}) will not result in reducing the number of required workers more than the case of $\lambda=z$ as the main benefit of creating gaps in powers of $C_B(x)$ is that it allows us to choose the powers of secret terms from the gaps that will be created in powers of $C_A(x)C_B(x)$ without interfering with the important powers. It is worth recalling that the total number of the powers of secret terms is equal to the number of colluding workers, $z$, \ie $|\mathbf{P}(S_{A}(x))|=|\mathbf{P}(S_{B}(x))|=z$, therefore considering more than $z$ number of gaps in powers of coded terms is not beneficial and just results in increasing the powers of coded and secret terms, and consequently increasing the required number of workers. In the following, we provide the mathematical proof.
}
{\begin{lemma}\label{lemma:non-zero-coeff-AGE-lambda>=z}
 For $z \leq \lambda \leq z+s-1, t\neq 1$, we have
\begin{align}\label{eq:hxlambdacase1}
    |\mathbf{P}({H}(x))|=2ts+(ts+\lambda)(t-1)+2z-1,
\end{align}
and for $ \lambda > z+s-1, t\neq 1$, we have the following 
\begin{align}\label{eq:hxlambdacase2}
    |\mathbf{P}({H}(x))|&=2ts+(ts+z+s-1)(t-1)+2z-1.
\end{align}
\end{lemma}
 {\em Proof:} 
 To prove this lemma, we first calculate $\mathbf{D}_3$ from (\ref{eq:AGE-p(CB)-th}) and (\ref{eq:S-A}):
 \begin{align}\label{eq:d3-AGE- lambda>=z}
      \mathbf{D}_3 = & \mathbf{P}(S_{A}(x))+\mathbf{P}(C_B(x)) \nonumber \\
      =& \mathbf{P}(S_{A_2}(x))+\mathbf{P}(C_B(x)) \nonumber \\
      = & \{ts+u: 0 \leq u \leq z-1\} \nonumber \\
      + & \{s-1-k+\theta l:0 \leq l \leq t-1,\; 0 \leq k \leq s-1,\}\nonumber \\
     = & \{ts,\dots,ts+z+s-2\}+\{\theta l: 0 \leq l \leq t-1\} \nonumber \\
     = & \bigcup_{l=0}^{t-1} \{\theta l+ts,\dots,\theta l +ts+z+s-2\}
     .
\end{align}
$\mathbf{D}_1$ defined in (\ref{eq:D1age}) is calculated in (\ref{eq:d1-AGE}). Therefore, we have:
\begin{align}
    \mathbf{D}_{13}=&\mathbf{D}_1 \cup \mathbf{D}_3\nonumber\\
    =&\bigcup\limits_{l=0}^{t-1}\{\theta l,\ldots,ts+s-2+\theta l\}\nonumber\\
    &\cup\bigcup_{l=0}^{t-1} \{\theta l+ts,\dots,\theta l +ts+z+s-2\}\nonumber\\
   = &\bigcup_{l=0}^{t-1} \{\theta l,\ldots,\theta l +ts+z+s-2\}\label{eq:d13AGE11-lambda>=z}
\end{align}
where (\ref{eq:d13AGE11-lambda>=z}) comes from the fact that $\theta l < \theta l+ts \leq ts+s-2+\theta l+1 \leq \theta l +ts+z+s-2$. To calculate $\mathbf{D}_{13}\cup\mathbf{D}_{2}$, we consider two different cases based on the value of $\lambda$.}

{Case 1: $z \leq \lambda \leq z+s-1$.
In this case, $\theta l+ts+z+s-2+1 \geq \theta l+ts+\lambda = \theta (l+1)$ and thus there is no gap between each two consecutive subsets of $\mathbf{D}_{13}$, \ie $\bigcup_{l=0}^{t-1} \{\theta l,\ldots,\theta l +ts+z+s-2\}$. Therefore, we have
\begin{align}\label{eq:d13AGE12>=z}
\mathbf{D}_{13}=\{0,\ldots,\theta (t-1) +ts+z+s-2\}.
\end{align}
Next, we calculate $\mathbf{D}_{123}=\mathbf{D}_{13} \cup \mathbf{D}_{2}$ from (\ref{eq:d13AGE12>=z}) and (\ref{eq:d2-AGE}).
\begin{align}\label{eq:d123AGE1>=z}
    \mathbf{D}_{123}=&\mathbf{D}_{1} \cup \mathbf{D}_{3} \cup \mathbf{D}_{2}\nonumber\\
    =&\{0,\ldots,\theta (t-1) +ts+z+s-2\} \nonumber\\
    &\cup\{ts+\theta(t-1),\ldots,2ts+\theta(t-1)+z-2\}\nonumber\\
    =&\{0,\ldots,2ts+\theta(t-1) +z-2\},
\end{align}
where the last equality comes from the fact that $0 < ts+\theta(t-1) \leq \theta(t-1)+ts+z+s-2 < 2ts+\theta(t-1)+z-2$. Next, we first calculate $\mathbf{D}_4$, and then its union with $\mathbf{D}_{123}$.
From (\ref{eq:S-A}) and (\ref{eq:S-B}), we have
 \begin{align}\label{eq:d4-AGE- lambda>=z}
      \mathbf{D}_4 = & \mathbf{P}(S_{A}(x))+\mathbf{P}(S_B(x)) \nonumber \\
      = & \{ts,\dots,ts+z-1\} \nonumber \\
      + & \{ts+\theta(t-1),\dots,ts+\theta(t-1)+z-1\} \nonumber \\
      = & \{2ts+\theta(t-1),\dots,2ts+\theta(t-1)+2z-2\}.
\end{align}
From (\ref{eq:PHx-AGE}), (\ref{eq:d123AGE1>=z}) and (\ref{eq:d4-AGE- lambda>=z}), we have
\begin{align}
    \mathbf{P}({H}(x)) = & \mathbf{D}_{123} \cup \mathbf{D}_4 \nonumber\\
    &=\{0,\ldots,2ts+\theta(t-1) +z-2\} \cup\nonumber\\ &\{2ts+\theta(t-1),\dots,2ts+\theta(t-1)+2z-2\}\nonumber\\
    =&\{0,\ldots,2ts+\theta(t-1)+2z-2\}.
\end{align}
Therefore, in case 1, $|\mathbf{P}({H}(x))|=2ts+\theta(t-1)+2z-2+1=2ts+(ts+\lambda)(t-1)+2z-1$. This proves (\ref{eq:hxlambdacase1}) in Lemma \ref{lemma:non-zero-coeff-AGE-lambda>=z}.}

{Case 2: $\lambda > z+s-1$. In this case from (\ref{eq:d13AGE11-lambda>=z}), we have 
\begin{align}\label{eq:D13-case2}
\mathbf{D}_{13} =& \bigcup_{l=0}^{t-1} \{\theta l,\ldots,\theta l +ts+z+s-2\} \nonumber \\
= & \{0,\ldots,ts+z+s-2\} \nonumber \\
\cup& \{\theta,\ldots,\theta+ts+z+s-2\} 
\cup \dots \nonumber \\
\cup& \{\theta(t-1),\ldots,\theta(t-1)+ts+z+s-2\}.
\end{align}
Now let us calculate $\mathbf{D}_{123}=\mathbf{D}_{13} \cup \mathbf{D}_{2}$ from (\ref{eq:D13-case2}) and (\ref{eq:d2-AGE})
\begin{align}\label{eq:d123AGE1>=z-case2}
    \mathbf{D}_{123}=&\mathbf{D}_{1} \cup \mathbf{D}_{3} \cup \mathbf{D}_{2}\nonumber\\
    =&\bigcup_{l=0}^{t-1} \{\theta l,\ldots,\theta l +ts+z+s-2\} \nonumber\\
    &\cup\{ts+\theta(t-1),\ldots,2ts+\theta(t-1)+z-2\}\nonumber\\
    =& \bigcup_{l=0}^{t-2} \{\theta l,\ldots,\theta l +ts+z+s-2\} \nonumber \\
    \cup& \{\theta(t-1),\ldots,2ts+\theta(t-1)+z-2\},
\end{align}
where (\ref{eq:d123AGE1>=z-case2}) comes from the fact that $\theta(t-1) < ts+\theta(t-1) < \theta(t-1)+ts+z+s-2 < 2ts+\theta(t-1)+z-2$. Next, we calculate $\mathbf{P}({H}(x))=\mathbf{D}_{123} \cup \mathbf{D}_4$ from (\ref{eq:d4-AGE- lambda>=z}) and (\ref{eq:d123AGE1>=z-case2}).  
\begin{align}\label{eq:p(H(x))-case2}
    \mathbf{P}({H}(x)) = & \mathbf{D}_{123} \cup \mathbf{D}_4 \nonumber \\
    &=\bigcup_{l=0}^{t-2} \{\theta l,\ldots,\theta l +ts+z+s-2\}\nonumber \\
    \cup& \{\theta(t-1),\ldots,2ts+\theta(t-1)+z-2\} \nonumber\\ \cup& \{2ts+\theta(t-1),\dots,2ts+\theta(t-1)+2z-2\}\nonumber\\
    =& \bigcup_{l=0}^{t-2} \{\theta l,\ldots,\theta l +ts+z+s-2\} \nonumber \\
    \cup& \{\theta(t-1),\ldots,2ts+\theta(t-1)+2z-2\},
\end{align}
where (\ref{eq:p(H(x))-case2}) is resulted from the fact that $\theta(t-1)<2ts+\theta(t-1) \leq 2ts+\theta(t-1)+z-2+1 \leq 2ts+\theta(t-1)+2z-2$. In the above equation, there exist $\theta(l+1)-(\theta l+ts+z+s-2)-1=\lambda - z-s+1$ gaps between each two consecutive subsets of $\bigcup_{l=0}^{t-2} \{\theta l,\ldots,\theta l +ts+z+s-2\}$. Therefore, in case 2 we have 
\begin{align}
&|\mathbf{P}({H}(x))|\\
&=2ts+\theta(t-1)+2z-2+1-(t-1)(\lambda-z-s+1) \nonumber \\
&= 2ts+(ts+\lambda-\lambda+z+s-1)(t-1)+2z-1 \nonumber \\
&= 2ts+(ts+z+s-1)(t-1)+2z-1.
\end{align}
This proves (\ref{eq:hxlambdacase2}) in Lemma \ref{lemma:non-zero-coeff-AGE-lambda>=z}.} 

This completes the proof of Lemma \ref{lemma:non-zero-coeff-AGE-lambda>=z}. \hfill $\Box$

From Lemma \ref{lemma:non-zero-coeff-AGE-lambda>=z}, 
$|\mathbf{P}({H}(x))|$ in (\ref{eq:hxlambdacase1}) is always less than or equal to $|\mathbf{P}({H}(x))|$ in (\ref{eq:hxlambdacase2}) because of the fact that in (\ref{eq:hxlambdacase1}) we have $\lambda \leq z+s-1$ and as a result $ 2ts+(ts+\lambda)(t-1)+2z-1 \leq 2ts+(ts+z+s-1)(t-1)+2z-1$. On the other hand, $|\mathbf{P}({H}(x))|$ in (\ref{eq:hxlambdacase1}) is an increasing function of $\lambda$. Therefore, the choice of $\lambda=z$ results is the minimum required number of workers in the range of $z \leq \lambda$. This completes the proof. \hfill $\Box$

\end{document}